\documentclass[12pt,a4paper,fleqn]{article}

\usepackage{lscape,exscale,amsthm}

\usepackage[intlimits]{amsmath}
\usepackage{algorithm}
\usepackage{rawfonts}
\usepackage{latexsym}
\usepackage[cp850]{inputenc}
\usepackage{epsfig}
\usepackage{bbm}
\usepackage{enumerate,dsfont}
\usepackage{natbib}
\renewcommand{\baselinestretch}{1.4}
\newcommand{\bep}{\mbox{\boldmath$\varepsilon$}}
\newcommand{\bmu}{\mbox{\boldmath$\mu$}}

\newcommand{\blambda}{\mbox{\boldmath$\lambda$}}

\newcommand{\bPsi}{\mathbf{\Psi}}
\newcommand{\bOmega}{\mathbf{\Omega}}
\newcommand{\bx}{\mathbf{x}}

\newcommand{\bX}{\mathbf{X}}
\newcommand{\bU}{\mathbf{U}}

\newcommand{\bZ}{\mathbf{Z}}

\newcommand{\bS}{\mathbf{S}}

\newcommand{\bi}{\mathbf{1}}
\newcommand{\bI}{\mathbf{I}}

\newcommand{\bSigma}{\mathbf{\Sigma}}

\newcommand{\E}{\mathds{E}}

\usepackage{xr} %

\usepackage{authblk}
\title{Gibbs sampler approach for objective Bayeisan inference in elliptical multivariate random effects model}

\author[1]{Olha Bodnar}
\author[2]{Taras Bodnar}

\affil[1]{National Institute of Standards and Technology, Gaithersburg, MD 20899-8980, USA and Unit of Statistics, School of Business, \"Orebro University, SE-70182 \"Orebro, Sweden}
\affil[2]{Department of Mathematics, Stockholm University, SE-10691 Stockholm, Sweden}
\date{}
\providecommand{\keywords}[1]
{
\small	
\textbf{\textit{Keywords}:} #1
}
\begin{document}

%
\maketitle
\begin{abstract}
In this paper, we present the Bayesian inference procedures for the parameters of the multivariate random effects model derived under the assumption of an elliptically contoured distribution when the Berger and Bernardo reference and the Jeffreys priors are assigned to the model parameters. We develop a new numerical algorithm for drawing samples from the posterior distribution, which is based on the hybrid Gibbs sampler. The new approach is compared to the two Metropolis-Hastings algorithms, which were previously derived in the literature, via an extensive simulation study. The results are implemented in practice by considering ten studies about the effectiveness of hypertension treatment for reducing blood pressure where the treatment effects on both the systolic blood pressure and diastolic blood pressure are investigated.
\end{abstract}


\newpage
\section{Introduction}\label{sec:intro}

Multivariate random effects model presents a classical quantitative tool used to combine the measurements of individual studies into a single consensus vector. The approach is widely used in multivariate meta-analysis in medicine, psychology, chemistry, among others (see, e.g.,  \citet{gasparrini2012multivariate}, \citet{wei2013bayesian}, \citet{jackson2014refined}, \citet{liu2015multivariate}, \citet{noma2019efficient}, \cite{negeri2020robust}, \citet{jackson2020multivariate}) as well as in interlaboratory comparison study in physics and metrology (see, e.g., \cite{rukhin2007estimating}, \cite{TomanFischerElster2012}, \cite{Rukhin2013}, \cite{BodnarEriksson2023}).

In most of applications, the random effects model is based on the assumption that the individual measurements are normally distributed with unknown common mean vector and known covariance matrices which are provided together with the measurement results by each participant (cf., \cite{lambert2005vague}, \cite{Viechtbauer2007}, \cite{SuttonHiggins2008}, \cite{riley2010meta}, \cite{StrawdermanRukhin2010}, \cite{Guolo2012}, \cite{Turner2015}, \cite{bodnar2017bayesian}, \cite{wynants2018random}, \cite{michael2019exact}, \cite{veroniki2019methods}, \cite{BodnarEriksson2023}). In the multivariate case, the model is determined by the following equation
\begin{equation}\label{mult-rem-nor}
\bx_i=\bmu+ \blambda_i + \bep_i \quad \text{with} \quad \blambda_i  \sim \mathcal{N}_p(\mathbf{0},\bPsi) \quad \text{and} \quad \bep_i \sim \mathcal{N}_p(\mathbf{0},\bU_i) ,
\end{equation}
where $\{\blambda_i\}_{i=1,...,n}$ and $\{\bep_i\}_{i=1,...,n}$ are mutually independent. Both $\blambda_i$ and $\bep_i$ are normally distributed random vectors in the stochastic representation \eqref{mult-rem-nor} of $\mathbf{x}_i$.

The aim of a meta-analysis and/or of an interlaboratory comparison study is to use the observation vectors $\bx_i$, $i=1,...,n$ to produce a single estimate of the common mean vector $\bmu$. Usually, it appears that the constructed estimator is more volatile as one would expect by using the uncertainties reported by each individual study $\bU_i$, $i=1,...,n$. This effect is known as the dark uncertainty of between-study variability (cf., \cite{thompson2011dark}) in statistical literature and it is explained by the fact that the individual studies may be performed in different places and in different times resulting in extra viability. Thus the random effects $\blambda_i$, $i=1,...,n$ are introduced in \eqref{mult-rem-nor}, while their covariance matrix $\bPsi$, known also the between-study covariance matrix, represents the scope of dark uncertainty.

An extension of the multivariate random effects model was considered by \cite{BodnarLinkElster2015} and \cite{bodnar2019} in the univariate case and by \cite{bodnar2023objective} in the multivariate case. It is based on the assumption of elliptically contoured distributions to model the stochastic behaviour of measurement results, which include the normal distribution as a special case (see, \cite{GuptaVargaBodnar}). Let $\bX=(\bx_1,...,\bx_n)$ denote the observation matrix. Then, the model is given by
\begin{equation}\label{mult-rem}
p(\bX | \bmu, \bPsi) = \frac{1}{\sqrt{\text{det}(\bPsi \otimes \bI+\bU)}} f\left(\text{vec}(\bX - \bmu \bi^\top)^\top (\bPsi \otimes \bI+\bU)^{-1}\text{vec}(\bX - \bmu \bi) \right),
\end{equation}
where $\bi$ is the vector of ones, $\bI$ is the identity matrix of an appropriate order, and $\bU=\text{diag}(\bU_1,...,\bU_n)$ is a block-diagonal matrix consisting of the reported covariance matrices $\bU_i$, $i=1,...,n$, as its diagonal blocks. The symbol $\otimes$ denotes the Kronecker product, while $\text{vec}$ stands for the $vec$ operator (see, e.g., \cite{Harville97}). The function $f(.)$ is density generator which determines the type of an elliptical distribution used in the model assumption.

Statistical inference procedures for the parameters of the multivariate random effects model were derived from both viewpoints of the frequentist and Bayesian statistics. Under the assumption of normality, \citet{jackson2010extending} extended the DerSimonian and Laird approach to the multivariate data. \citet{chen2012method} developed the method based on the restricted maximum likelihood. These two procedures together with the maximum likelihood estimator constitute the most commonly used methods of the frequentist statistics (cf., \citet{jackson2013matrix}, \citet{schwarzer2015meta}, \citet{jackson2020multivariate}). While \citet{nam2003multivariate} and \citet{paul2010bayesian} derived Bayesian inference procedures under the assumption of the multivariate normal random effects model by assigning informative priors to model parameters, recently \cite{bodnar2023objective} developed a full objective Bayesian analysis of the generalized multivariate random effects model. They also presented two Metropolis-Hastings algorithms for drawing samples from the derived posterior distribution. In the current paper, we extend the previous findings by deriving a Gibbs sampler algorithm, which will be compared to the existent approaches for both simulated and real data.

The rest of the paper is structured as follows. In the next section, the objective Bayesian inference procedures for the parameters of the generalized random effects model are present, which are derived by using the Berger and Bernardo reference prior (see, \cite{BergerBernardo1992c}, \cite{berger2009formal}) and the Jeffreys prior (see, \cite{Jeffreys1946}). The Gibbs sampler methods for drawing samples from the posterior distribution of the parameters of the generalized random effects model is introduced in Section \ref{sec:gibbs}. Within an extensive simulation study we compare the new approach to the existent ones in Section \ref{sec:sim}, where split-$\hat{R}$ estimates based on the rank normalization of \cite{vehtari2021rank} and the coverage probabilities of the constructed credible intervals are used as performance measures. In Section \ref{sec:emp} the results of the empirical illustration are provided, while Section \ref{sec:sum} summarizes the findings obtained in the paper.

\section{Bayesian inferences in generalized multivariate random effects model}\label{sec:ba-MREM}
In many practical situations no information or only vague information is available for model parameters. In such a case, it is preferable to construct Bayesian inference procedures by assigning a noninformative prior. The Laplace prior, the Jeffreys prior and the Berger and Bernardo reference prior present the most commonly used noninformative priors in statistical literature (see, \cite{Laplace1812}, \cite{Jeffreys1946}, \cite{BergerBernardo1992c}, \cite{berger2009formal}).

The Berger and Bernardo reference prior and the Jeffreys prior for the parameters $\bmu$ and $\bPsi$ of the generalized multivariate random effects model were derived in \cite{bodnar2023objective}. They are given by
\begin{itemize}
\item The Berger and Bernardo reference prior
\begin{eqnarray}\label{prior_R}
&&\pi_R(\bmu,\bPsi)=\pi_R(\bPsi)\propto \Bigg(\text{det}\Bigg[\mathbf{G}_p^\top\Bigg[ \frac{2J_2}{2pn+p^2n^2}\sum_{i=1}^{n}\left((\bPsi + \bU_i)^{-1} \otimes (\bPsi + \bU_i)^{-1}\right) \nonumber\\
&+&\left( \frac{J_2}{2pn+p^2n^2} -\frac{1}{4}\right) \text{vec}\left(\sum_{i=1}^{n}(\bPsi + \bU_i)^{-1}\right)\text{vec}\left(\sum_{j=1}^{n}(\bPsi + \bU_j)^{-1}\right)^\top
\Bigg]\mathbf{G}_p\Bigg]\Bigg)^{1/2}
\end{eqnarray}
with
\begin{eqnarray}\label{Ji}
J_2=\E\left((R^2)^2\left(\frac{f^\prime\left(R^2\right)}
{f\left(R^2\right)} \right)^2 \right),
\end{eqnarray}
where $R^2=\text{vec}(\bZ)^\top\text{vec}(\bZ)$ with $\bZ \sim E_{p,n}(\mathbf{O}_{p,n},\bI_{p\times n},f)$ (matrix-variate elliptically contoured distribution with zero location matrix $\mathbf{O}_{p,n}$, identity dispersion matrix $\bI_{p\times n}$ and density generator $f(.)$). The symbol $\mathbf{G}_p$ denotes for the duplication matrix (see, \cite{magnus2019matrix}).
\item The Jeffreys prior
\begin{eqnarray}\label{prior_J}
\pi_J(\bmu,\bPsi)=\pi_J(\bPsi)\propto \pi_R(\bPsi) \sqrt{\text{det}\left[\sum_{i=1}^{n}(\bPsi + \bU_i)^{-1}\right]}.
\end{eqnarray}
\end{itemize}

Both priors in \eqref{prior_R} and \eqref{prior_J} depend only on $\bPsi$ and assign the constant prior to $\bmu$, which is not surprising since $\bmu$ is the location parameter of the model. As such, the choice between the Berger and Bernardo reference prior and the Jeffreys prior has impact on the marginal posterior distribution of $\bPsi$, while the conditional posterior for $\bmu$ given $\bPsi$ is same and it is given by
\begin{eqnarray}\label{con_posterior_mu}
\pi(\bmu|\bPsi, \bX) &\propto& \sqrt{\text{det}\left(\sum_{i=1}^{n}(\bPsi + \bU_i)^{-1}\right)} f_{\bPsi,\bX}\left((\bmu-\tilde{\bx}(\bPsi))^\top\left(\sum_{i=1}^{n}(\bPsi+ \bU_i)^{-1}\right)(\bmu-\tilde{\bx}(\bPsi))\right)\!,
\end{eqnarray}
where
\begin{equation}\label{g_sig-lam_bx}
f_{\bPsi,\bX}\left(u\right)=f\left(\sum_{i=1}^{n} (\bx_i-\tilde{\bx}(\bPsi))^\top (\bPsi+ \bU_i)^{-1}(\bx_i-\tilde{\bx}(\bPsi))+u\right) \qquad u \ge 0\,,
\end{equation}
with
\begin{equation}\label{tilde_x}
\tilde{\bx}(\bPsi)=\left(\sum_{i=1}^{n}(\bPsi+ \bU_i)^{-1}\right)^{-1}\sum_{i=1}^{n}(\bPsi+ \bU_i)^{-1}\bx_i.
\end{equation}
Furthermore, a similar statement for the conditional posterior of $\bmu$ holds for any prior of $\bmu$ and $\bPsi$, which is a function of $\bPsi$ only. Using a generic notation $\pi(\bPsi)$, we get the marginal posterior for $\bPsi$ expressed as
\begin{eqnarray}\label{marg_posterior_Psi}
\pi(\bPsi| \bX) &\propto& \frac{\pi(\bPsi)}{\sqrt{\text{det}(\sum_{i=1}^{n}(\bPsi + \bU_i)^{-1})}\prod_{i=1}^{n}\sqrt{\text{det}(\bPsi + \bU_i)}}\nonumber\\
&\times&\int_{0}^{\infty} u^{p-1} f\left(u^2+\sum_{i=1}^{n} (\bx_i-\tilde{\bx}(\bPsi))^\top (\bPsi+ \bU_i)^{-1}(\bx_i-\tilde{\bx}(\bPsi))\right) \mathbf{d}u.
\end{eqnarray}

From \eqref{con_posterior_mu} one concludes that the conditional posterior of $\bmu$ belongs to the class of elliptically contoured distribution with density generator $f_{\bPsi,\bX}(.)$ which is related to $f(.)$ as shown in \eqref{g_sig-lam_bx}. In some special cases, the density generator $f_{\bPsi,\bX}(.)$ can be expressed analytically. For instance, the conditional posterior of $\bmu$ is a multivariate normal distribution under the normal multivariate random effects model, while $\bmu$ conditionally on $\bPsi$ and $\bX$ is $t$-distributed under the assumption of the $t$ multivariate random effects model (see, Section 5 in \cite{bodnar2023objective}).

The situation is more challenging in the case with the marginal posterior of $\bPsi$. Even though the integral in \eqref{marg_posterior_Psi} can be analytically evaluated for some families of elliptically contoured distribution, like the normal distribution and the $t$-distribution, the resulting expression of the marginal posterior of $\bPsi$ remains a complicated function of $\bPsi$ which makes it impossible to derive in closed-form Bayesian inference procedures neither for $\bPsi$ nor for $\bmu$. As such, two Metropolis-Hastings algorithms for drawing samples from the joint posterior of $\bmu$ and $\bPsi$ were proposed in \cite{bodnar2023objective}. In the next section, we suggest a further approach that is based on the Gibbs sampler and utilizes the fact that the conditional posterior of $\bmu$ is an elliptically contoured distribution.

It is also remarkable that the assumption of known covariance matrices $\bU_i$, $i=1,...,n$ reported by individual studies can be weakened. Namely, one can extend the generalized random effects model \eqref{mult-rem} by assuming that
\begin{equation}\label{mult-rem-con1}
p(\bX | \bmu, \bPsi,\bU_1,...,\bU_n) = \frac{1}{\sqrt{\text{det}(\bPsi \otimes \bI+\bU)}} f\left(\text{vec}(\bX - \bmu \bi^\top)^\top (\bPsi \otimes \bI+\bU)^{-1}\text{vec}(\bX - \bmu \bi) \right),
\end{equation}
and
\begin{equation}\label{mult-rem-con2}
p(\bU_1,...,\bU_n| \bSigma_1,...,\bSigma_n) = g\left(\bU_1,...,\bU_n; \bSigma_1,...,\bSigma_n \right),
\end{equation}
where $g(.)$ is a density parameterized by $\bSigma_1,...,\bSigma_n$. Such a model under the assumption of multivariate normal distribution was considered in \cite{rukhin2007estimating} and \cite{zhao2018some}. If the parameter space for $\bmu,\bPsi$ and $\bSigma_1,...,\bSigma_n$ can be presented as a Cartesian product, then the statistics $\bU_1,...,\bU_n$ are ancillary and the conditional principle can be used (\cite{barndorff1994inference}, \cite{reid1995roles}, \cite{fraser2004ancillaries}, \cite{ghosh2010ancillary}, \cite{sundberg2019statistical}). As such, the conditional likelihood $p(\bX | \bmu, \bPsi,\bU_1,...,\bU_n)$ is used in the derivation of Bayesian inference procedures and the realizations of $\bU_1,...,\bU_n$ are treated as known quantities. This approach is widely used in Bayesian regression analysis among others where explanatory factors are usually assumed to be ancillary (cf., \cite{gelman2013bayesian}, \cite{norets2015bayesian}).

\section{Gibbs sampler}\label{sec:gibbs}

It is shown in \eqref{con_posterior_mu} that the conditional posterior of $\bmu$ given $\bPsi$ belong to the family of elliptically contoured distributions with location parameter $\tilde{\bx}(\bPsi)$, dispersion matrix $\left(\sum_{i=1}^{n}(\bPsi+ \bU_i)^{-1}\right)^{-1}$ and density generator $f_{\bPsi,\bX}(.)$. As such, using the last generated value of the between-study covariance matrix $\bPsi^{(b-1)}$, the new $\bmu^{(b)}$ is drawn from the conditional distribution
\begin{equation*}
\pi(\bmu|\bPsi=\bPsi^{(b-1)}, \bX) \propto f_{\bPsi^{(b-1)},\bX}\left((\bmu-\tilde{\bx}(\bPsi^{(b-1)}))^\top\left(\sum_{i=1}^{n}(\bPsi^{(b-1)}+ \bU_i)^{-1}\right)(\bmu-\tilde{\bx}(\bPsi^{(b-1)}))\right).
\end{equation*}
In some special cases, like under the normal multivariate random effects model and the $t$ multivariate random effects model, this step of the Gibbs sampler algorithm can further be simplified as presented in detail in Section \ref{sec:gibbs-nor} and \ref{sec:gibbs-t}.

Next, we discuss the way how a new value of $\bPsi$ can be drawn from the conditional posterior distribution of $\bPsi$ given $\bmu$ and $\bX$. Under the assumption $\frac{J_2}{2pn+p^2n^2} -\frac{1}{4} \le 0$ where $J_2$ is defined in \eqref{Ji}, we get from the proof of Theorem 4 in \cite{bodnar2023objective} that
\begin{equation}\label{ineq_R}
 \frac{\pi(\bPsi)}{\prod_{i=1}^{n}\sqrt{\text{det}(\bPsi + \bU_i)}} \le \text{det}(\bPsi)^{-(n+p+1)/2}
\end{equation}
under the Berger and Bernardo reference prior and
\begin{equation}\label{ineq_J}
\frac{\pi(\bPsi)}{\prod_{i=1}^{n}\sqrt{\text{det}(\bPsi + \bU_i)}} \le \text{det}(\bPsi)^{-(n+p+2)/2}
\end{equation}
under the Jeffreys prior. Furthermore, assuming that the density generator $f(u)$ is a non-increasing function in $u \ge 0$ it holds that
\begin{eqnarray}\label{ineq_f}
 f\left(\text{vec}(\bX - \bmu \bi^\top)^\top (\bPsi \otimes \bI+\bU)^{-1}\text{vec}(\bX - \bmu \bi) \right)&\le&  f\left(\text{vec}(\bX - \bmu \bi^\top)^\top \bPsi^{-1}\text{vec}(\bX - \bmu \bi)\right) \nonumber\\
 &=&f\left(\text{tr}(\bPsi^{-1} \bS(\bmu))\right),
\end{eqnarray}
where
\begin{equation}\label{bS_bmu}
\bS(\bmu)=\sum_{i=1}^{n}(\bx_i-\bmu)(\bx_i-\bmu)^\top.
\end{equation}

In using \eqref{ineq_R}-\eqref{ineq_f} and the fact that the conditional posterior of $\bPsi$ is proportional to the joint posterior of $\bmu$ and $\bPsi$ with the proportionality constant being the function of $\bmu$, we get that the conditional posterior for $\bPsi$ given $\bmu$ is bounded by
\begin{equation}\label{q_R}
q_R(\bPsi|\bmu)=\text{det}(\bPsi)^{-(n+p+1)/2} f\left(\text{tr}(\bPsi^{-1} \bS(\bmu))\right)
\end{equation}
under the Berger and Bernardo reference prior and by
\begin{equation}\label{q_J}
q_J(\bPsi|\bmu)=\text{det}(\bPsi)^{-(n+p+2)/2} f\left(\text{tr}(\bPsi^{-1} \bS(\bmu))\right)
\end{equation}
under the Jeffreys prior. The two expressions $q_R(\bPsi|\bmu)$ and $q_R(\bPsi|\bmu)$ are the kernels of the generalized $p$-dimensional inverse Wishart distribution (see, e.g., \citet{sutradhar1989generalization}) with scale matrix $\sum_{i=1}^n(\bx_i - \bmu)(\bx_i - \bmu)^\top$, density generator $f$, and $(n+p+1)$ degrees of freedom when the Berger and Bernardo reference prior and with $(n+p+2)$ degrees of freedom for the Jeffreys prior. We will denote these assertions by $\bPsi|\bmu, \bX \sim GIW_{p}(n+p+1,\bS(\bmu),f)$ and $\bPsi|\bmu, \bX \sim GIW_{p}(n+p+2,\bS(\bmu),f)$, respectively. Finally, it is noted that the assumption of $\frac{J_2}{2pn+p^2n^2} -\frac{1}{4} \le 0$ and of a non-increasing density generator $f(.)$, which are used in the derivation of \eqref{q_R} and \eqref{q_J}, are quite general and they are fulfilled for many families of elliptically contoured distributions, like for the normal distribution and $t$-distribution as discussed in Sections \ref{sec:gibbs-nor} and \ref{sec:gibbs-t} below.

The derived results provide a motivation of choosing the generalized Wishart distribution as a proposal used in drawing samples from the conditional posterior of $\bPsi$ given $\bmu$. It is remarkable that it is simple to draw samples from the generalized Wishart distribution. Also, the generated matrices are positive definite by construction. In the general case, the resulting hybrid Gibbs sampler approach is performed under the Berger and Bernardo reference prior is performed in the following way:

\begin{algorithm}[H]
{\small\caption{Gibbs sampler for drawing realizations from the joint posterior distribution of $\bmu$ and $\bPsi$ under the generalized multivariate random effects model \eqref{mult-rem} and the Berger and Bernardo reference prior \eqref{prior_R}}\label{algorithmGibbs-gen}
\begin{enumerate}[(1)]
\item \textbf{Initialization:} Choose the initial values $\bmu^{(0)}$ and $\bPsi^{(0)}$ for $\bmu$ and $\bPsi$ and set $b =0$.
\item \textbf{Given the previous value $\bPsi^{(b-1)}$ and data $\bX$, generate $\bmu^{(b)}$ from the conditional posterior \eqref{con_posterior_mu}:}
\begin{eqnarray}\label{algC-gen}
\bmu|\bPsi=\bPsi^{(b-1)},\bX \sim E_{p,1}\left(\tilde{\bx}(\bPsi^{(b-1)}),\left(\sum_{i=1}^{N}(\bPsi^{(b-1)}+ \bU_i)^{-1}\right)^{-1},f_{\bPsi^{(b-1)},\bX}\right),
\end{eqnarray}
where $\tilde{\bx}(\bPsi^{(b-1)})$ is given in \eqref{tilde_x} and $f_{\bPsi^{(b-1)},\bX}$ is defined in \eqref{g_sig-lam_bx}.
\item \textbf{Given the previous value $\bmu^{(b)}$ and data $\bX$, generate new value of $\bPsi^{(b)}$:}
\begin{enumerate}[(i)]
\item Using $\bmu^{(b)}$ and $\bX$, generate $\bPsi^{(w)}$ from
\begin{equation}\label{algC-iW}
\bPsi|\bmu=\bmu^{(b)}, \bX  \sim GIW_{p}\left(n+p+1,\bS(\bmu^{(b)}),f)\right),
\end{equation}
where $\bS(\bmu^{(b)})$ is defined in \eqref{bS_bmu}.
\item Compute of the Metropolis-Hastings ratio:
\begin{equation}\label{MHratio}
MH^{(b)}=\frac{\pi(\bPsi^{(w)}|\bmu^{(b)},\bX )q_R(\bPsi^{(b-1)}| \bmu^{(b)})}{\pi(\bPsi^{(b-1)}|\bmu^{(b)},\bX) q_R(\bPsi^{(w)}| \bmu^{(b)})},
\end{equation}
where $q_R(\bPsi|\bmu)$ is the conditional density of $\bPsi$ given $\bmu$, i.e., the density of the generalized inverse Wishart distribution \eqref{q_R}.
\item Moving to the next state of the Markov chain:\\
Generate $U^{(b)}$ from the uniform distribution on $[0,1]$. If $U^{b}<\min\left\{1,MH^{(b)}\right\}$, then set $\bmu^{(b)}=\bmu^{(w)}$ and $\bPsi^{(b)}=\bPsi^{(w)}$. Otherwise, set $\bmu^{(b)}=\bmu^{(b-1)}$ and $\bPsi^{(b)}=\bPsi^{(b-1)}$.
\end{enumerate}
\item \textbf{Return to step (2), increase $b$ by 1, and repeat until the sample of size $B$ is accumulated.}
\end{enumerate}
}
\end{algorithm}

Since the same value of $\bmu$ is used in the conditional posterior for $\bPsi$ when the Metropolis-Hastings ratio is computed, the conditional posterior for $\bPsi$ can be replaced by the joint posterior for $\bmu$ and $\bPsi$ in \eqref{MHratio}. Under the Jeffreys prior, step (3.i) in the algorithm should be replaced by
\begin{enumerate}[(i)]
\item Using $\bmu^{(b)}$ and $\bX$, generate $\bPsi^{(w)}$ from
\begin{equation}\label{algC-iW-J}
\bPsi|\bmu=\bmu^{(b)}, \bX \sim GIW_{p}\left(n+p+2,\bS(\bmu^{(b)})\right),
\end{equation}
\end{enumerate}
while $q_J(\bPsi|\bmu)$ should be used instead of $q_R(\bPsi|\bmu)$ in the computation of the  Metropolis-Hastings ratio in \eqref{MHratio}.

In the next two subsections we present the Gibbs sampler in two partial cases of the normal distribution and $t$-distribution.

\subsection{Gibbs sampler under the normal multivariate random effects model}\label{sec:gibbs-nor}

Under the assumption of the normal multivariate random effects model, it holds that
\begin{equation}\label{f_nor}
f(u)=K_{p,n}\exp(-u/2), ~ u\ge 0~~ \text{with} ~~ K_{p,n}=(2\pi)^{-pn/2},
\end{equation}
which is a decreasing function. Moreover, under the assumption of normality it holds that
$$\frac{J_2}{2pn+p^2n^2} -\frac{1}{4}=0.$$
Thus, the two conditions present in the derivation of the Gibbs sampler algorithm are fulfilled.

Using \eqref{f_nor}, we get that
\begin{eqnarray}\label{con_posterior_mu_nor}
\bmu|\bPsi,\bX \sim \mathcal{N}_p\left(\left(\sum_{i=1}^{n}(\bPsi+ \bU_i)^{-1}\right)^{-1}\sum_{i=1}^{n}(\bPsi+ \bU_i)^{-1}\bx_i,\left(\sum_{i=1}^{n}(\bPsi+ \bU_i)^{-1}\right)^{-1}\right),
\end{eqnarray}
and the marginal posterior for $\bPsi$ is given by
\begin{eqnarray}\label{marg_posterior_Psi_nor}
\pi(\bPsi|\bX)&\propto& \frac{\pi(\bPsi)}{\sqrt{\text{det}(\sum_{i=1}^{n}(\bPsi + \bU_i)^{-1})}\prod_{i=1}^{n}\sqrt{\text{det}(\bPsi + \bU_i)}}\nonumber\\
&\times& \exp\left(-\frac{1}{2}\sum_{i=1}^{n} (\bx_i-\tilde{\bx}(\bPsi))^\top (\bPsi+ \bU_i)^{-1}(\bx_i-\tilde{\bx}(\bPsi))\right)\,.
\end{eqnarray}

Hence, the hybrid Gibbs sampler algorithm derived under the Berger and Bernardo reference prior simplifies to
\begin{algorithm}[H]
{\small\caption{Gibbs sampler for drawing realizations from the joint posterior distribution of $\bmu$ and $\bPsi$ under the normal multivariate random effects model and the Berger and Bernardo reference prior \eqref{prior_R}}\label{algorithmGibbs-nor}
\begin{enumerate}[(1)]
\item \textbf{Initialization:} Choose the initial values $\bmu^{(0)}$ and $\bPsi^{(0)}$ for $\bmu$ and $\bPsi$ and set $b =0$.
\item \textbf{Given the previous value $\bPsi^{(b-1)}$ and data $\bX$, generate $\bmu^{(b)}$ from:}
\begin{eqnarray}\label{algC-nor}
\bmu|\bPsi=\bPsi^{(b-1)},\bX \sim \mathcal{N}_p\left(\tilde{\bx}(\bPsi^{(b-1)}),\left(\sum_{i=1}^{N}(\bPsi^{(b-1)}+ \bU_i)^{-1}\right)^{-1}\right),
\end{eqnarray}
where $\tilde{\bx}(\bPsi^{(b-1)})$ is given in \eqref{tilde_x}.
\item \textbf{Given the previous value $\bmu^{(b)}$ and data $\bX$, generate new value of $\bPsi^{(b)}$:}
\begin{enumerate}[(i)]
\item Using $\bmu^{(b)}$ and $\bX$, generate $\bPsi^{(w)}$ from
\begin{equation}\label{algC-iW-nor}
\bPsi|\bmu=\bmu^{(b)}, \bX  \sim IW_{p}\left(n+p+1,\bS(\bmu^{(b)}),f)\right),
\end{equation}
where $\bS(\bmu^{(b)})$ is defined in \eqref{bS_bmu} and the symbol $IW_p$ denotes the inverse Wishart distribution (see, e.g., \cite{gupta2000matrix}).
\item Compute of the Metropolis-Hastings ratio:
\begin{equation}\label{MHratio-nor}
MH^{(b)}=\frac{\pi(\bPsi^{(w)}|\bmu^{(b)},\bX )q_R(\bPsi^{(b-1)}| \bmu^{(b)})}{\pi(\bPsi^{(b-1)}|\bmu^{(b)},\bX) q_R(\bPsi^{(w)}| \bmu^{(b)})},
\end{equation}
where $q_R(\bPsi|\bmu)$ is given in \eqref{q_R} with $f(.)$ as in \eqref{f_nor}.
\item Moving to the next state of the Markov chain:\\
Generate $U^{(b)}$ from the uniform distribution on $[0,1]$. If $U^{b}<\min\left\{1,MH^{(b)}\right\}$, then set $\bmu^{(b)}=\bmu^{(w)}$ and $\bPsi^{(b)}=\bPsi^{(w)}$. Otherwise, set $\bmu^{(b)}=\bmu^{(b-1)}$ and $\bPsi^{(b)}=\bPsi^{(b-1)}$.
\end{enumerate}
\item \textbf{Return to step (2), increase $b$ by 1, and repeat until the sample of size $B$ is accumulated.}
\end{enumerate}
}
\end{algorithm}

Under the Jeffreys prior, step (3.i) in the algorithm becomes
\begin{enumerate}[(i)]
\item Using $\bmu^{(b)}$ and $\bX$, generate $\bPsi^{(w)}$ from
\begin{equation}\label{algC-iW-J-nor}
\bPsi|\bmu=\bmu^{(b)}, \bX \sim IW_{p}\left(n+p+2,\bS(\bmu^{(b)})\right).
\end{equation}
\end{enumerate}
Finally, $q_R(\bPsi|\bmu)$ should be replaced by $q_J(\bPsi|\bmu)$ in the computation of the Metropolis-Hastings ratio in \eqref{MHratio-nor}.

\subsection{Gibbs sampler under the $t$ multivariate random effects model}\label{sec:gibbs-t}

Under the assumption of the $t$ multivariate random effects model with $d$ degrees of freedom, we get
\begin{equation}\label{f_t}
f(u)=K_{p,n,d}(1+u/d)^{-(pn+d)/2}~~ \text{with} ~~ K_{p,n,d}=(\pi d)^{-pn/2}\dfrac{\Gamma\left((d+pn)/2\right)}{\Gamma\left(d/2\right)},
\end{equation}
which  is a decreasing function. Moreover, $J_2=\frac{pn(pn+2)(pn+d)}{4(pn+2+d)}$ (see, Section 3.2 in \cite{bodnar2019}) and, consequently,
\[\frac{J_2}{2pn+p^2n^2}=\frac{pn+d}{4(pn+d+2)}<\frac{1}{4}.\]
Hence, the two conditions used in the derivation of the Gibbs sampler algorithm are fulfilled.

Then, it holds that (see, Section 5.2 in \cite{bodnar2023objective})
\begin{eqnarray}\label{con_posterior_mu_t}
\pi(\bmu|\bPsi,\bX) &\propto& \Bigg(1+\frac{1}{pn+d-p}
\frac{pn+d-p}{d+\sum_{i=1}^{n} (\bx_i-\tilde{\bx}(\bPsi))^\top (\bPsi+ \bU_i)^{-1}(\bx_i-\tilde{\bx}(\bPsi))}\nonumber\\
&\times&(\bmu-\tilde{\bx}(\bPsi))^\top\left(\sum_{i=1}^{n}(\bPsi+ \bU_i)^{-1}\right)(\bmu-\tilde{\bx}(\bPsi))\Bigg)^{-(pn+d)/2},
\end{eqnarray}
i.e., $\bmu$ conditionally on $\bPsi$ and $\bX$ has a $p$-dimensional $t$-distribution with $pn+d-p$ degrees of freedom, location parameter $\tilde{\bx}(\bPsi)$ and dispersion matrix
\begin{equation}\label{disp-t}
\frac{d+\sum_{i=1}^{n} (\bx_i-\tilde{\bx}(\bPsi))^\top (\bPsi+ \bU_i)^{-1}(\bx_i-\tilde{\bx}(\bPsi))}{pn+d-p}\left(\sum_{i=1}^{n}(\bPsi+ \bU_i)^{-1}\right)^{-1}.
\end{equation}

The marginal posterior for $\bPsi$ is expressed as
\begin{eqnarray}\label{marg_posterior_Psi_t}
\pi(\bPsi|\bX)&\propto& \frac{\pi(\bPsi)}{\sqrt{\text{det}(\sum_{i=1}^{n}(\bPsi + \bU_i)^{-1})}\prod_{i=1}^{n}\sqrt{\text{det}(\bPsi + \bU_i)}}\nonumber\\
&\times&\left(1+\frac{1}{d}\sum_{i=1}^{n} (\bx_i-\tilde{\bx}(\bPsi))^\top (\bPsi+ \bU_i)^{-1}(\bx_i-\tilde{\bx}(\bPsi))\right)^{-(pn+d)/2}\,.
\end{eqnarray}
Finally, using the properties of the generalized inverse Wishart distribution with density generator corresponding to \eqref{f_t}, a random matrix from this distribution can be drawn by generating a random matrix form the inverse Wishart distribution which should be multiply by a random draw from the $\chi^2$-distribution with $d$ degrees of freedom divided by $d$. We summarize this approach in the algorithm derived for the $t$ multivariate random effect model when the Berger and Bernardo reference prior is employed.

\begin{algorithm}[H]
{\small\caption{Gibbs sampler for drawing realizations from the joint posterior distribution of $\bmu$ and $\bPsi$ under the $t$ multivariate random effects model and the Berger and Bernardo reference prior \eqref{prior_R}}\label{algorithmGibbs-t}
\begin{enumerate}[(1)]
\item \textbf{Initialization:} Choose the initial values $\bmu^{(0)}$ and $\bPsi^{(0)}$ for $\bmu$ and $\bPsi$ and set $b =0$.
\item \textbf{Given the previous value $\bPsi^{(b-1)}$ and data $\bX$, generate $\bmu^{(b)}$ from the $t$-distribution with $pn+d-p$ degrees of freedom, location parameter $\tilde{\bx}(\bPsi^{(b-1)})$ and dispersion matrix as in \eqref{disp-t} with $\bPsi$ replaced by $\bPsi^{(b-1)}$.}
\item \textbf{Given the previous value $\bmu^{(b)}$ and data $\bX$, generate new value of $\bPsi^{(b)}$:}
\begin{enumerate}[(i)]
\item Using $\bmu^{(b)}$ and $\bX$, generate $\bOmega^{(w)}$ from
\begin{equation}\label{algC-iW-t}
\bOmega|\bmu=\bmu^{(b)}, \bX  \sim IW_{p}\left(n+p+1,\bS(\bmu^{(b)}),f)\right),
\end{equation}
where $\bS(\bmu^{(b)})$ is defined in \eqref{bS_bmu}.
\item Generate $\xi^{(w)}$ from $\chi^2$-distribution with $d$ degrees of freedom independently of $\bOmega^{(w)}$ and compute
\[\bPsi^{(w)}=\frac{\xi^{(w)}}{d}\bOmega^{(w)}.\]
\item Compute of the Metropolis-Hastings ratio:
\begin{equation}\label{MHratio-t}
MH^{(b)}=\frac{\pi(\bPsi^{(w)}|\bmu^{(b)},\bX )q_R(\bPsi^{(b-1)}| \bmu^{(b)})}{\pi(\bPsi^{(b-1)}|\bmu^{(b)},\bX) q_R(\bPsi^{(w)}| \bmu^{(b)})},
\end{equation}
where $q_R(\bPsi|\bmu)$ is given in \eqref{q_R} with $f(.)$ as in \eqref{f_t}.
\item Moving to the next state of the Markov chain:\\
Generate $U^{(b)}$ from the uniform distribution on $[0,1]$. If $U^{b}<\min\left\{1,MH^{(b)}\right\}$, then set $\bmu^{(b)}=\bmu^{(w)}$ and $\bPsi^{(b)}=\bPsi^{(w)}$. Otherwise, set $\bmu^{(b)}=\bmu^{(b-1)}$ and $\bPsi^{(b)}=\bPsi^{(b-1)}$.
\end{enumerate}
\item \textbf{Return to step (2), increase $b$ by 1, and repeat until the sample of size $B$ is accumulated.}
\end{enumerate}
}
\end{algorithm}

When the Jeffreys prior is used, step (3.i) in the algorithm is changed to
\begin{enumerate}[(i)]
\item Using $\bmu^{(b)}$ and $\bX$, generate $\bOmega^{(w)}$ from
\begin{equation}\label{algC-iW-J-t}
\bOmega|\bmu=\bmu^{(b)}, \bX \sim IW_{p}\left(n+p+2,\bS(\bmu^{(b)})\right).
\end{equation}
\end{enumerate}
and $q_R(\bPsi|\bmu)$ should be replaced by $q_J(\bPsi|\bmu)$ in the computation of the Metropolis-Hastings ratio in \eqref{MHratio-t}.

\section{Simulation study}\label{sec:sim}

The aim of simulation study is two fold: (i) first, we study the convergence properties of the Markov chains constructed by the Gibbs sampler algorithm and compare it with those obtained by using two Metropolis-Hastings algorithms proposed in \cite{bodnar2023objective}; (ii) second, the coverage properties of the credible intervals, constructed for each parameter of the model are assessed. 

To study the convergence properties of Markov chains constructed by three algorithms we perform a simulation study for $p=2$ and $n=10$. The observation matrix $\bX$ is drawn from the normal multivariate random effects model and from the $t$ multivariate random effects model, where the elements of $\bmu$ are generated from the uniform distribution on $[1,5]$. We further set $\bPsi=\tau^2 \mathbf{\Xi}$, and $\bU=diag(\bU_1,...,\bU_n)$. The eigenvalues of $\mathbf{\Xi}$, $\bU_1$, ... , $\bU_{n}$, are drawn from the uniform distribution on $[1,4]$, while the eigenvectors are simulated from the Haar distribution (see, e.g., \cite{muirhead1982}). Finally, several values of $\tau$ are considered with $\tau^2 \in \{0.25, 0.5, 0.75, 1, 2\}$. We refer to two Metropolis-Hastings algorithms from \cite{bodnar2023objective} as Algorithm A and Algorithm B, while Algorithm C corresponds to the Gibbs sampler introduced in this paper. The average split-$\hat{R}$ estimates based on the rank normalization is used to analyze the convergence properties of the constructed Markov chains, while the empirical coverage probability is employed to assess the coverage properties of the credible intervals determined for each parameter of the model.

\begin{figure}[H]
\centering
\begin{tabular}{p{7.0cm}p{7.0cm}}
\includegraphics[width=6.5cm]{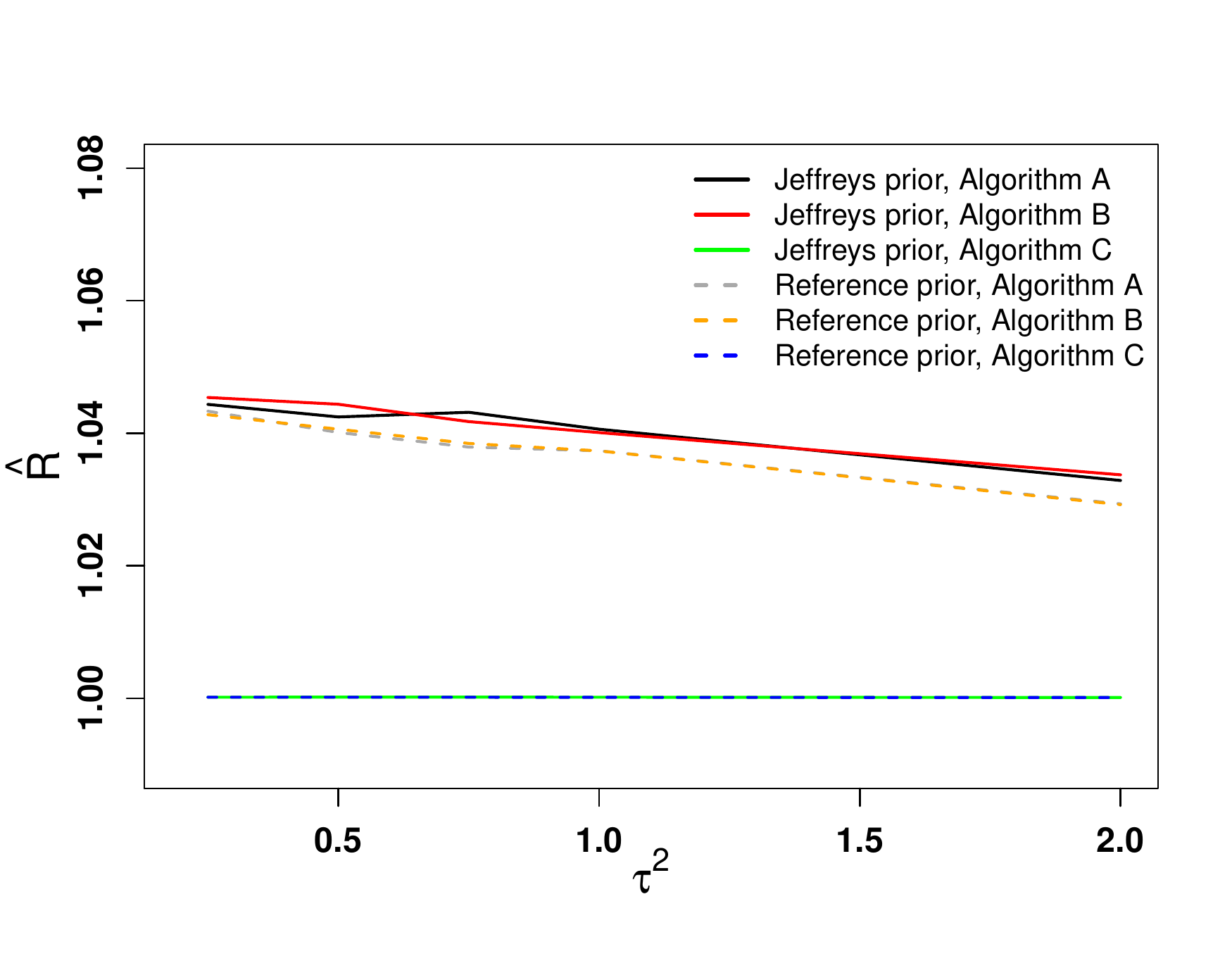}&\includegraphics[width=6.5cm]{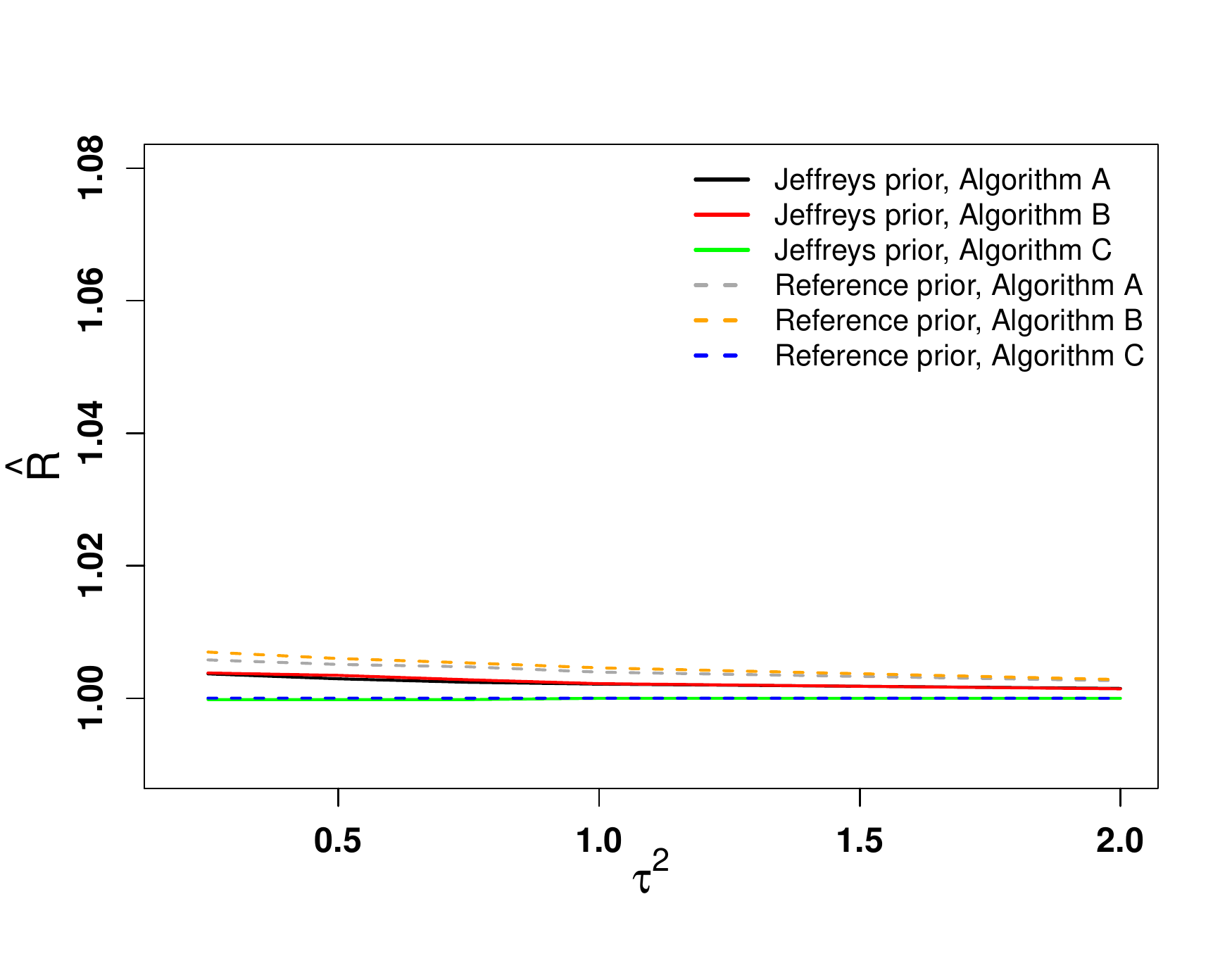}\\[-1cm]
\includegraphics[width=6.5cm]{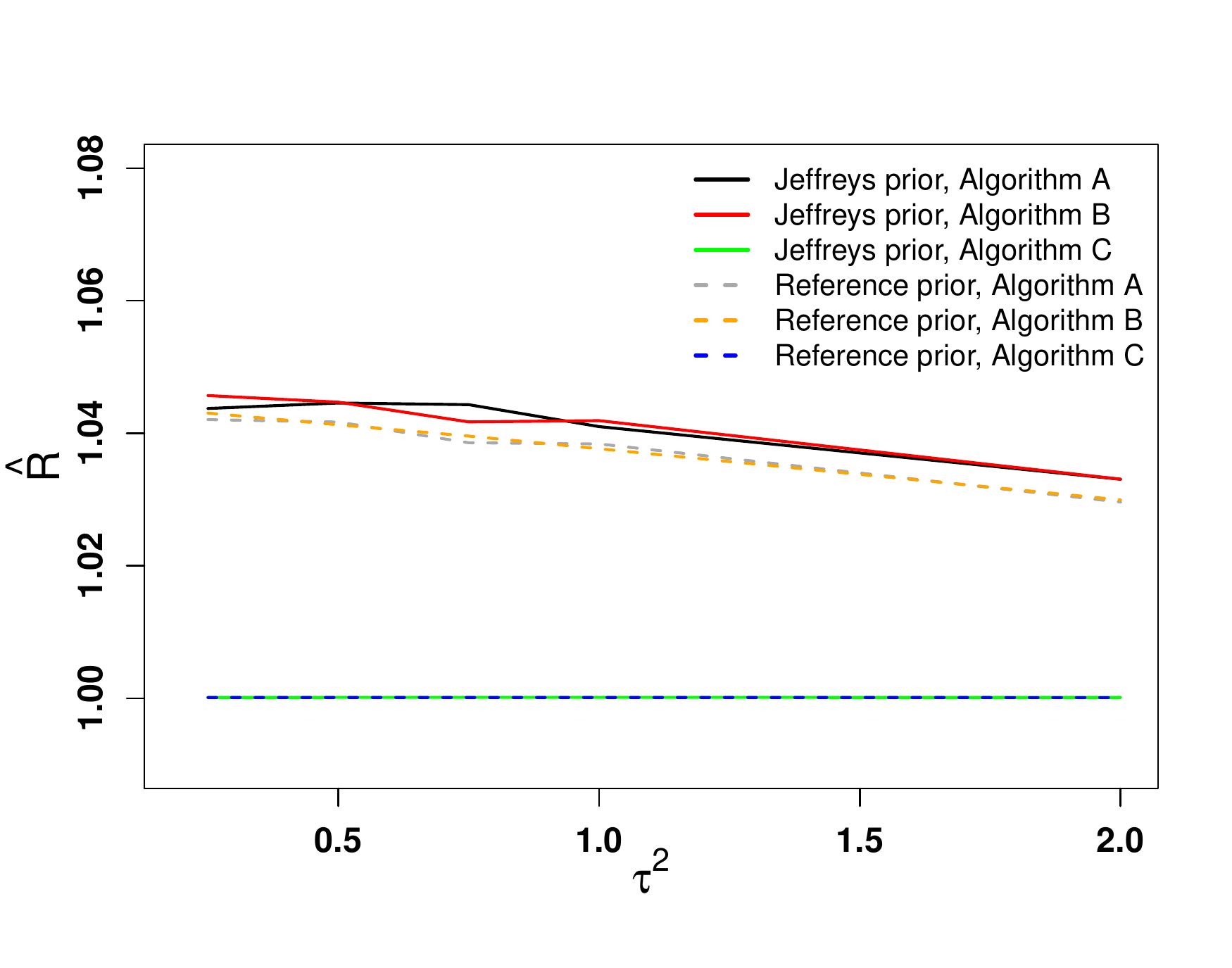}&\includegraphics[width=6.5cm]{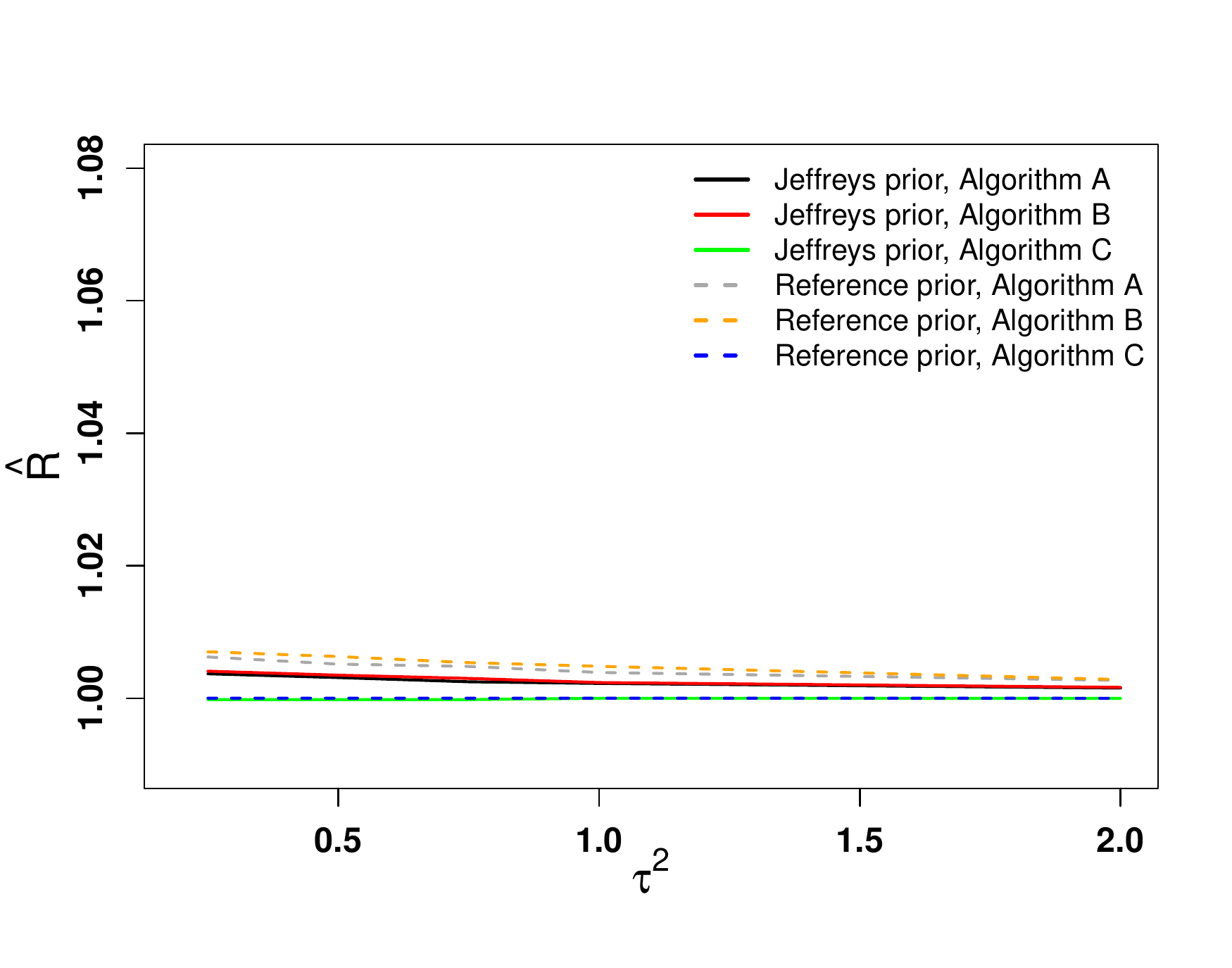}\\[-1cm]
\includegraphics[width=6.5cm]{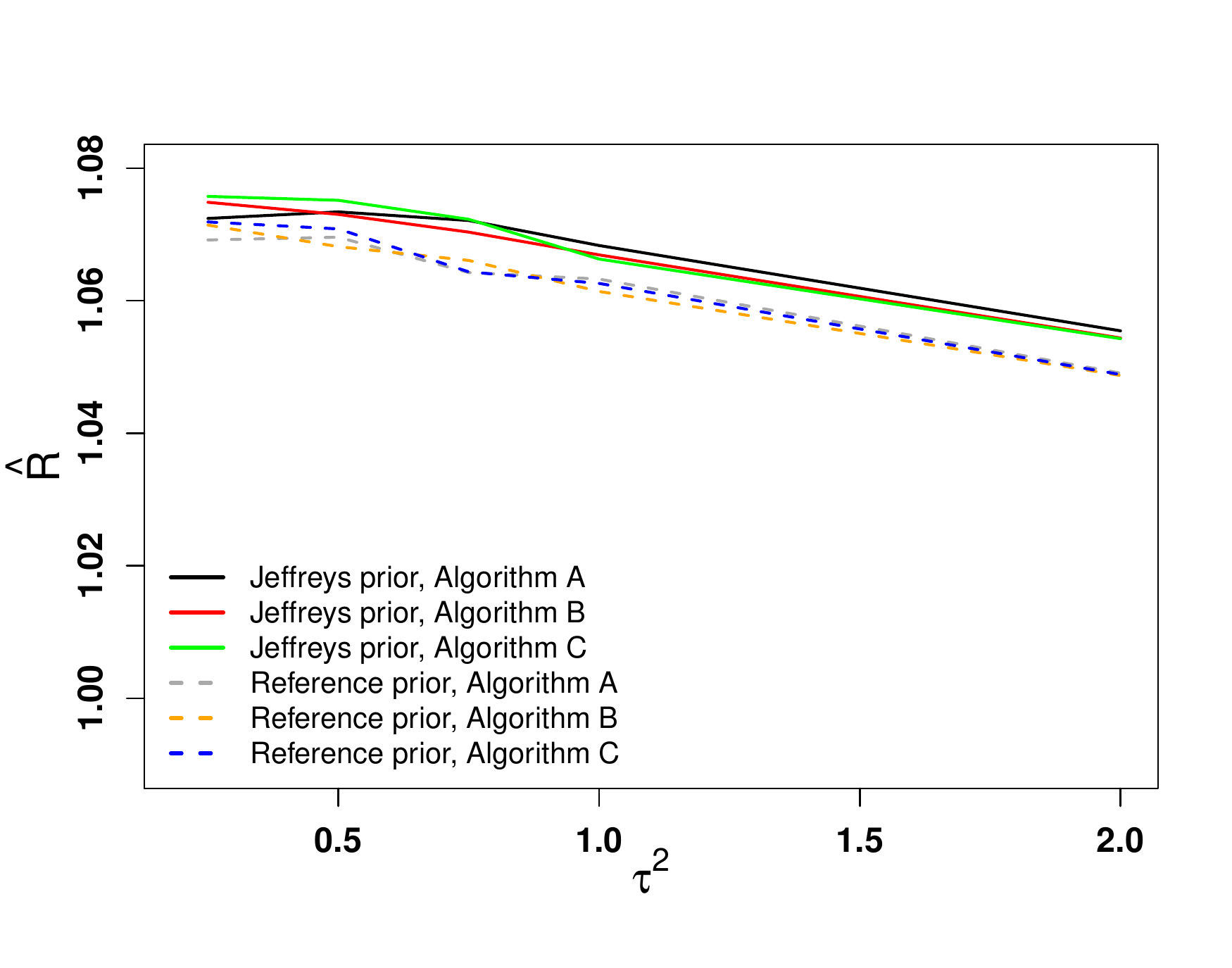}&\includegraphics[width=6.5cm]{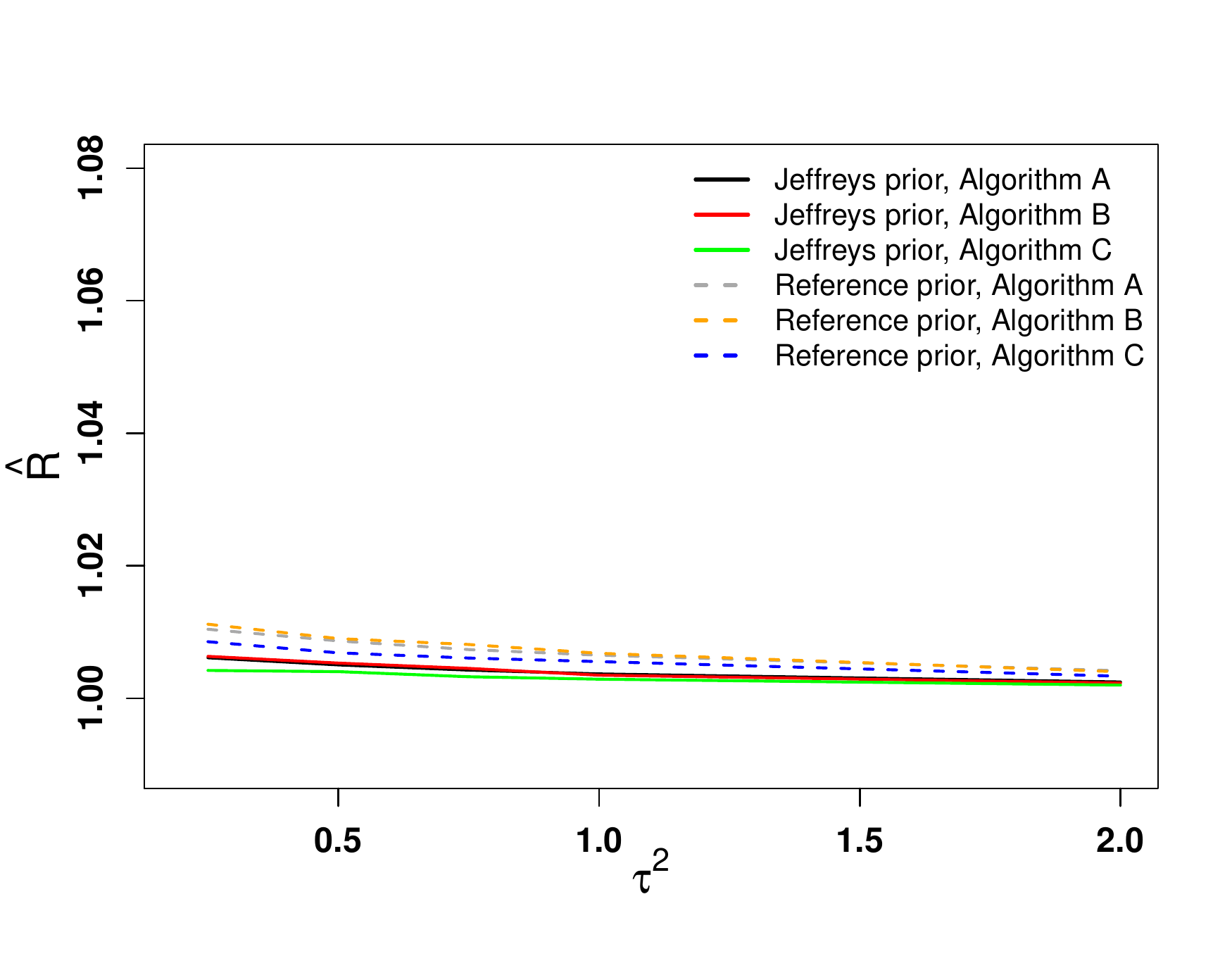}\\[-1cm]
\includegraphics[width=6.5cm]{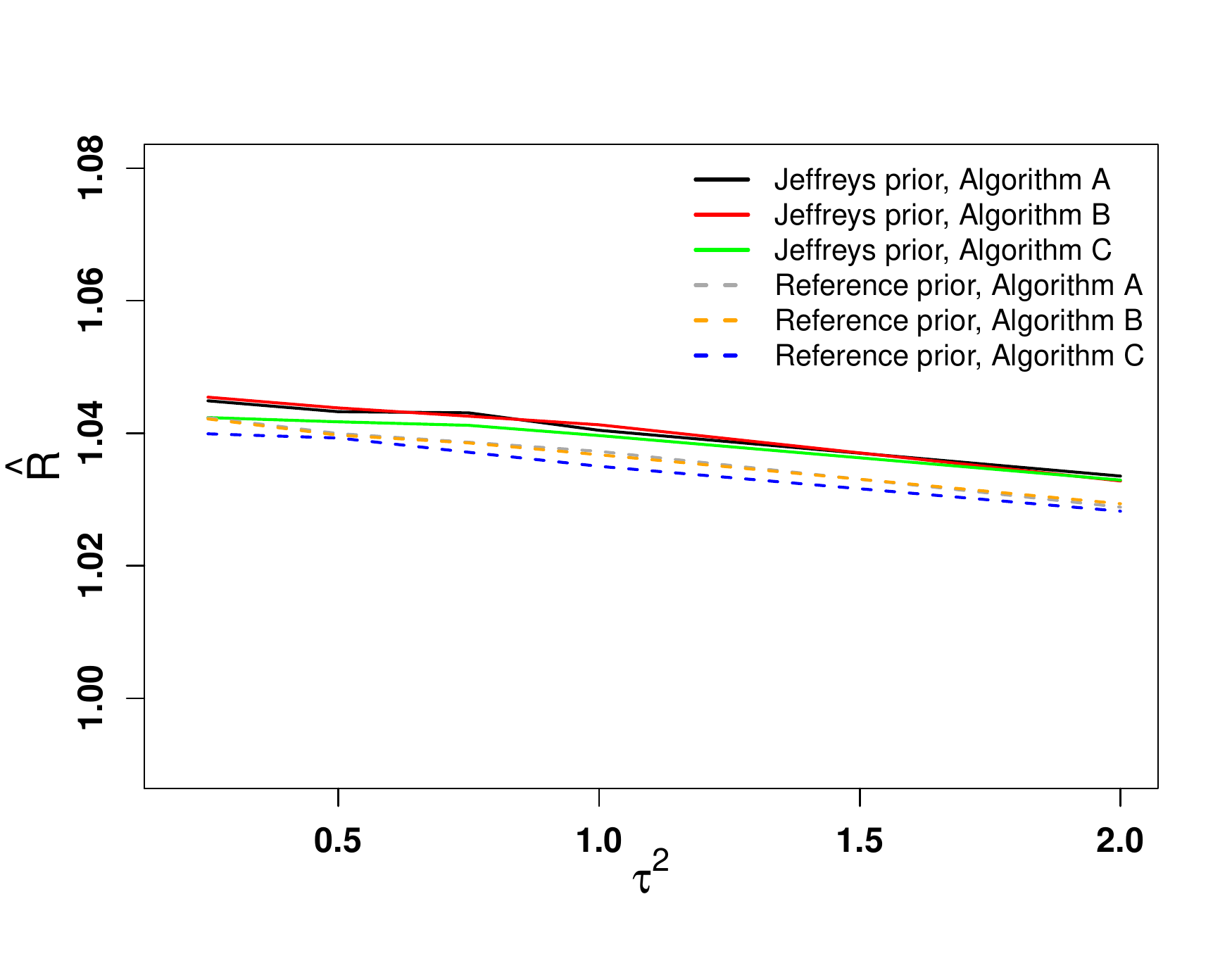}&\includegraphics[width=6.5cm]{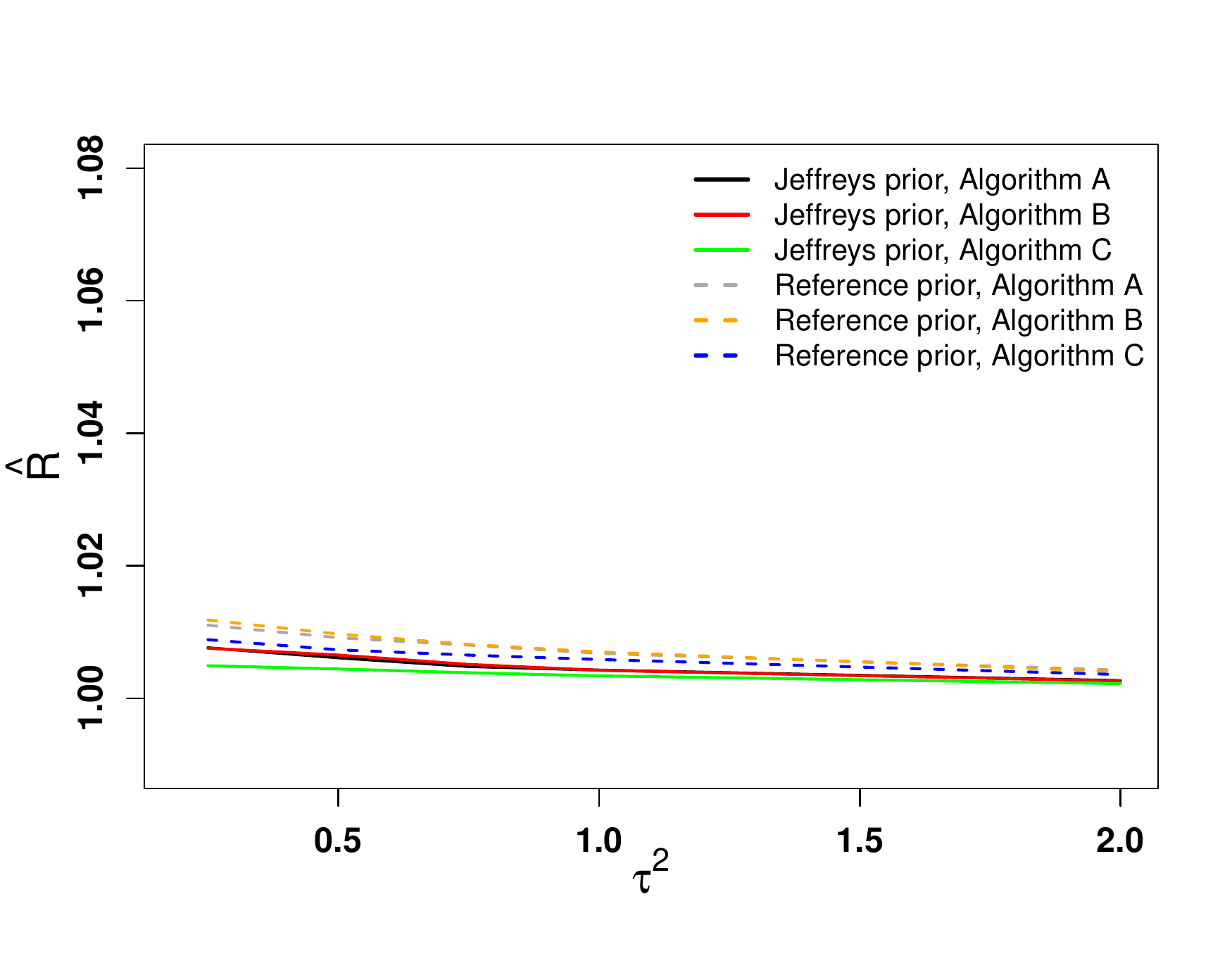}\\[-1cm]
\includegraphics[width=6.5cm]{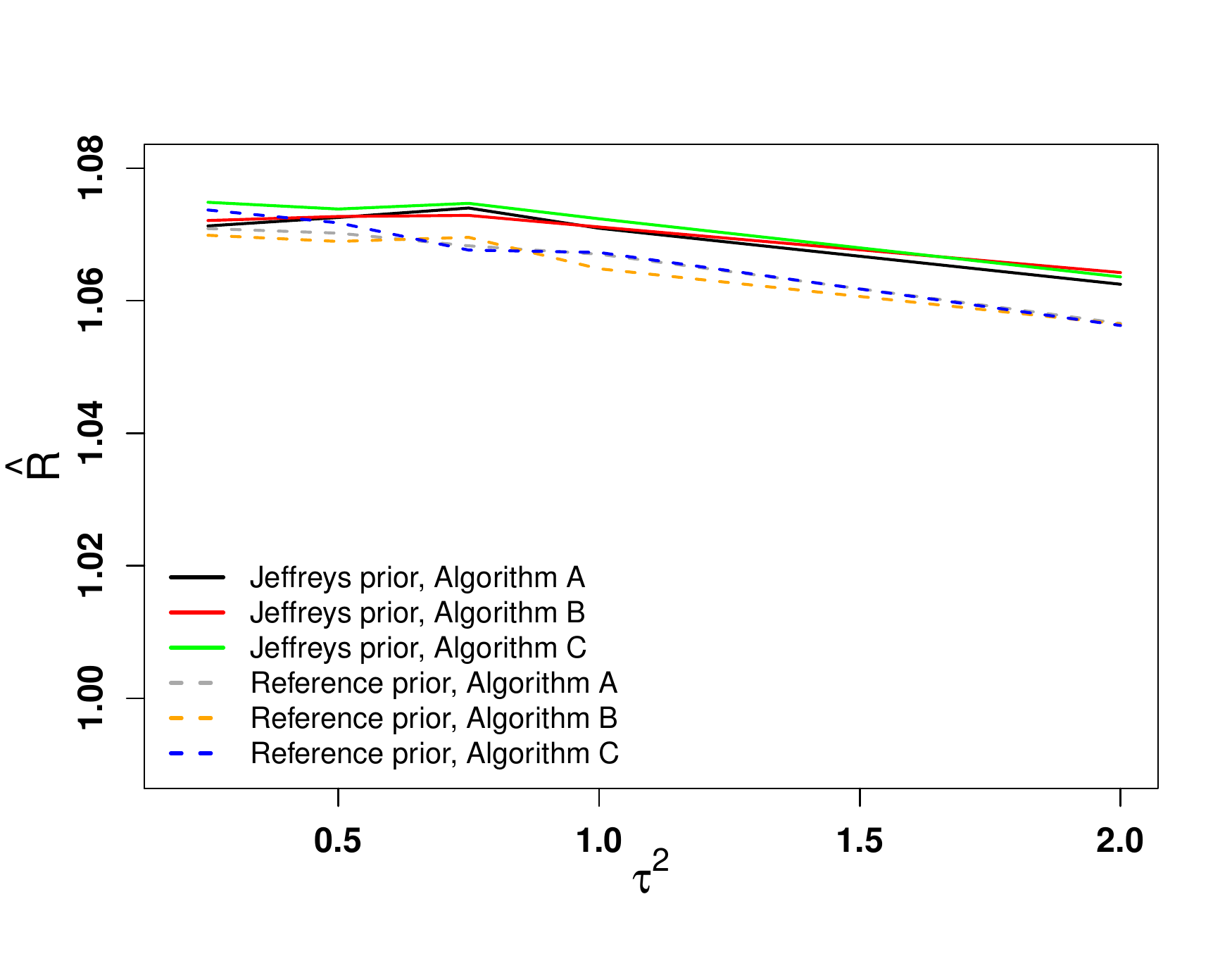}&\includegraphics[width=6.5cm]{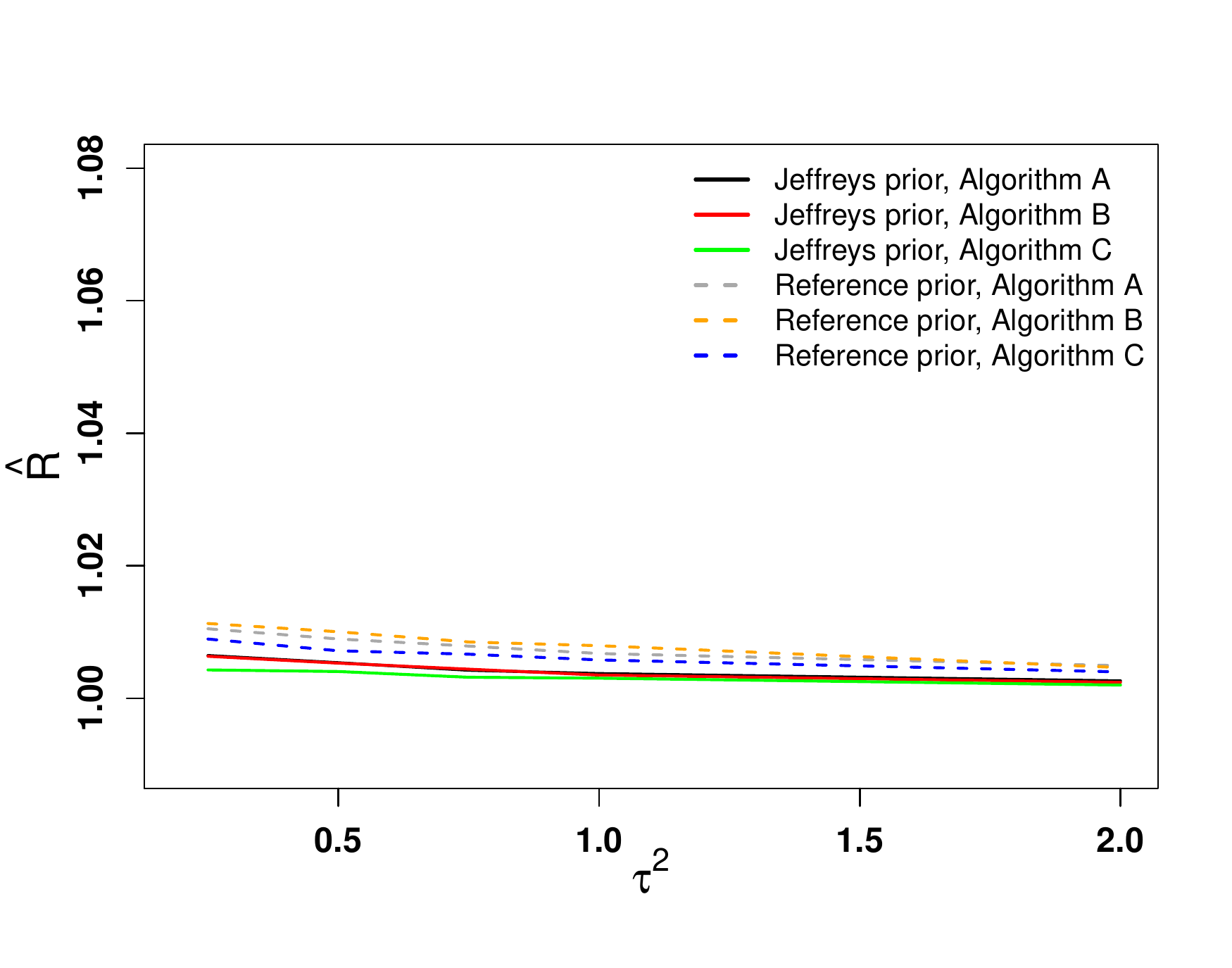}\\[-0.7cm]
\end{tabular}
 \caption{Average split-$\hat{R}$ estimates based on the rank normalization computed from four Markov chains constructed by using the joint posterior of $\bmu$ (first and second rows) and $\bPsi$ (third to fifth rows) given in Section \ref{sec:ba-MREM}. The observations are drawn from the normal multivariate random effects model (left column) and the $t$ multivariate random effects model (right column) with $p=2$ and $n=10$.}
\label{fig:sim-study-hatR}
 \end{figure}

Figure \ref{fig:sim-study-hatR} presents the average values of the split-$\hat{R}$ estimates based on the rank normalization as defined in \cite{vehtari2021rank}. The computations are based by constructing four Markov chains of length 10000 with burn-in period of 5000 and on $B=5000$ independent repetitions. This performance measure is suggested as a generalization of the $\hat{R}$ coefficient used in \cite{gelman2013bayesian} with the aim to define a measure which is robust to possible outliers that may be present in the constructed chains. In \cite{gelman2013bayesian}, it is recommended to conclude that the constructed Markov chain converges to its stationary distribution when the computed $\hat{R}$ coefficient is smaller than $1.1$. In the case of the split-$\hat{R}$ estimates based on the rank normalization, \cite{vehtari2021rank} proposed the usage of $1.01$ instead of $1.1$.

In Figure \ref{fig:sim-study-hatR} we observe that the Berger and Brenardo reference prior leads to smaller average values of the split-$\hat{R}$ estimates based on the rank normalization in comparison to the Jeffreys prior when the observations are drawn from the normal multivariate random effects model, while the reverse relation is present in the case of the $t$ multivariate random effects model. Moreover, independently of the employed prior, the algorithm used has only a minor impact on the split-$\hat{R}$ coefficients in the case of the between-study covariance matrix $\bPsi$. In contrast, the application of the Gibbs sample leads to the considerable improvements in the case of $\bmu$-estimation where the split-$\hat{R}$ coefficients are all close to one, independently of the value of $\tau$ and the distributional assumption of the multivariate random effects model. Finally, we note that the values obtained under the $t$ multivariate random effects model are considerably smaller than those computed under the normal multivariate random effects model.

\begin{figure}[H]
\centering
\begin{tabular}{p{7.0cm}p{7.0cm}}
\includegraphics[width=6.5cm]{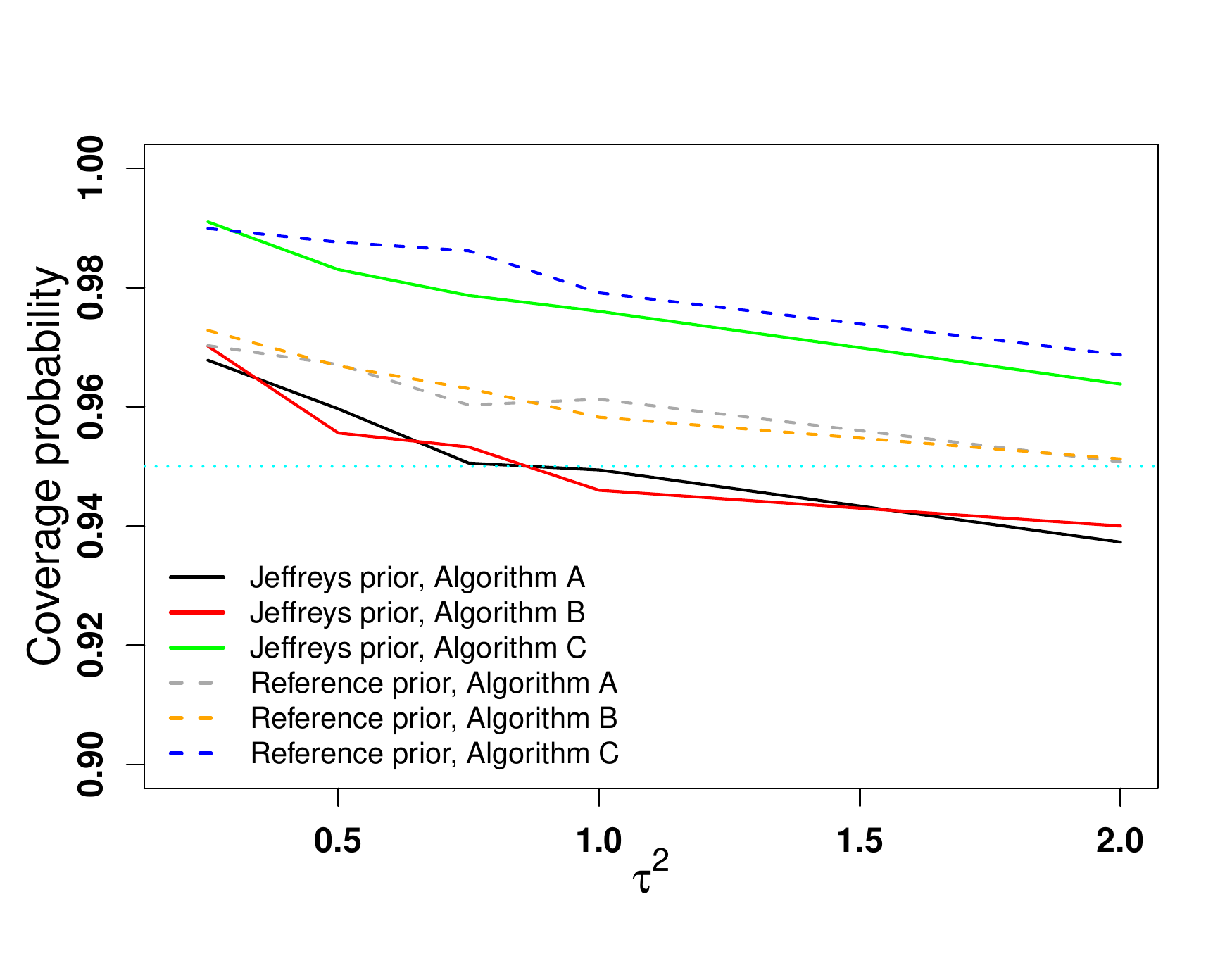}&\includegraphics[width=6.5cm]{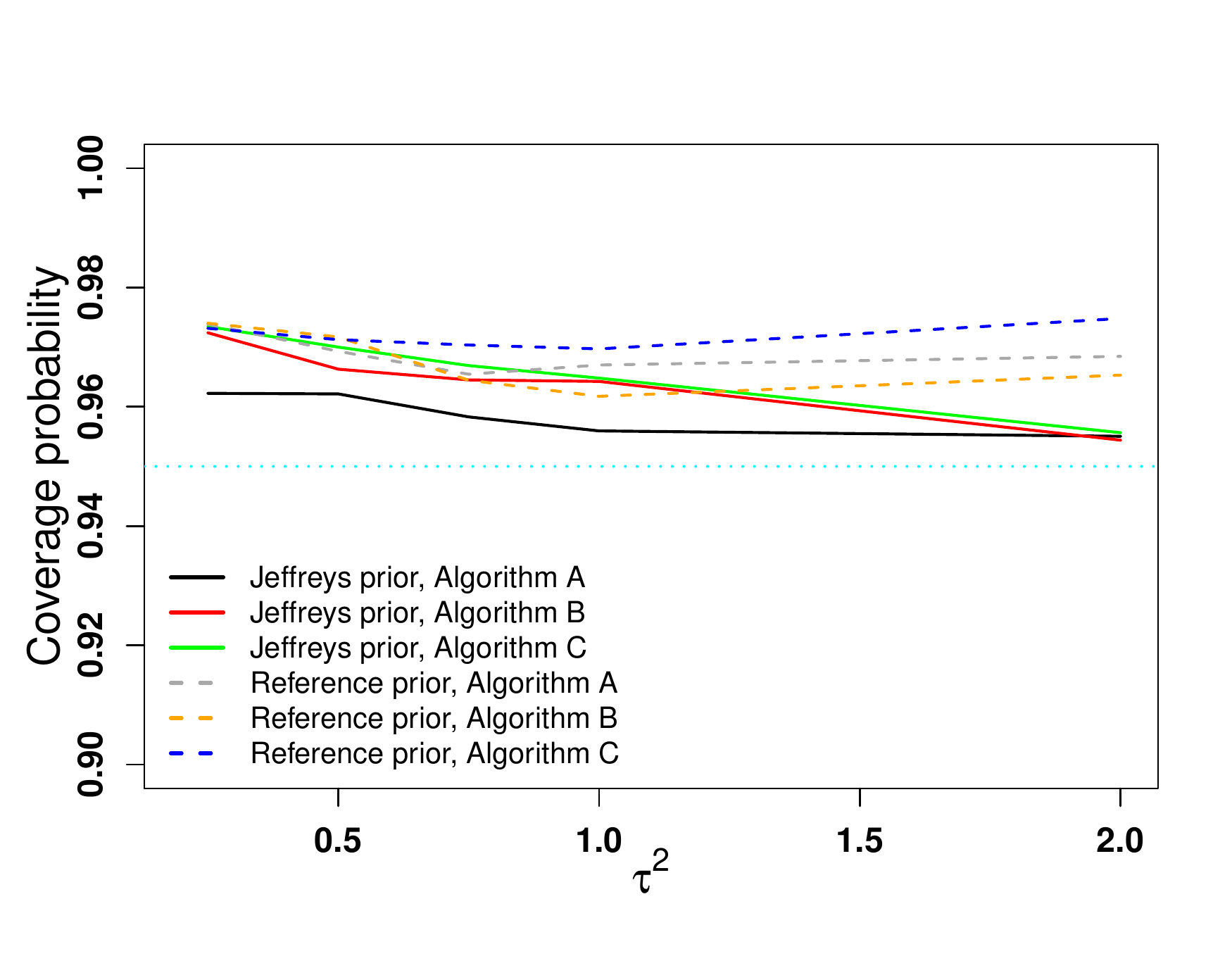}\\[-1cm]
\includegraphics[width=6.5cm]{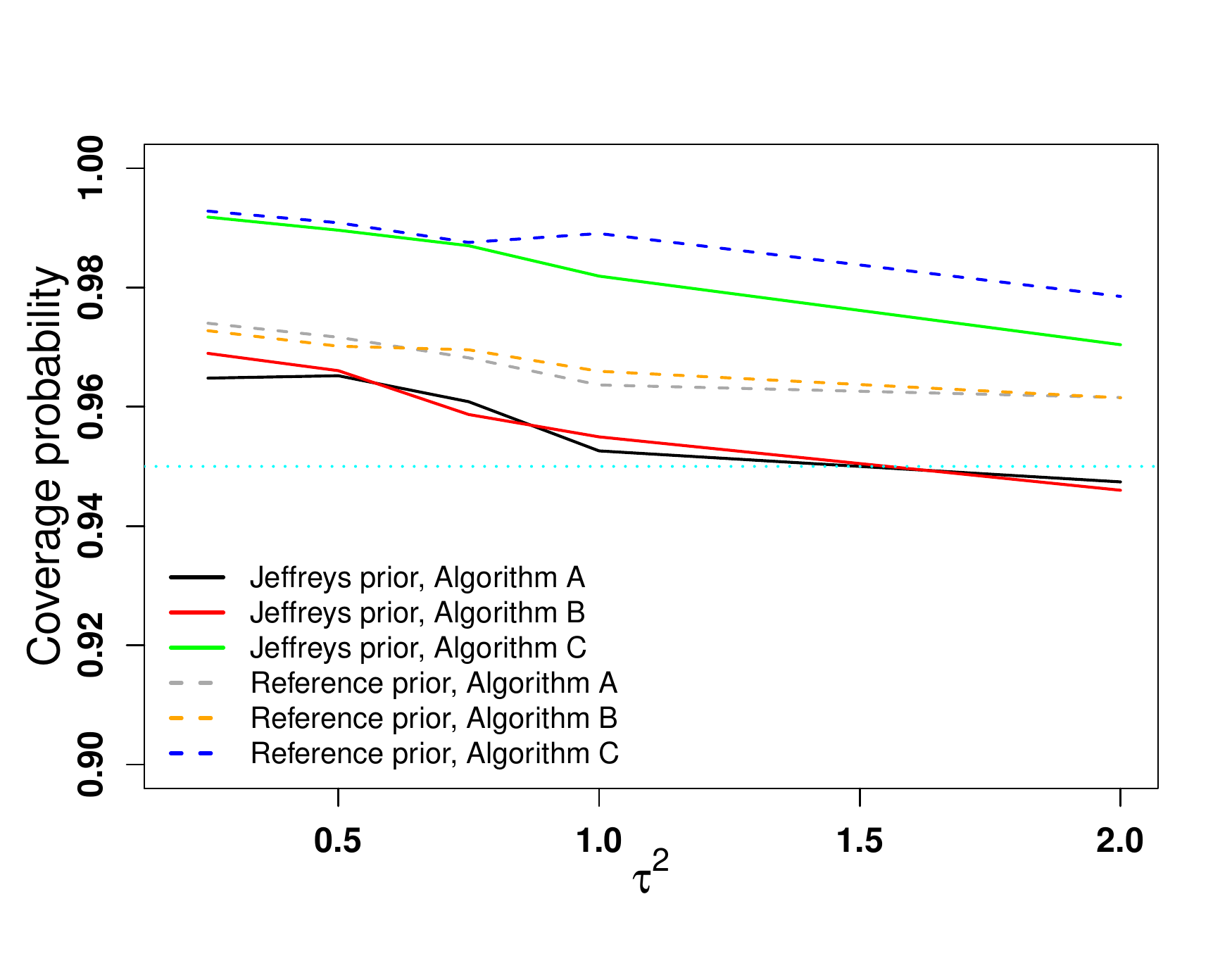}&\includegraphics[width=6.5cm]{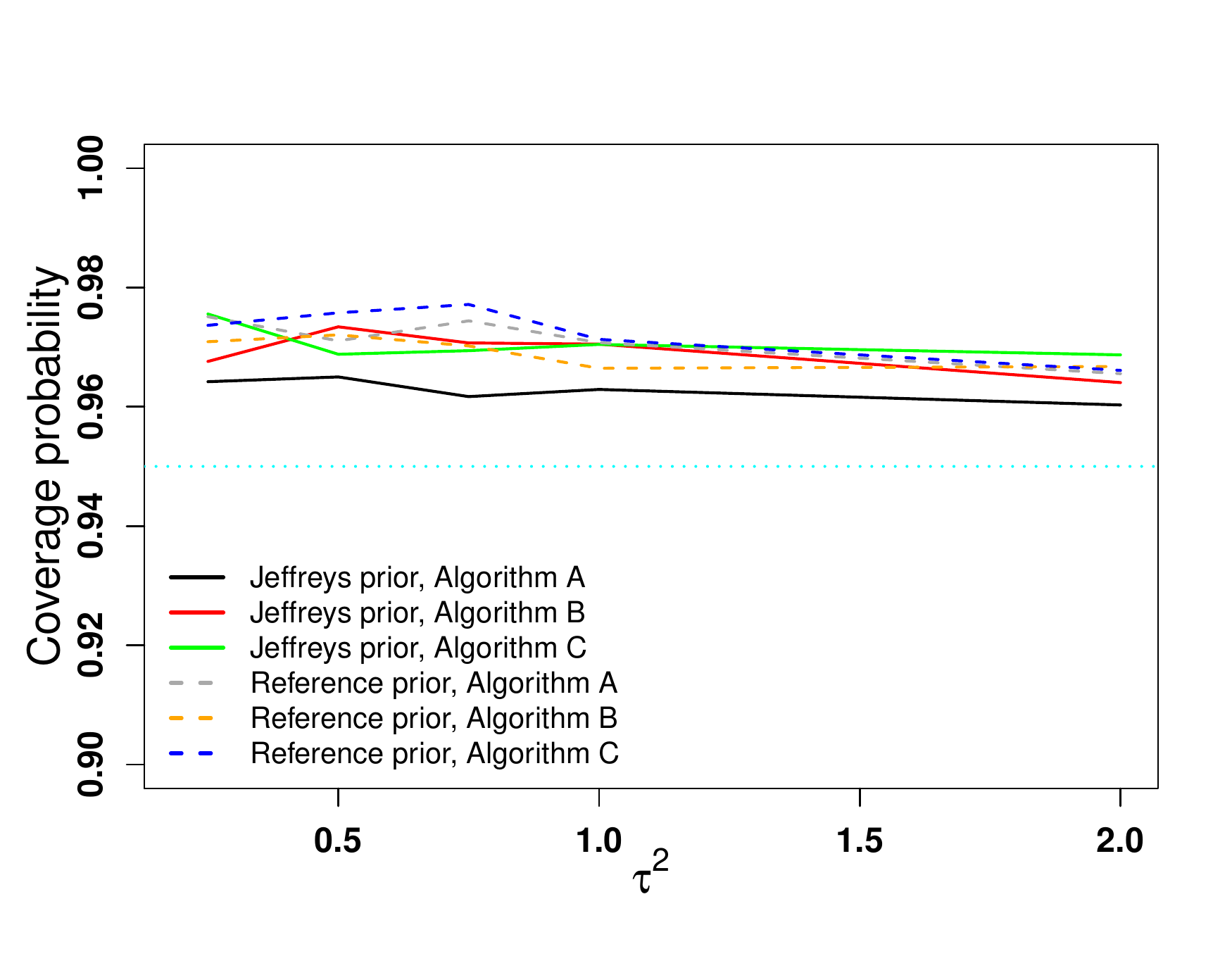}\\[-1cm]
\includegraphics[width=6.5cm]{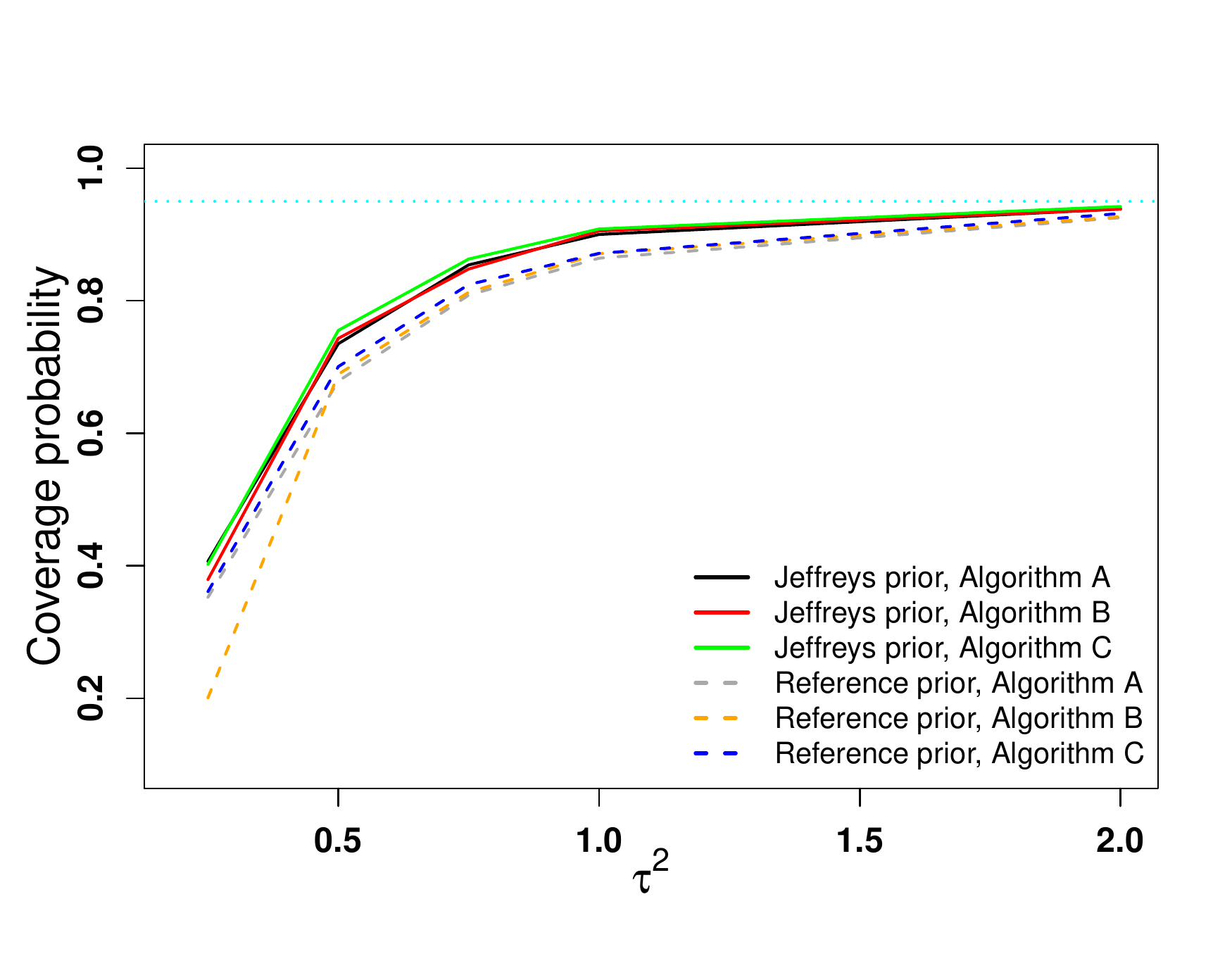}&\includegraphics[width=6.5cm]{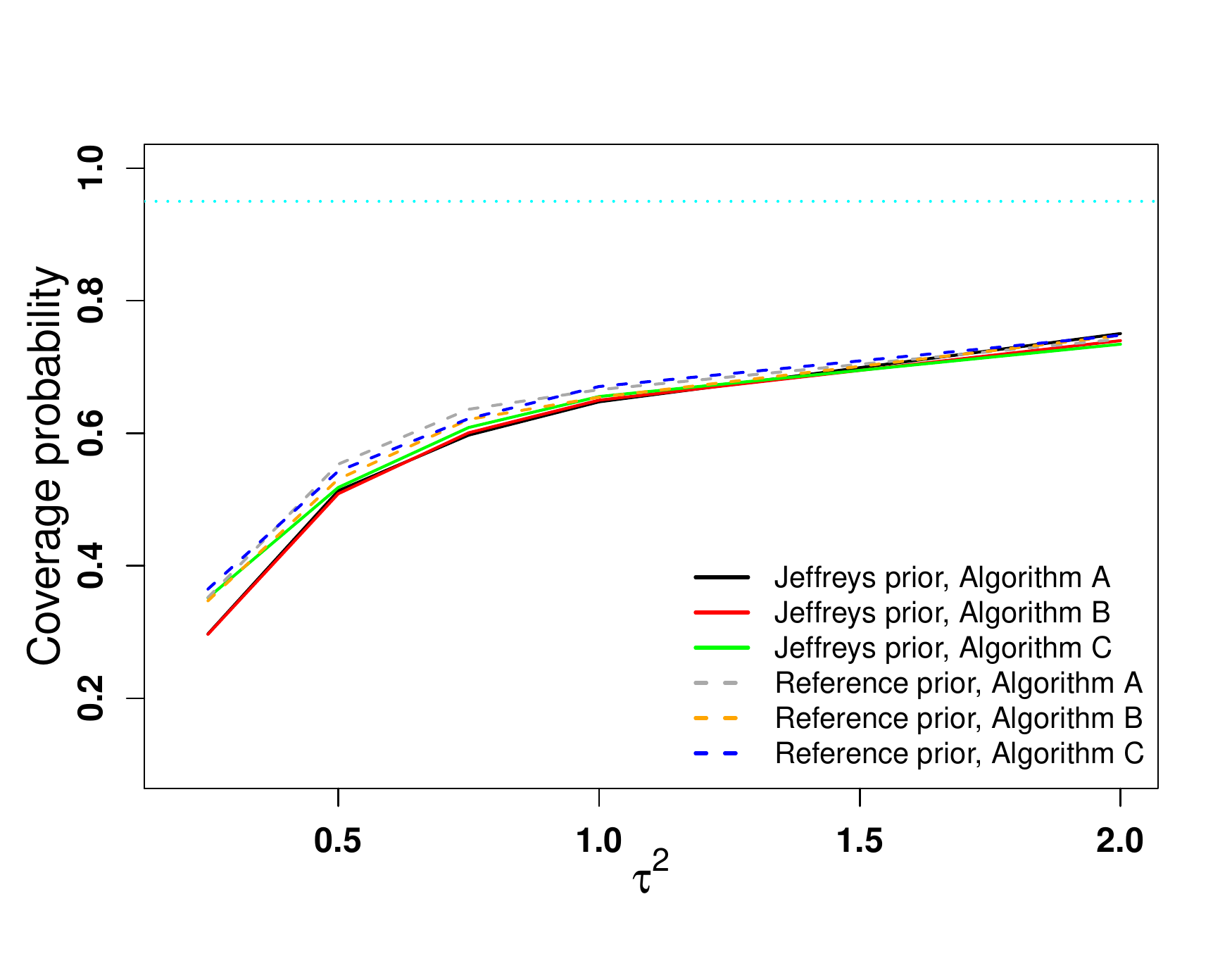}\\[-1cm]
\includegraphics[width=6.5cm]{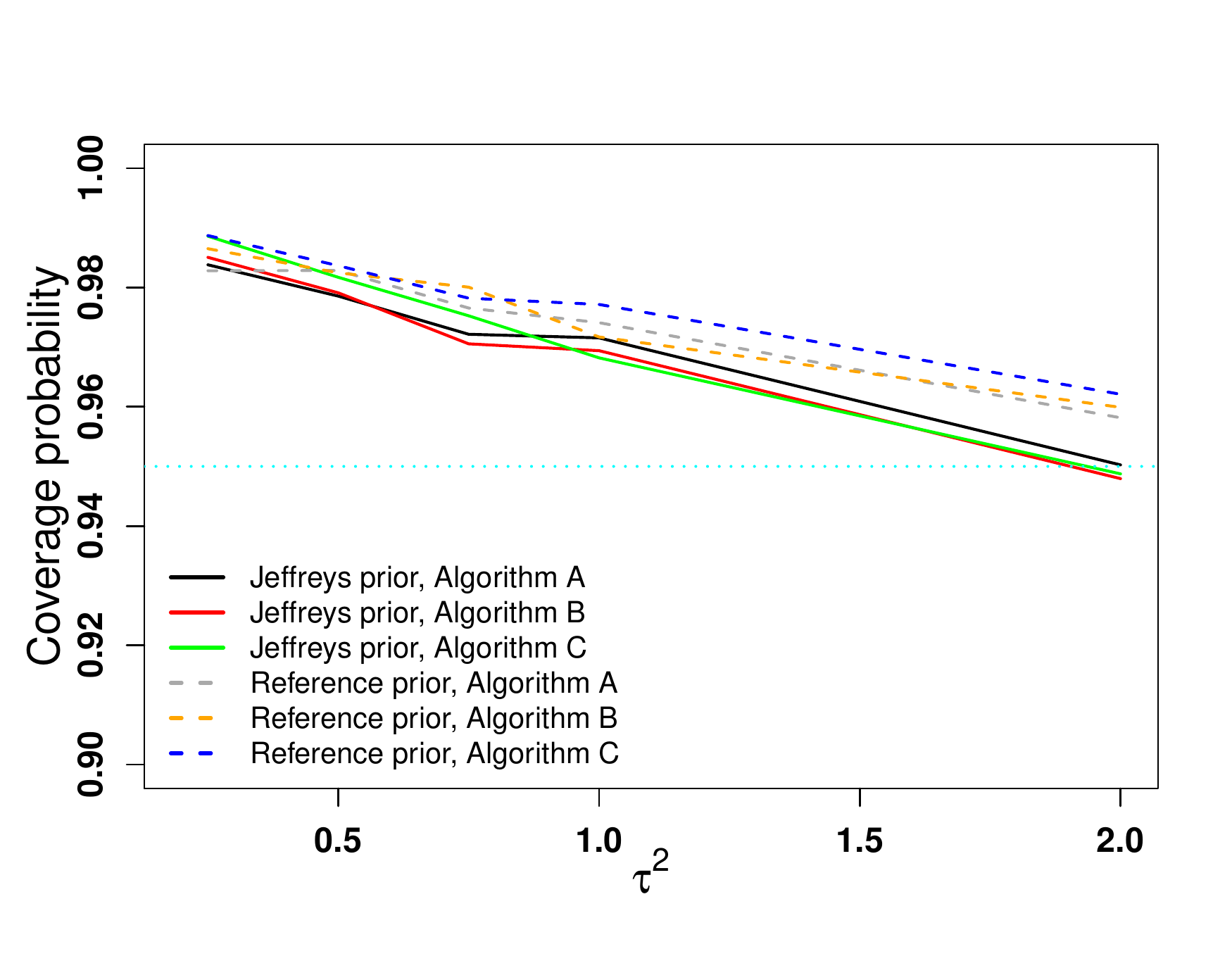}&\includegraphics[width=6.5cm]{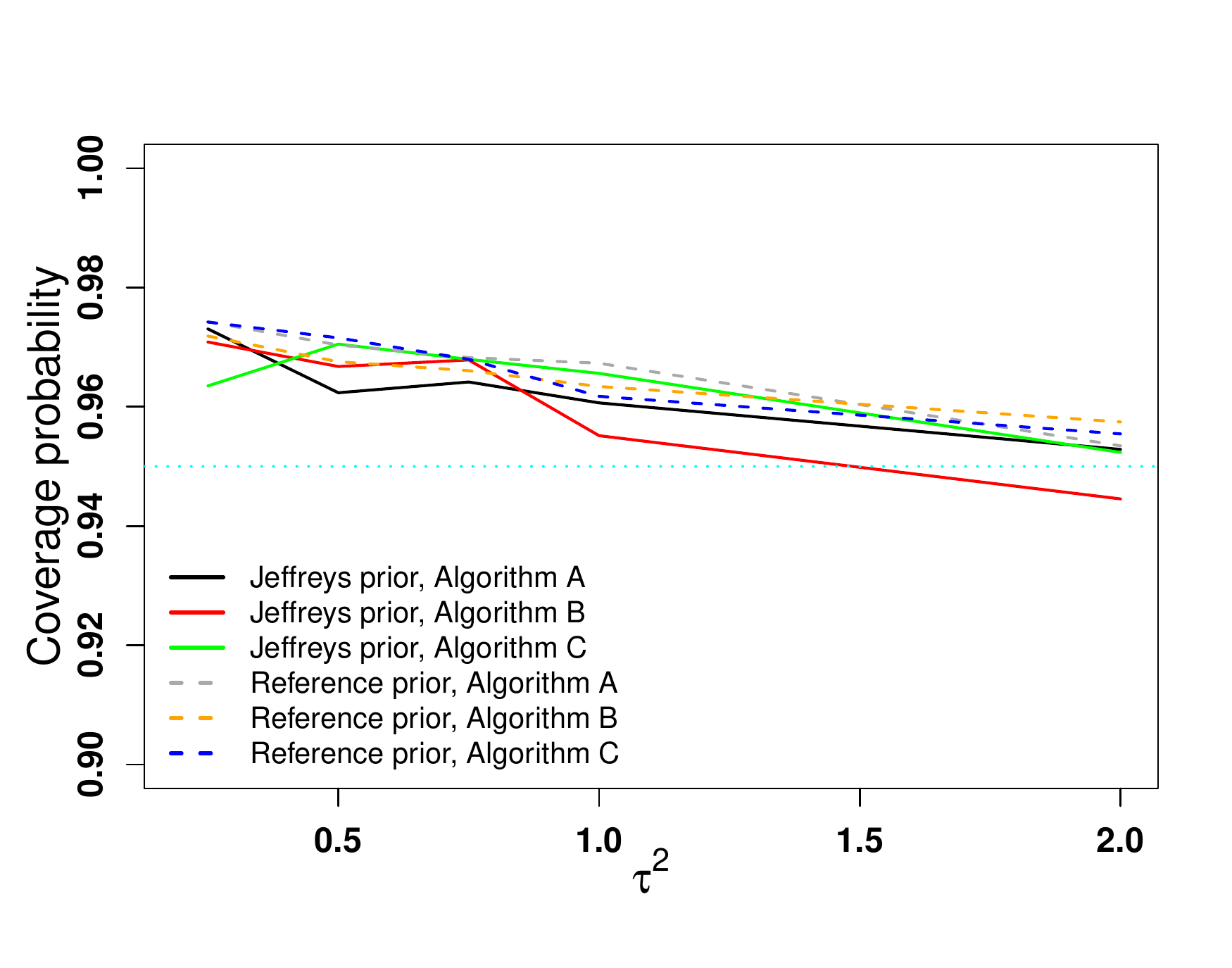}\\[-1cm]
\includegraphics[width=6.5cm]{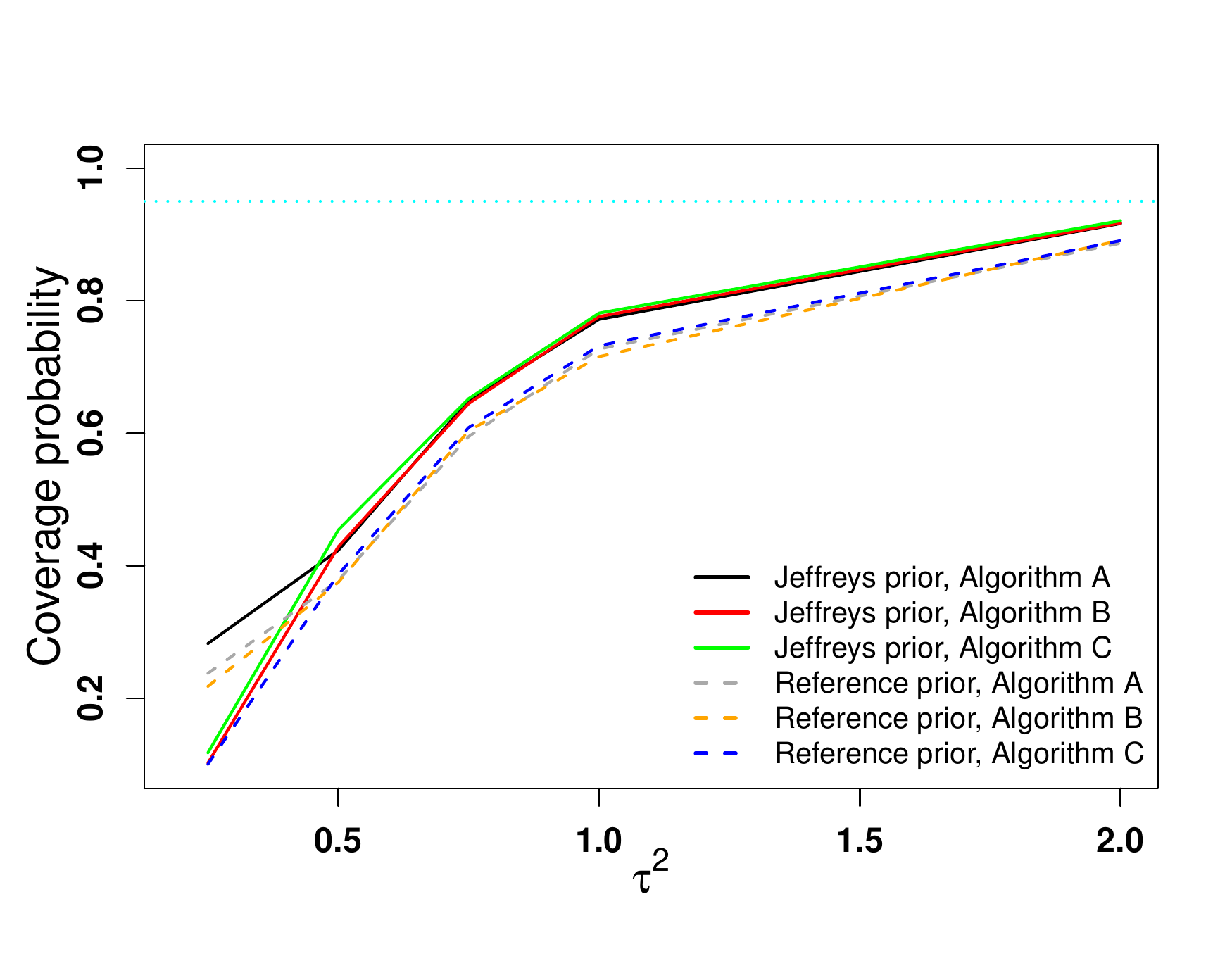}&\includegraphics[width=6.5cm]{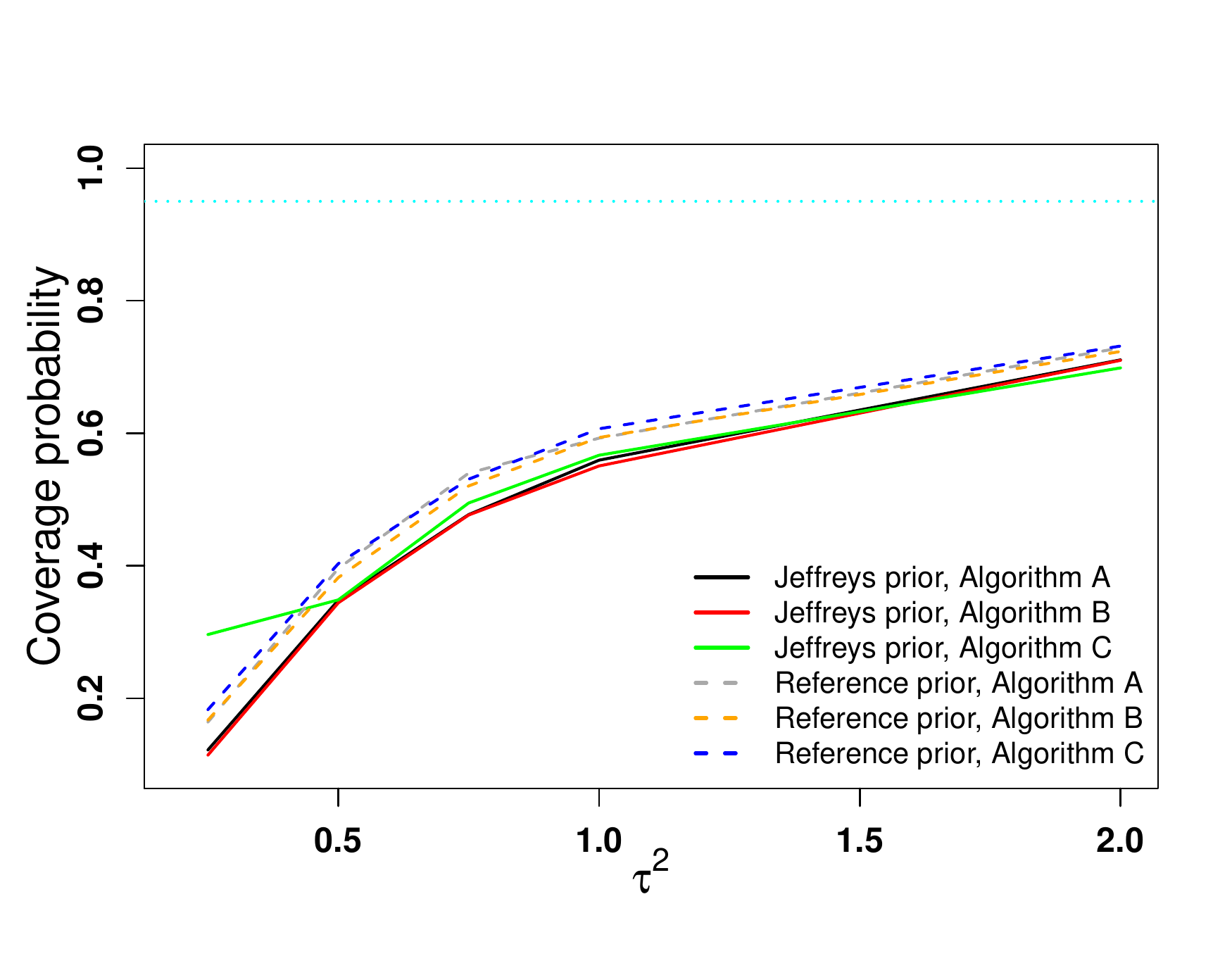}\\[-0.7cm]
\end{tabular}
 \caption{Empirical coverage probabilities of the $95\%$ probability symmetric univariate credible intervals determined for the parameters $\bmu$ (first and second rows) and $\bPsi$ (third to fifth rows) by employing the Berger and Bernardo reference prior and Jeffreys prior. The observations are drawn from the normal multivariate random effects model (left column) and the $t$ multivariate random effects model (right column) with $p=2$ and $n=10$.}
\label{fig:sim-study-cov-prob}
 \end{figure}

\begin{figure}[H]
\centering
\begin{tabular}{p{7.0cm}p{7.0cm}}
\includegraphics[width=6.5cm]{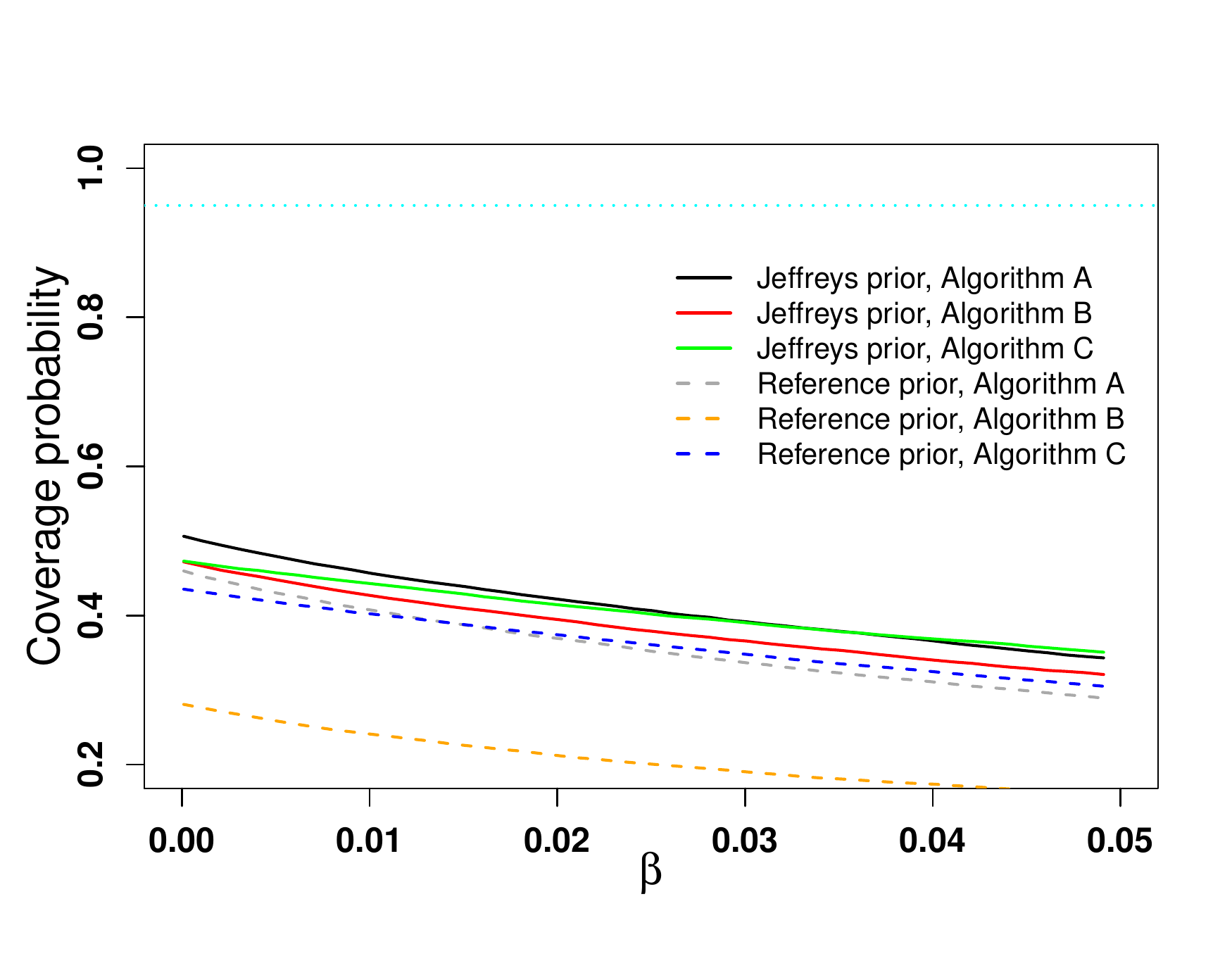}&\includegraphics[width=6.5cm]{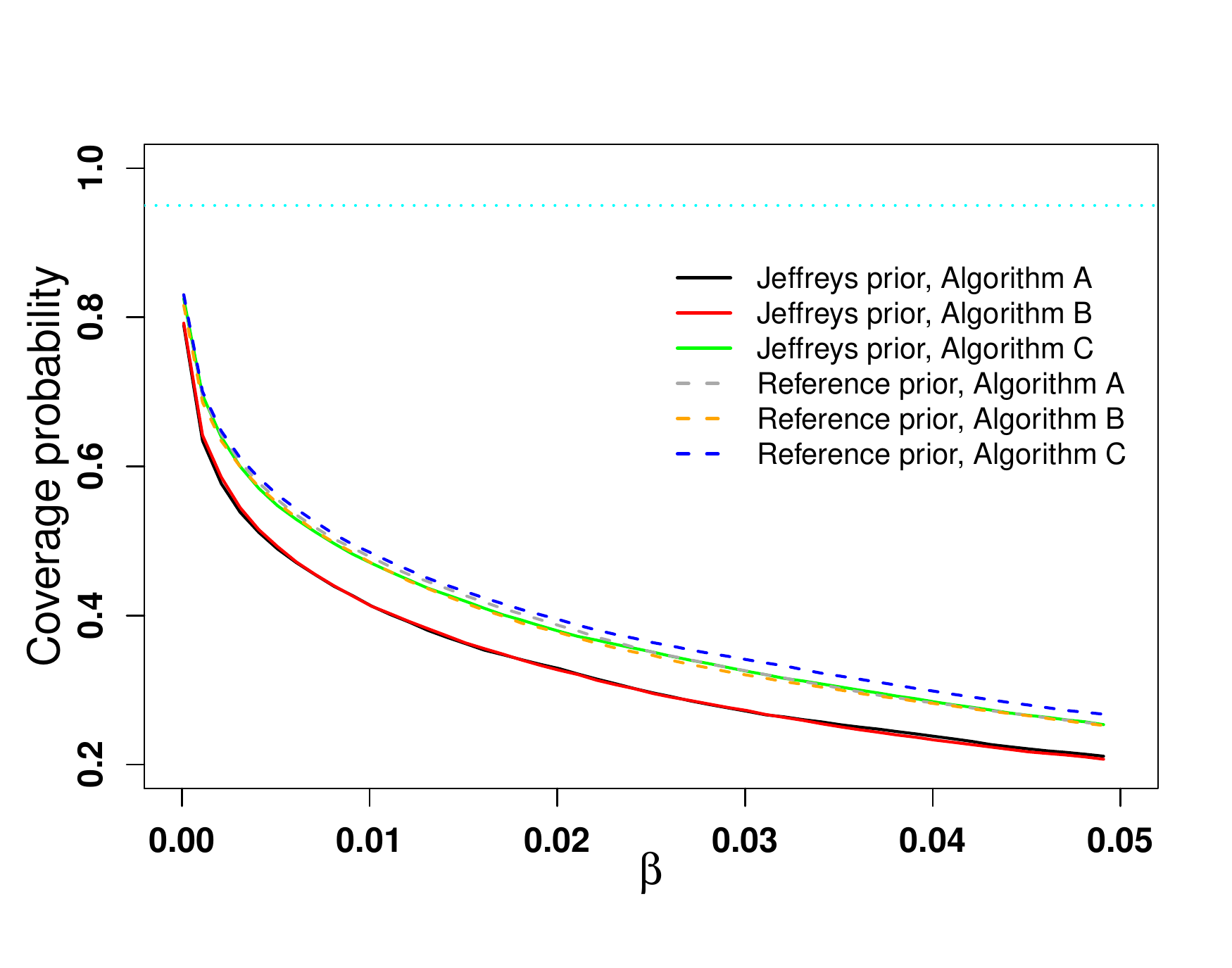}\\[-1cm]
\includegraphics[width=6.5cm]{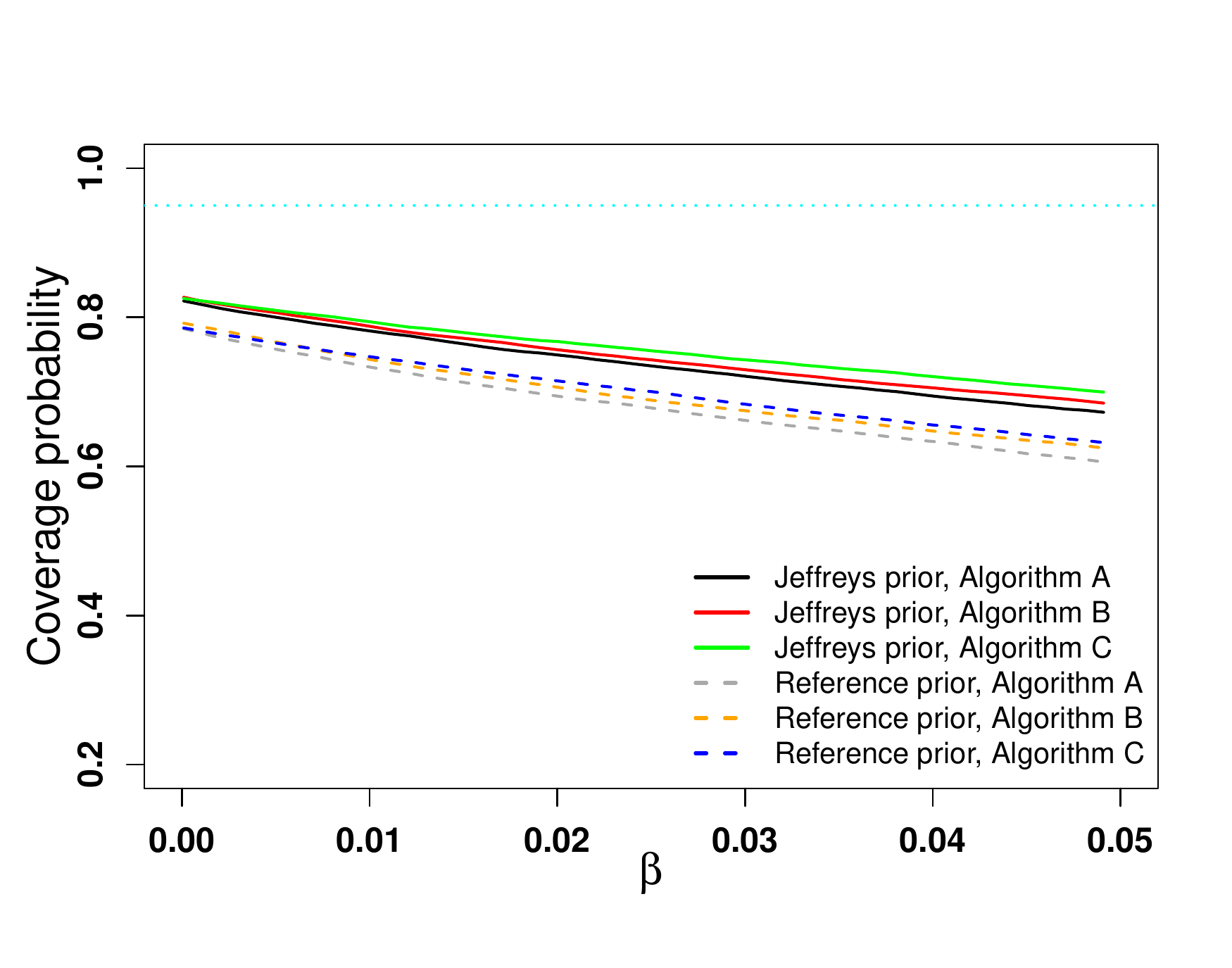}&\includegraphics[width=6.5cm]{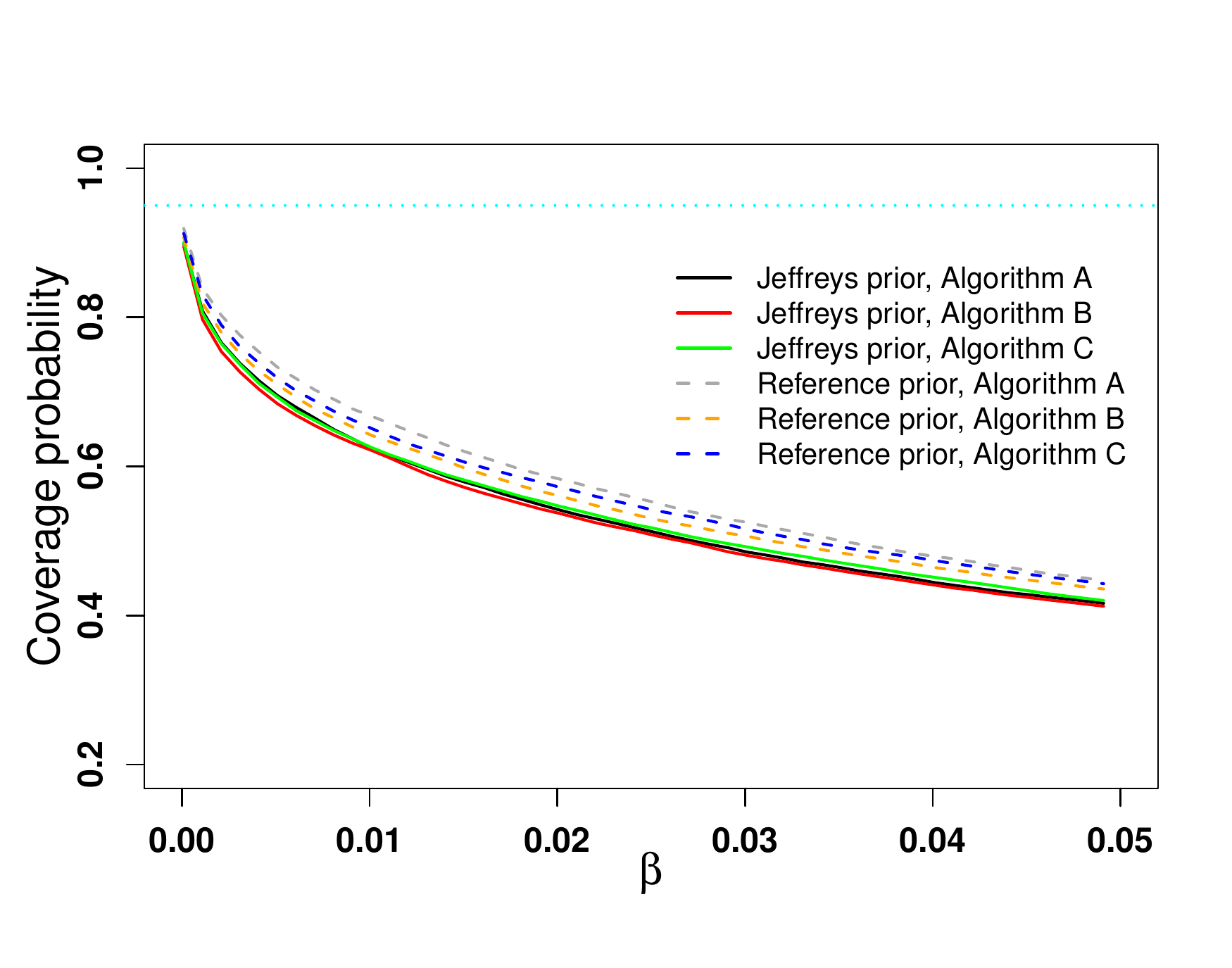}\\[-1cm]
\includegraphics[width=6.5cm]{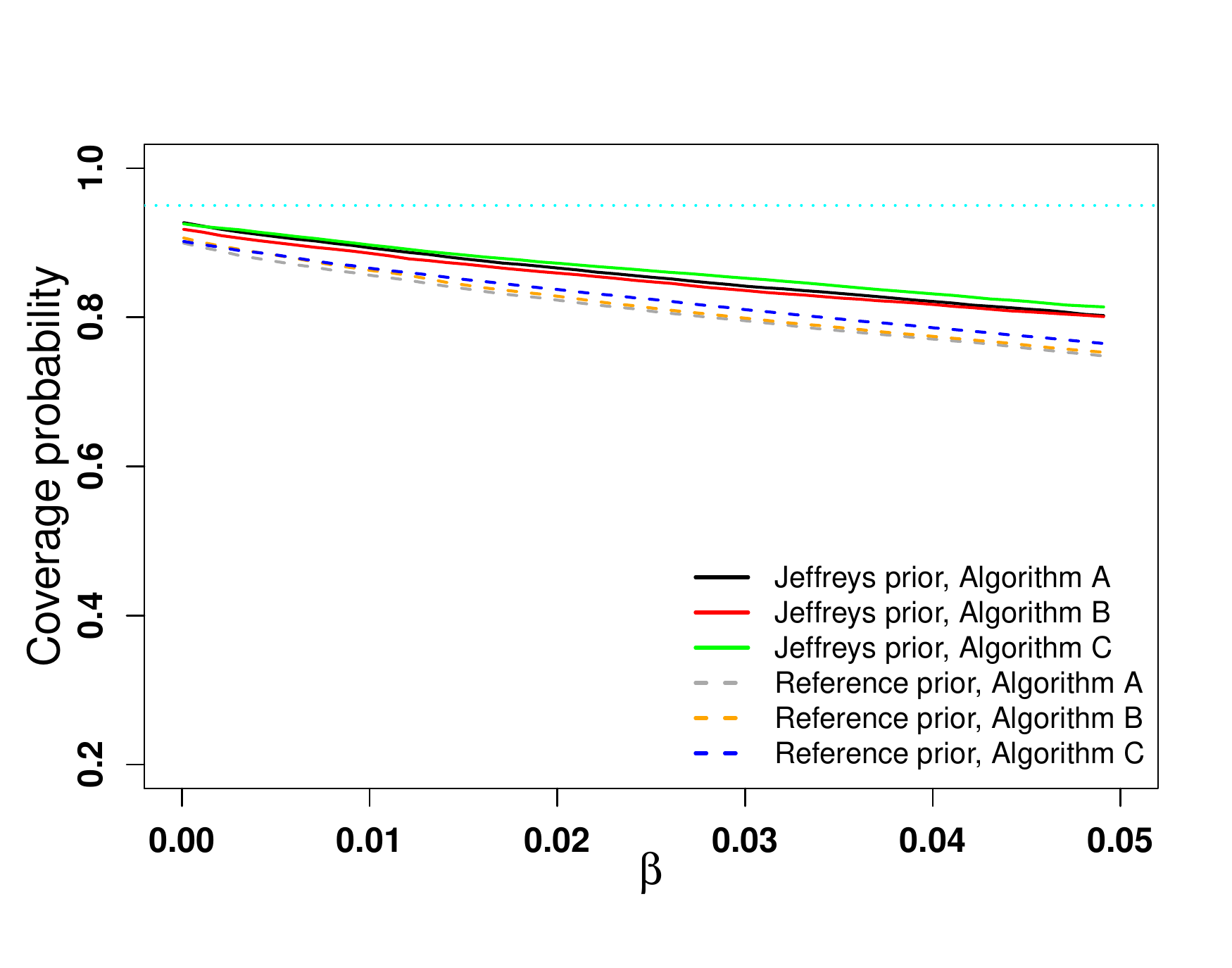}&\includegraphics[width=6.5cm]{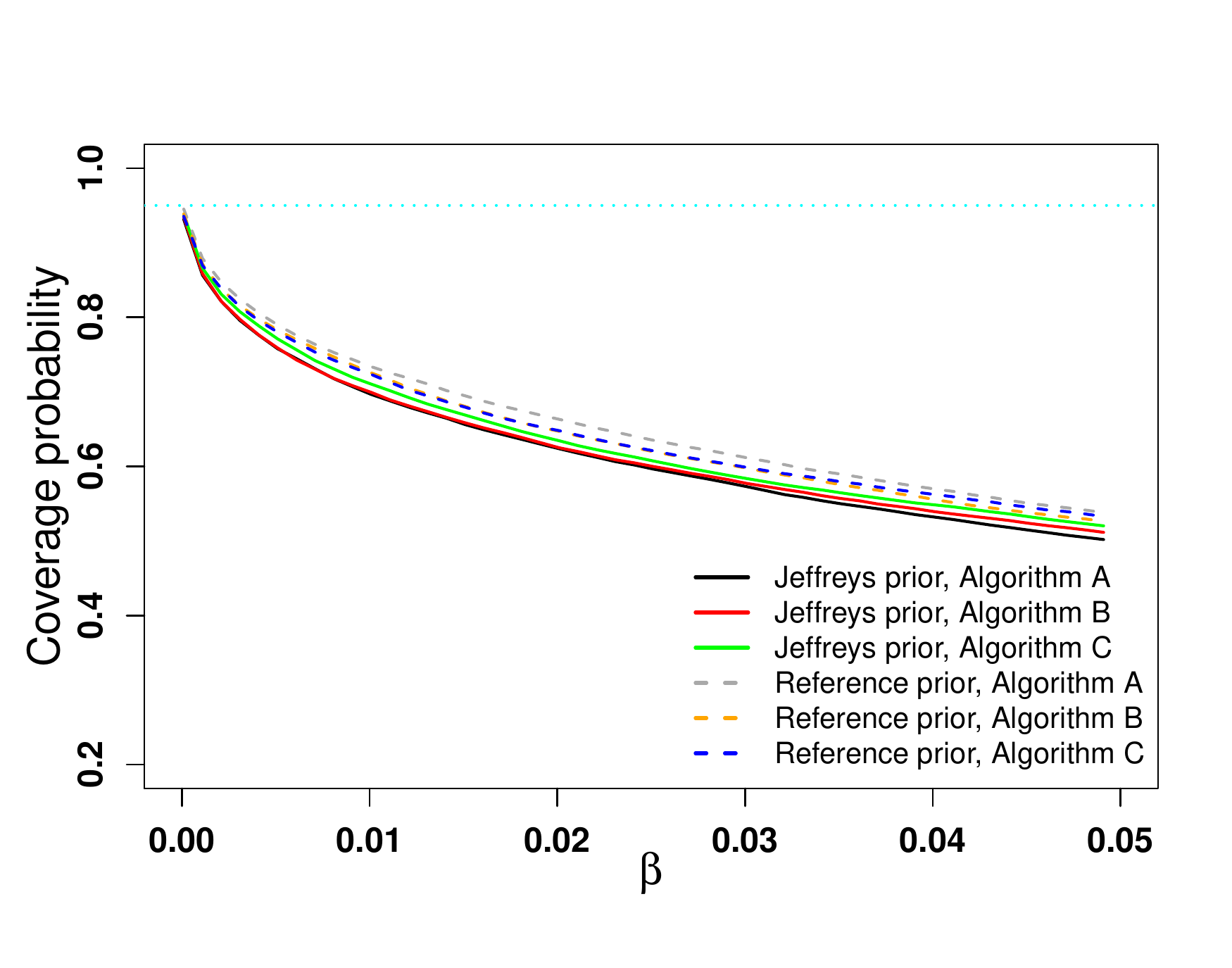}\\[-1cm]
\includegraphics[width=6.5cm]{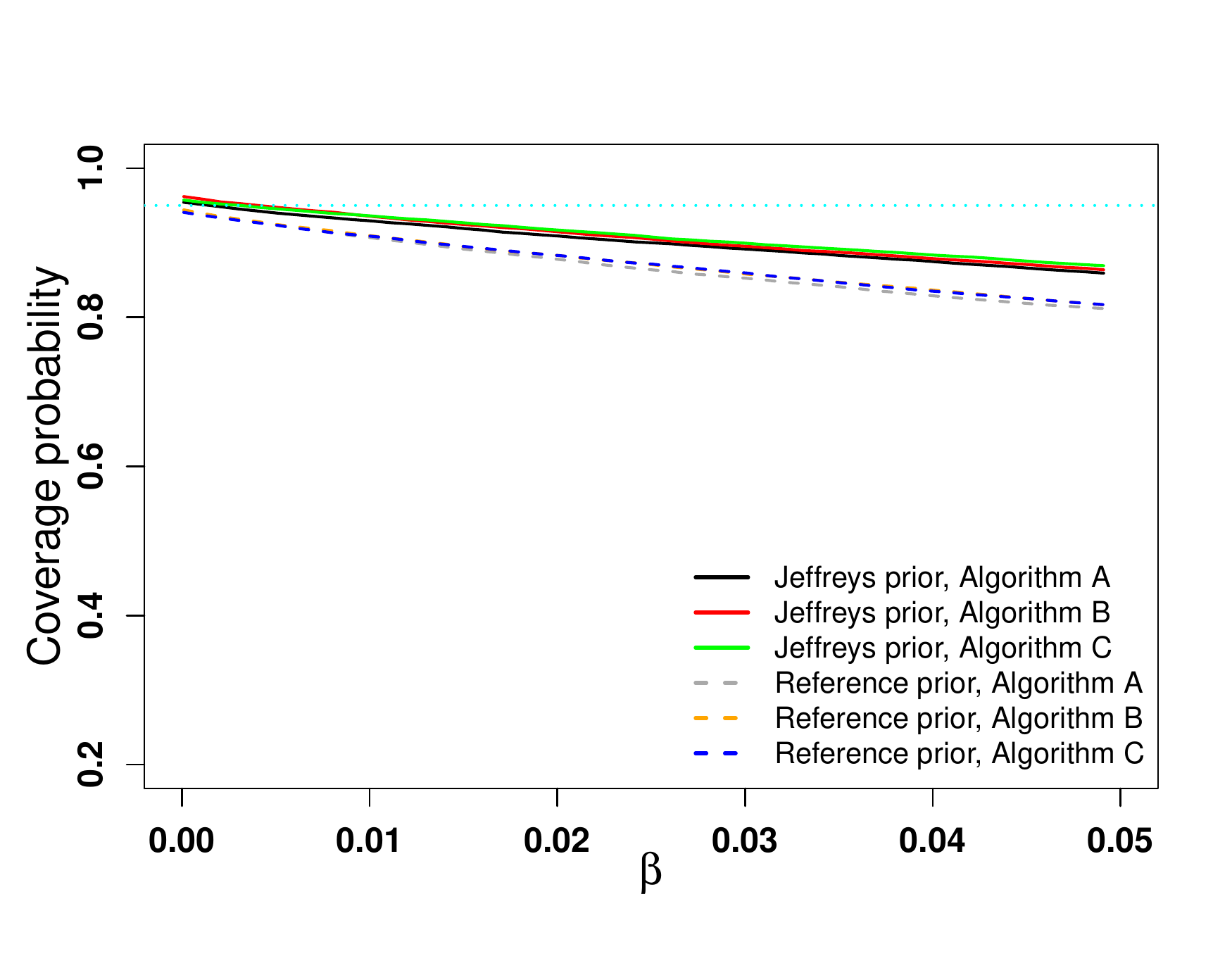}&\includegraphics[width=6.5cm]{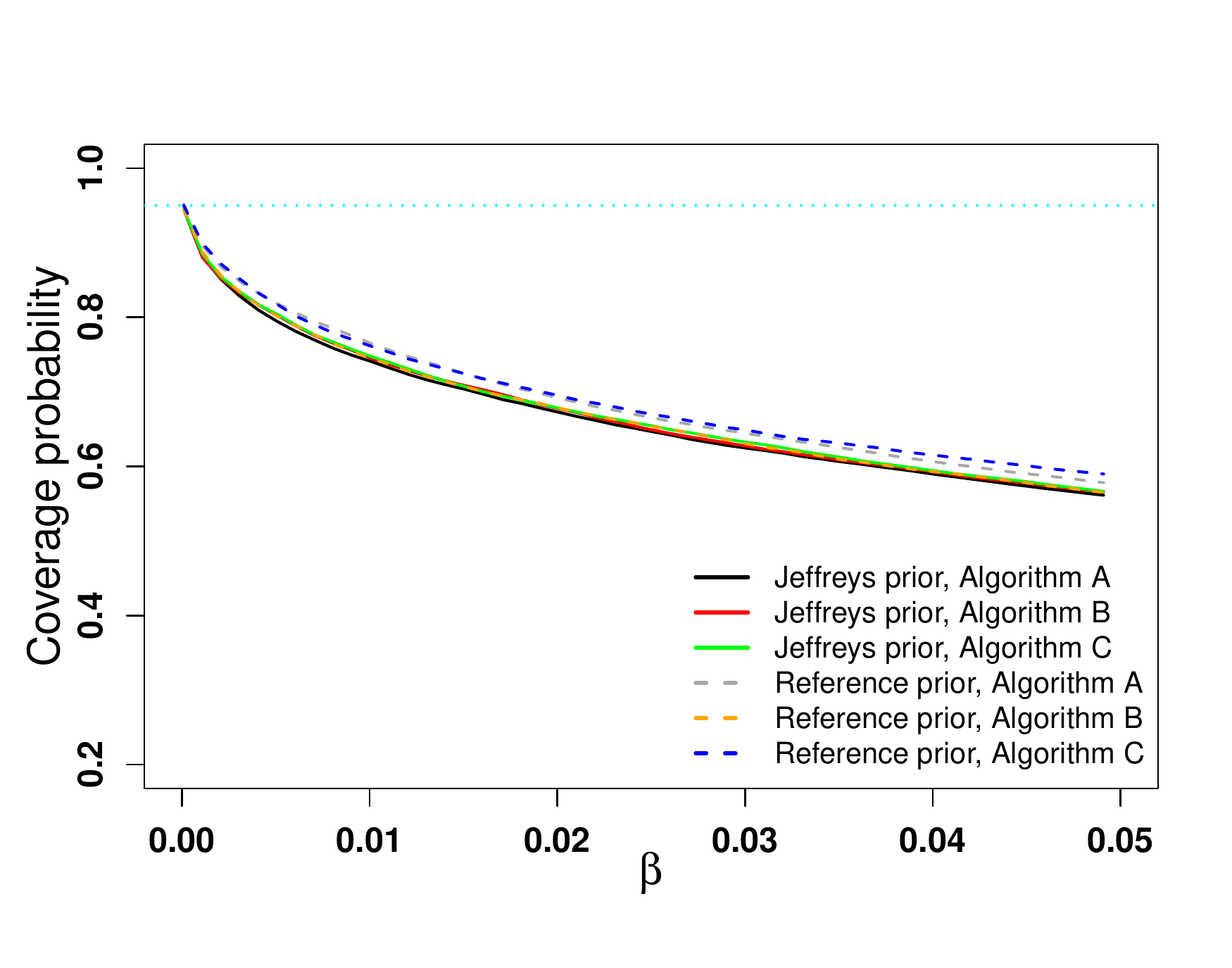}\\[-1cm]
\includegraphics[width=6.5cm]{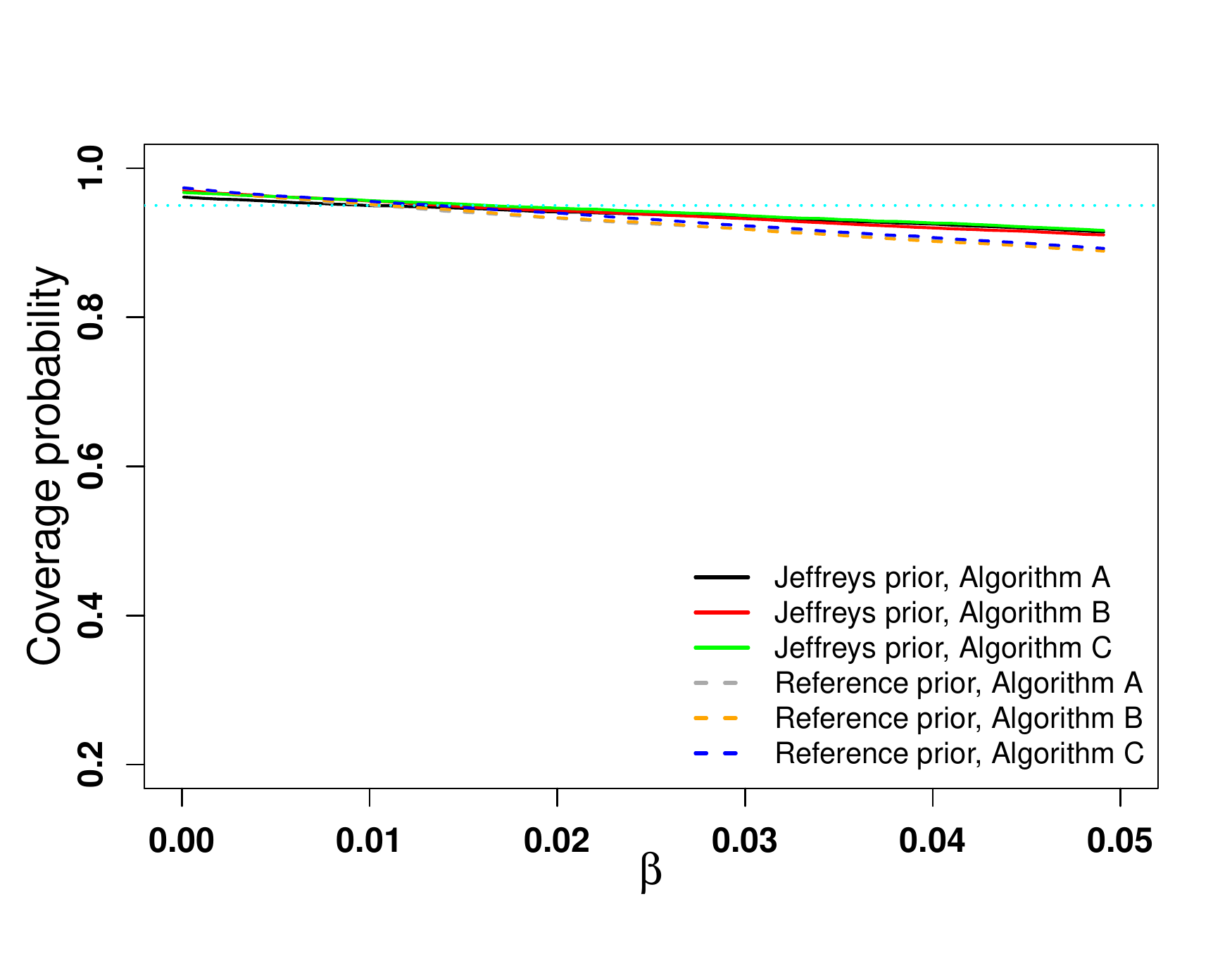}&\includegraphics[width=6.5cm]{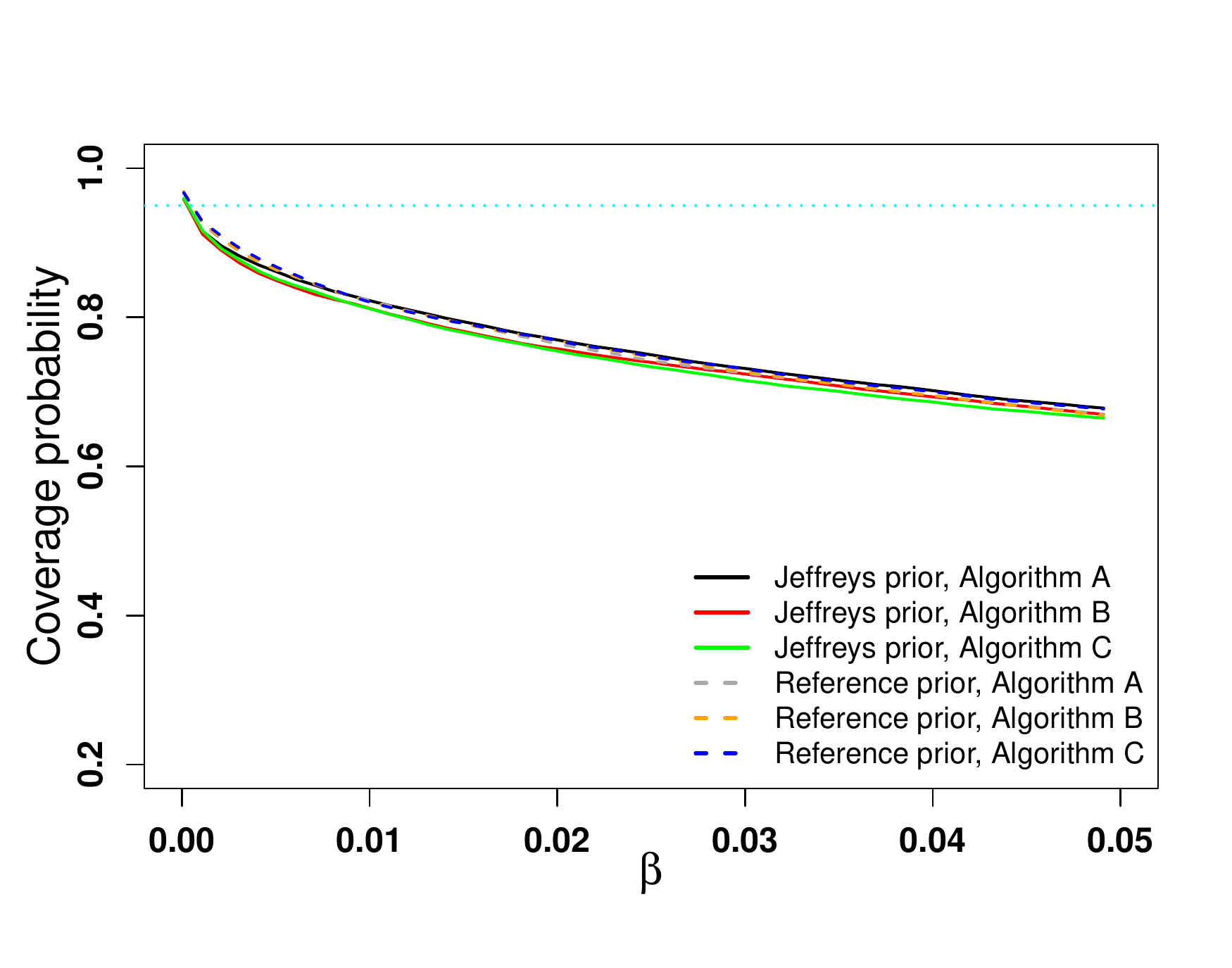}\\[-0.7cm]
\end{tabular}
 \caption{Empirical coverage probabilities of the $95\%$ credible intervals as function of $\beta$ determined for the parameter $\Psi_{11}$ by employing the Berger and Bernardo reference prior and Jeffreys prior. The observations are drawn from the normal multivariate random effects model (left column) and the $t$ multivariate random effects model (right column) with $p=2$, $n=10$ and $\tau^2=0.25$ (first row), $\tau^2=0.5$ (second row), $\tau^2=0.75$ (third row), $\tau^2=1$ (fourth row), $\tau^2=2$ (fifth row) .}
\label{fig:sim-study-cov-prob-beta}
 \end{figure}
 
Figure \ref{fig:sim-study-cov-prob} presents the empirical coverage probabilities of the probability symmetric credible intervals determined for the parameters of the multivariate random effects model. Despite the small sample size $n=10$, used in the construction of the credible intervals, the overall performance is good. Independently of the employed prior and the distributional assumption used in the data-generating model, the empirical coverage probabilities of the credible intervals for both components of the overall mean vector and the between-study covariance are always slightly above the desired level of $95\%$. Better results are present for the $t$ multivariate random effects model. Furthermore, the application of the Jeffreys prior leads the credible intervals whose coverage probabilities are smaller and, in general, closer to the target value.

The situation is different, when the credible intervals are determined for the between-study variances, the diagonal elements of $\bPsi$. Especially, when $\tau$ is small, then the empirical coverage probability is considerably smaller than the desired level of $95\%$. These finding are in line with the those documented in the univariate case. It is known that Bayesian procedures usually tend to overestimate the between-study variance, especially when the true between-study variance is small (see, e.g., \cite{bodnar2017bayesian}). Possible explanations of this effect could be the observation that the posterior of between-study variance is usually skewed to the right and it is heavy-tailed. Moreover, since the posterior is skewed to the right, the symmetric credible intervals may not be a good choice in such a situation.

The issues with the credible intervals constructed for the between-study variances are further investigated in Figure \ref{fig:sim-study-cov-prob-beta}, where the empirical coverage probabilities are reported for the credible intervals of the form $[q_{\beta},q_{1-\alpha+\beta}]$ for $\beta \in [0.0001,0.05]$ with $q_{\beta}$ being the $\beta$-quantile of the posterior distribution. When $\beta=\alpha/2$, the probability symmetric credible intervals are obtained. We observe that the coverage probabilities decrease as $\beta$ increases, independently of the chosen prior, employed algorithm or distributional assumption used in the specification of the multivariate random effects model. The best results are achieved for $\beta$ being close to its lower bound. These findings are again in line with the results obtained in the univariate case, namely that the Bayesian procedures tend to overestimate the true value of the between-study variance. Interestingly, the application of the Berger and Bernardo reference prior leads to smaller values of the coverage probability under the normal multivariate random effects model, while the reverse relationship is present when the data are drawn from the $t$ multivariate random effects model. 

\section{Empirical illustration}\label{sec:emp}

In this section we apply the three numerical algorithms to draw samples from the posterior distribution of $\bmu$ and $\bPsi$ to real data, which consist of results documented in ten studies designed to investigate the effectiveness of hypertension treatment for reducing blood pressure. The treatment effects on both the systolic blood pressure and diastolic blood pressure were analyzed where the negative values document beneficial effect of the treatment. The results of ten studies together with the reported within-study covariance matrices are provided in \citet{jackson2013matrix} and they are also summarized in Table \ref{tab:data}.

\begin{table}[h!t]
\centering
\begin{tabular}{c|c|c|c|c|c}
  \hline \hline
 Study & $X_{i;1}$ (SBP) & $X_{i;2}$ (DBP) & $\sqrt{U_{i;11}}$ (SBP) & $\rho_{i;12}=\dfrac{U_{i;12}}{\sqrt{U_{i;11}U_{i;22}}}$ & $\sqrt{U_{i;22}}$ (DBP)\\
\hline
 1&  -6.66& -2.99& 0.72 & 0.78& 0.27 \\
 2& -14.17& -7.87& 4.73 & 0.45& 1.44 \\
 3& -12.88& -6.01&10.31 & 0.59& 1.77 \\
 4& -8.71 & -5.11& 0.30 & 0.77& 0.10 \\
 5& -8.70 & -4.64& 0.14 & 0.66& 0.05 \\
 6& -10.60& -5.56& 0.58 & 0.49& 0.18 \\
 7& -11.36& -3.98& 0.30 & 0.50& 0.27 \\
 8& -17.93& -6.54& 5.82 & 0.61& 1.31 \\
 9& -6.55 & -2.08& 0.41 & 0.45& 0.11 \\
10& -10.26& -3.49& 0.20 & 0.51& 0.04 \\
\hline\hline
\end{tabular}
\caption{Data from 10 studies about the effectiveness of hypertension treatment with the aim to reduce blood pressure. The variables $X_{i;1}$ and $X_{i;2}$ denote the treatment effects on the systolic blood pressure (SBP) and the diastolic blood pressures (DBP) from the $i$th study, while $\mathbf{U}_i=(U_{i;lj})_{lj=1,2}$ is the corresponding within-study covariance matrix.}
\label{tab:data}
\end{table}

Bayesian multivariate meta-analysis is conducted for the data from Table \ref{tab:data} by assuming the normal multivariate random effects model and the $t$-multivariate random effects model and by assigning the Berger and Bernardo reference prior and the Jeffreys prior to the model parameters. Using the two the two Metropolis-Hastings algorithms (Algorithm A and Algorithm B) of \cite{bodnar2023objective} and the Gibbs sampler (Algorithm C) introduced in Section \ref{sec:gibbs}, four Markov chains are constructed of length $10^5$ with the burn-in period of the same length for each model assumption and the employed prior. 

\begin{figure}[h!t]
\centering
\begin{tabular}{p{4.0cm}p{4.0cm}p{4.0cm}p{4.0cm}}
\hspace{0.0cm}\includegraphics[width=4.0cm]{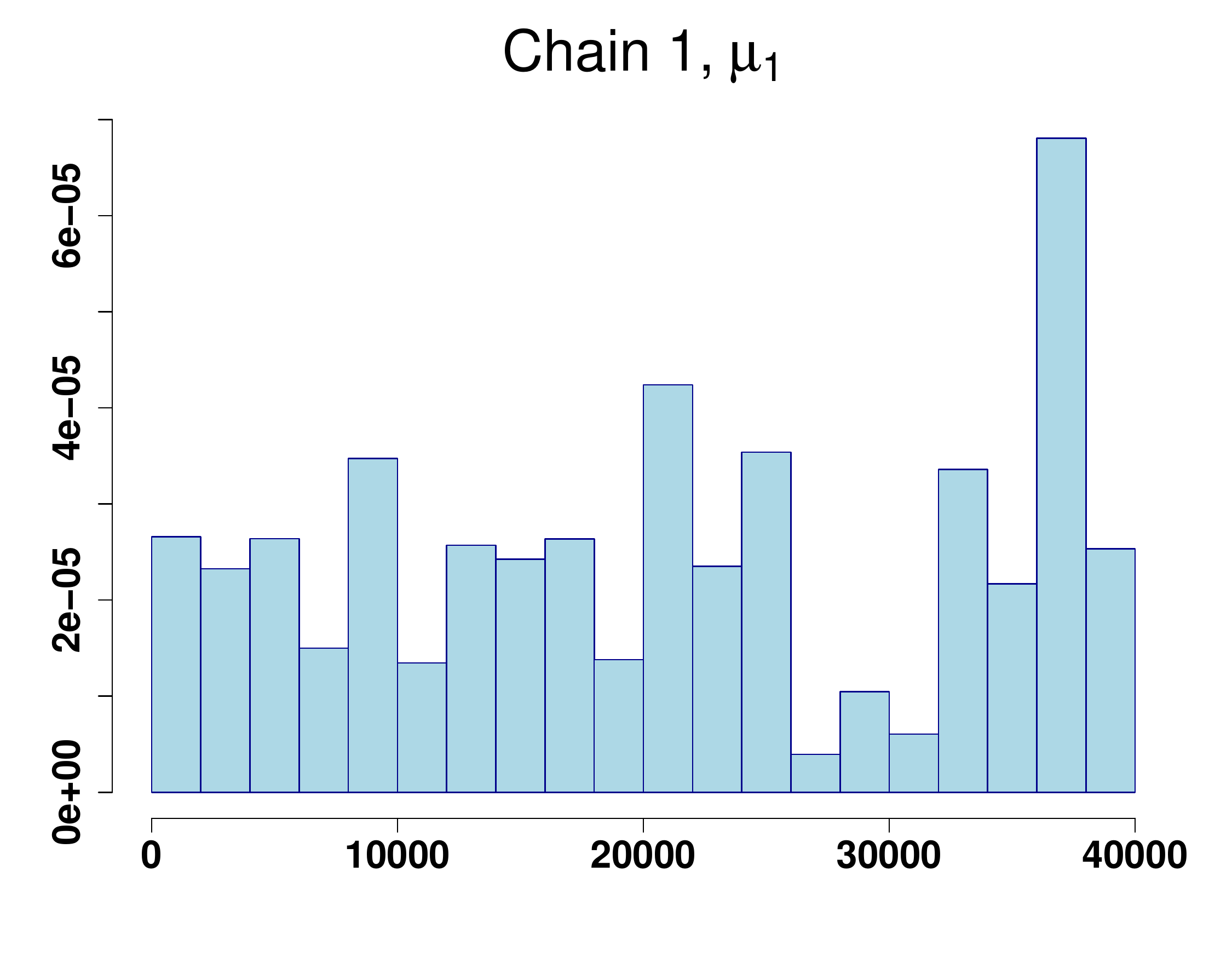}&\hspace{-0.5cm}\includegraphics[width=4.0cm]{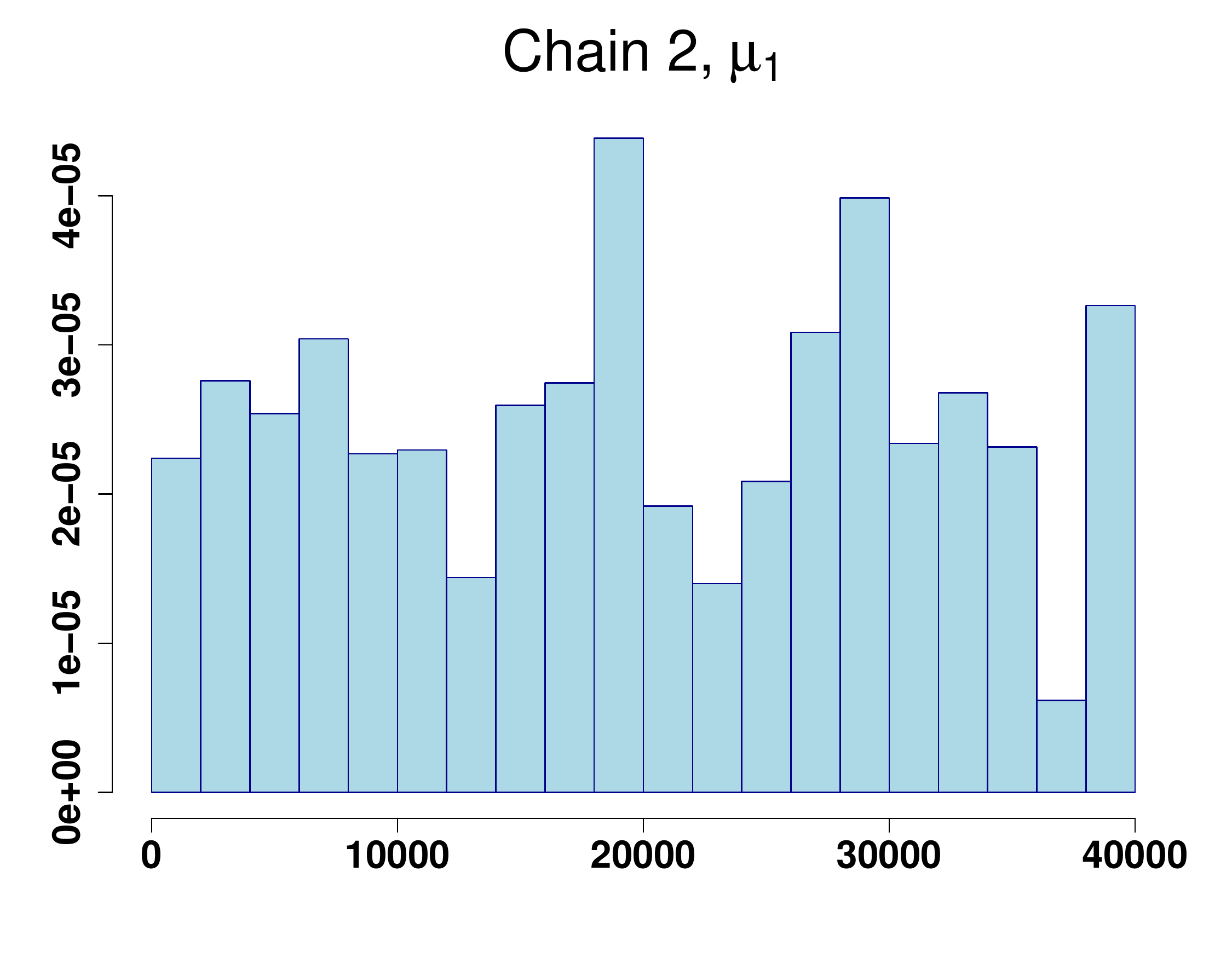}&
\hspace{-1.0cm}\includegraphics[width=4.0cm]{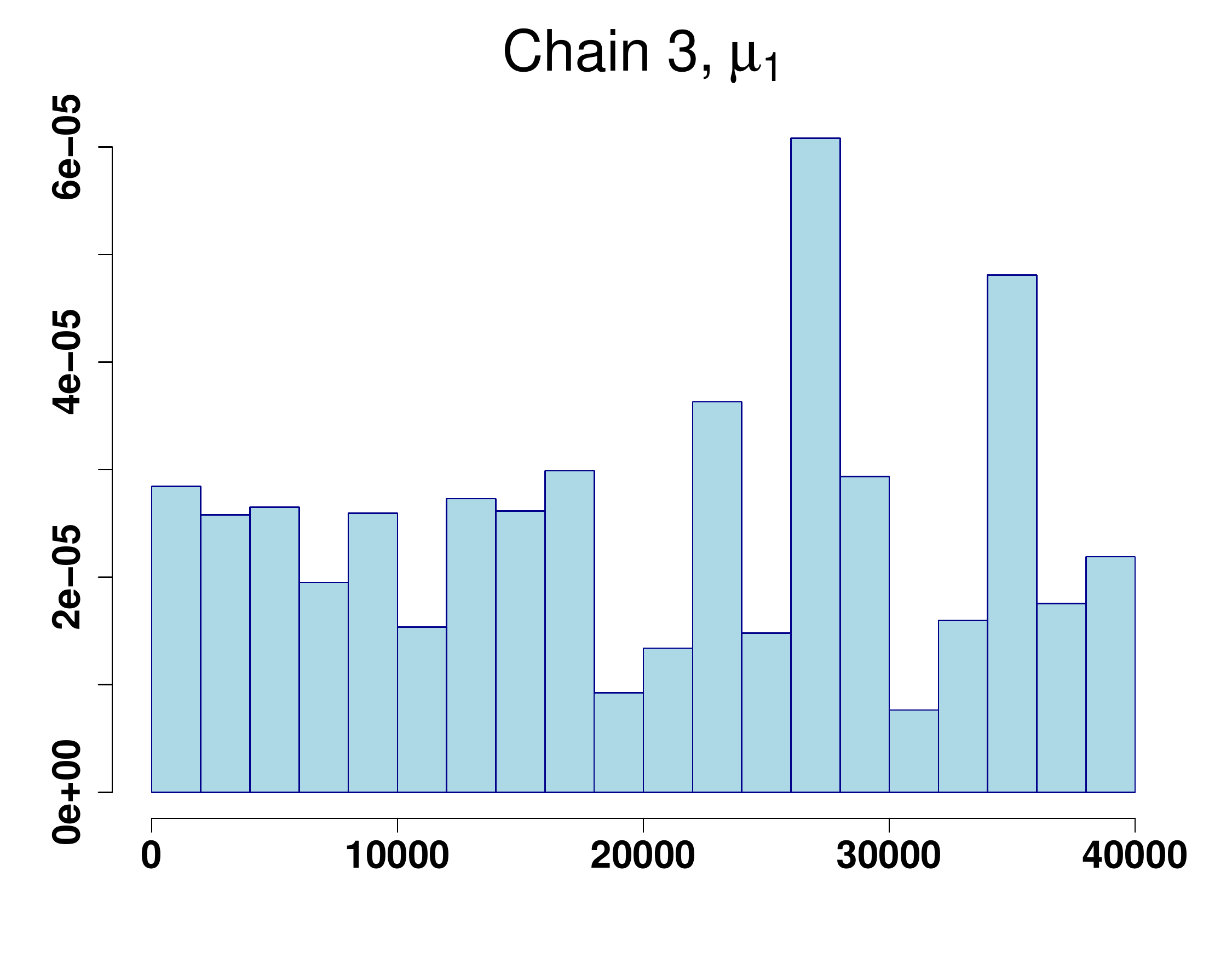}&\hspace{-1.5cm}\includegraphics[width=4.0cm]{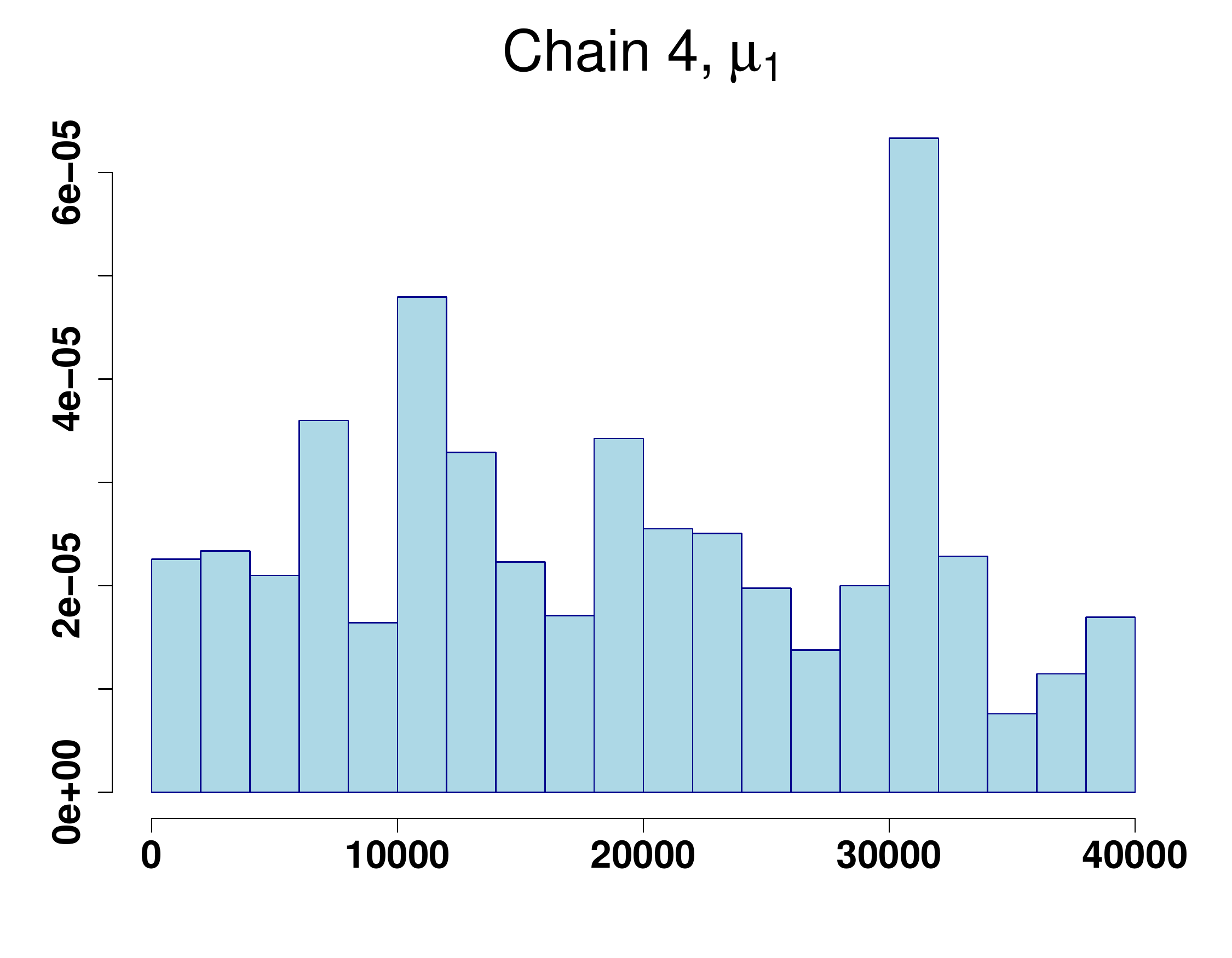}\\
\hspace{0.0cm}\includegraphics[width=4.0cm]{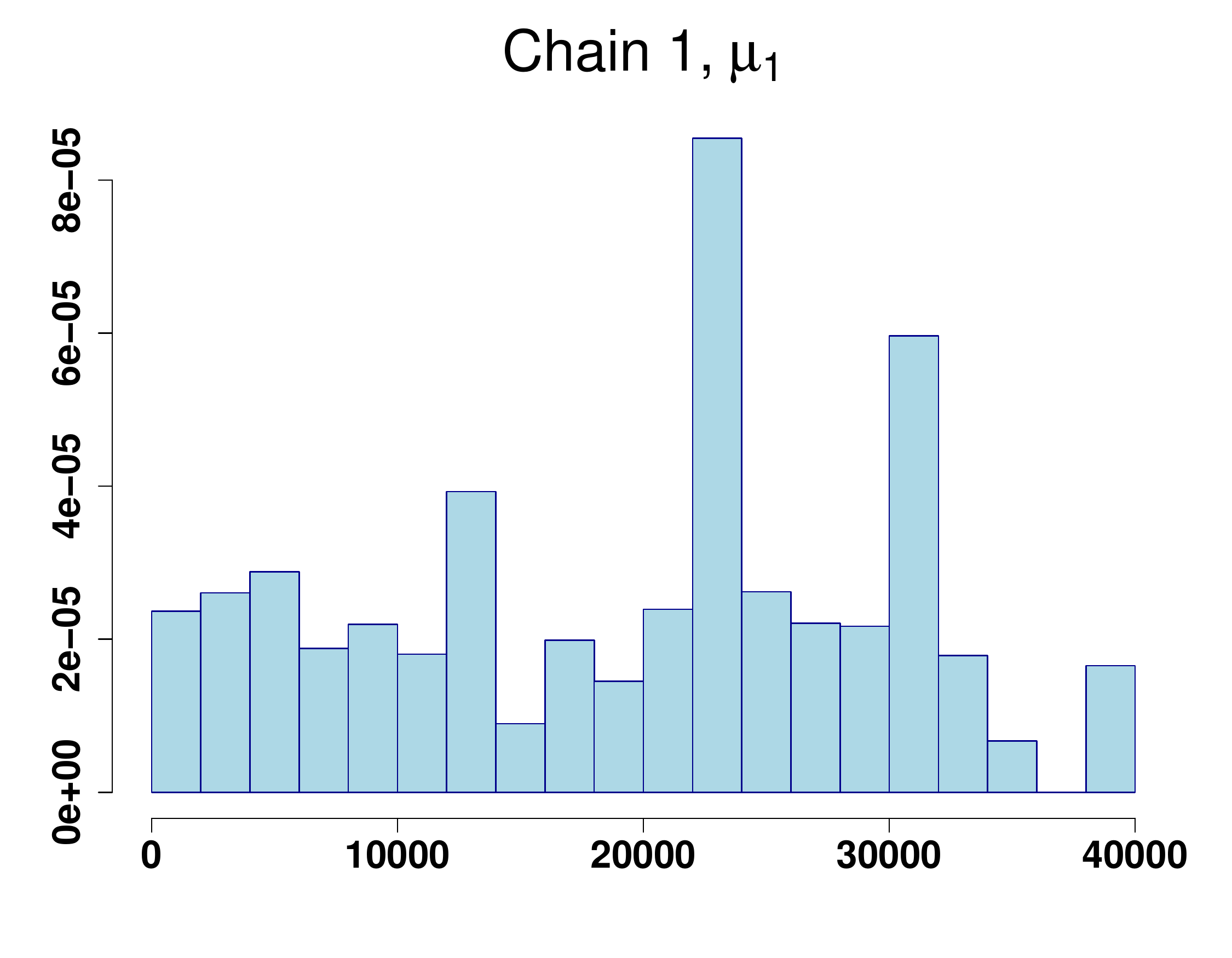}&\hspace{-0.5cm}\includegraphics[width=4.0cm]{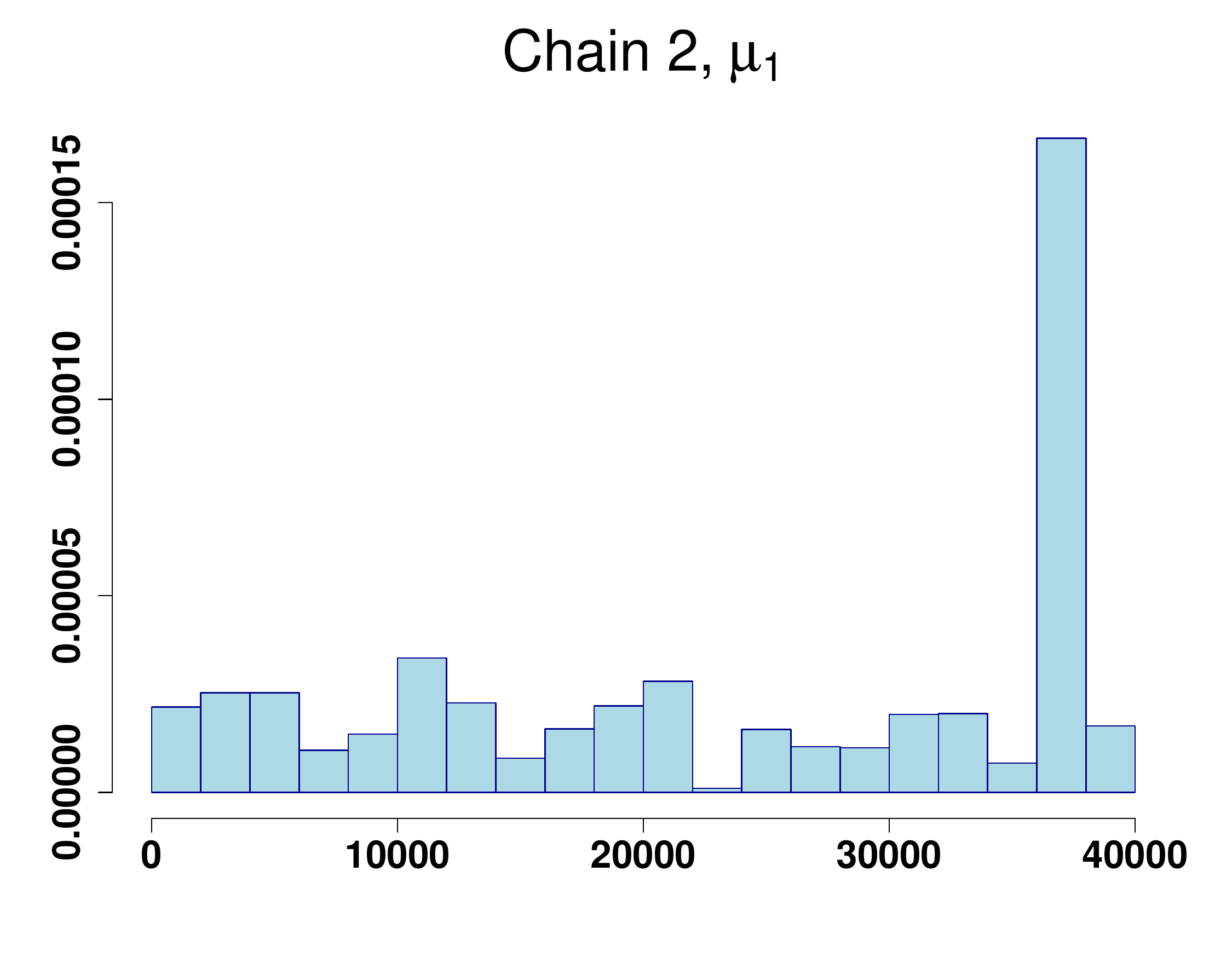}&
\hspace{-1.0cm}\includegraphics[width=4.0cm]{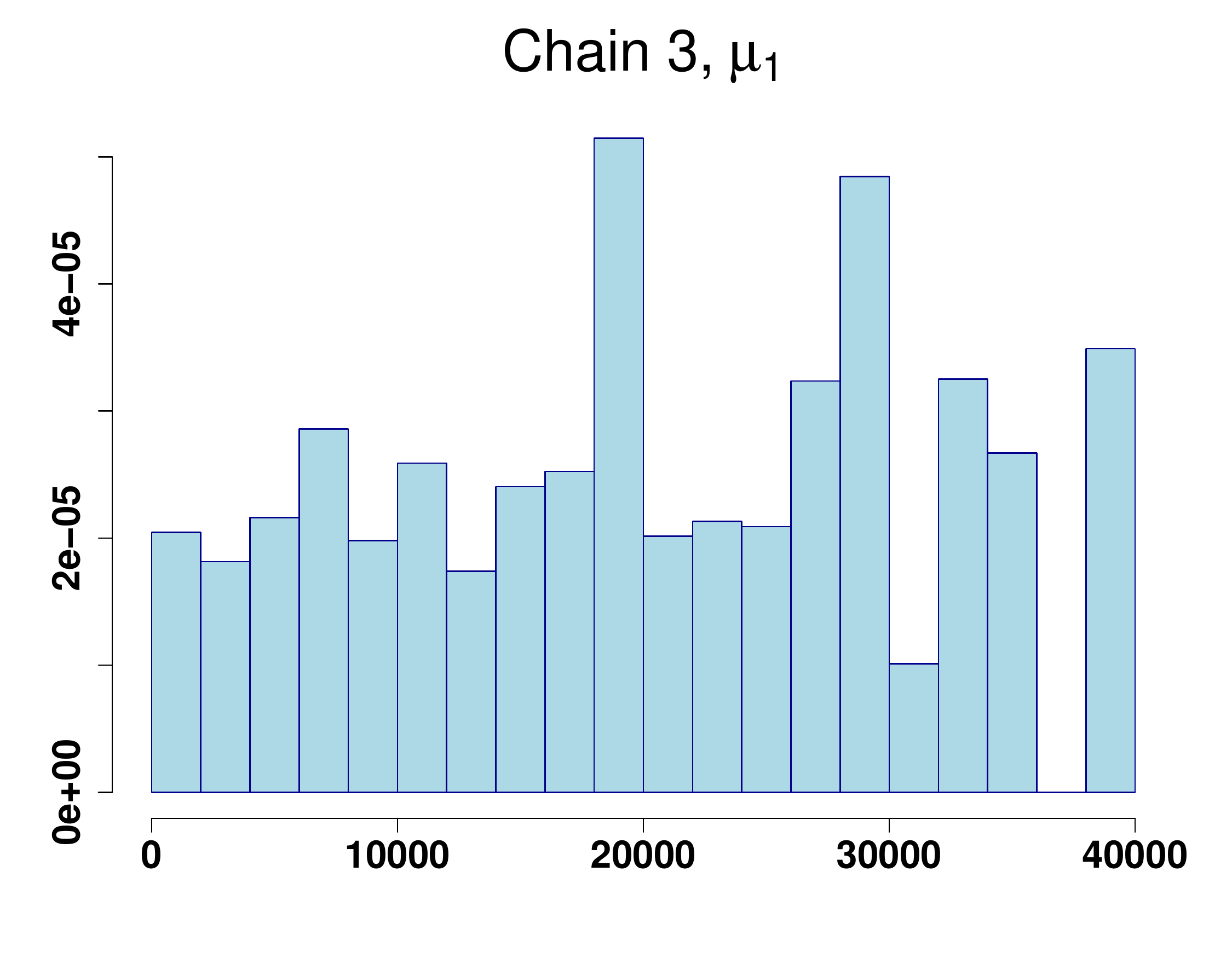}&\hspace{-1.5cm}\includegraphics[width=4.0cm]{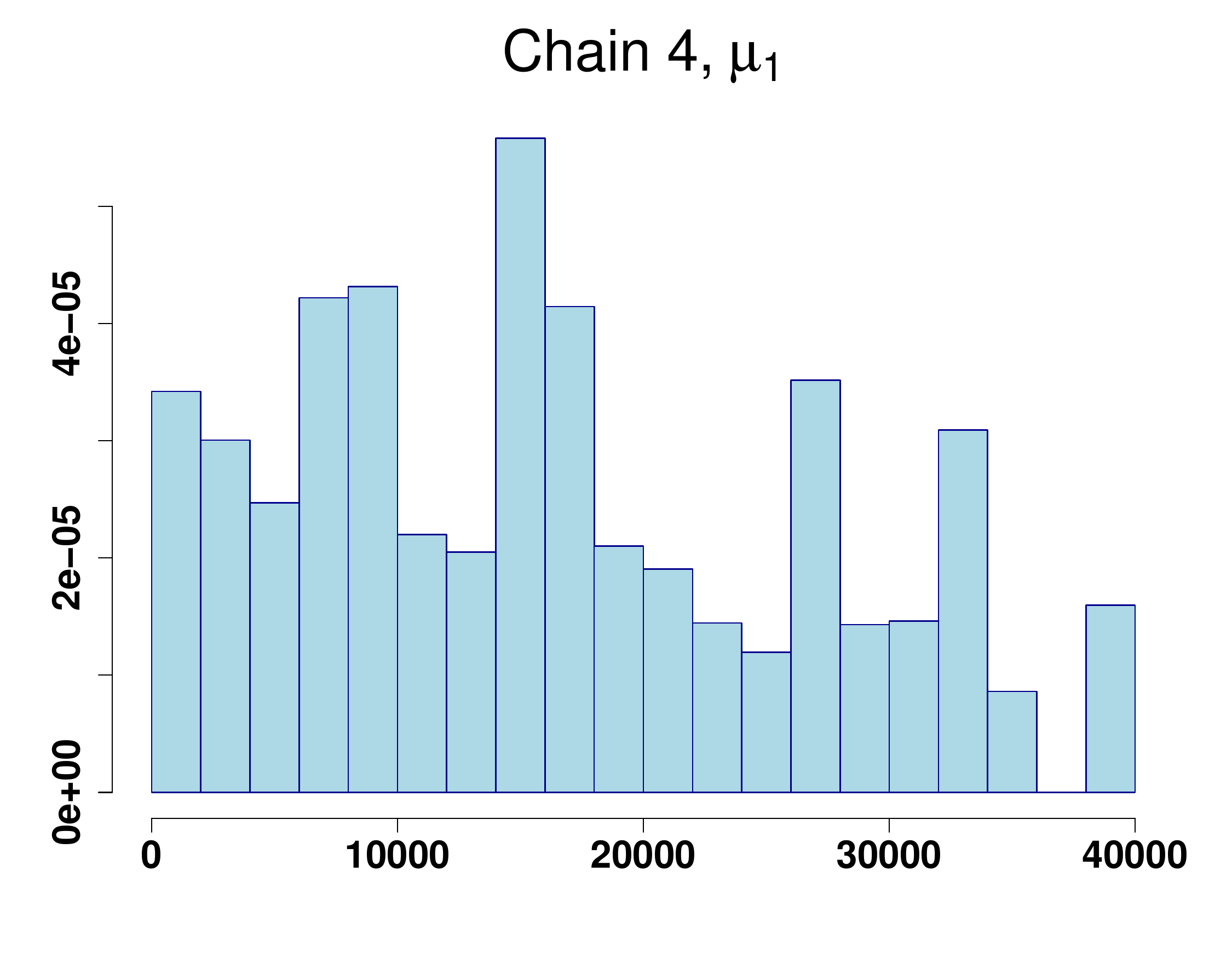}\\
\hspace{0.0cm}\includegraphics[width=4.0cm]{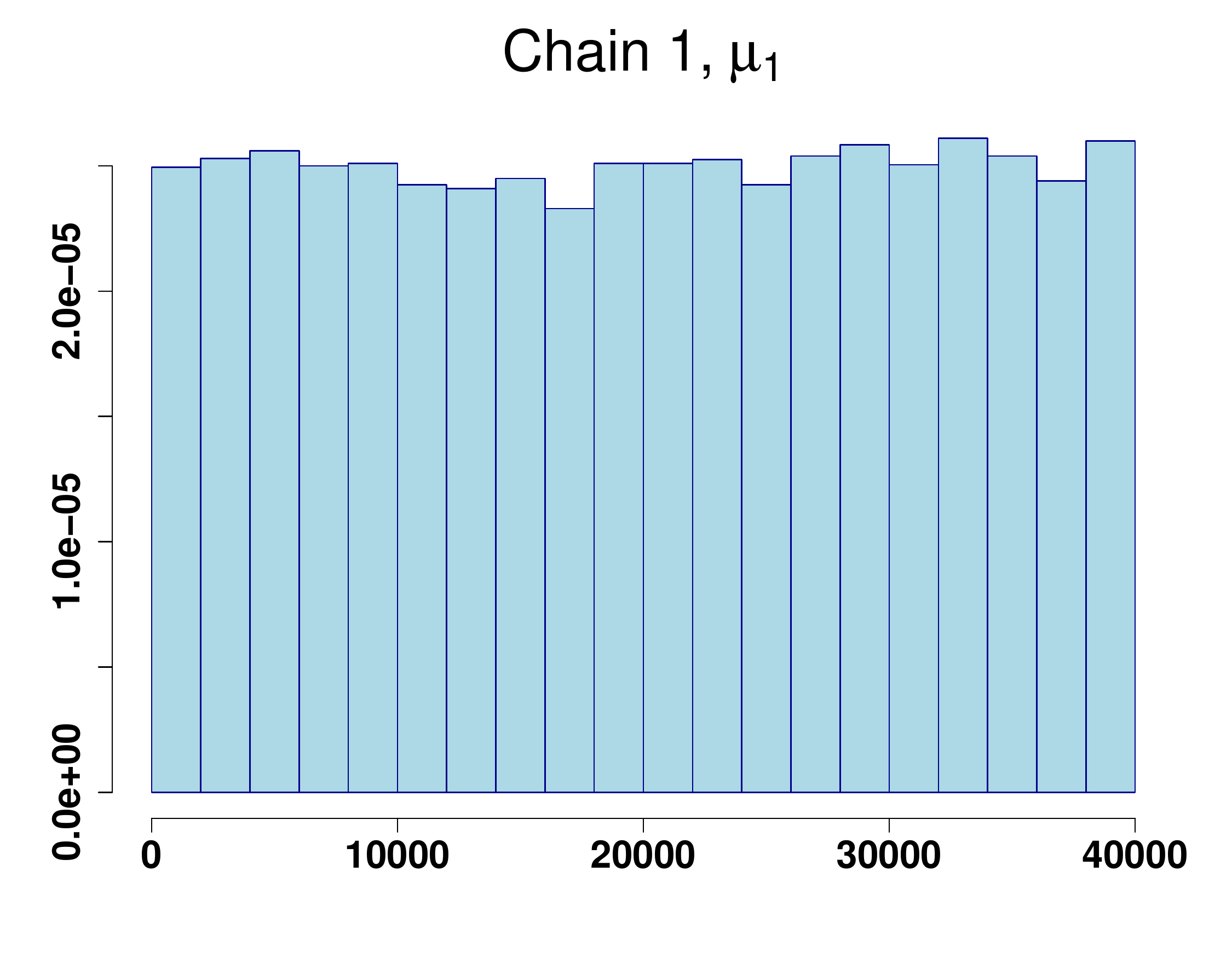}&\hspace{-0.5cm}\includegraphics[width=4.0cm]{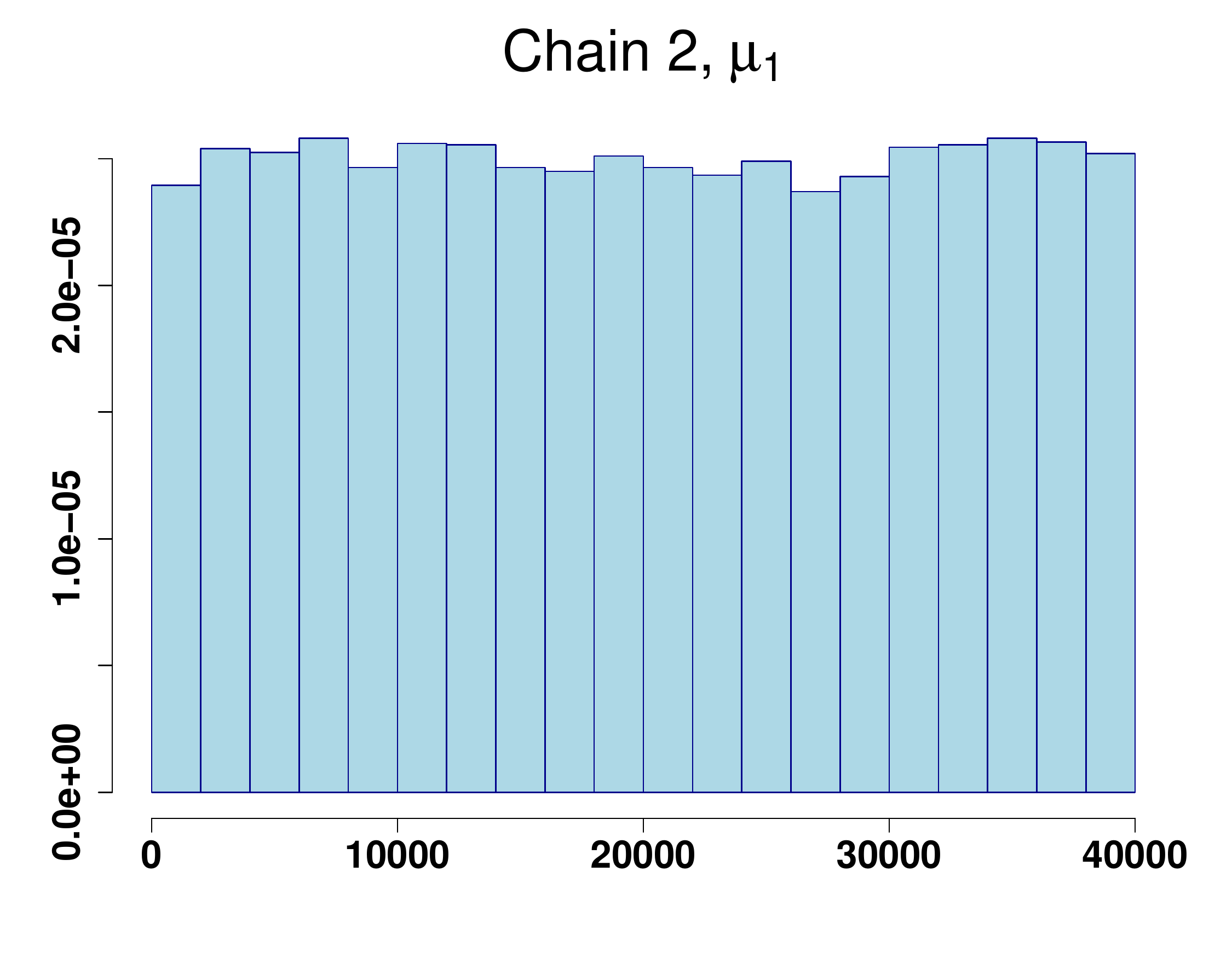}&
\hspace{-1.0cm}\includegraphics[width=4.0cm]{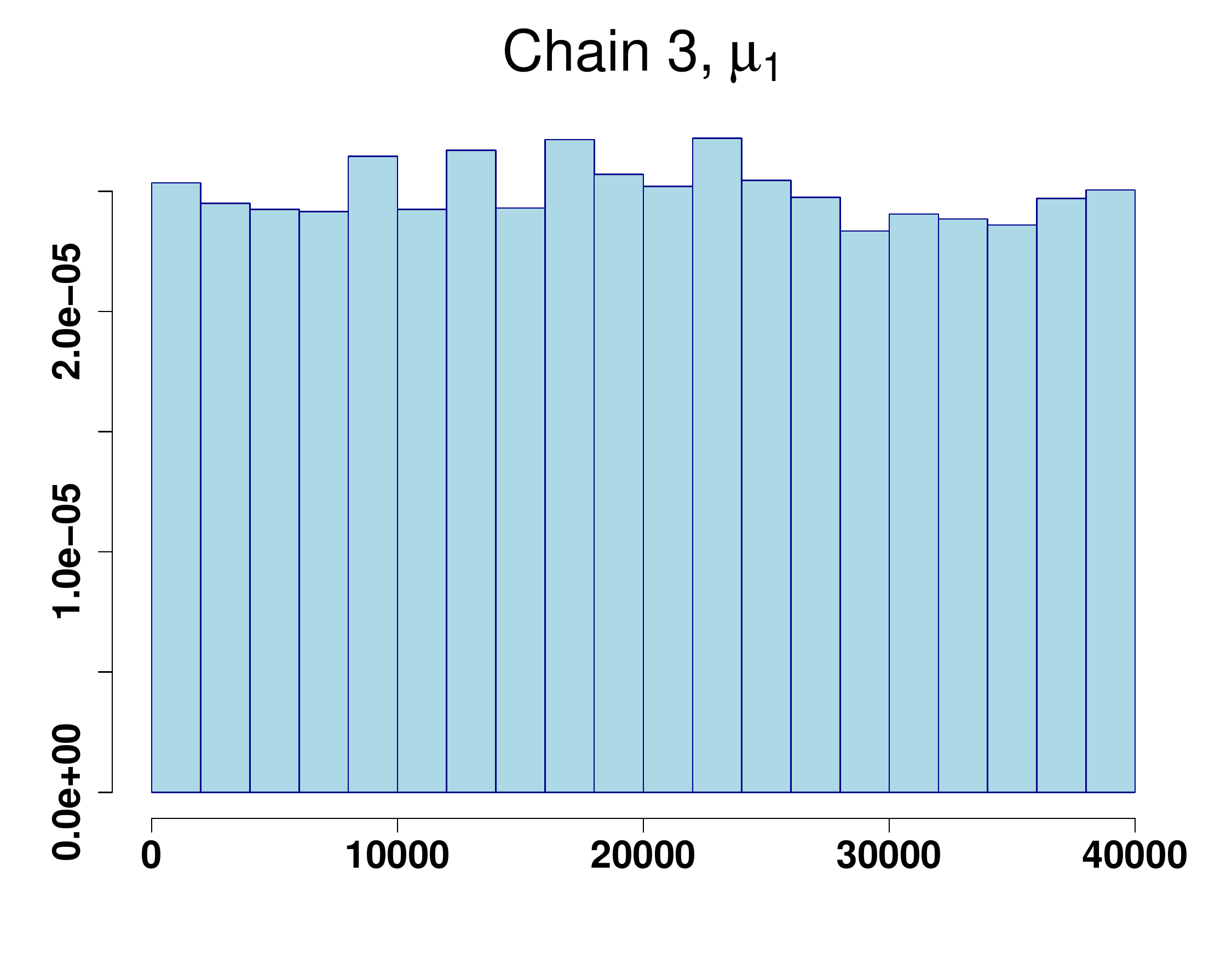}&\hspace{-1.5cm}\includegraphics[width=4.0cm]{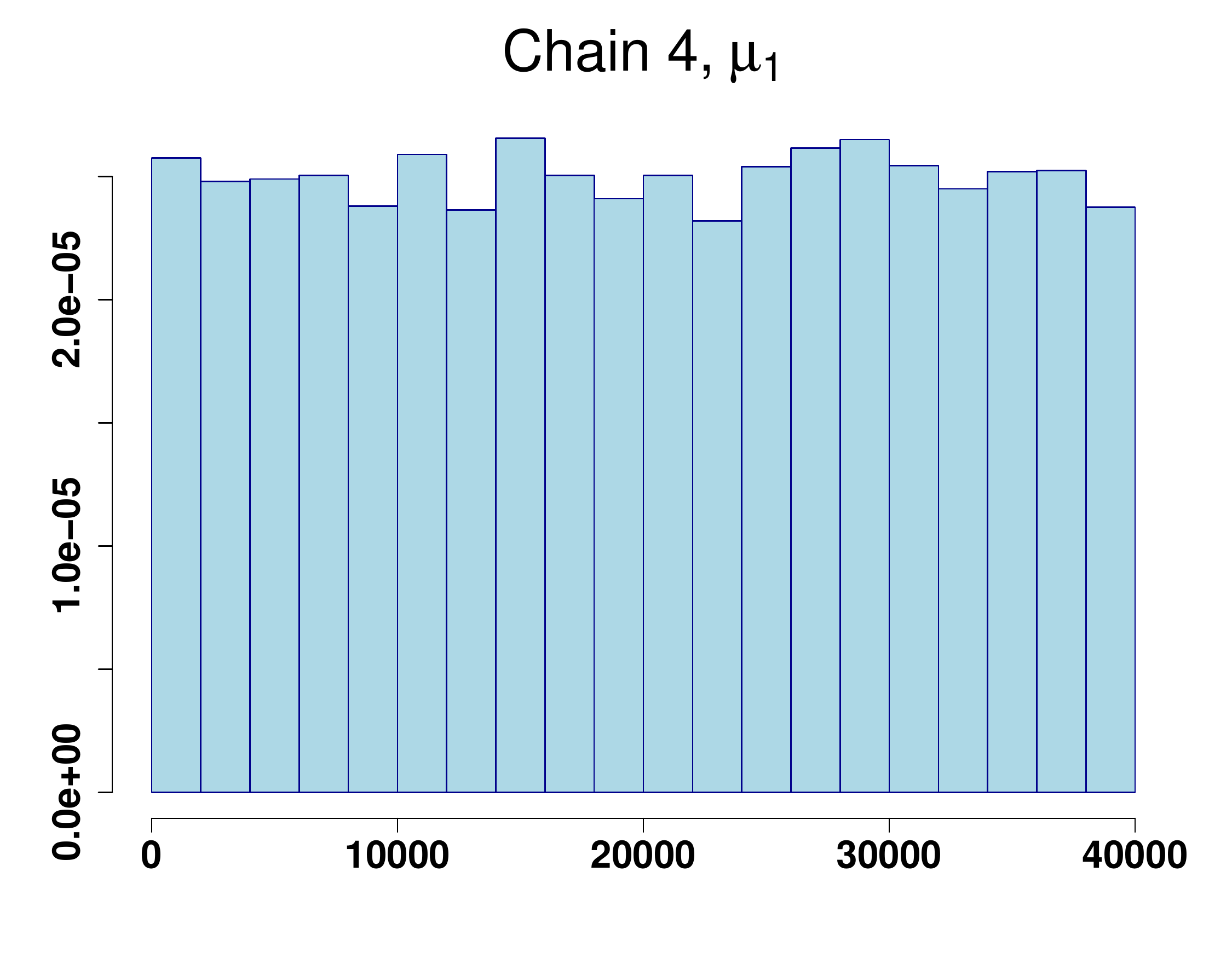}\\
\hspace{0.0cm}\includegraphics[width=4.0cm]{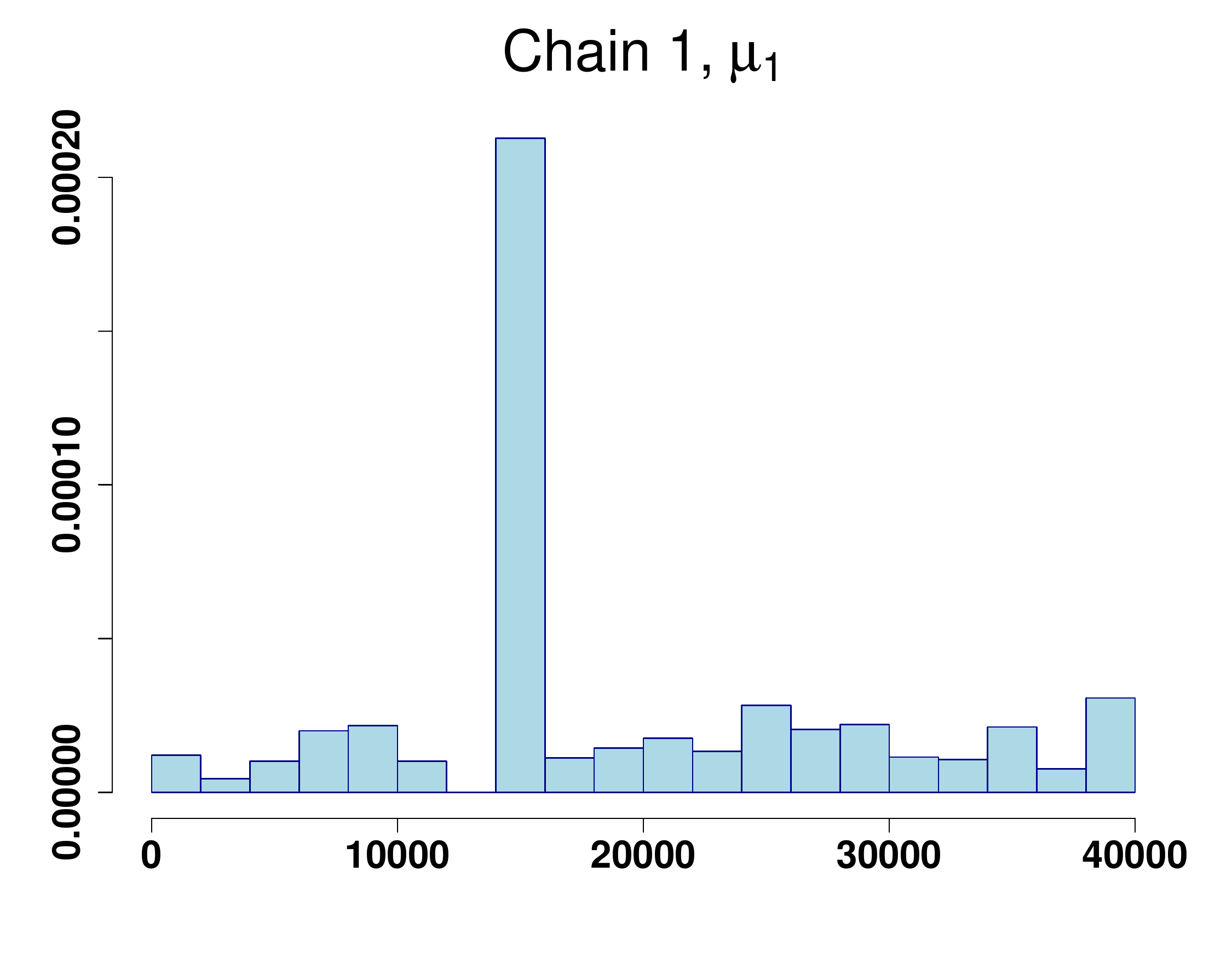}&\hspace{-0.5cm}\includegraphics[width=4.0cm]{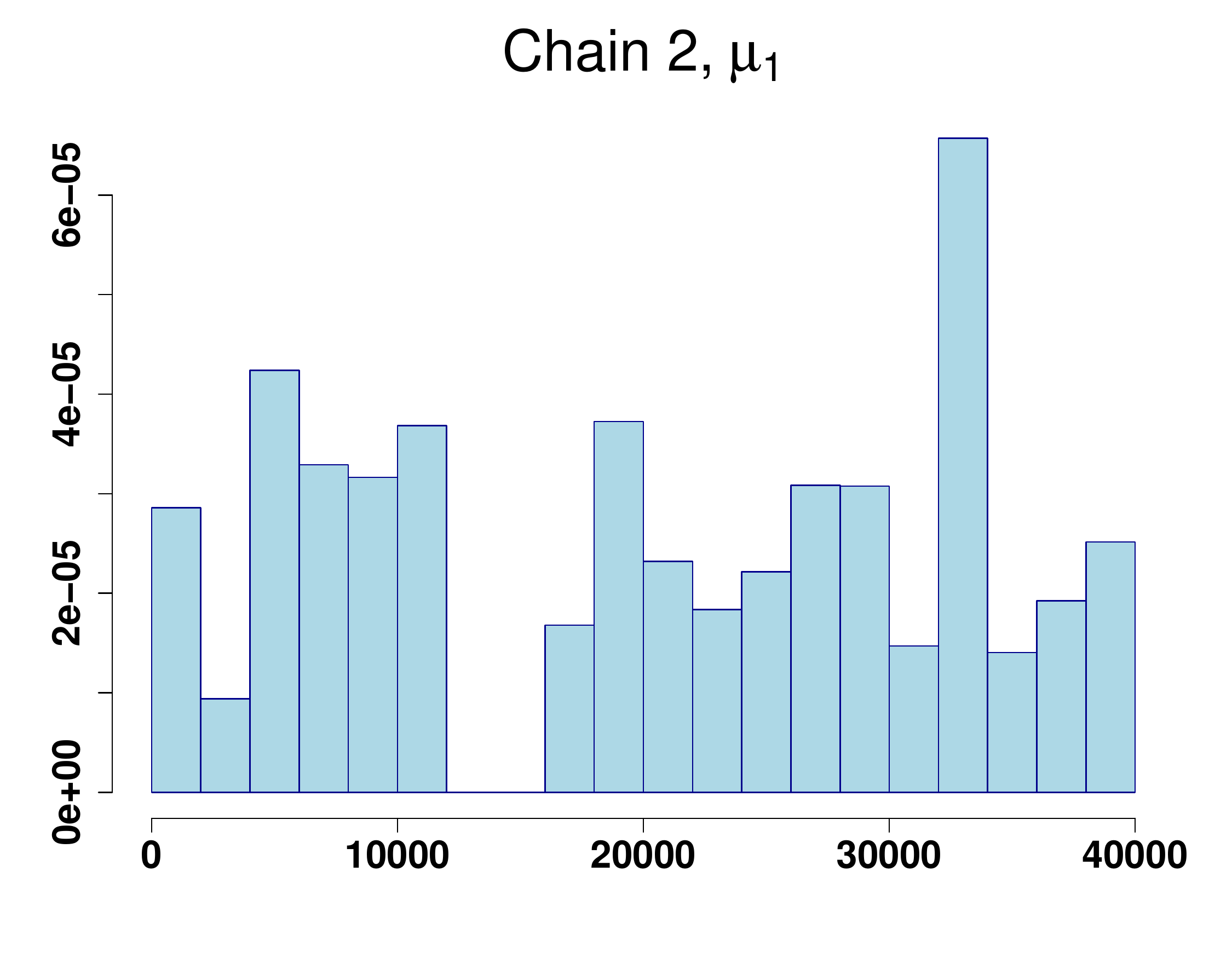}&
\hspace{-1.0cm}\includegraphics[width=4.0cm]{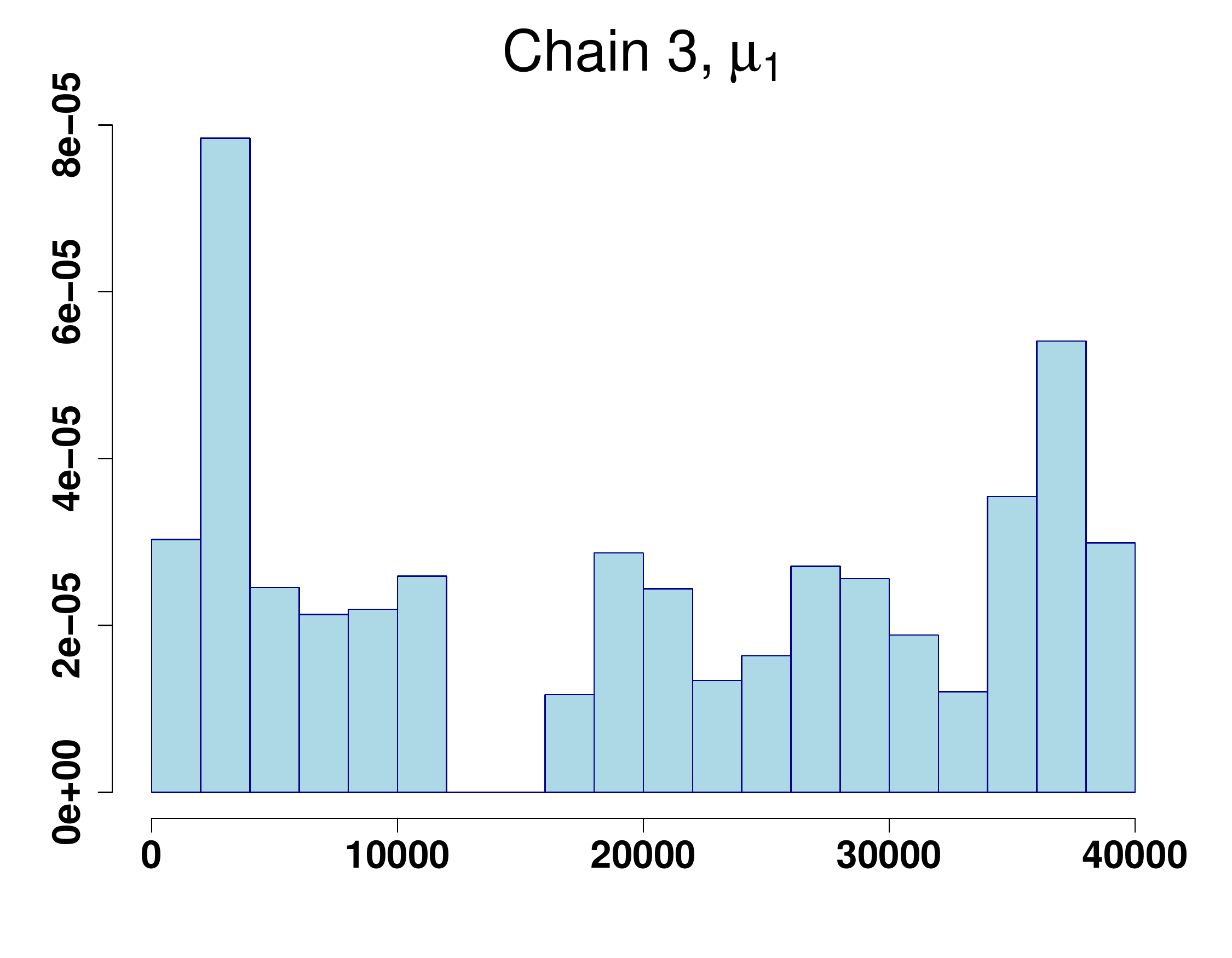}&\hspace{-1.5cm}\includegraphics[width=4.0cm]{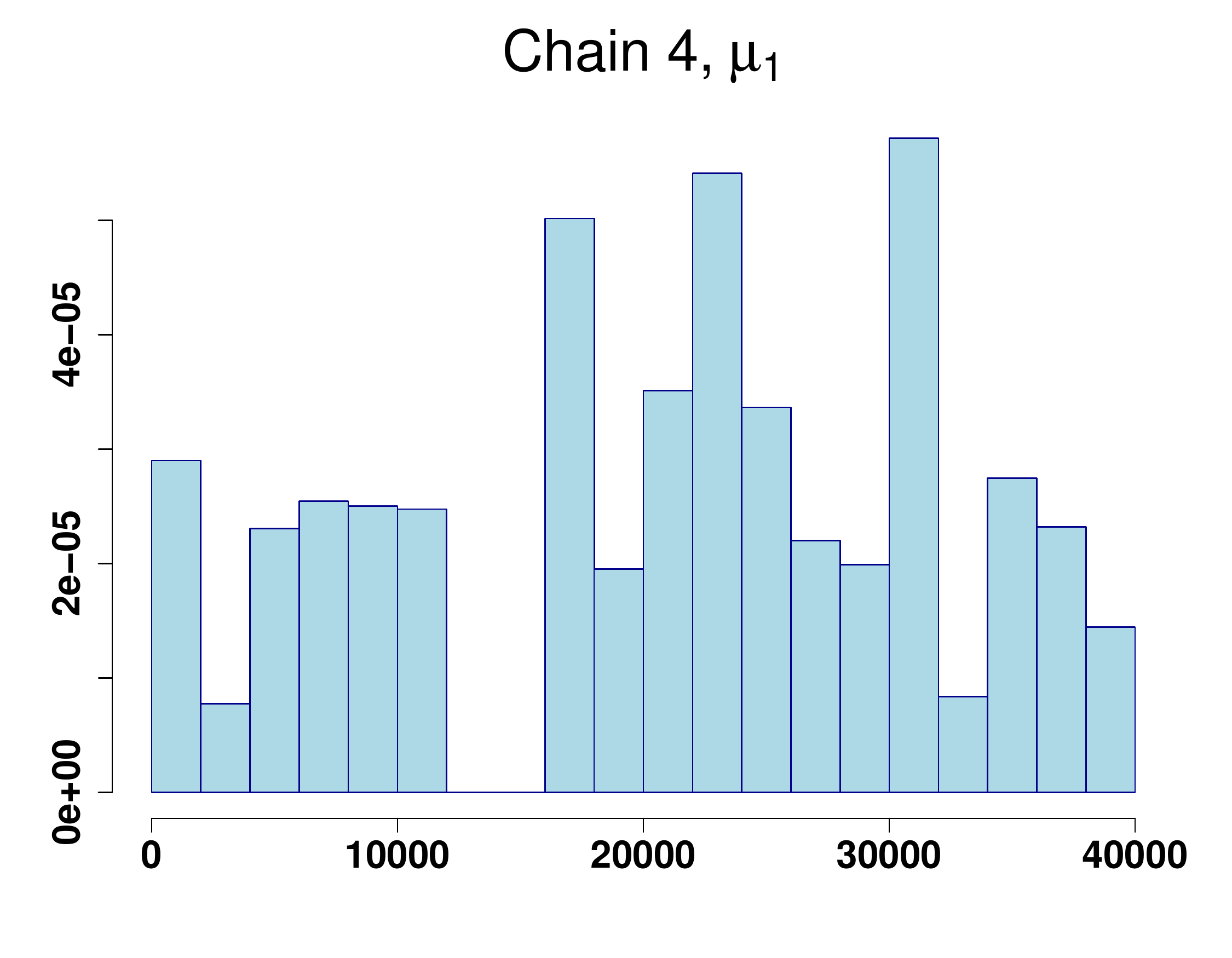}\\
\hspace{0.0cm}\includegraphics[width=4.0cm]{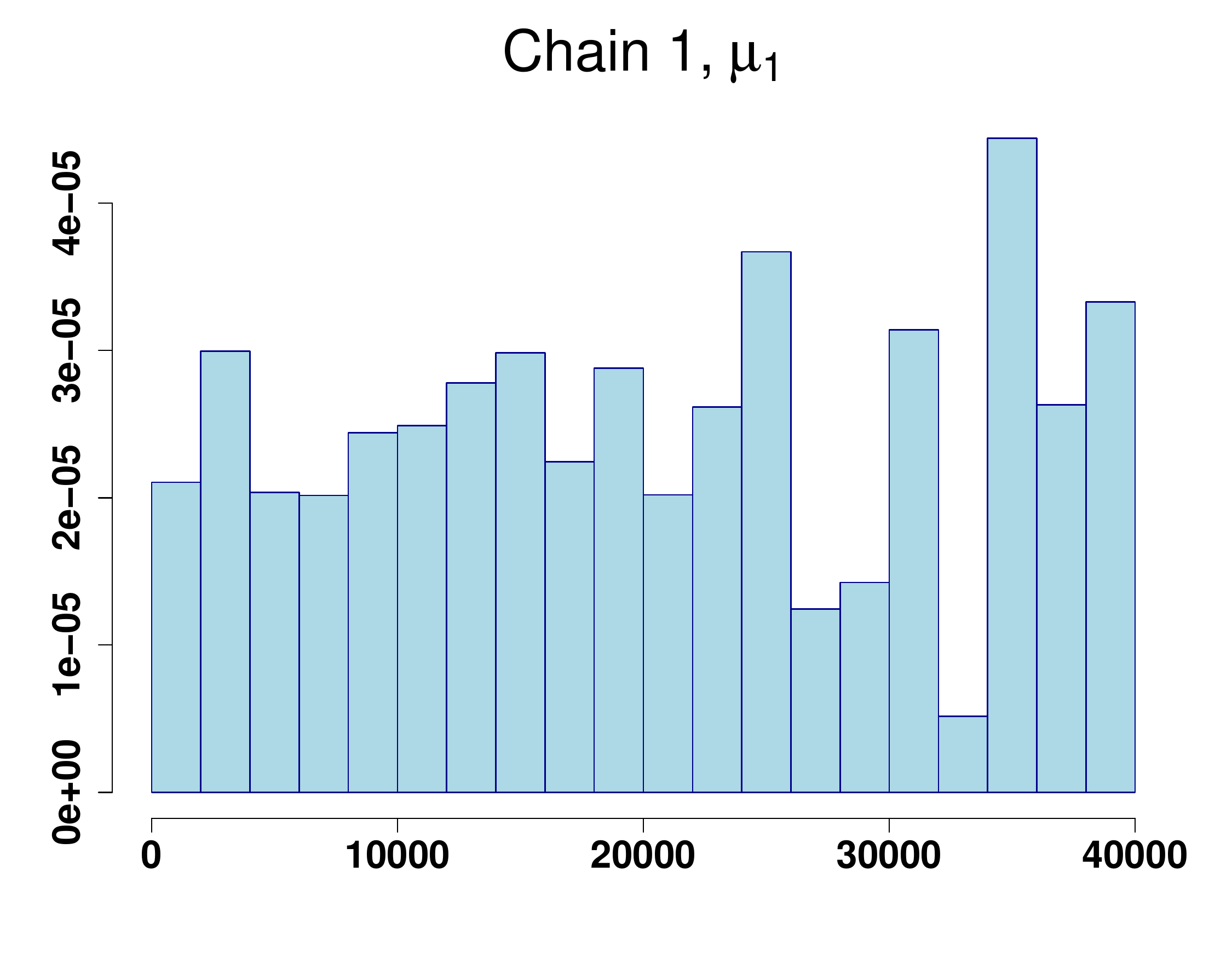}&\hspace{-0.5cm}\includegraphics[width=4.0cm]{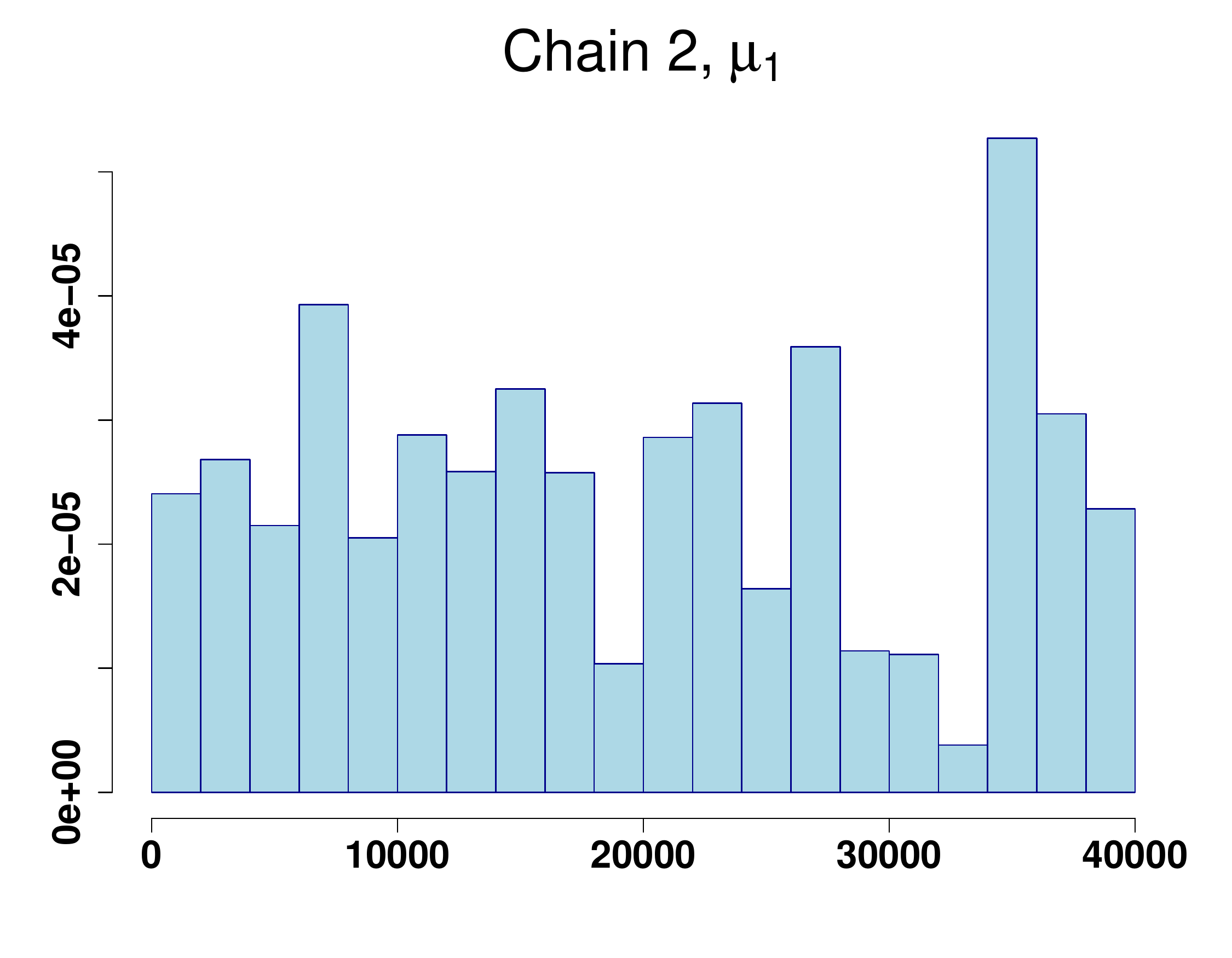}&
\hspace{-1.0cm}\includegraphics[width=4.0cm]{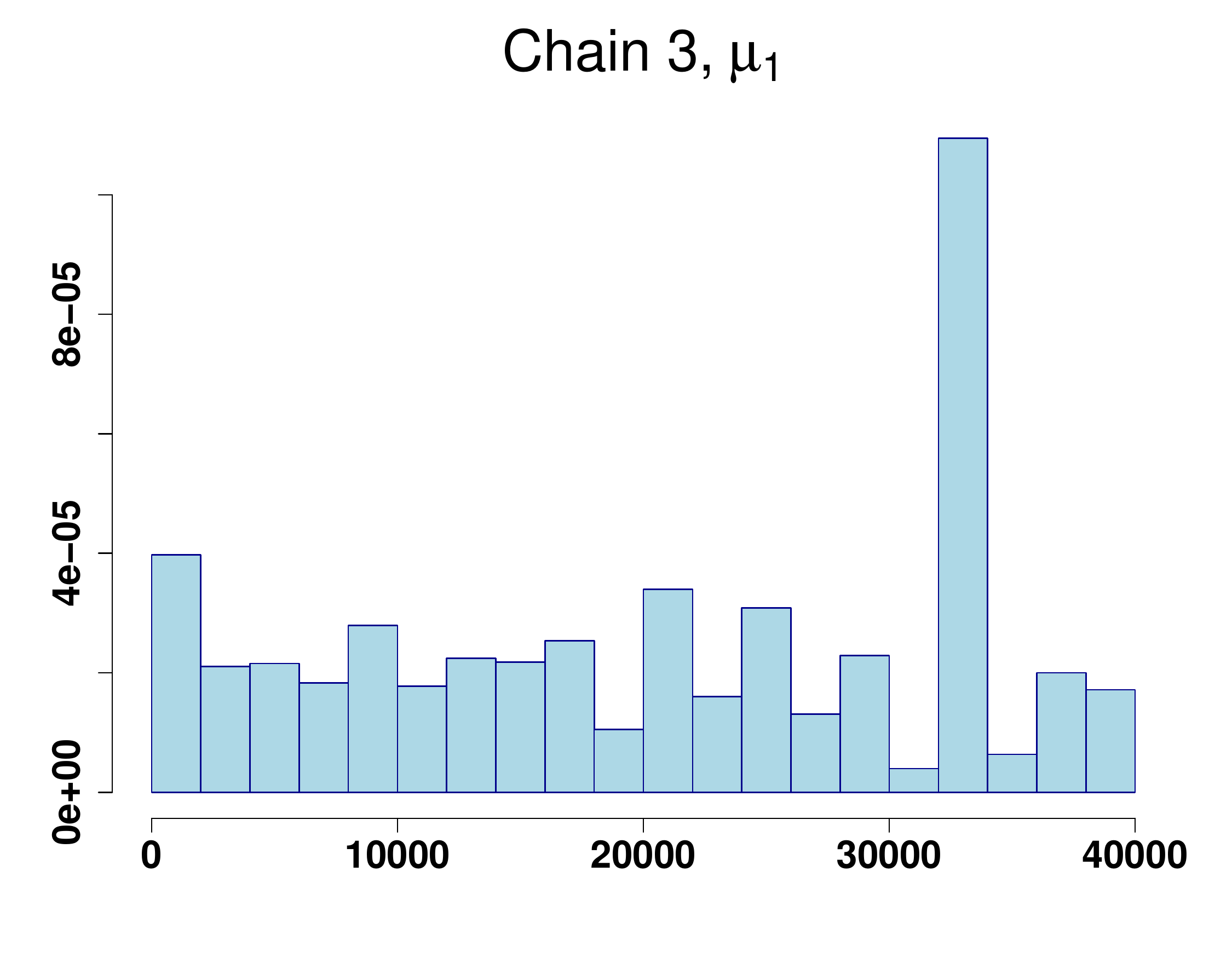}&\hspace{-1.5cm}\includegraphics[width=4.0cm]{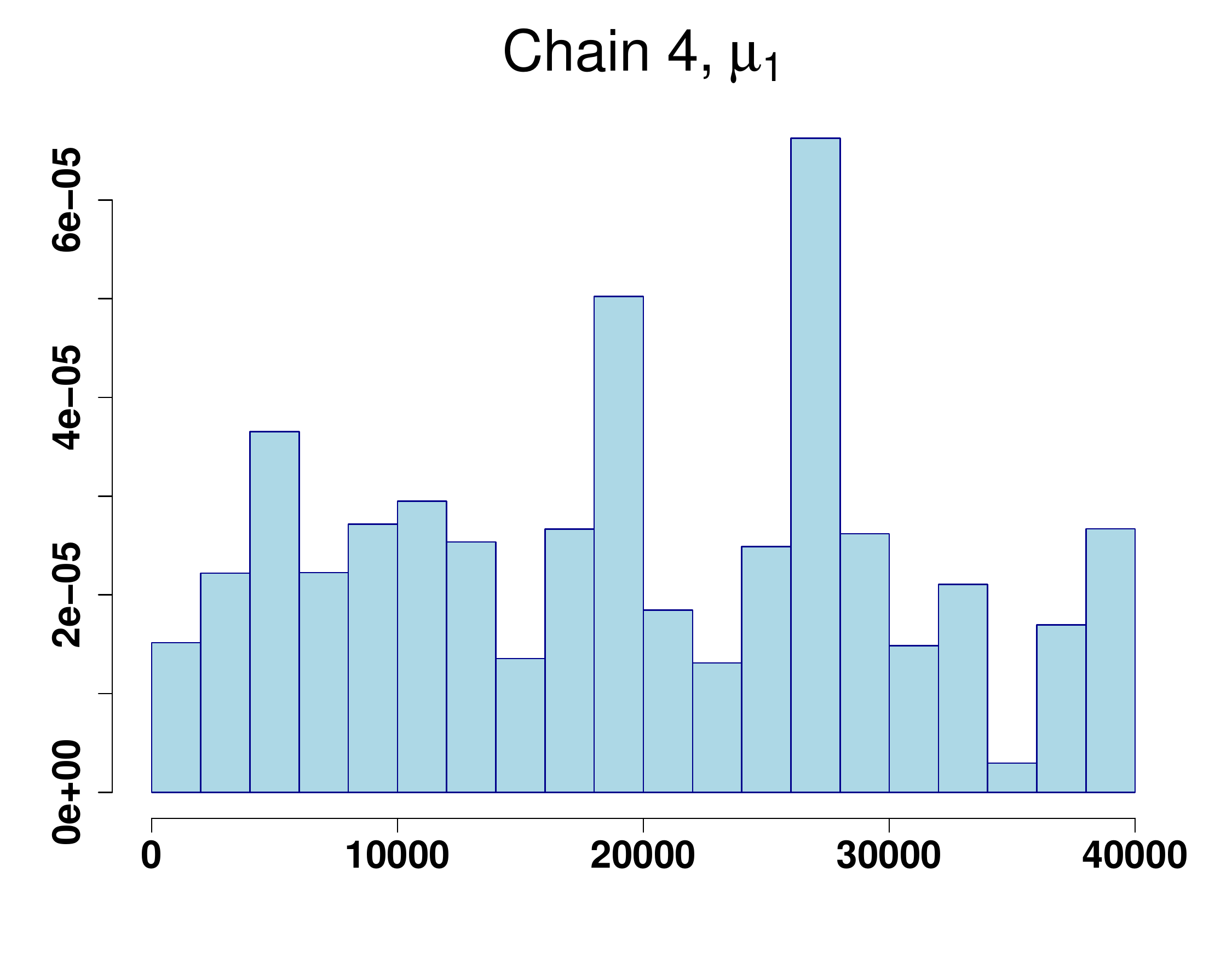}\\
\hspace{0.0cm}\includegraphics[width=4.0cm]{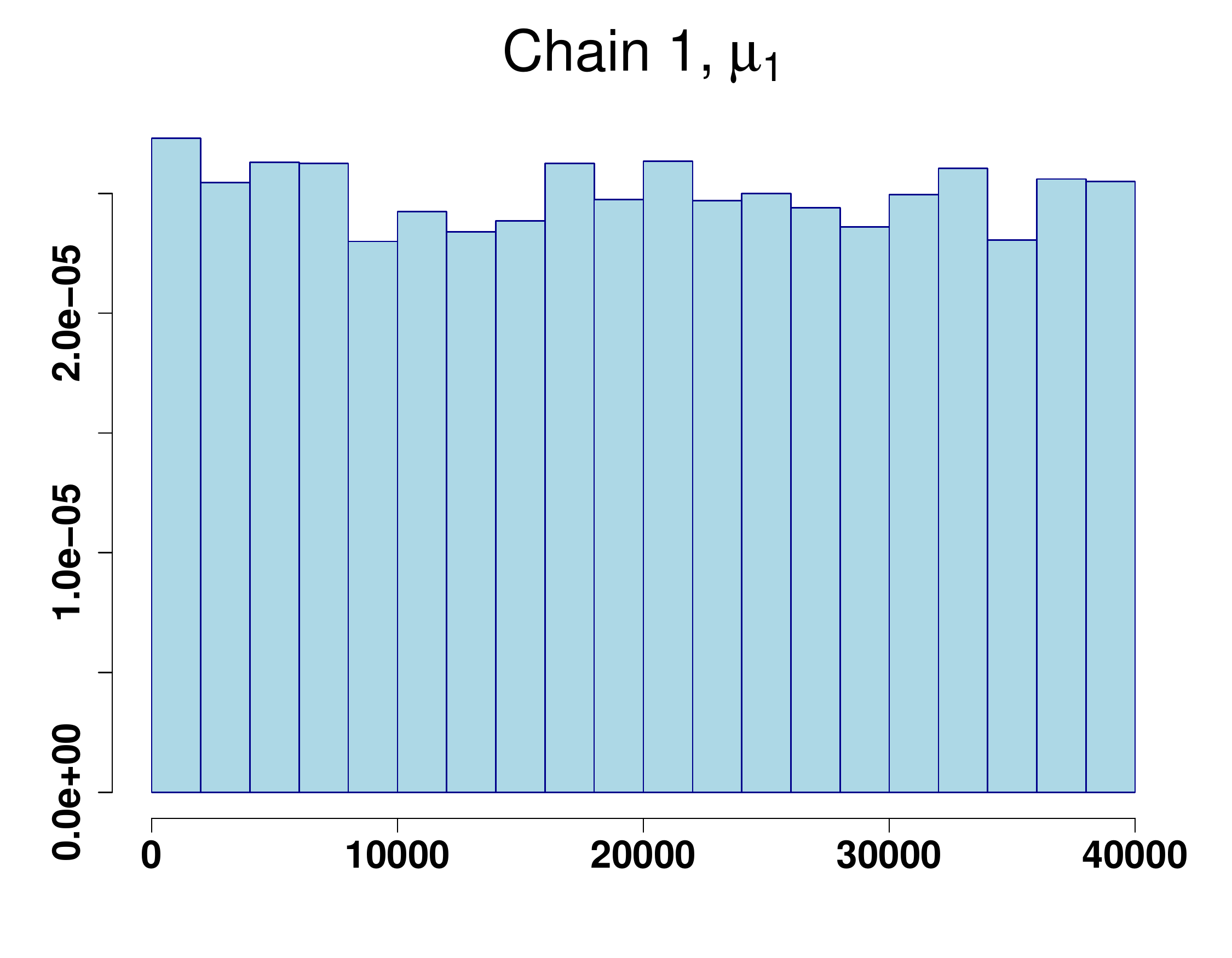}&\hspace{-0.5cm}\includegraphics[width=4.0cm]{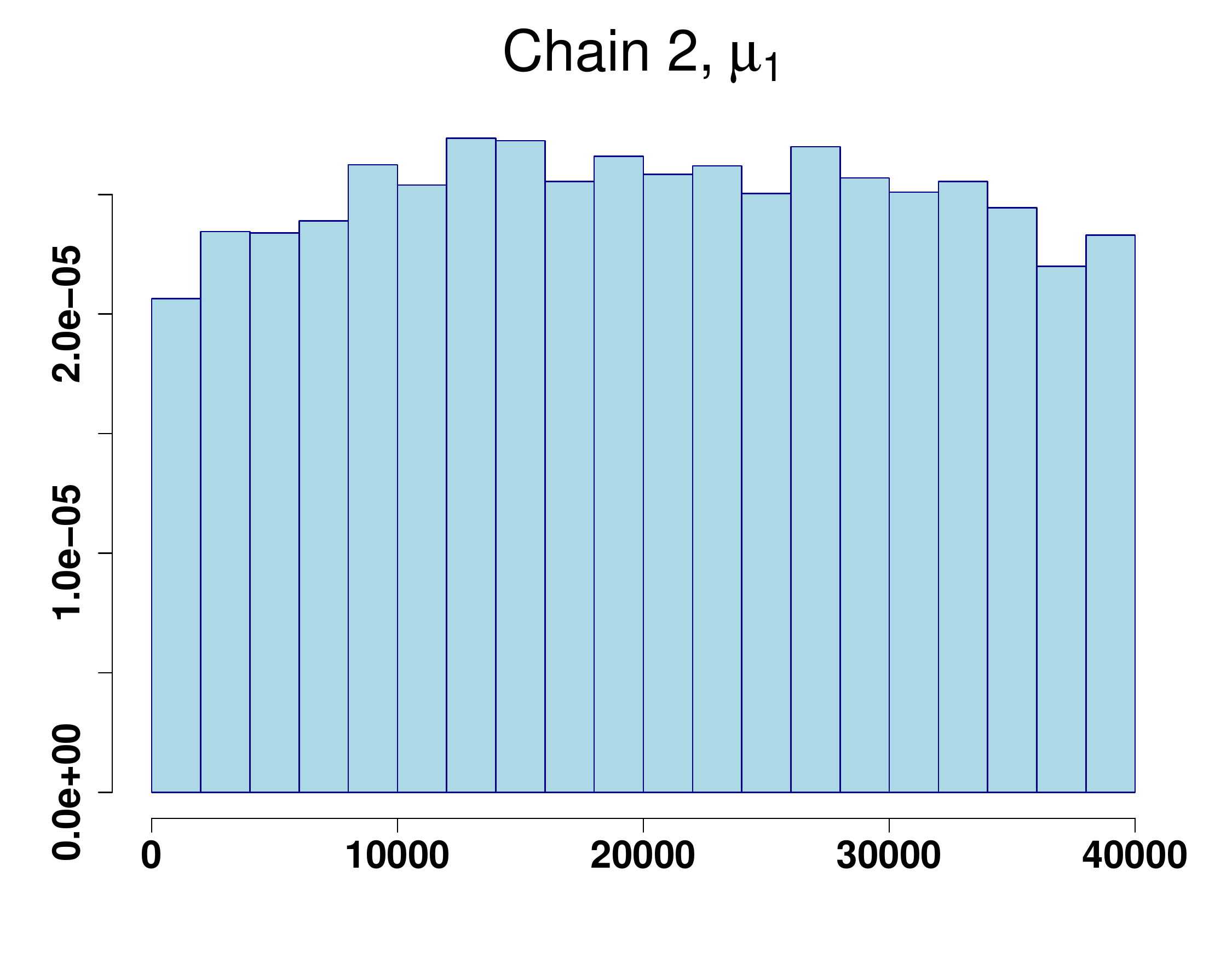}&
\hspace{-1.0cm}\includegraphics[width=4.0cm]{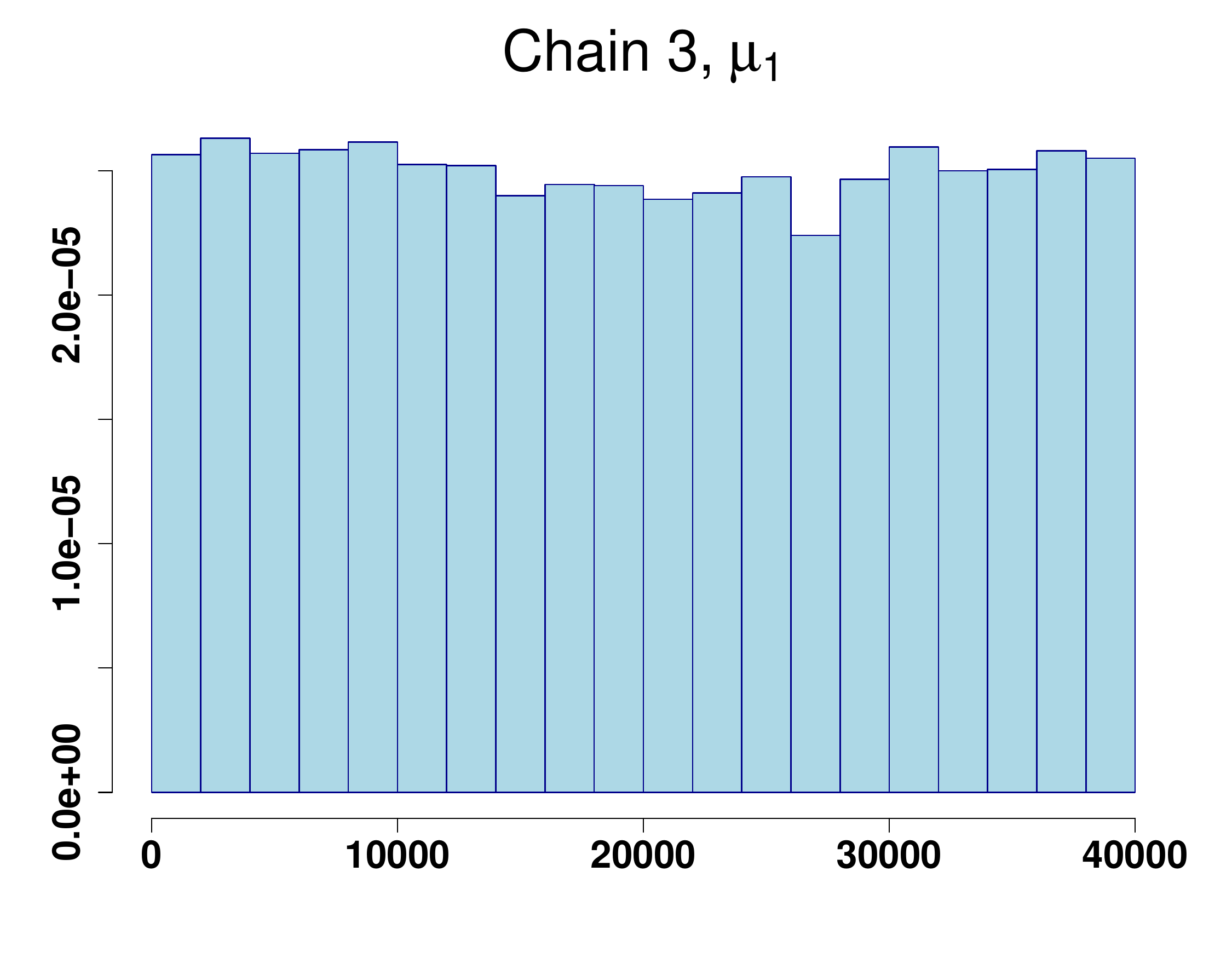}&\hspace{-1.5cm}\includegraphics[width=4.0cm]{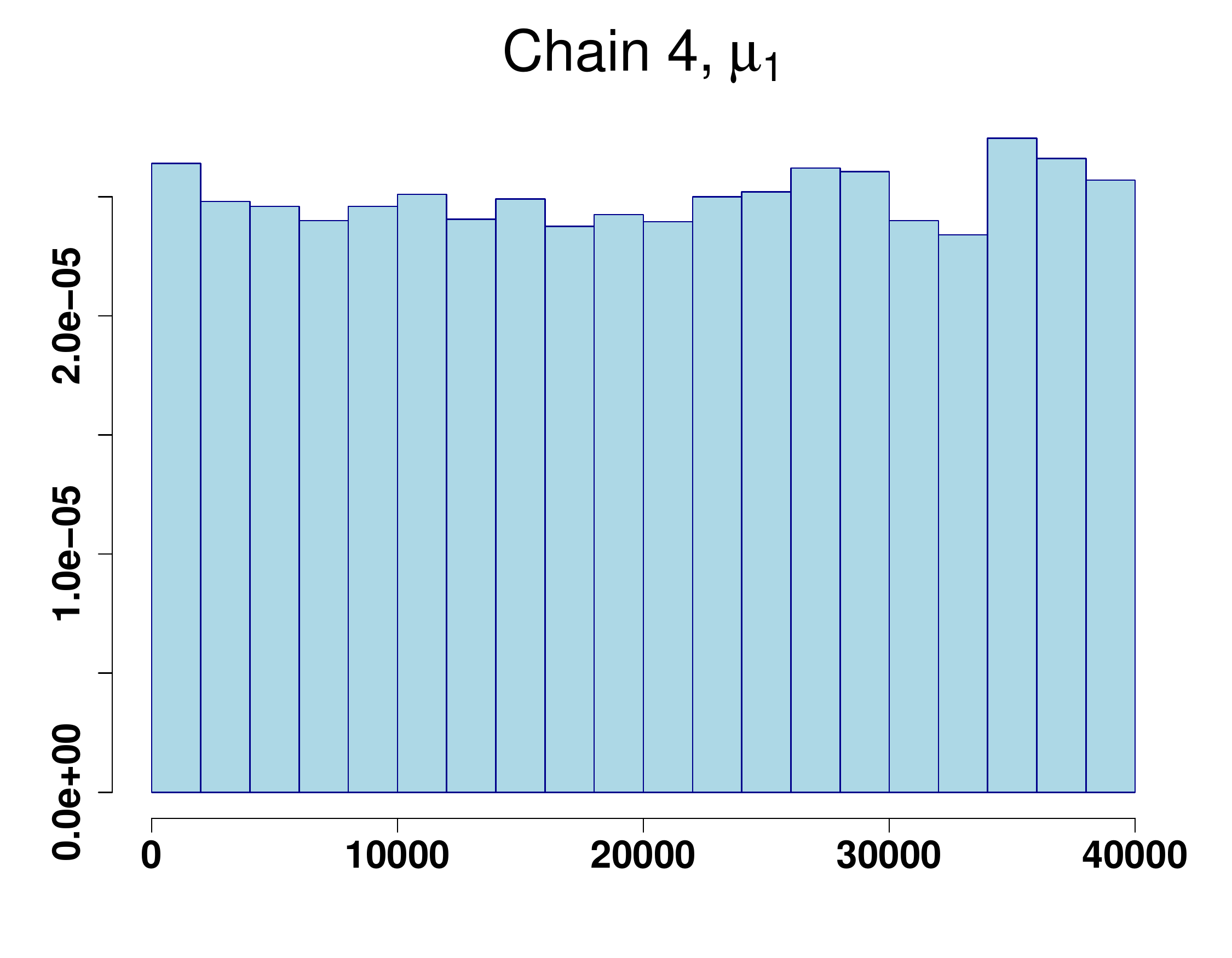}\\
\end{tabular}
 \caption{Rank plots of posterior draws from four chains in the case of the parameter $\mu_1$ (SBP) of the normal multivariate random effects model by employing the Jeffreys prior (first to third rows) and the Berger and Bernardo reference prior (fourth to sixth rows). The samples from the posterior distributions are drawn by Algorithm A (first and fourth rows), Algorithm B (second and fifth rows) and Algorithm C (third and sixth rows).}
\label{fig:emp-study-rank-mu1-nor}
 \end{figure}

\begin{figure}[h!t]
\centering
\begin{tabular}{p{4.0cm}p{4.0cm}p{4.0cm}p{4.0cm}}
\hspace{0.0cm}\includegraphics[width=4.0cm]{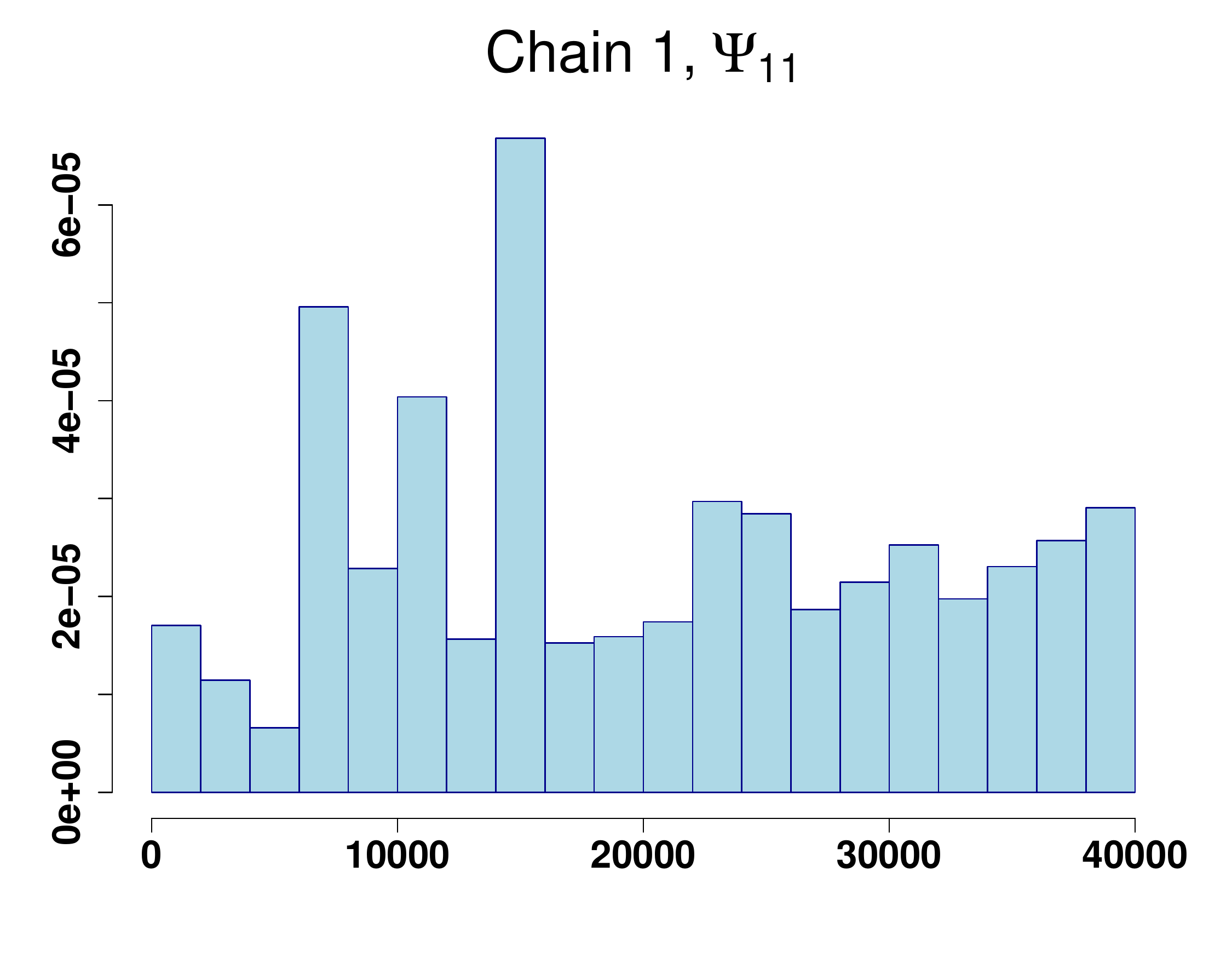}&\hspace{-0.5cm}\includegraphics[width=4.0cm]{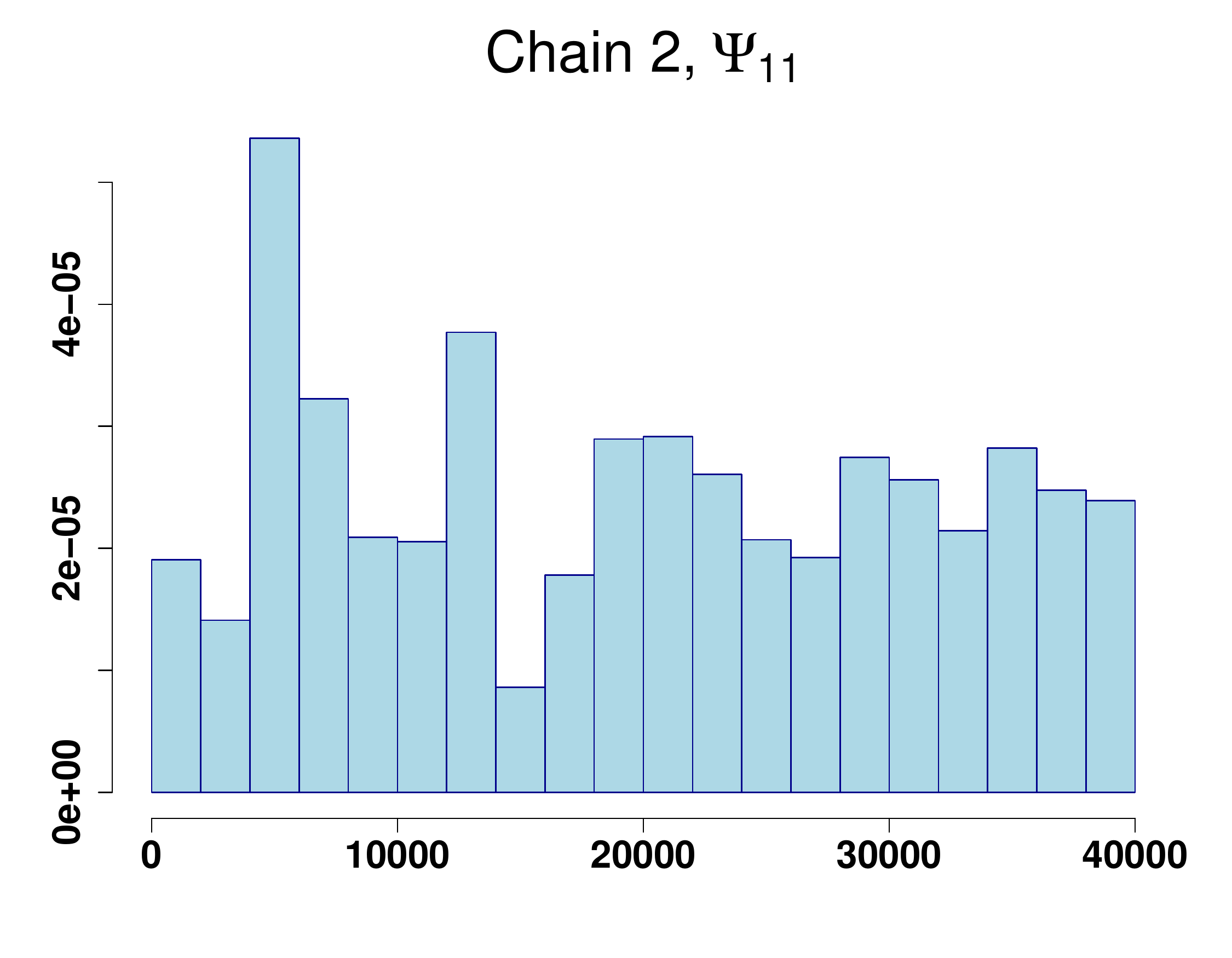}&
\hspace{-1.0cm}\includegraphics[width=4.0cm]{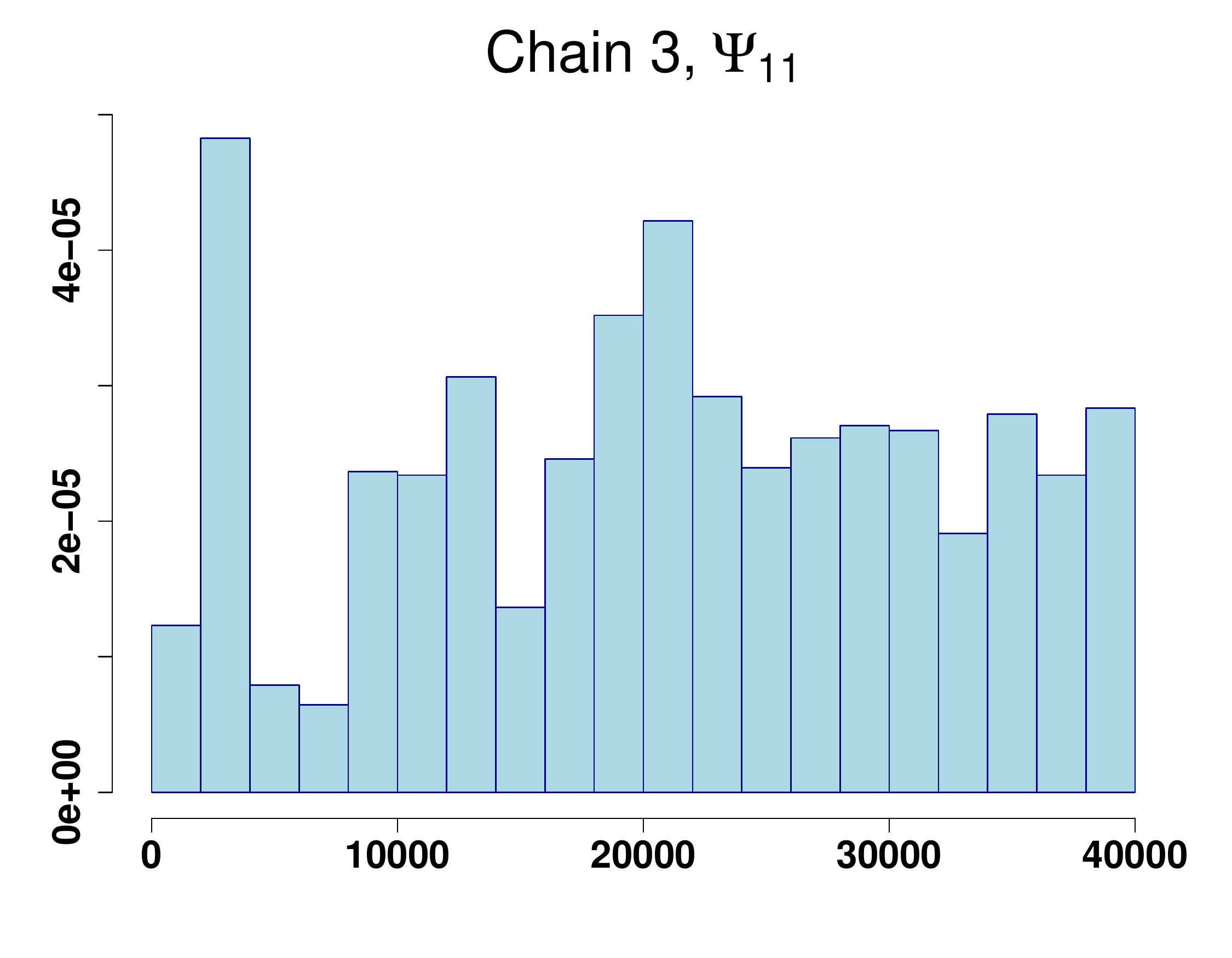}&\hspace{-1.5cm}\includegraphics[width=4.0cm]{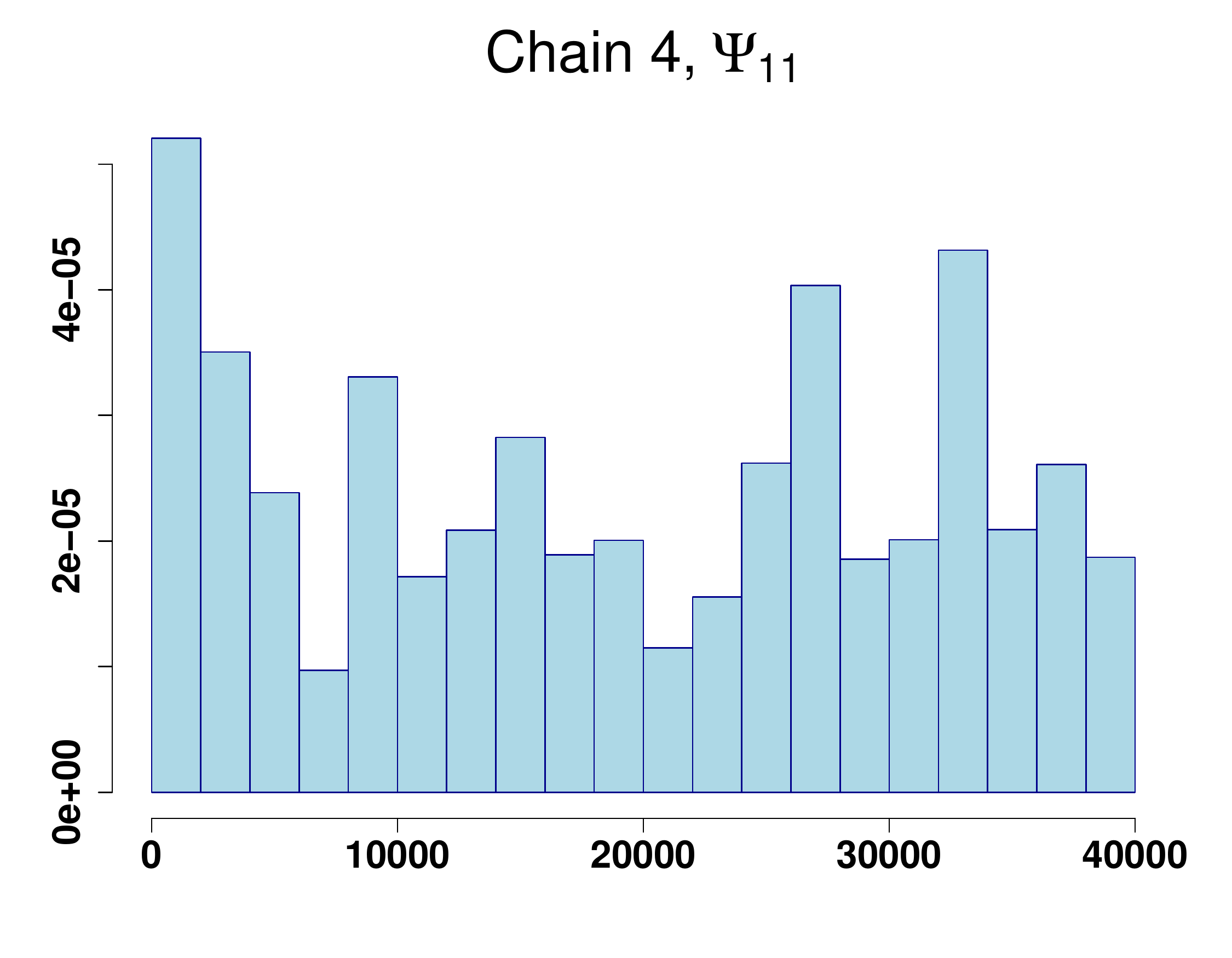}\\
\hspace{0.0cm}\includegraphics[width=4.0cm]{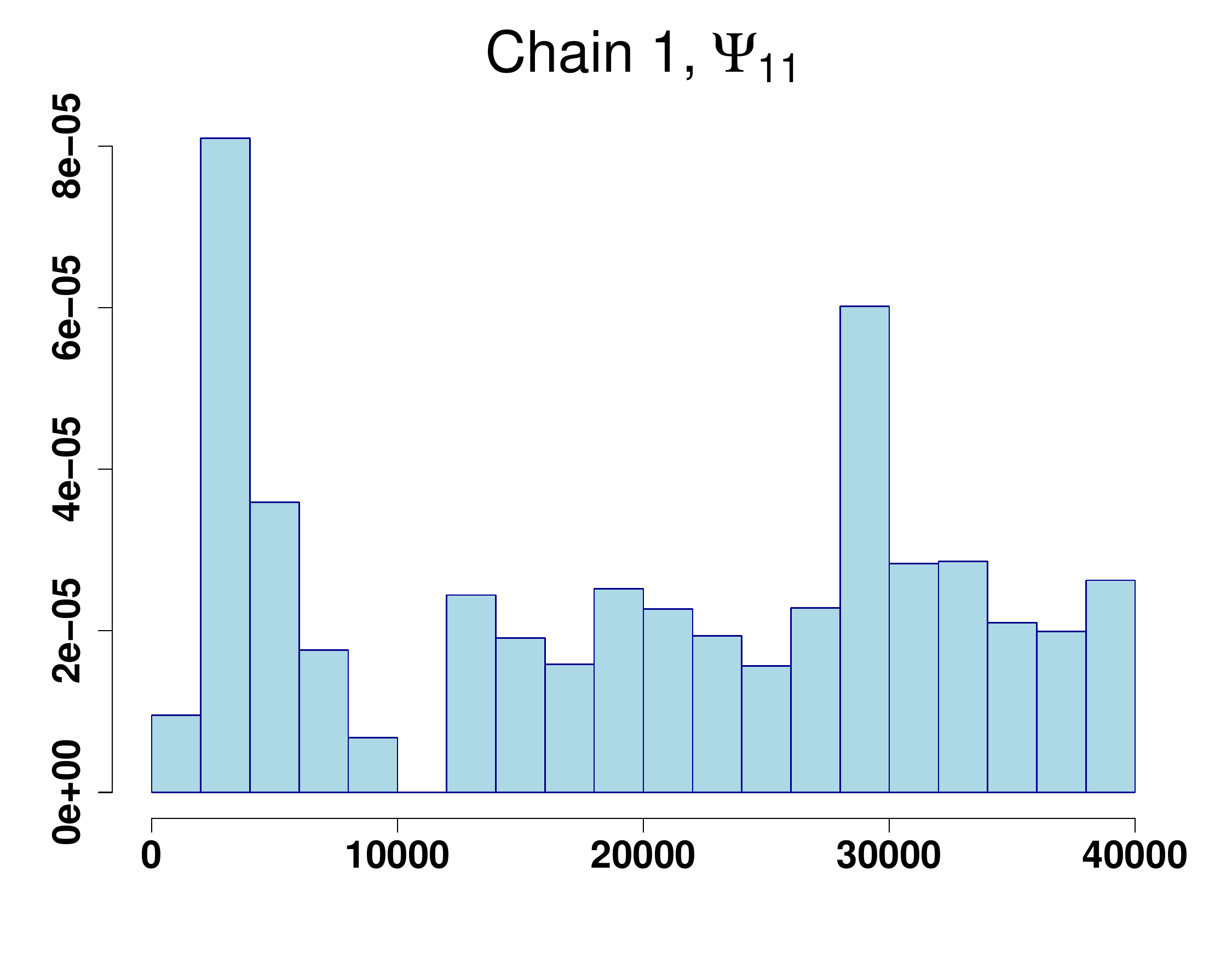}&\hspace{-0.5cm}\includegraphics[width=4.0cm]{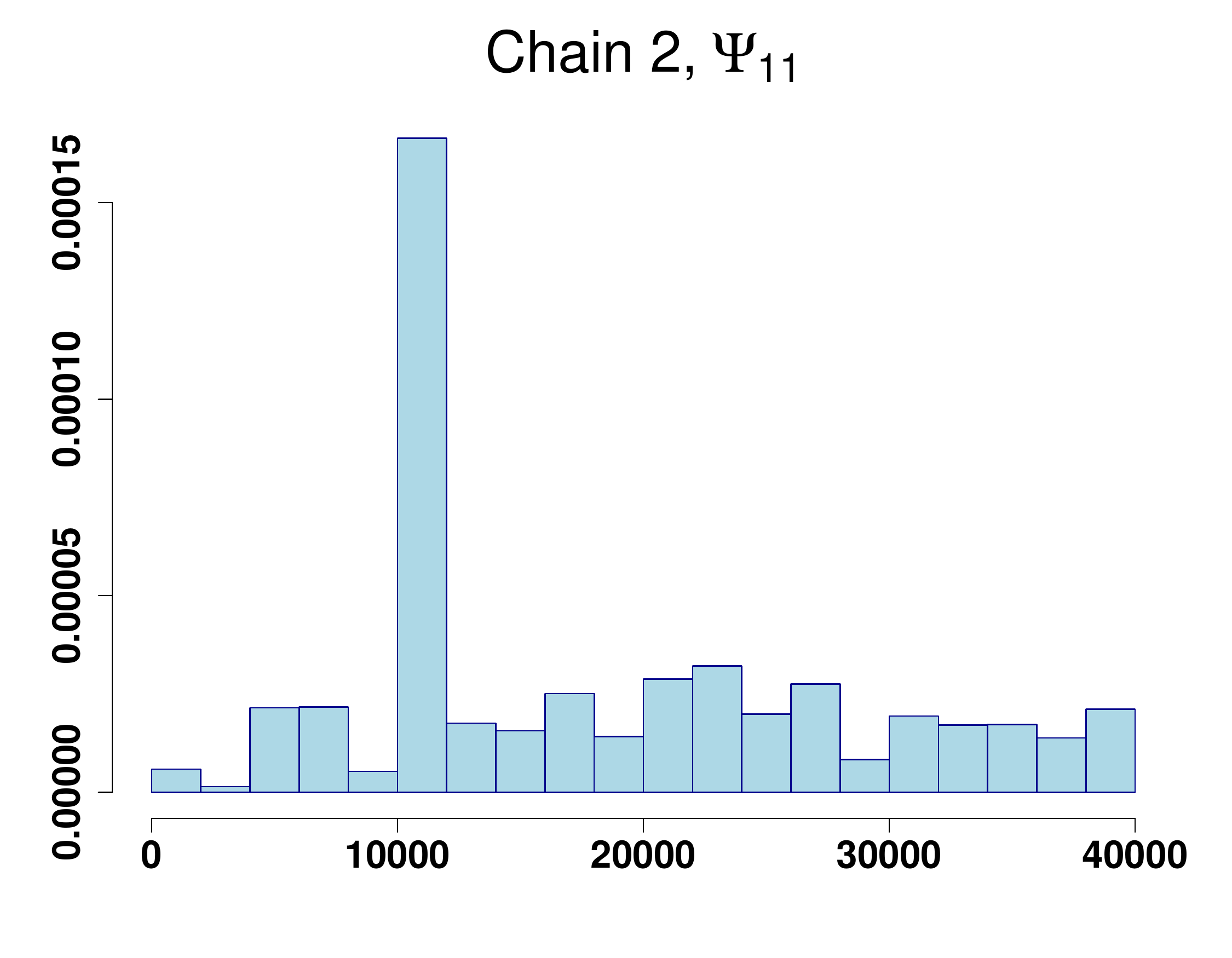}&
\hspace{-1.0cm}\includegraphics[width=4.0cm]{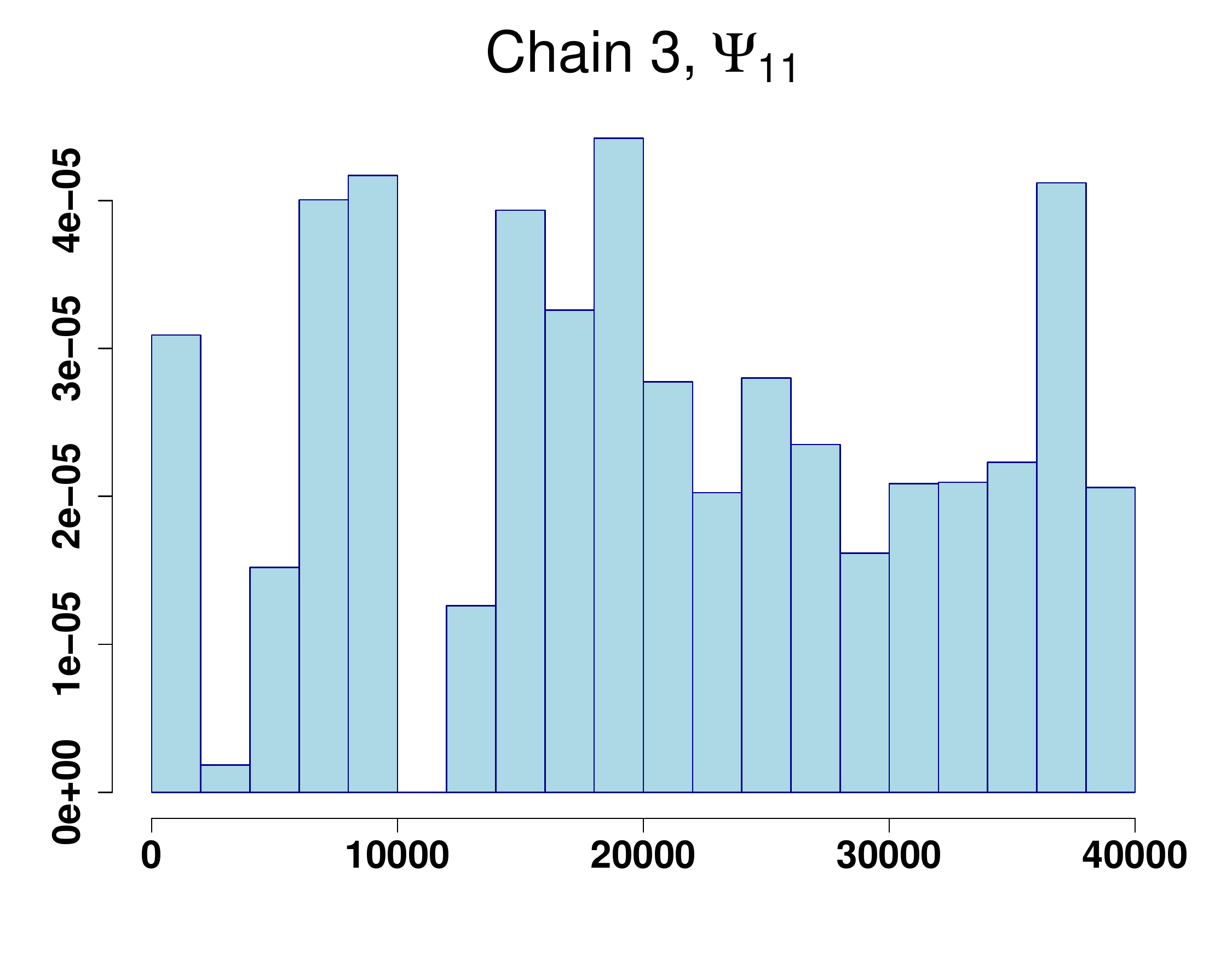}&\hspace{-1.5cm}\includegraphics[width=4.0cm]{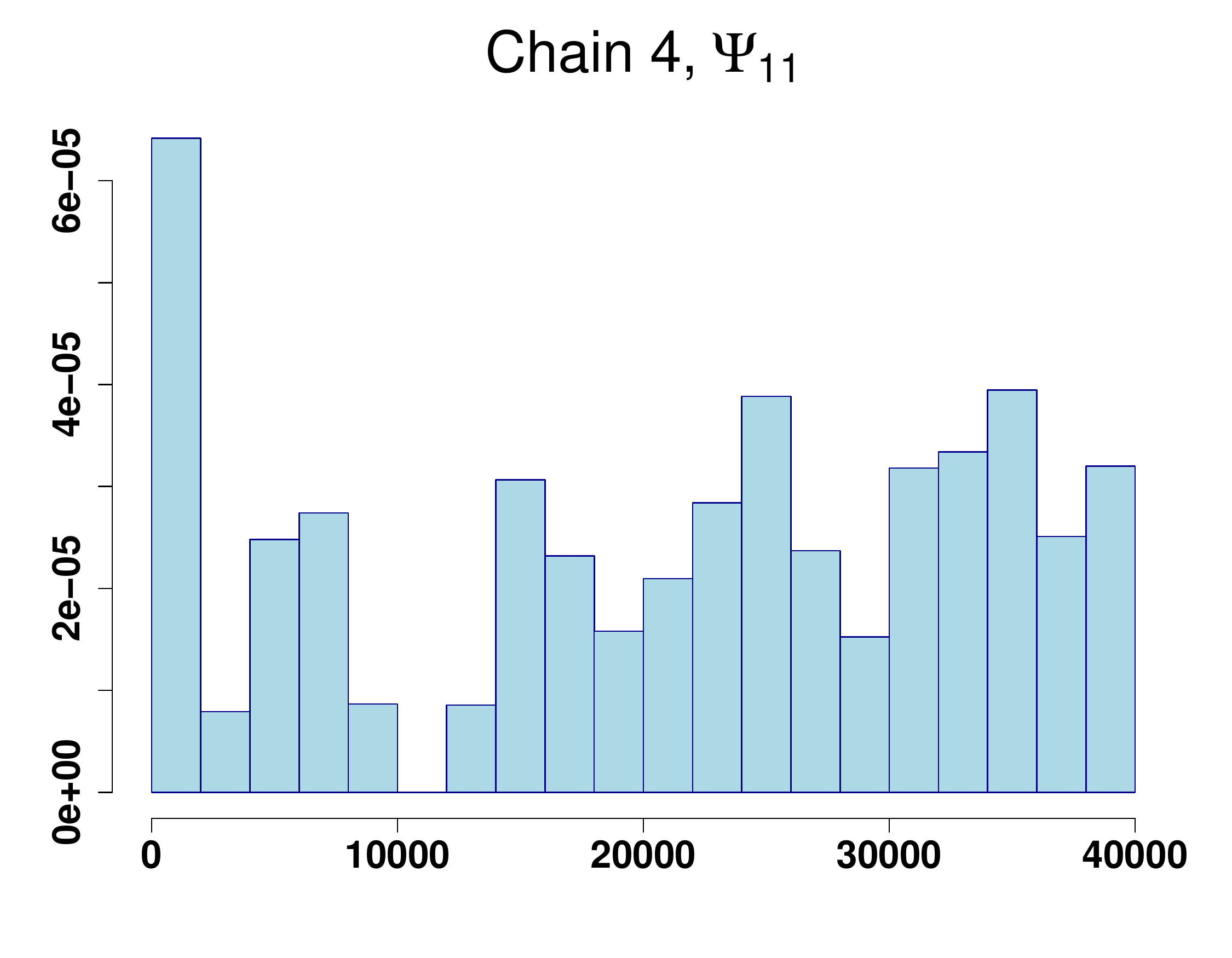}\\
\hspace{0.0cm}\includegraphics[width=4.0cm]{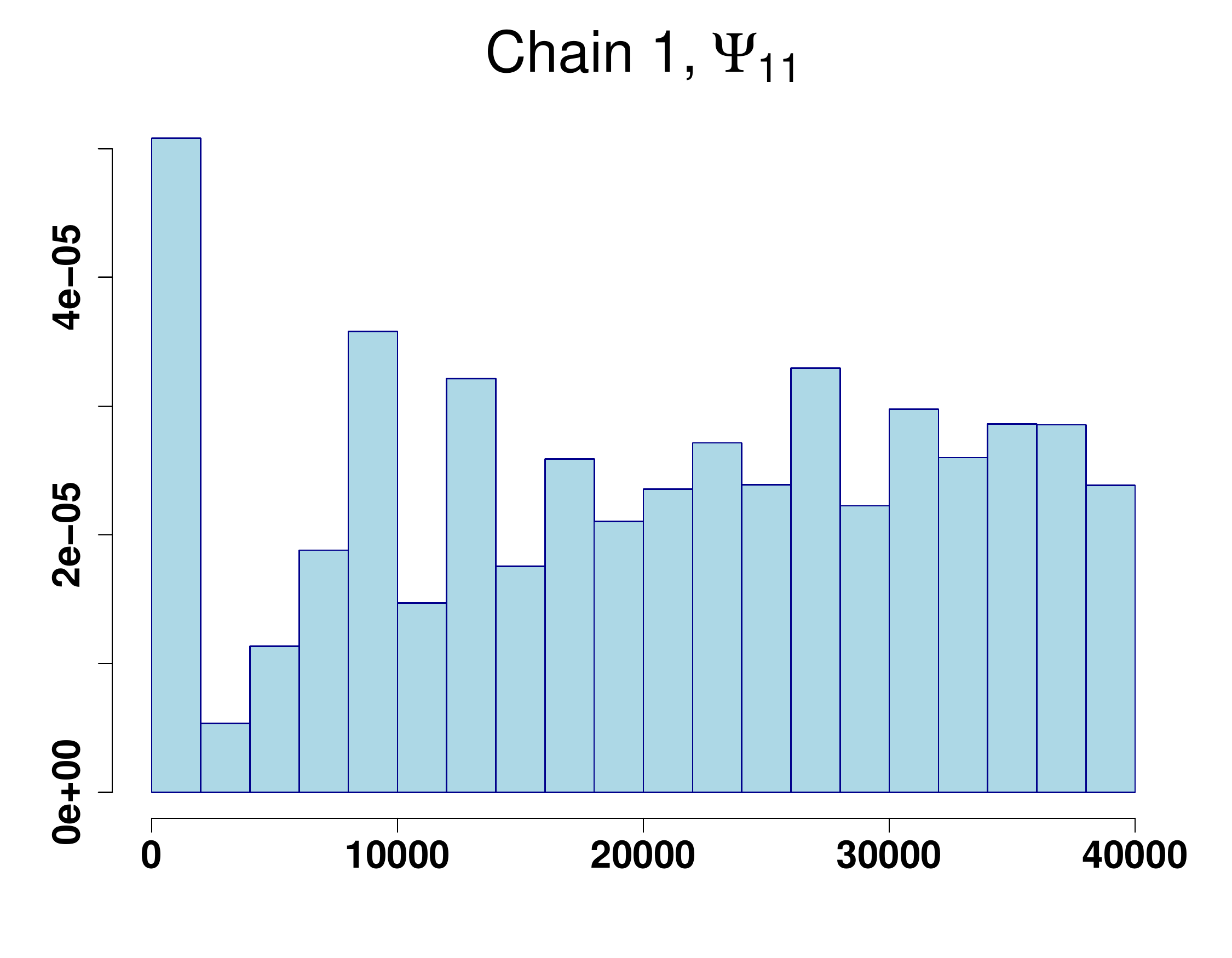}&\hspace{-0.5cm}\includegraphics[width=4.0cm]{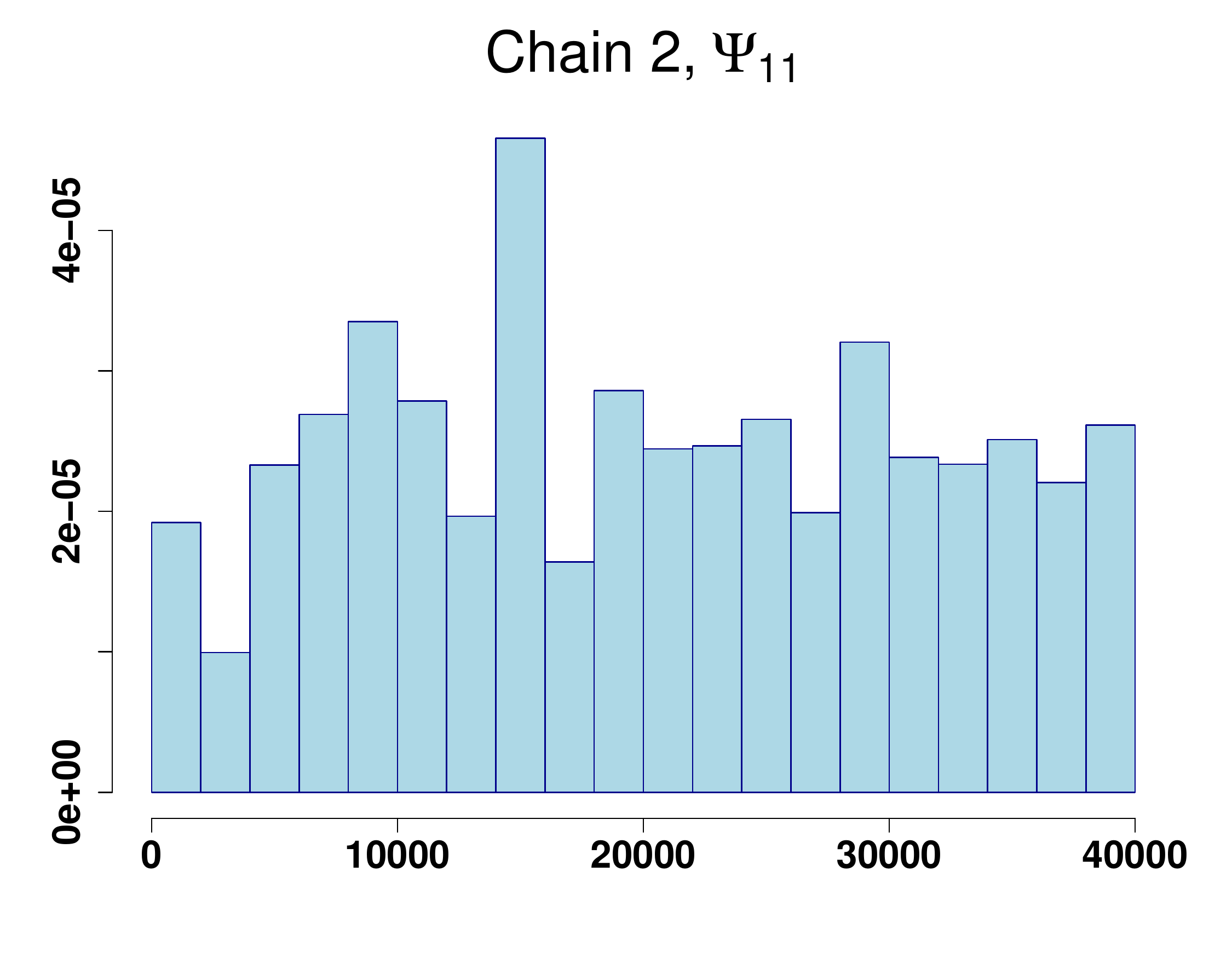}&
\hspace{-1.0cm}\includegraphics[width=4.0cm]{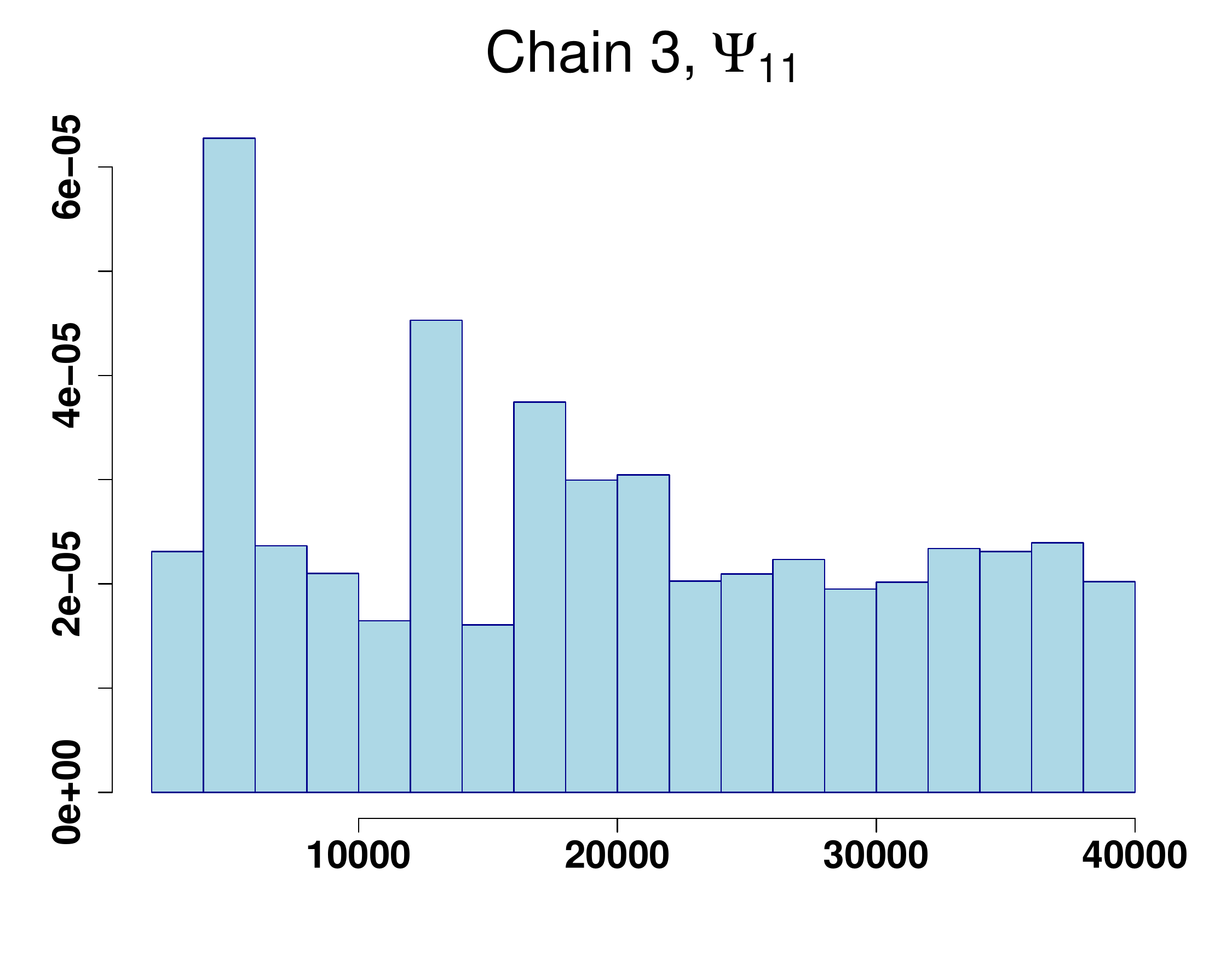}&\hspace{-1.5cm}\includegraphics[width=4.0cm]{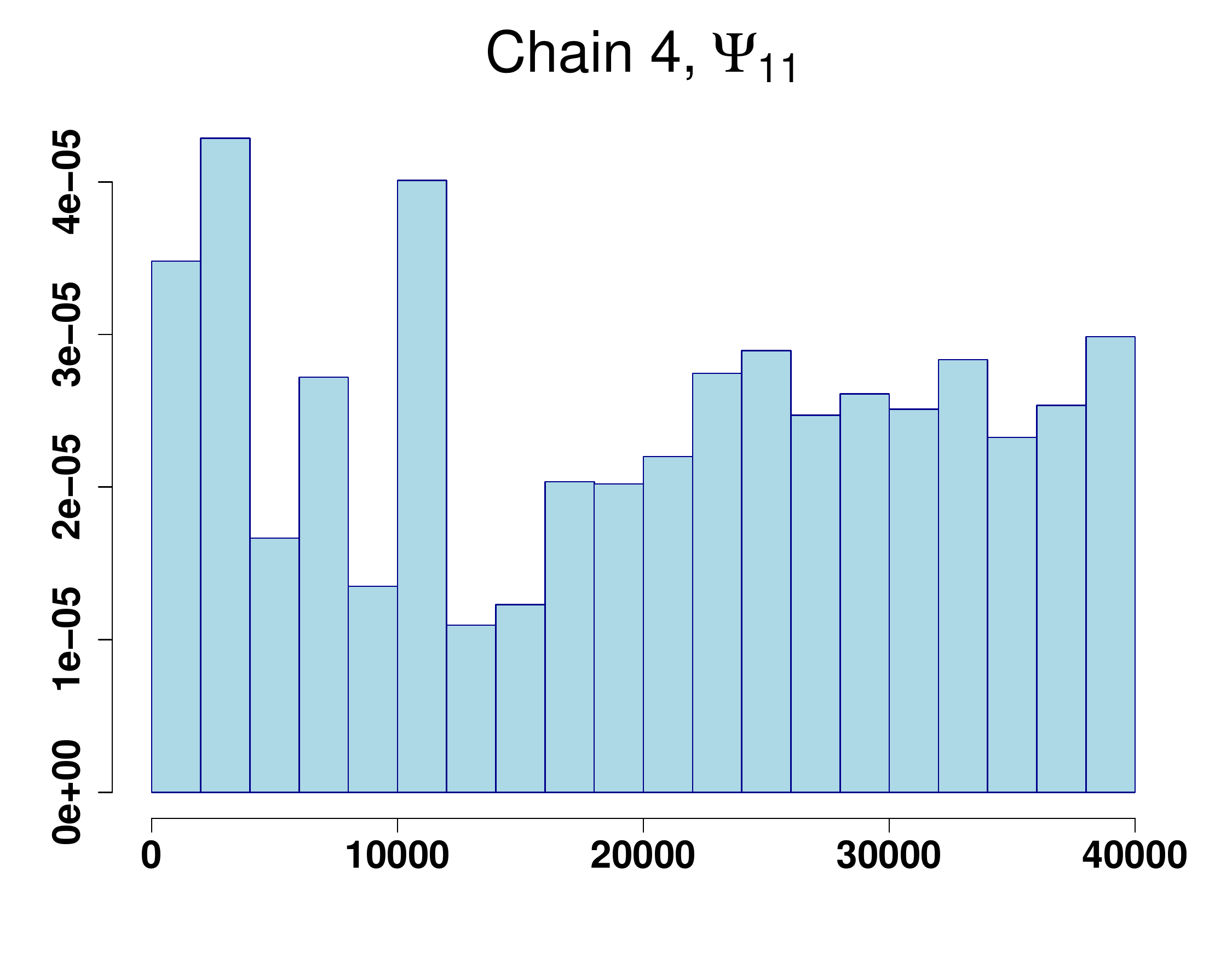}\\
\hspace{0.0cm}\includegraphics[width=4.0cm]{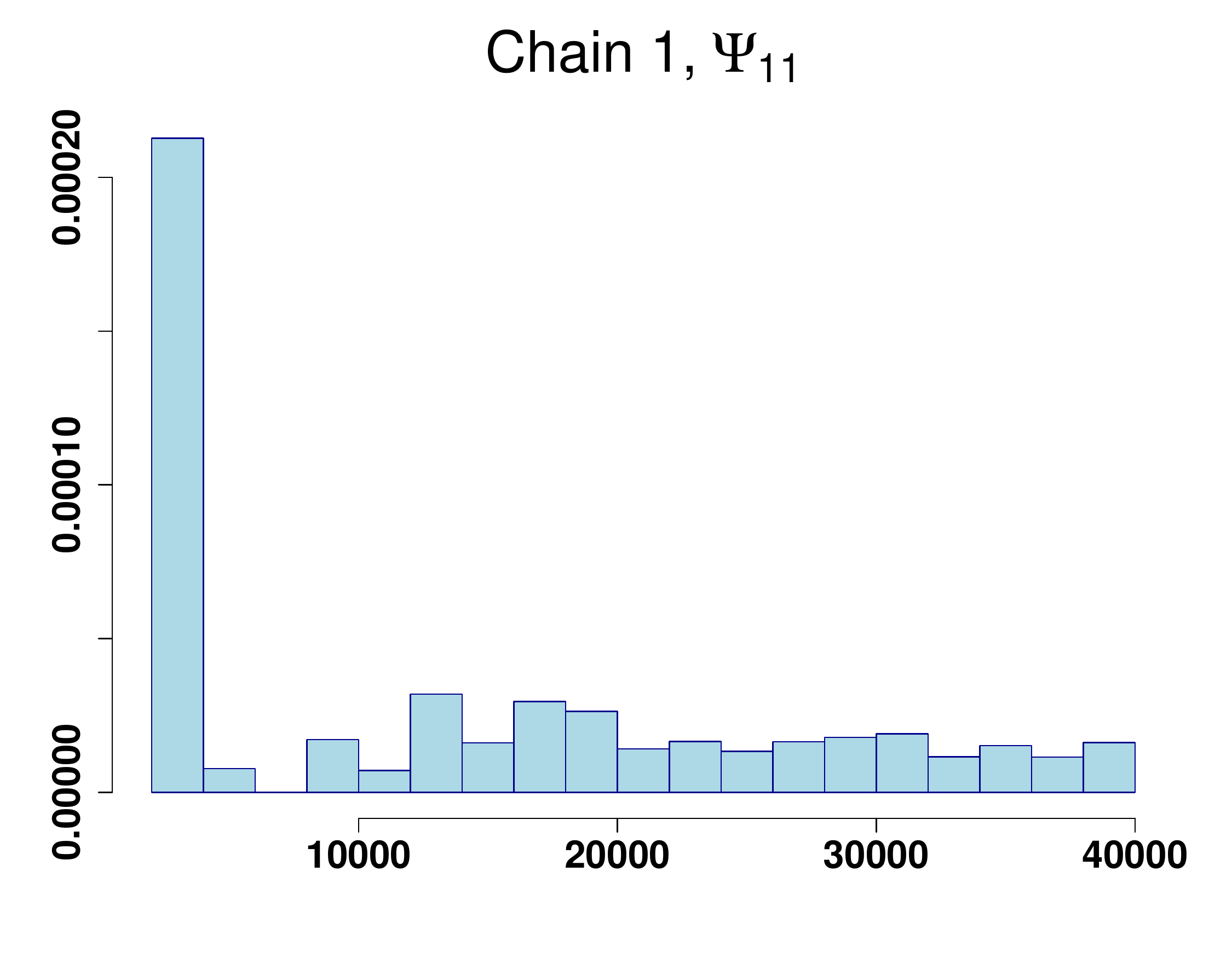}&\hspace{-0.5cm}\includegraphics[width=4.0cm]{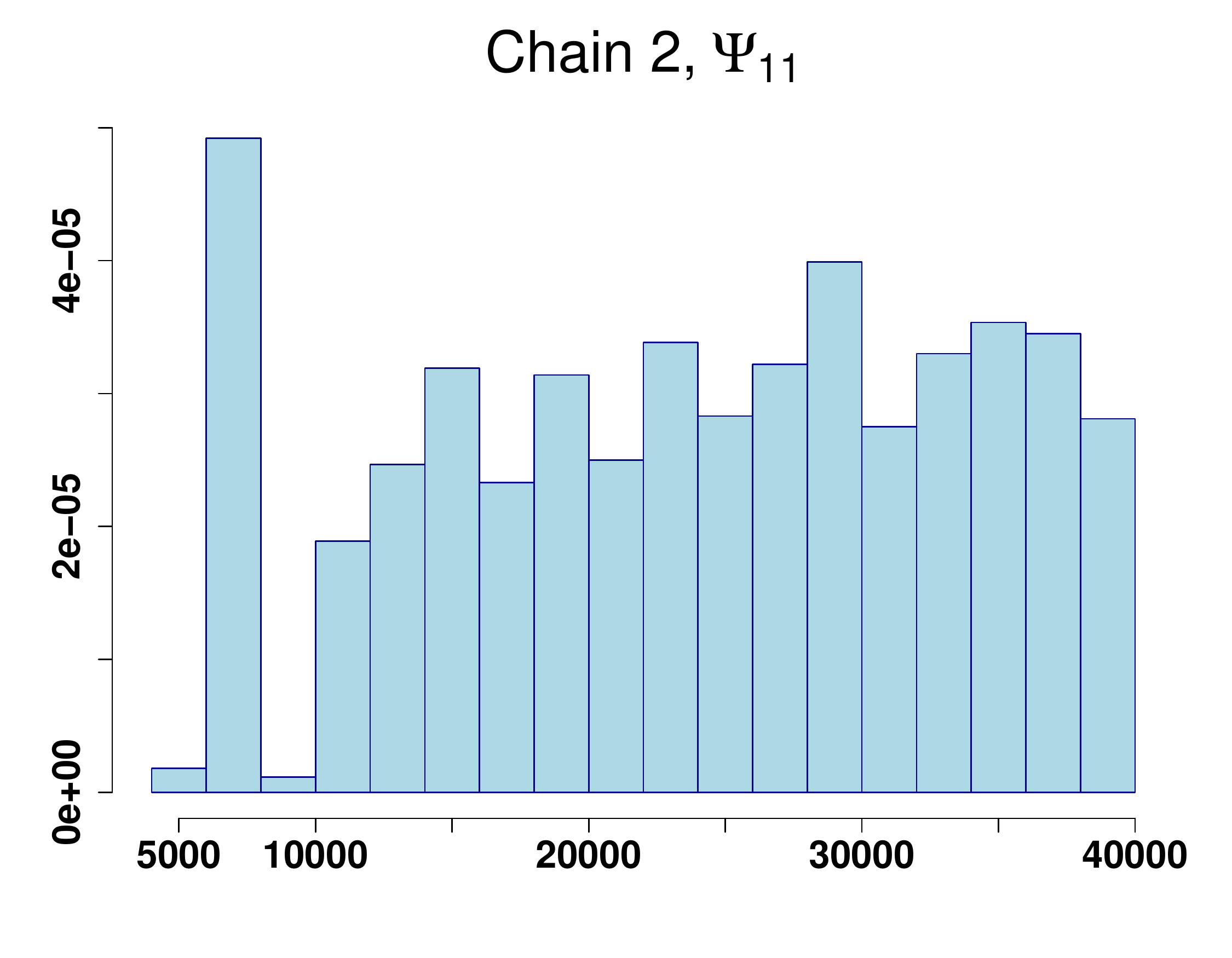}&
\hspace{-1.0cm}\includegraphics[width=4.0cm]{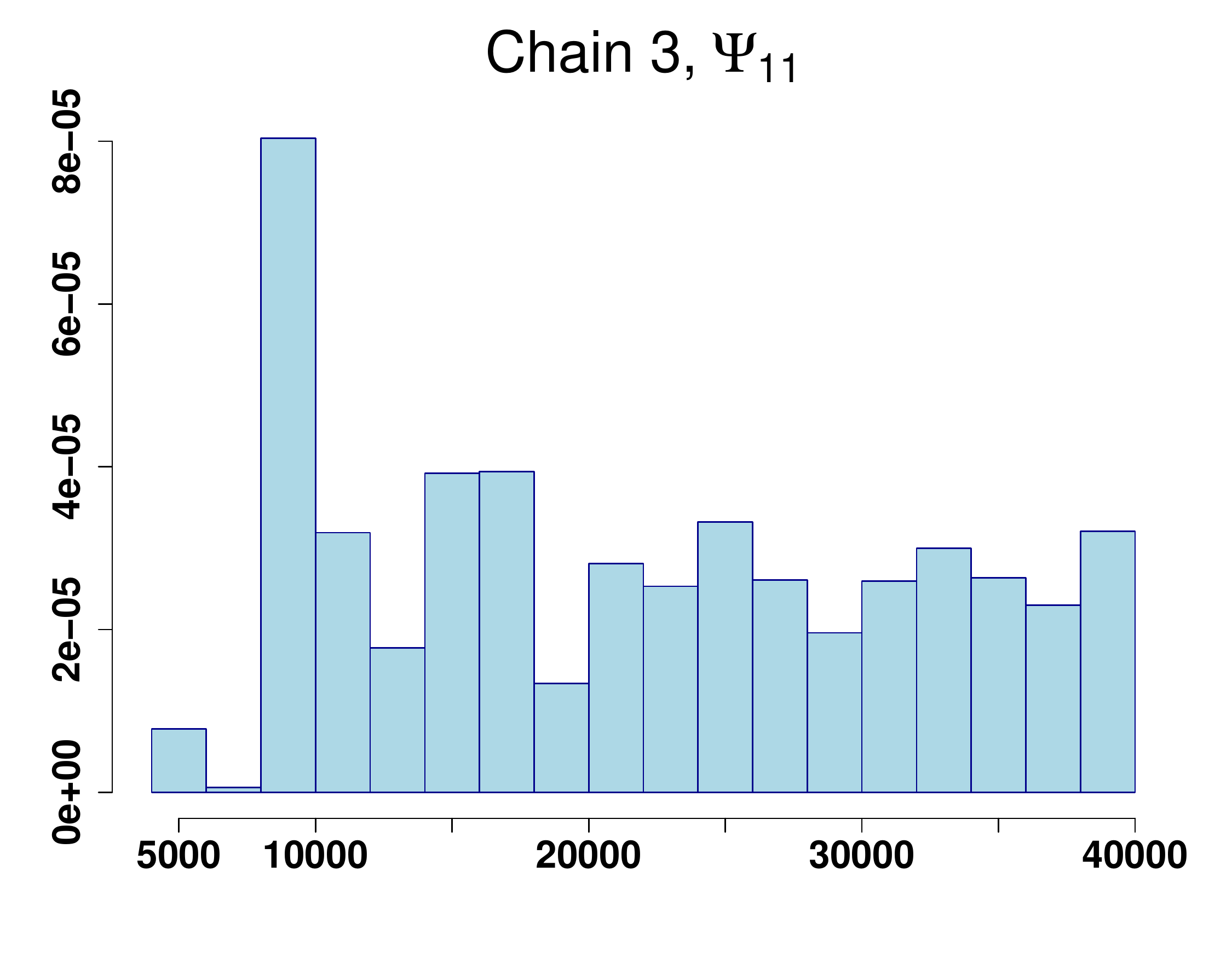}&\hspace{-1.5cm}\includegraphics[width=4.0cm]{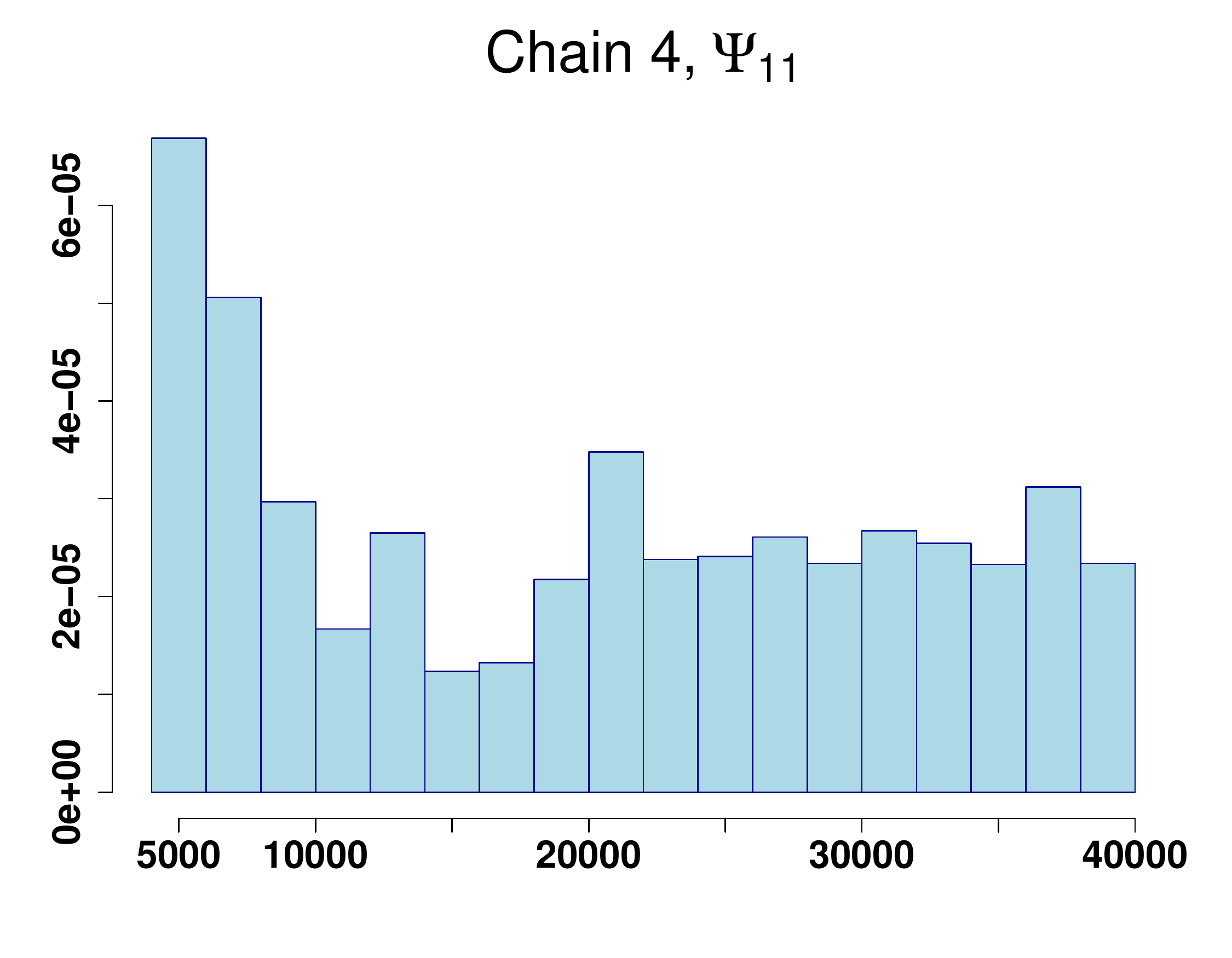}\\
\hspace{0.0cm}\includegraphics[width=4.0cm]{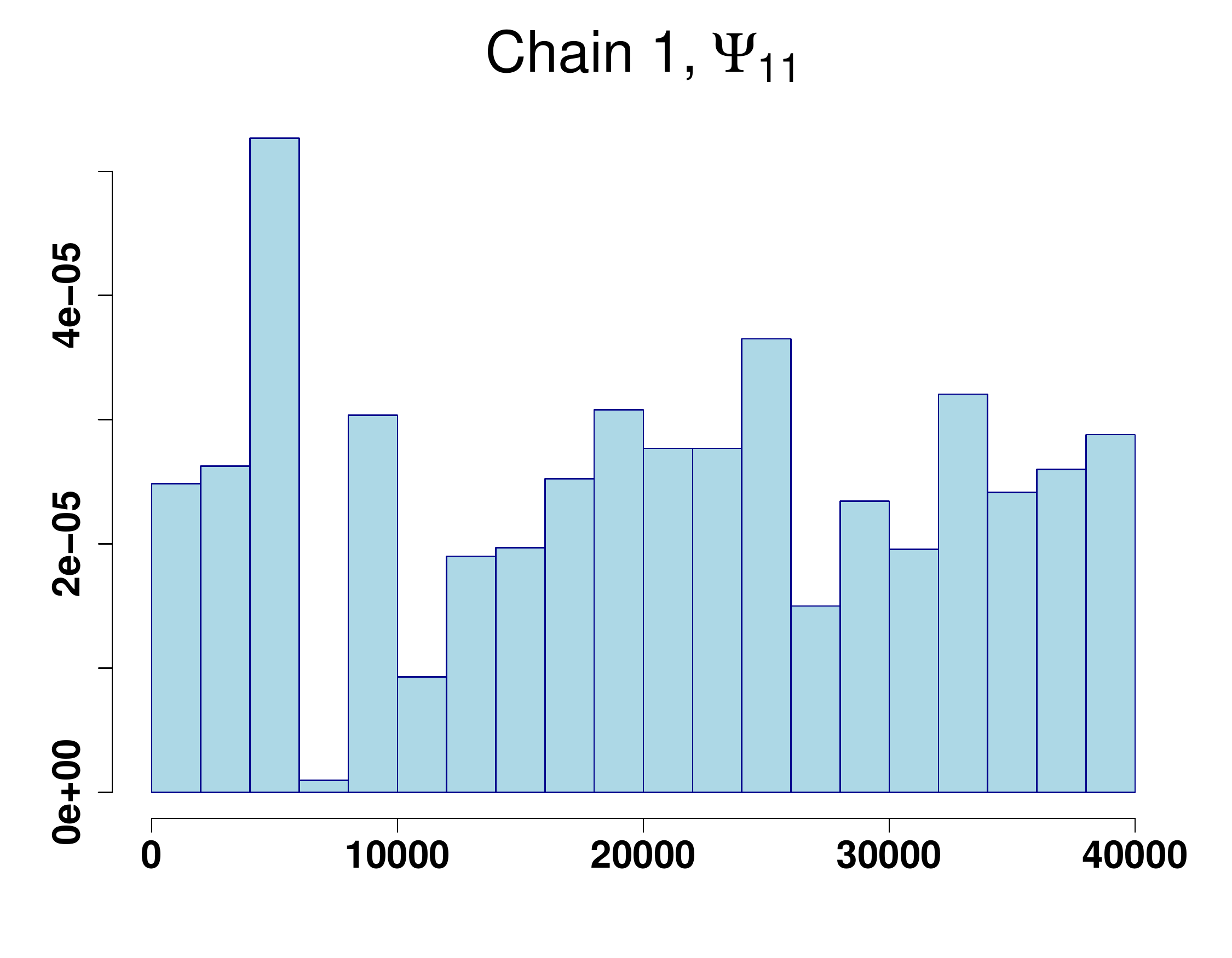}&\hspace{-0.5cm}\includegraphics[width=4.0cm]{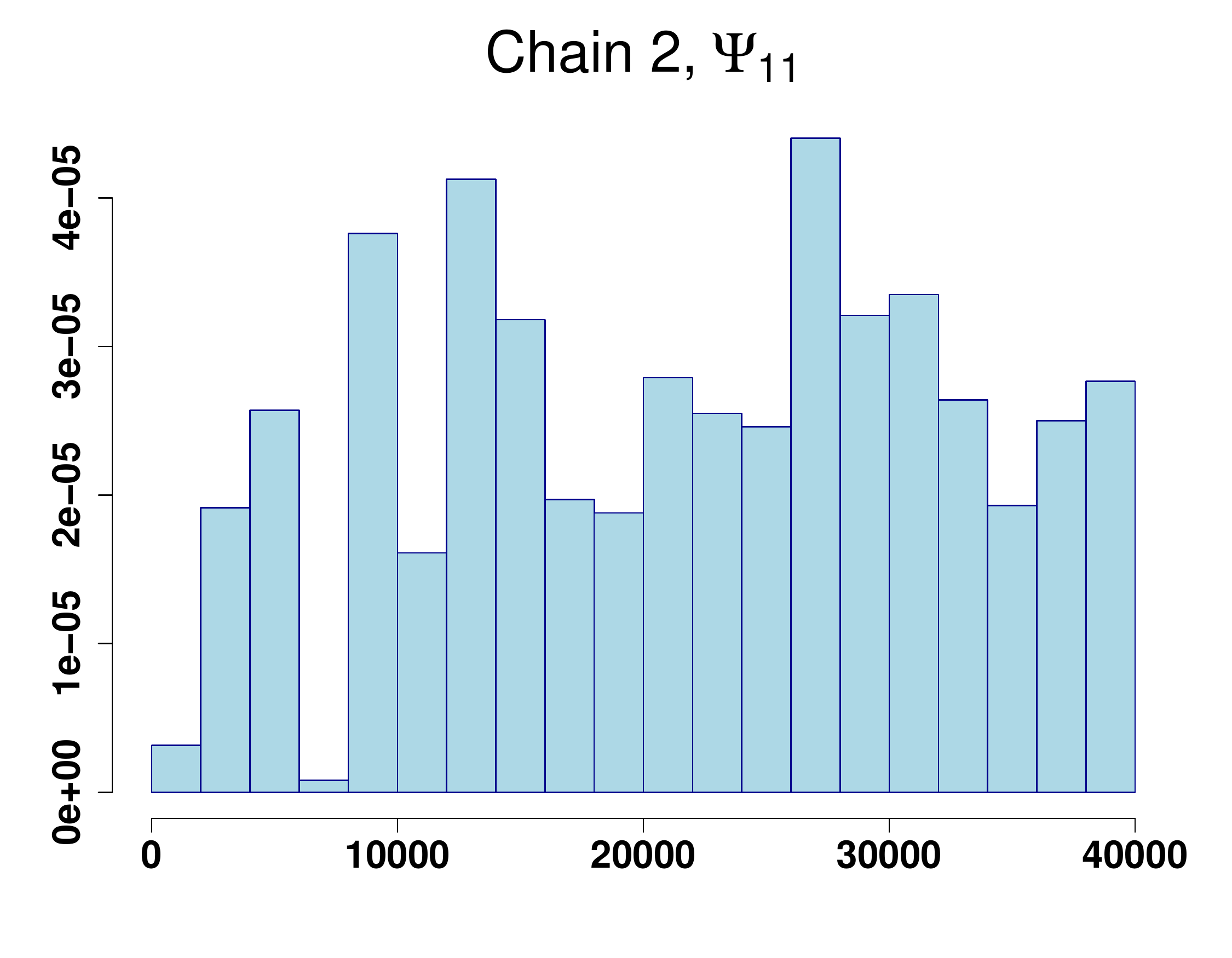}&
\hspace{-1.0cm}\includegraphics[width=4.0cm]{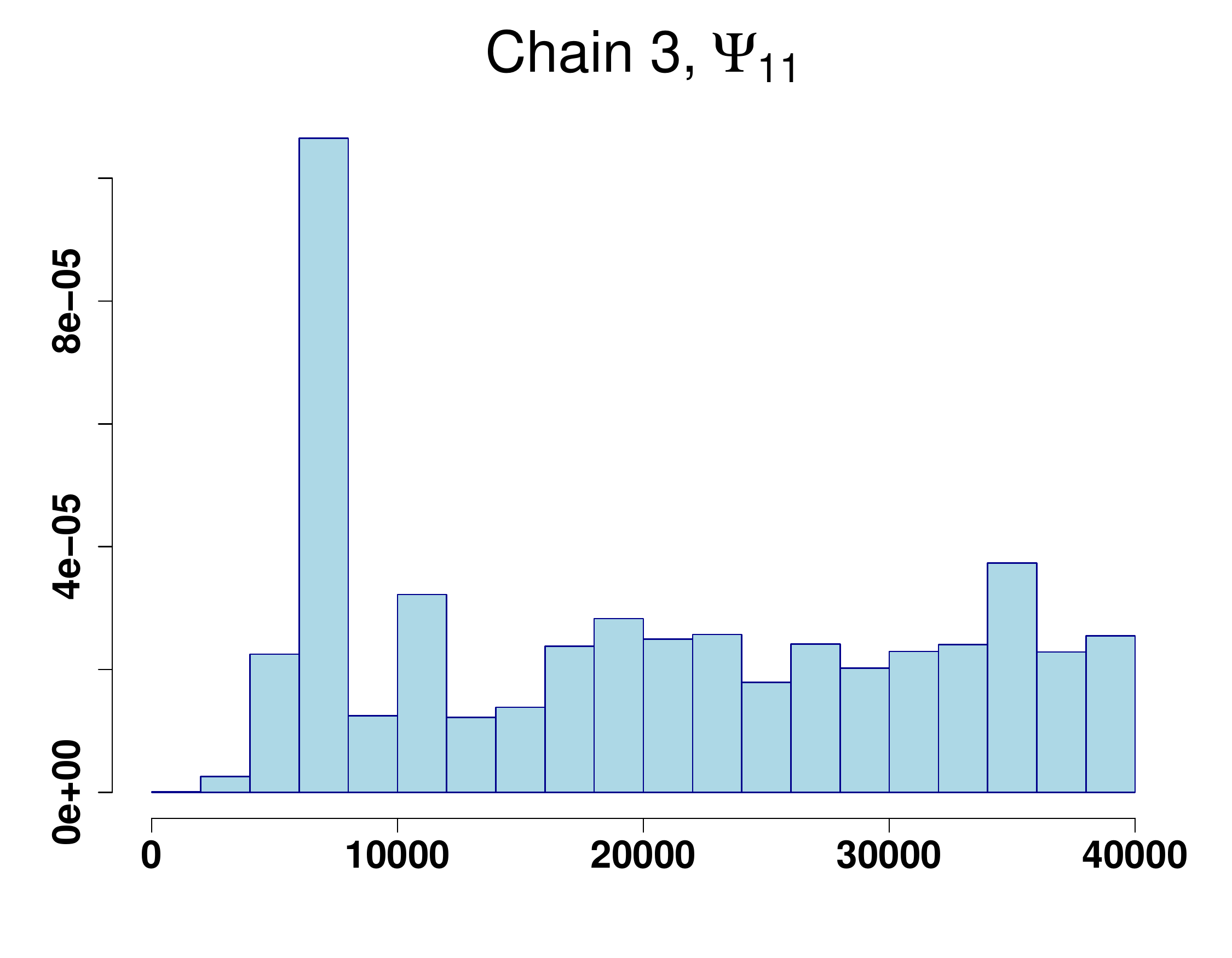}&\hspace{-1.5cm}\includegraphics[width=4.0cm]{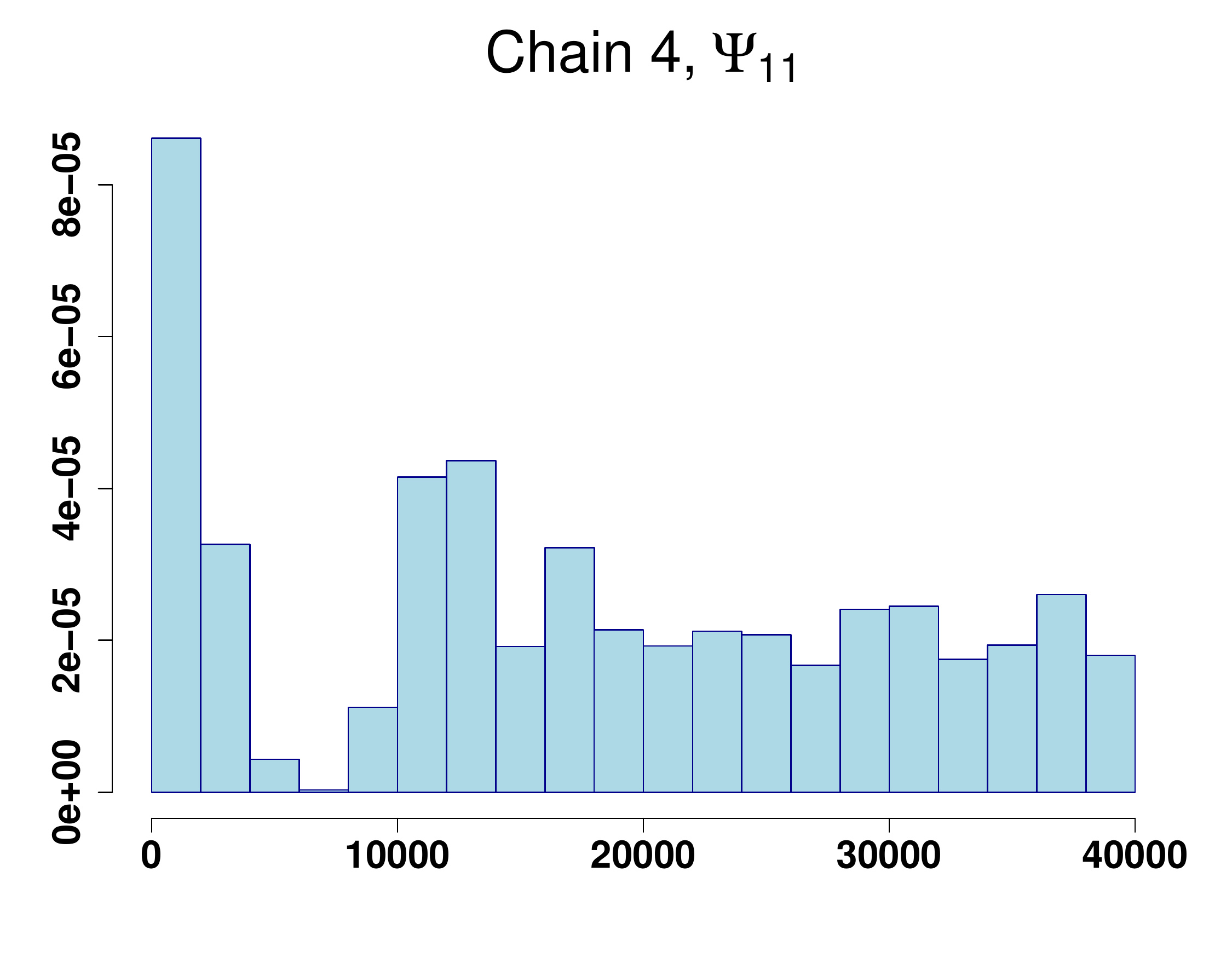}\\
\hspace{0.0cm}\includegraphics[width=4.0cm]{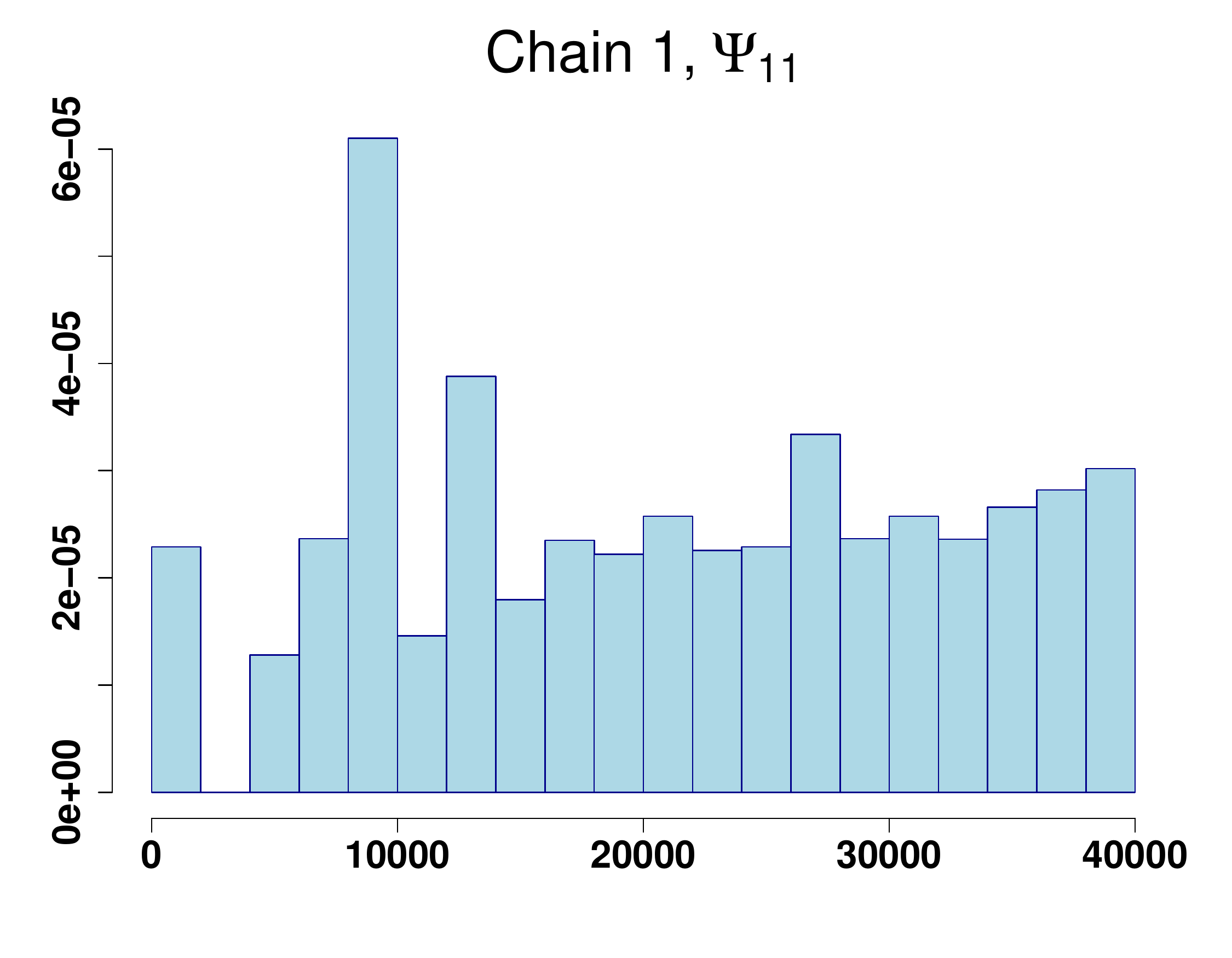}&\hspace{-0.5cm}\includegraphics[width=4.0cm]{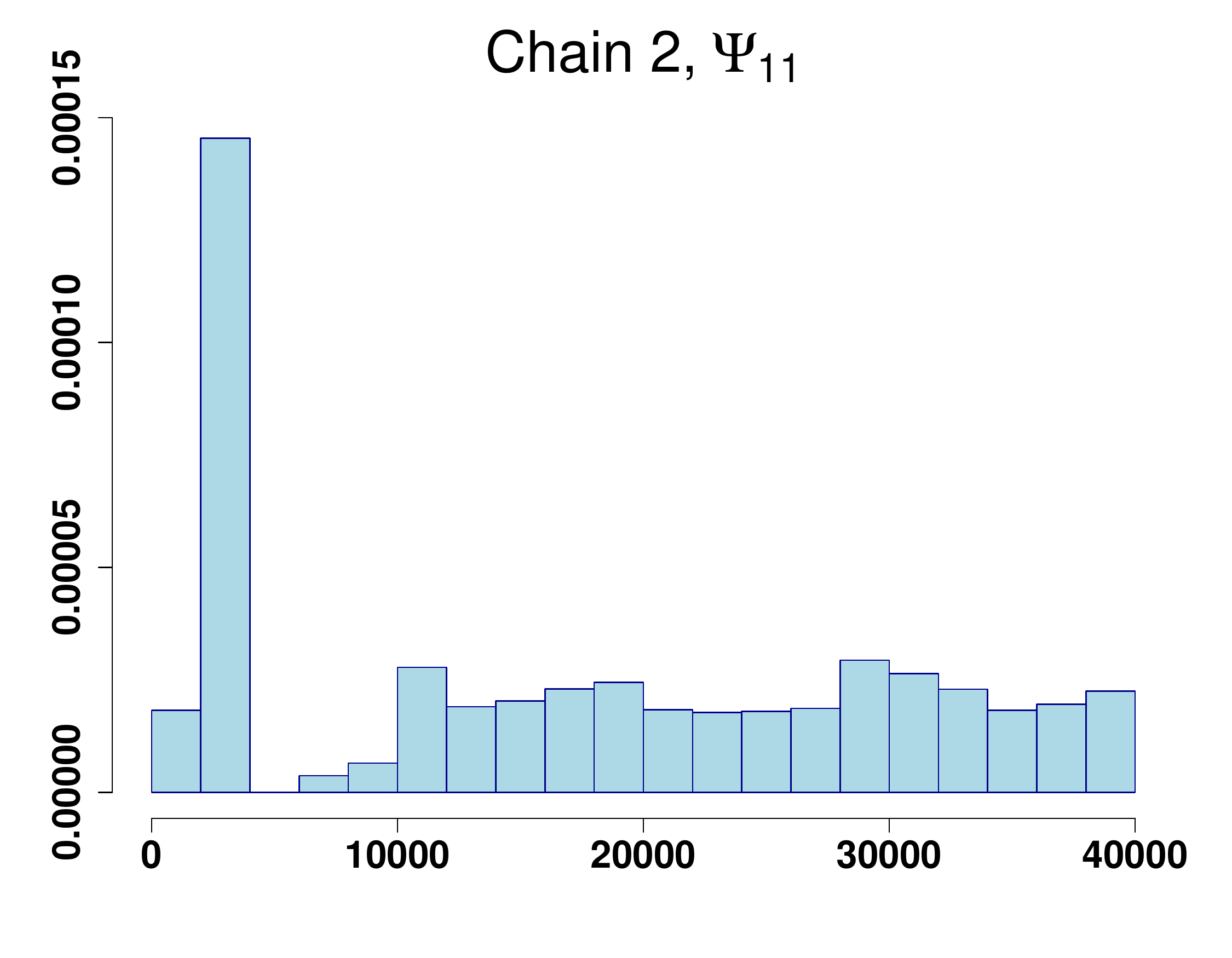}&
\hspace{-1.0cm}\includegraphics[width=4.0cm]{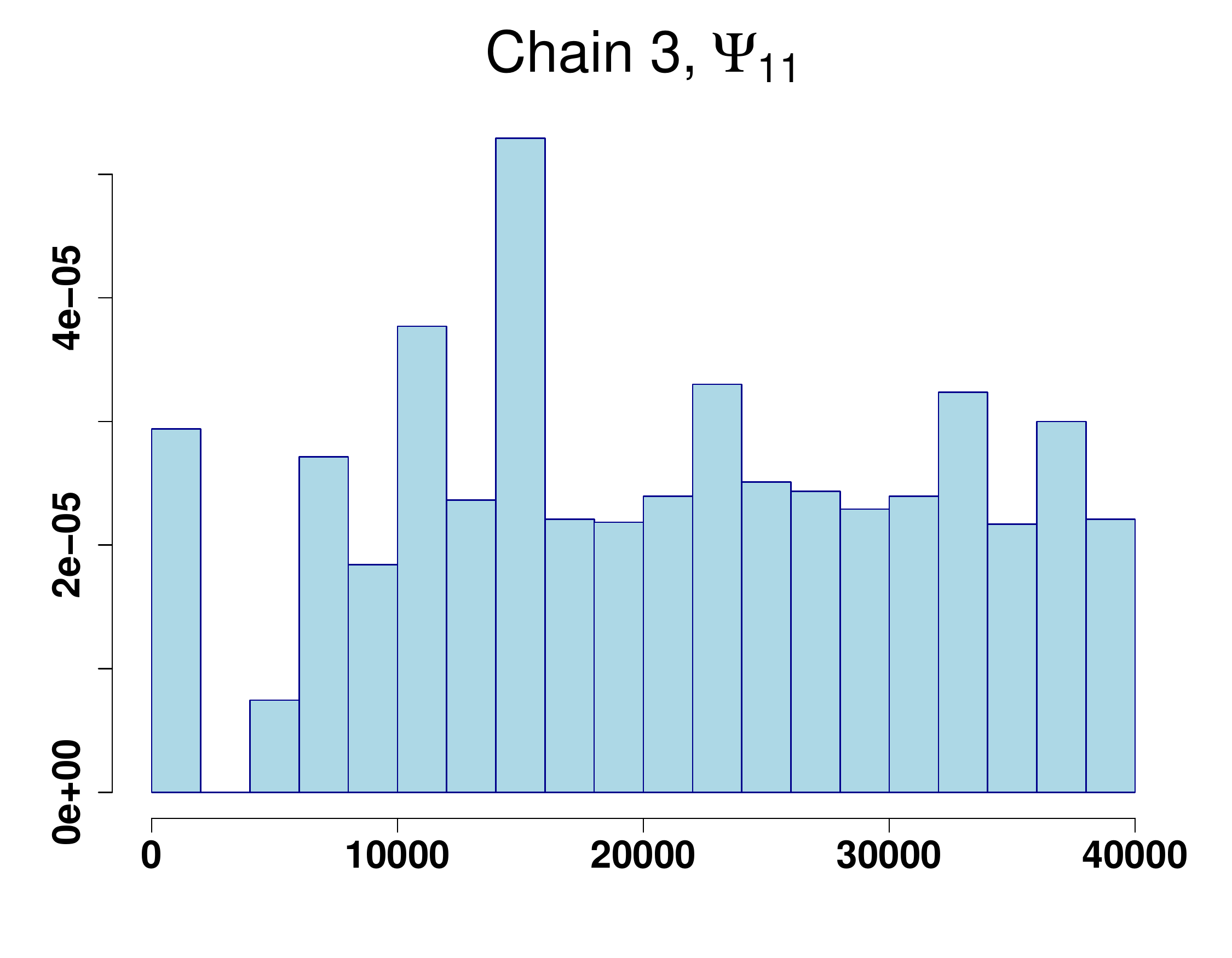}&\hspace{-1.5cm}\includegraphics[width=4.0cm]{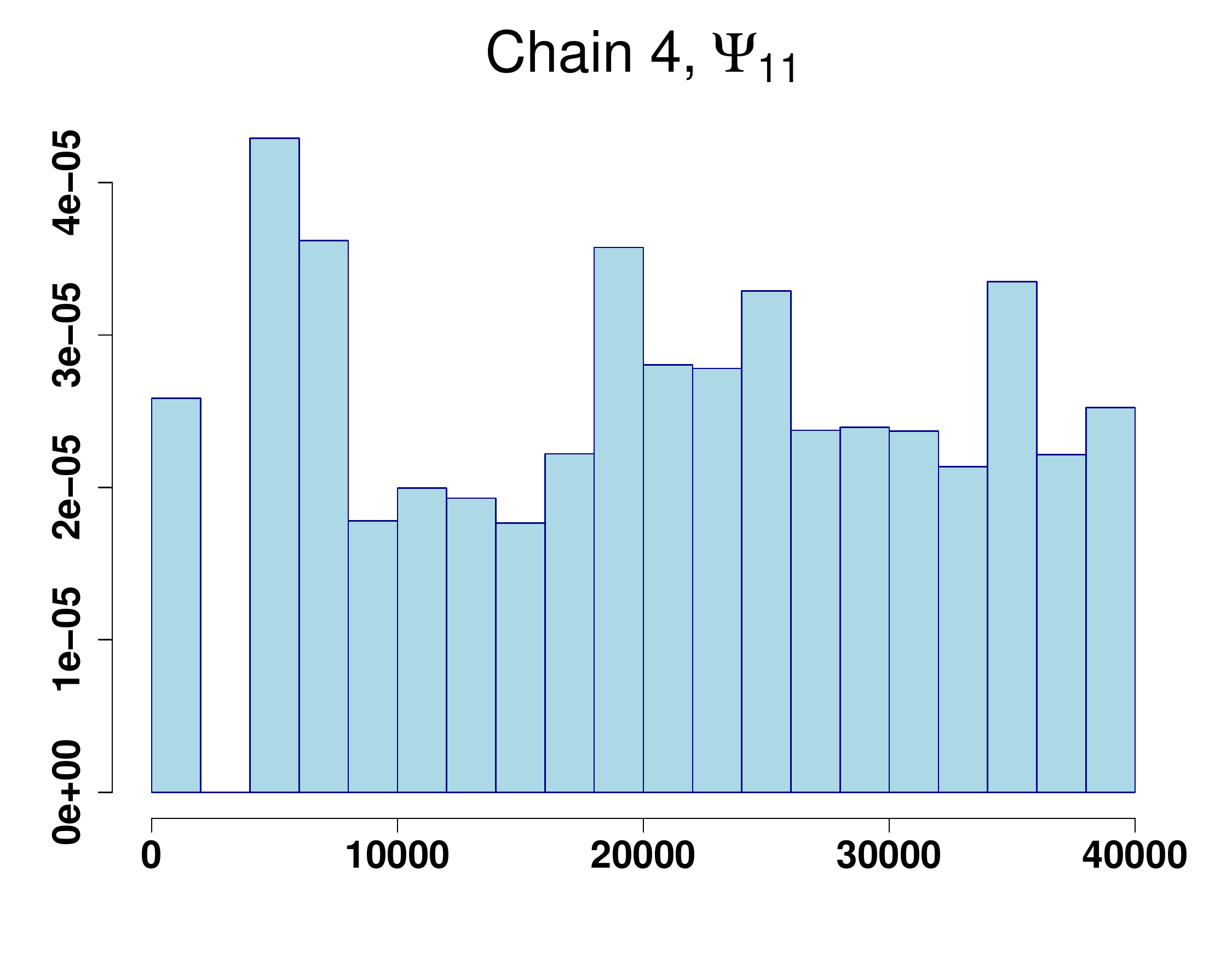}\\
\end{tabular}
 \caption{Rank plots of posterior draws from four chains in the case of the parameter $\Psi_{11}$ (SBP) of the normal multivariate random effects model by employing the Jeffreys prior (first to third rows) and the Berger and Bernardo reference prior (fourth to sixth rows). The samples from the posterior distributions are drawn by Algorithm A (first and fourth rows), Algorithm B (second and fifth rows) and Algorithm C (third and sixth rows).}
\label{fig:emp-study-rank-Psi11-nor}
 \end{figure}

\begin{figure}[h!t]
\centering
\begin{tabular}{p{4.0cm}p{4.0cm}p{4.0cm}p{4.0cm}}
\hspace{0.0cm}\includegraphics[width=4.0cm]{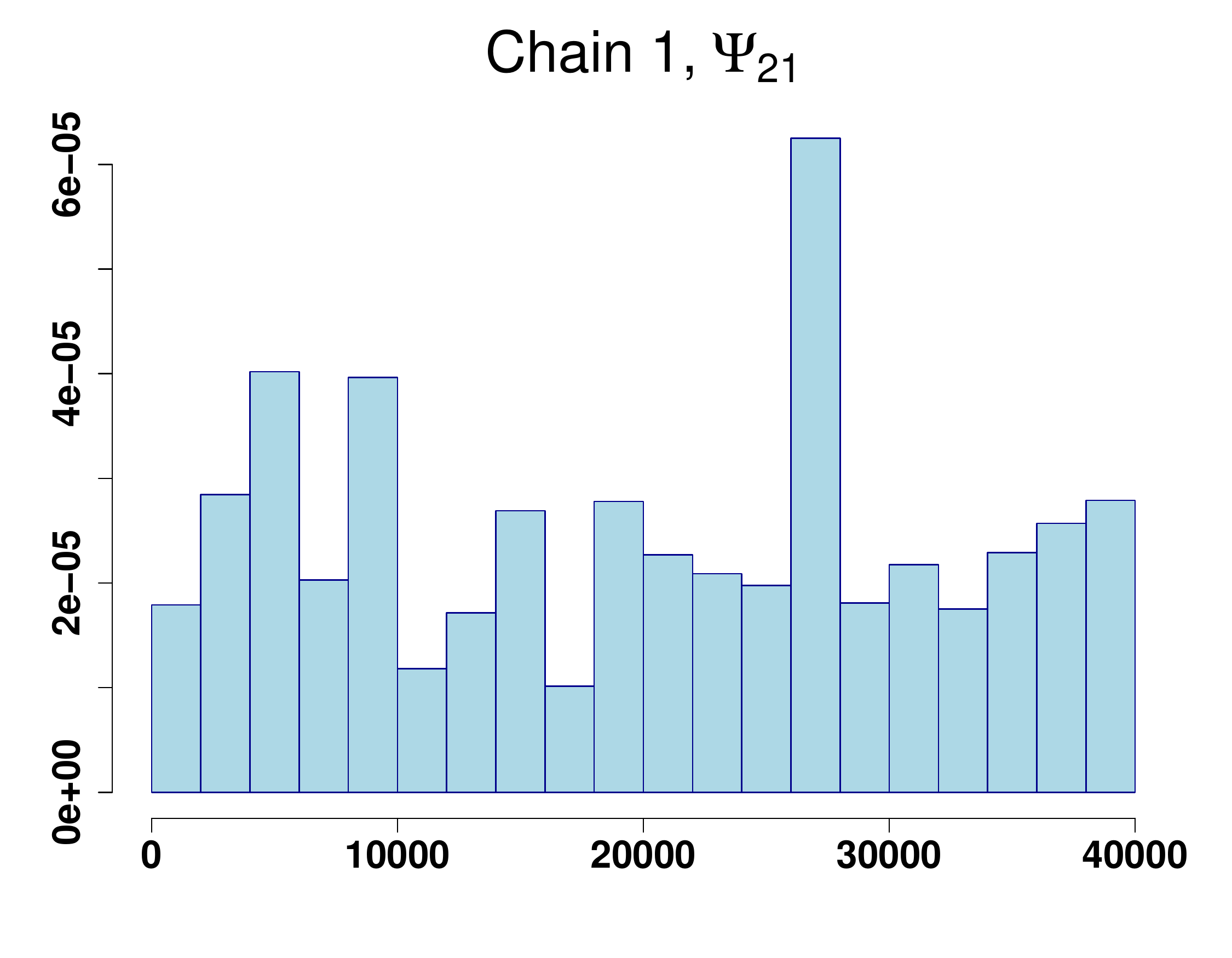}&\hspace{-0.5cm}\includegraphics[width=4.0cm]{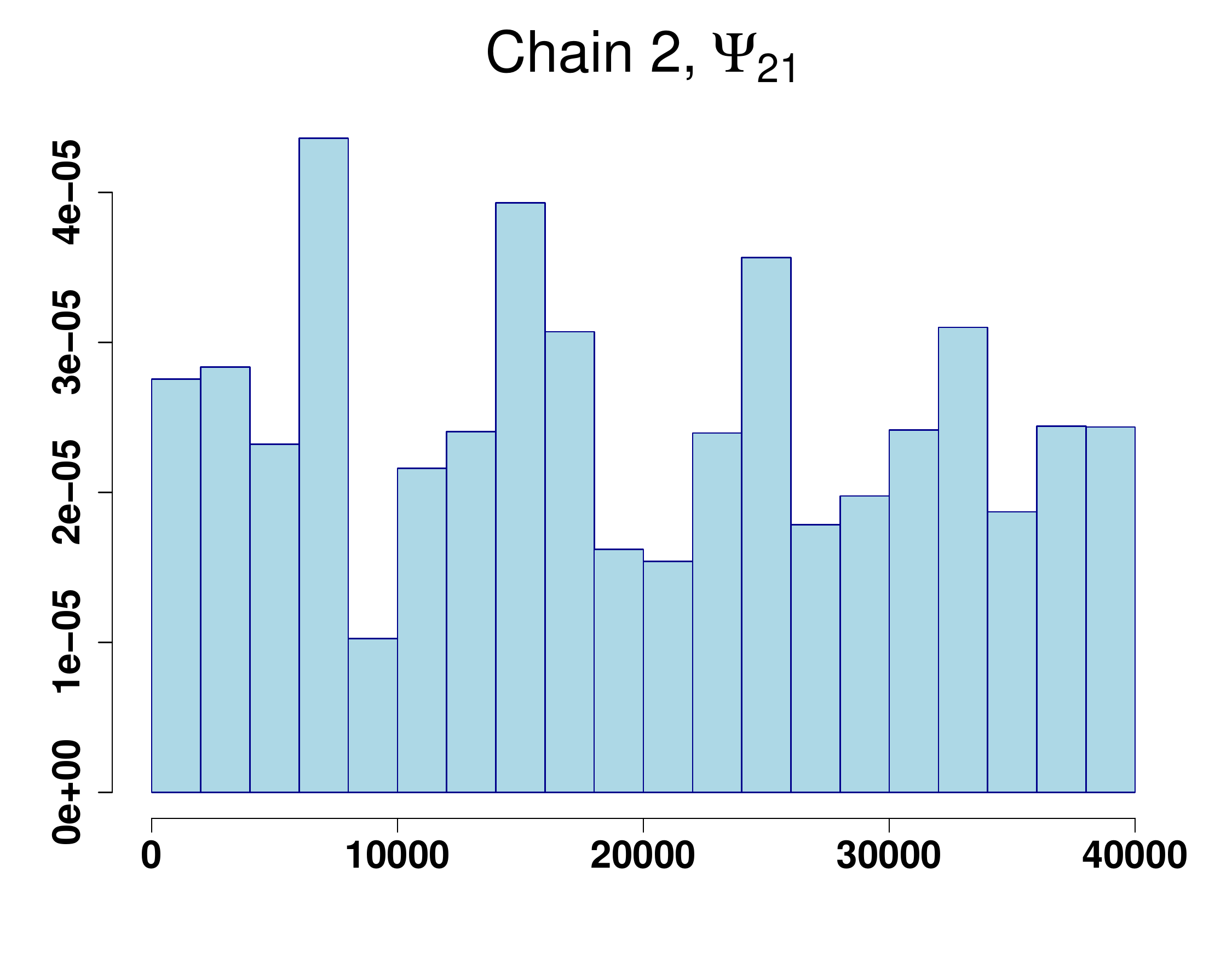}&
\hspace{-1.0cm}\includegraphics[width=4.0cm]{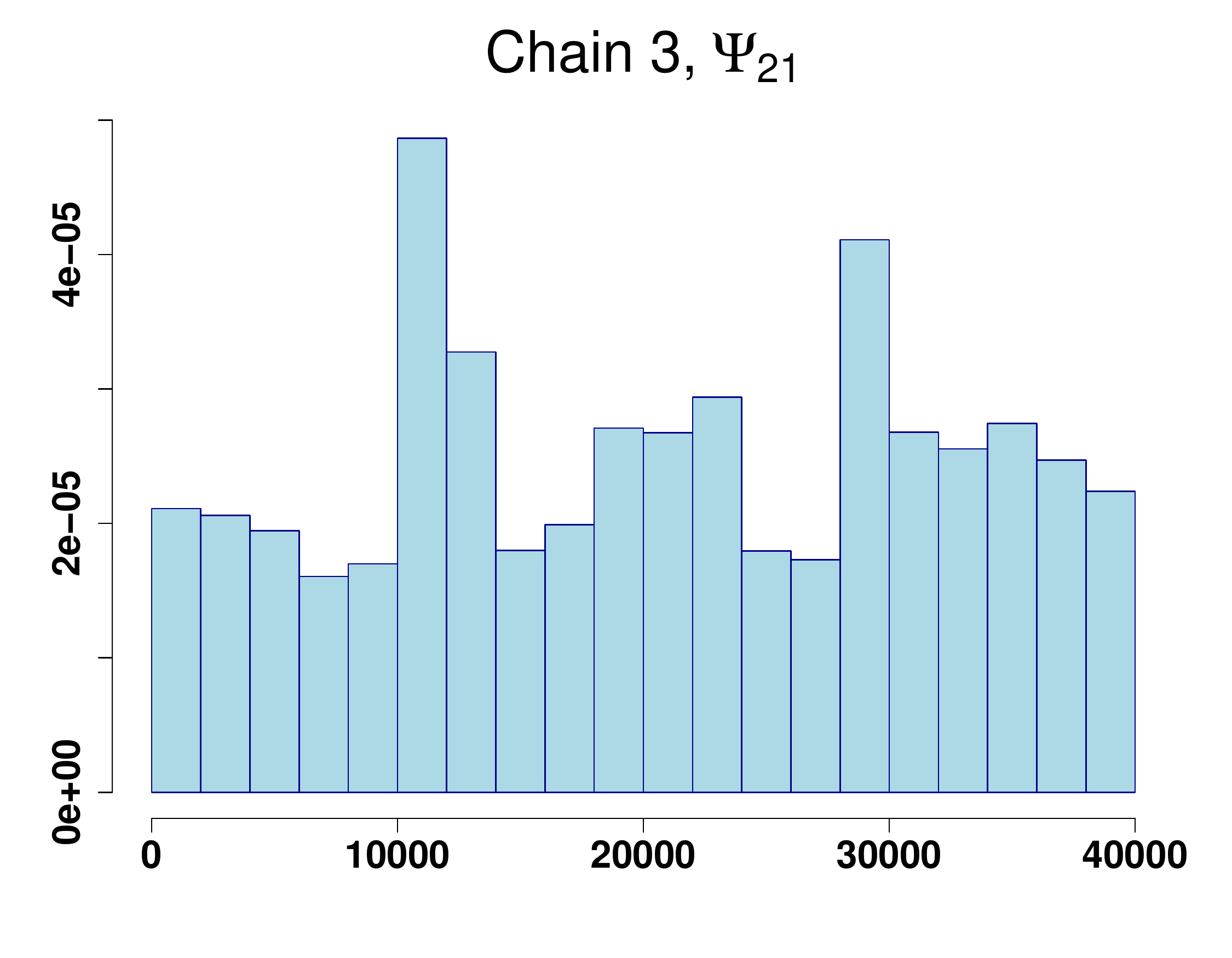}&\hspace{-1.5cm}\includegraphics[width=4.0cm]{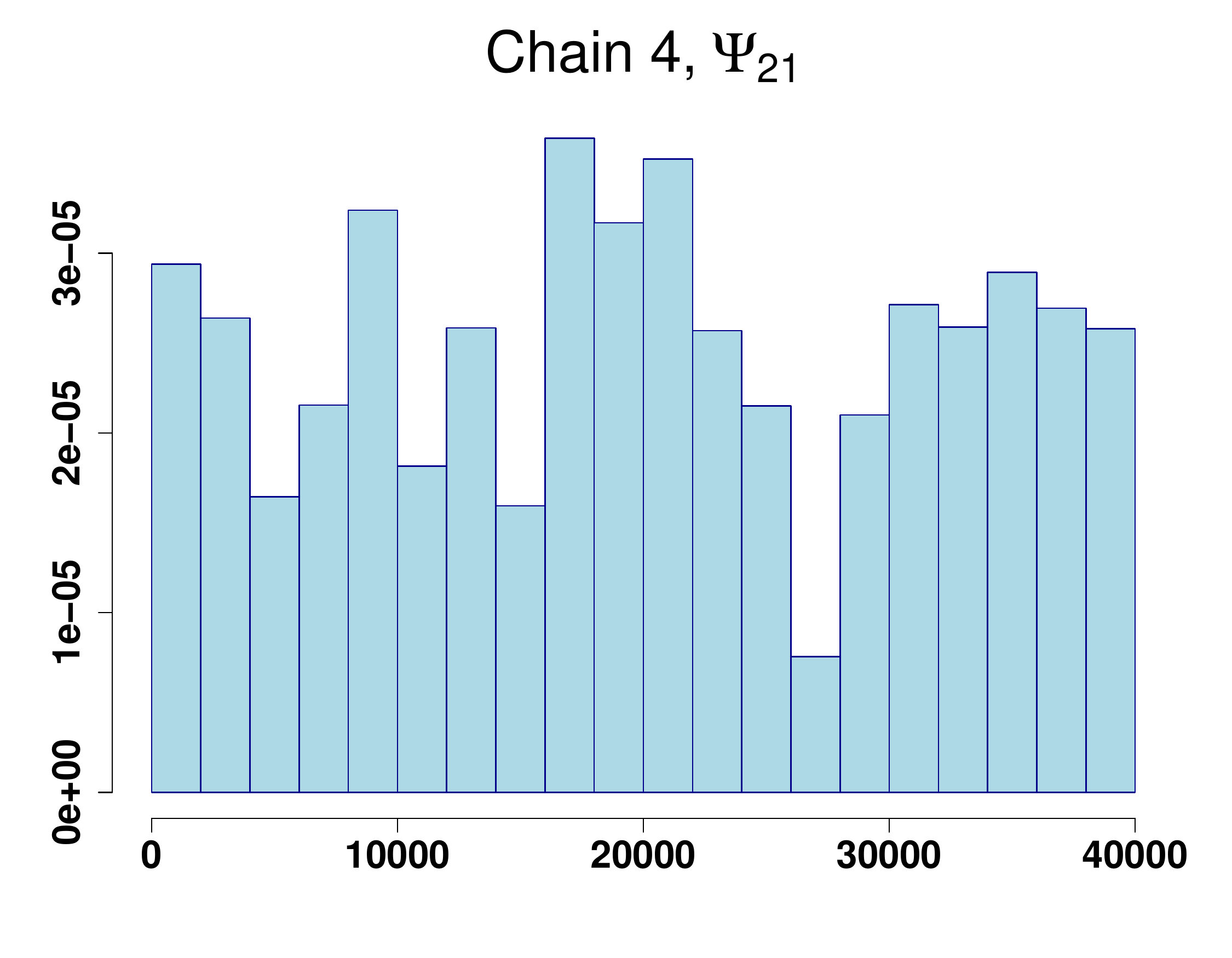}\\
\hspace{0.0cm}\includegraphics[width=4.0cm]{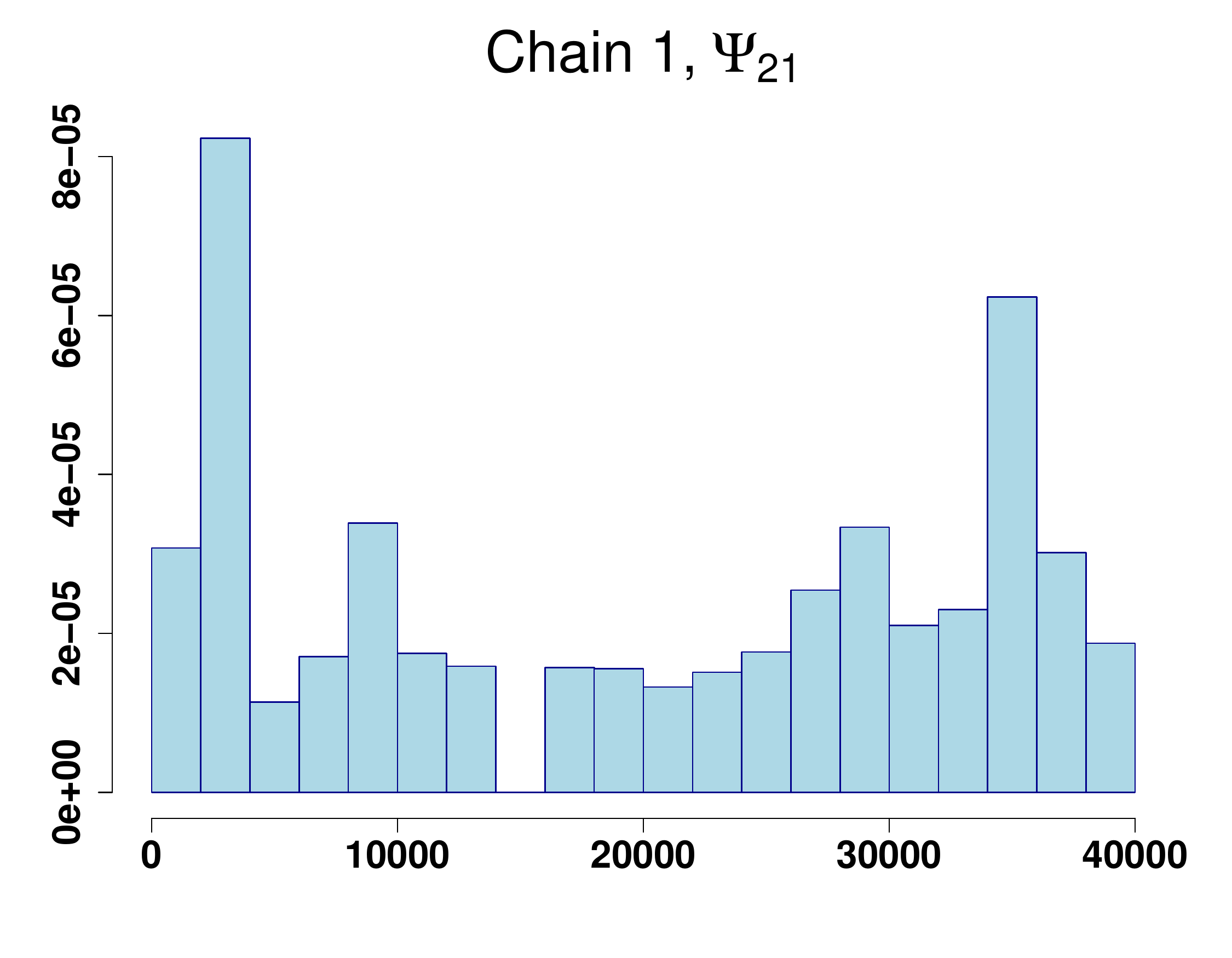}&\hspace{-0.5cm}\includegraphics[width=4.0cm]{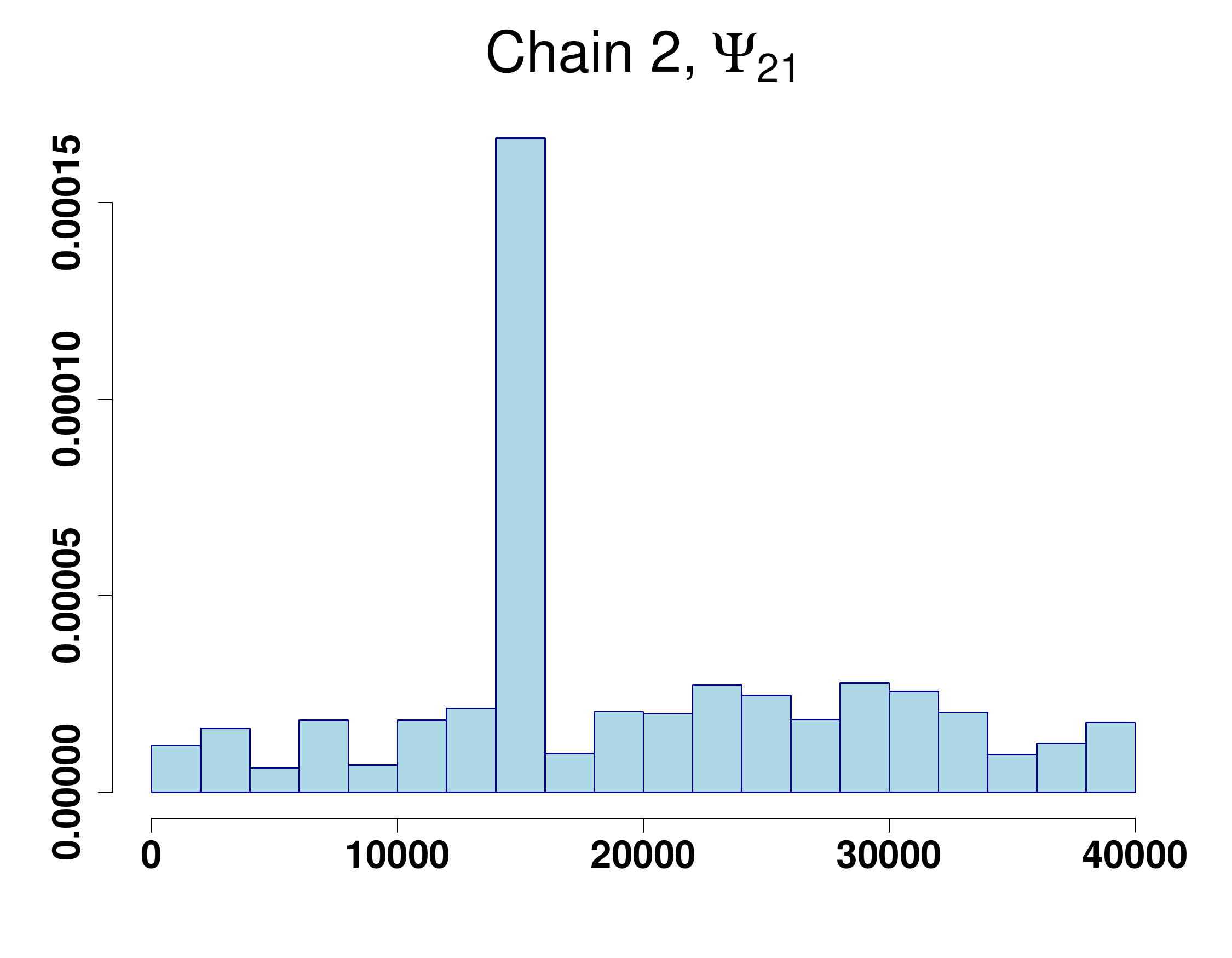}&
\hspace{-1.0cm}\includegraphics[width=4.0cm]{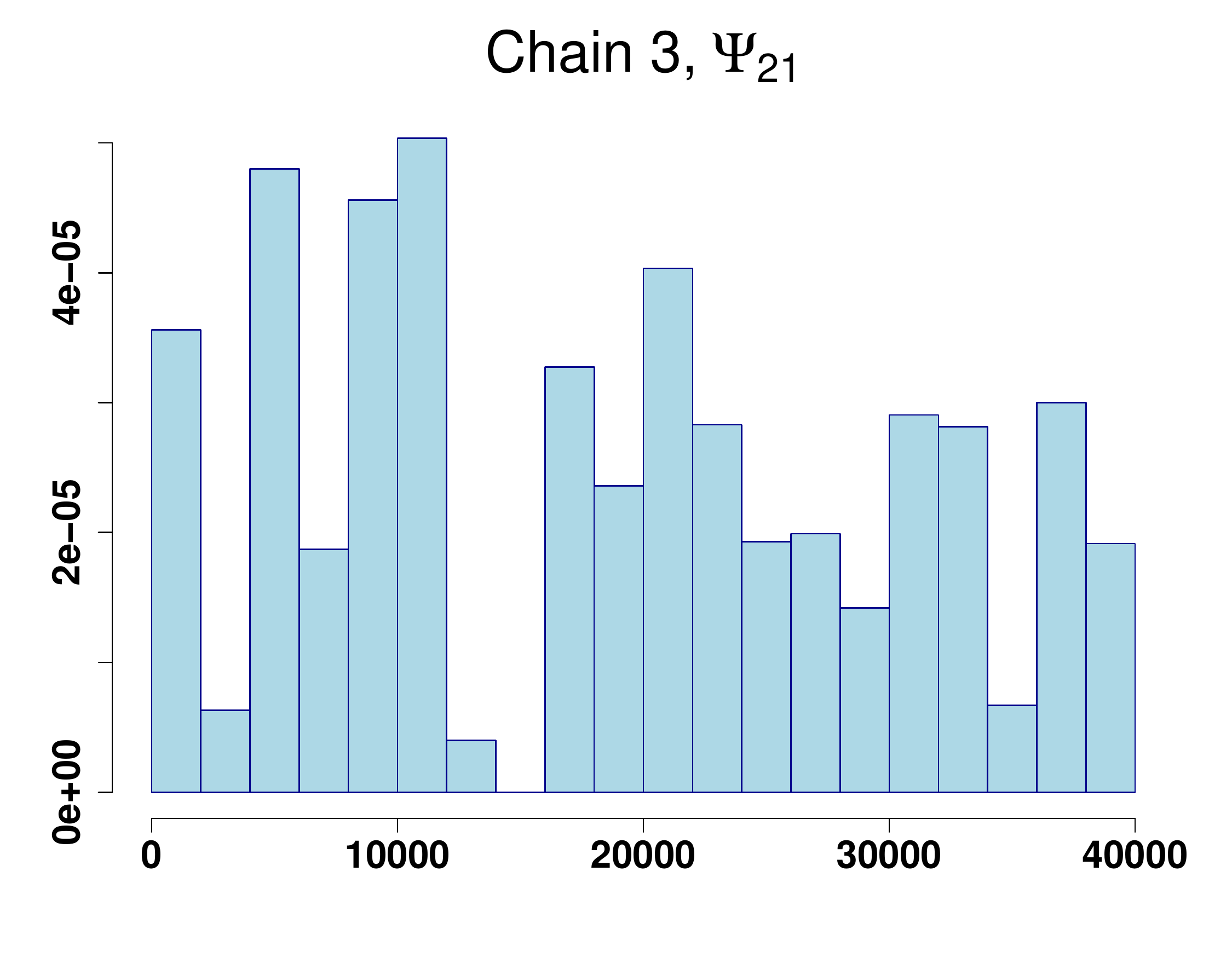}&\hspace{-1.5cm}\includegraphics[width=4.0cm]{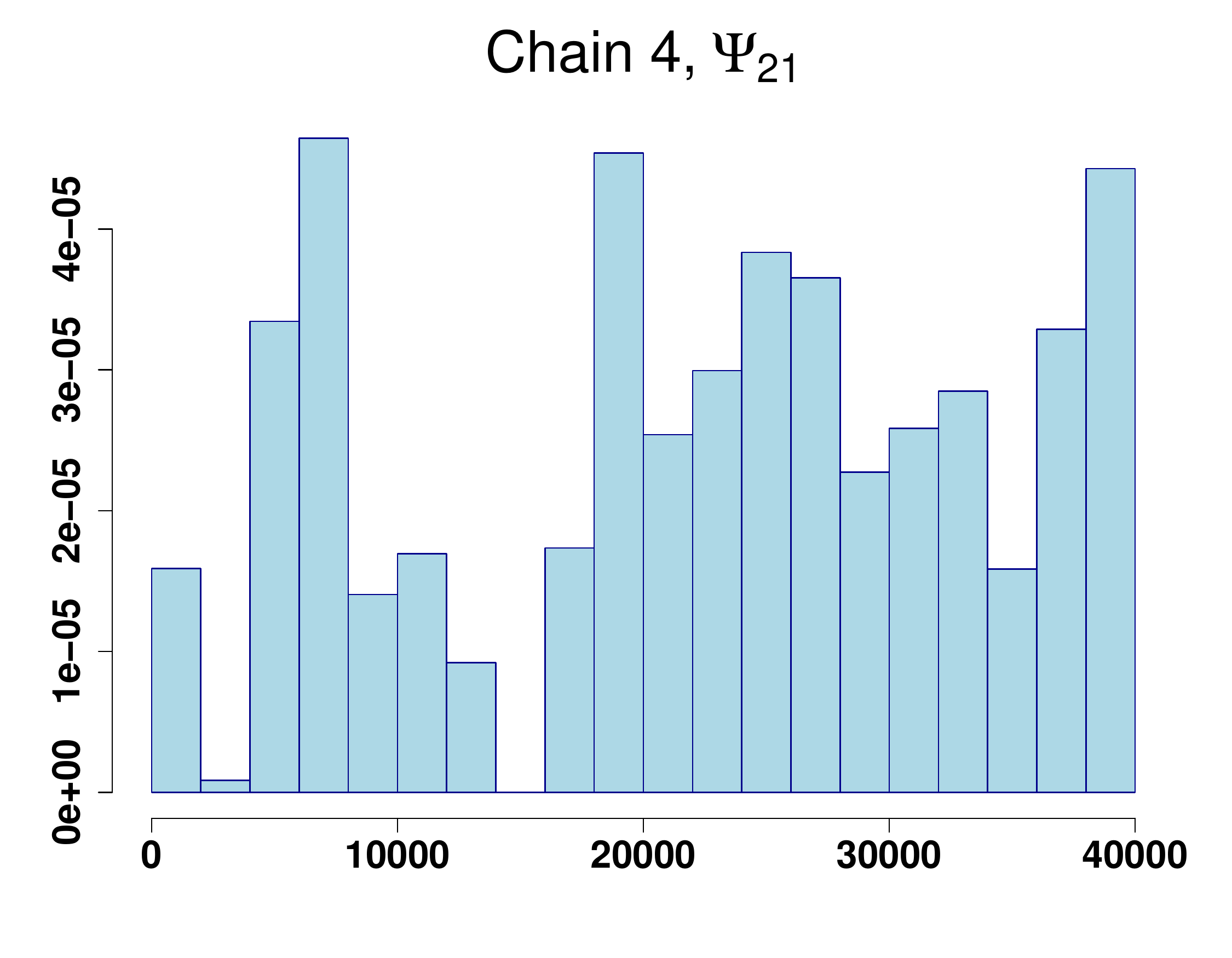}\\
\hspace{0.0cm}\includegraphics[width=4.0cm]{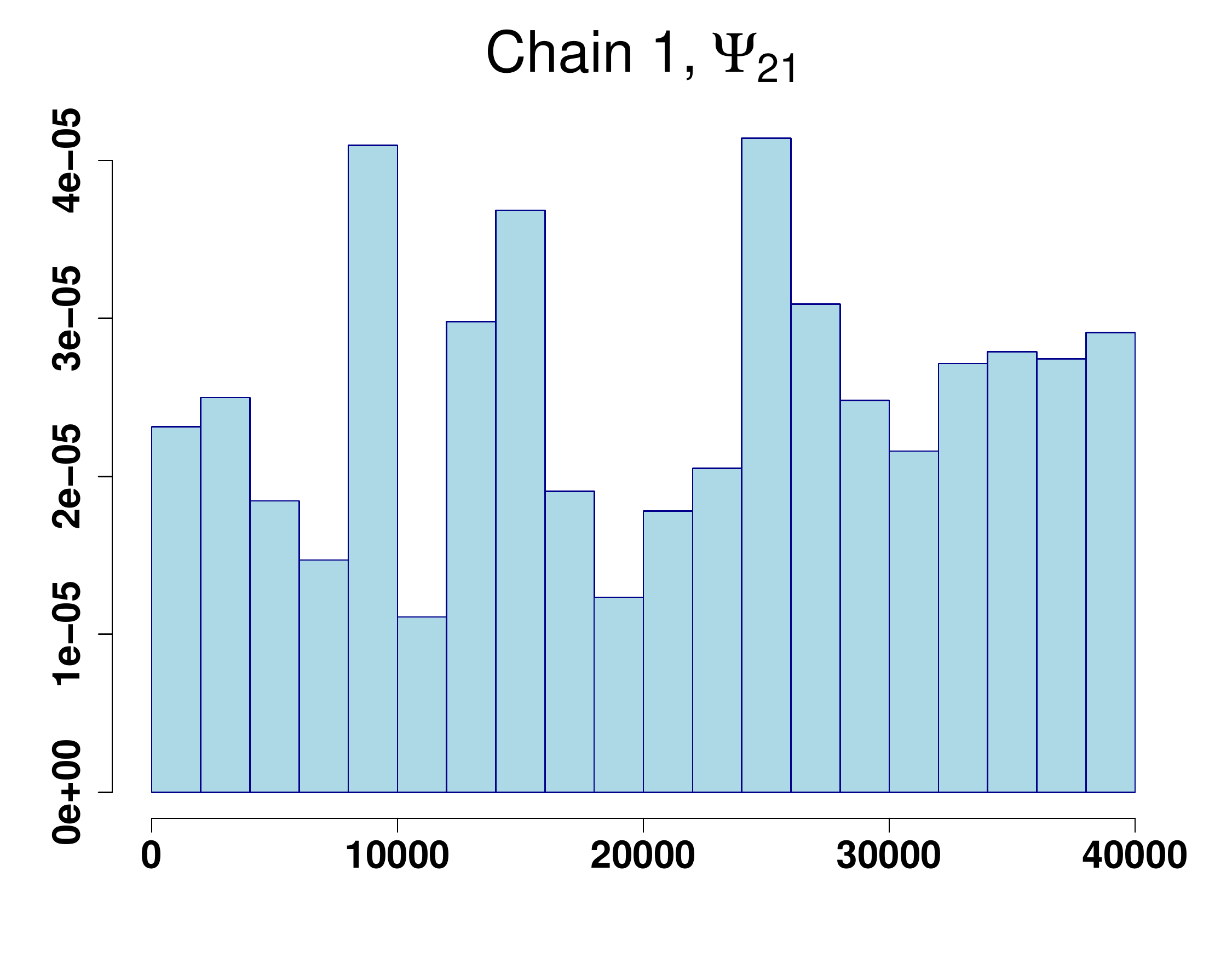}&\hspace{-0.5cm}\includegraphics[width=4.0cm]{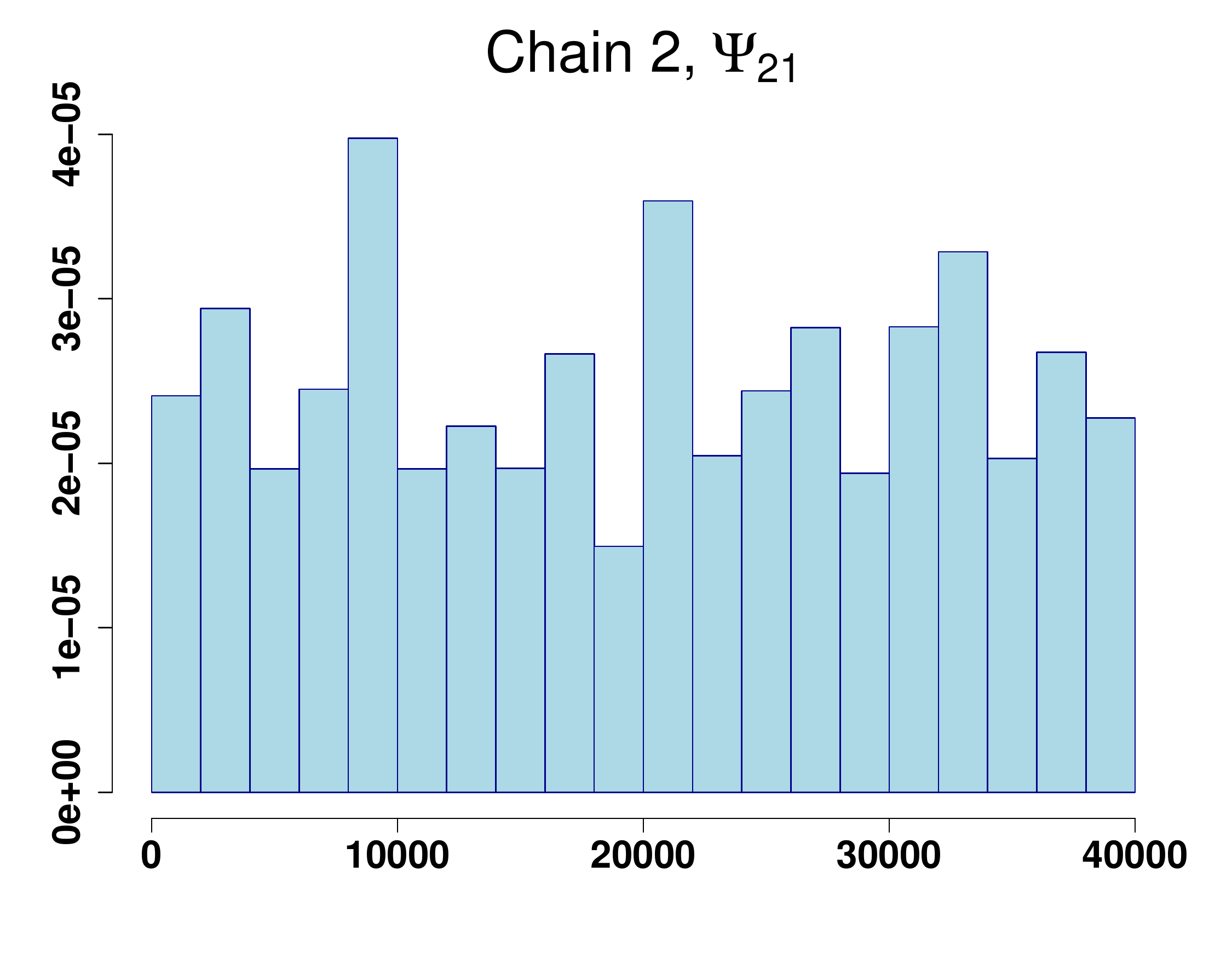}&
\hspace{-1.0cm}\includegraphics[width=4.0cm]{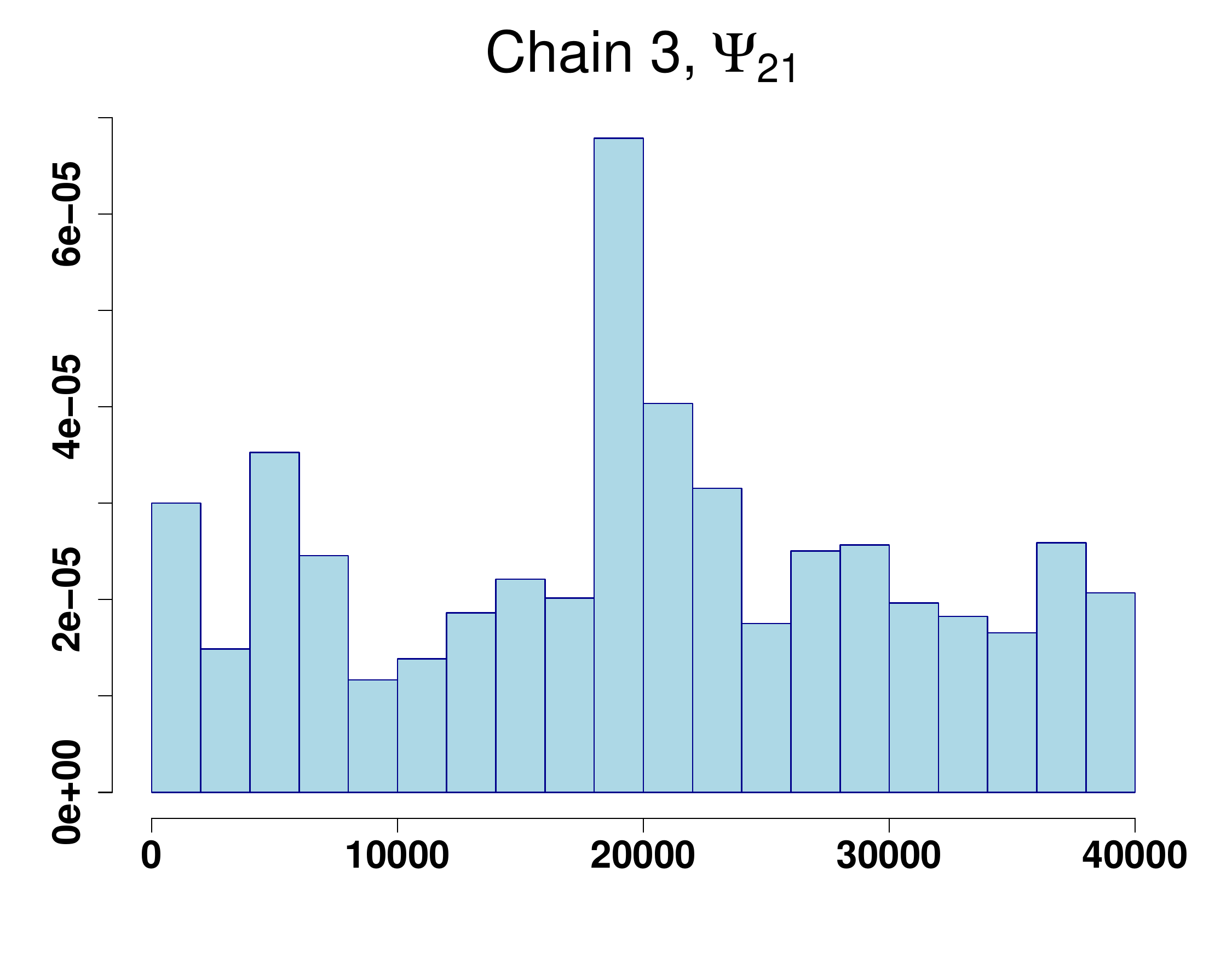}&\hspace{-1.5cm}\includegraphics[width=4.0cm]{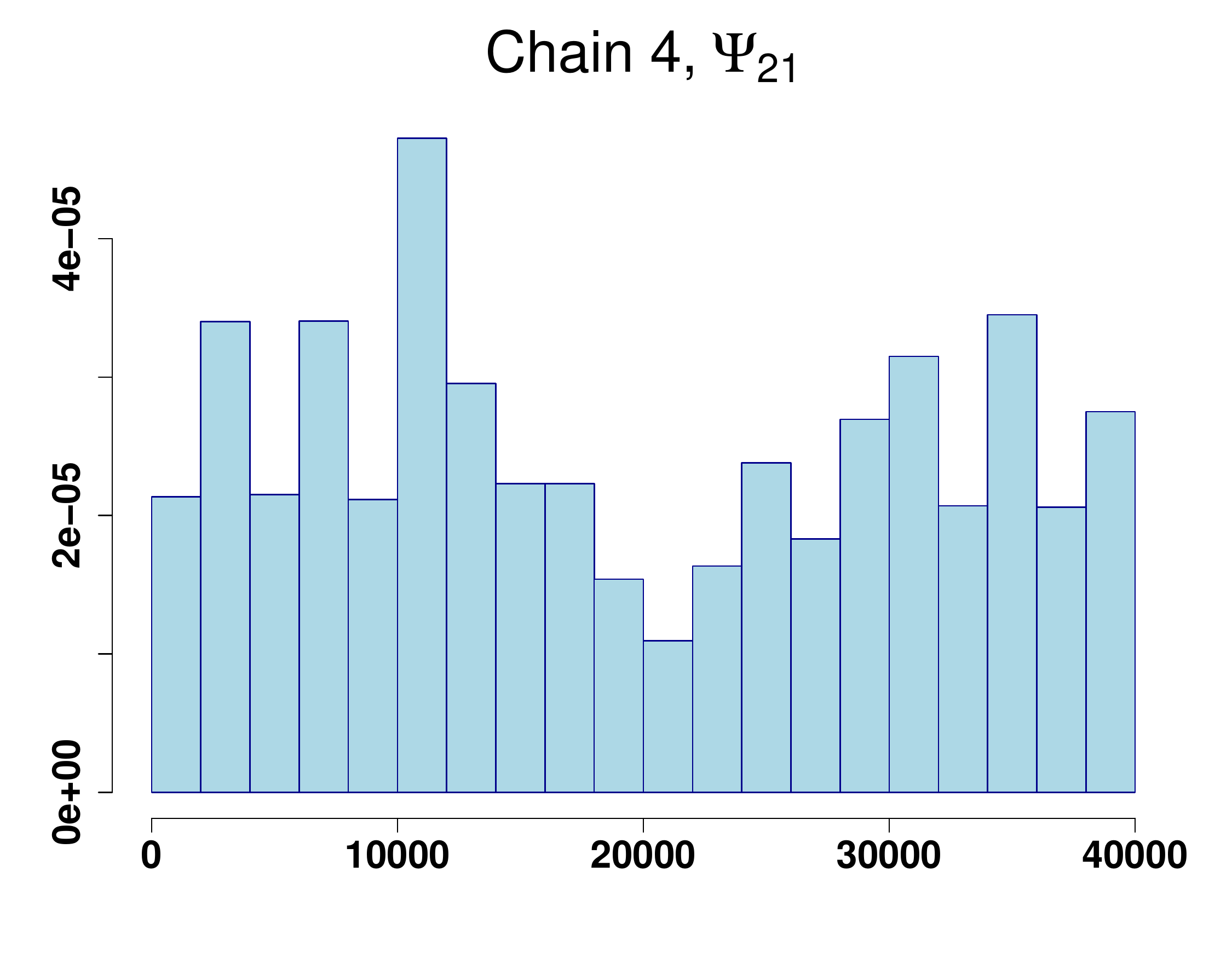}\\
\hspace{0.0cm}\includegraphics[width=4.0cm]{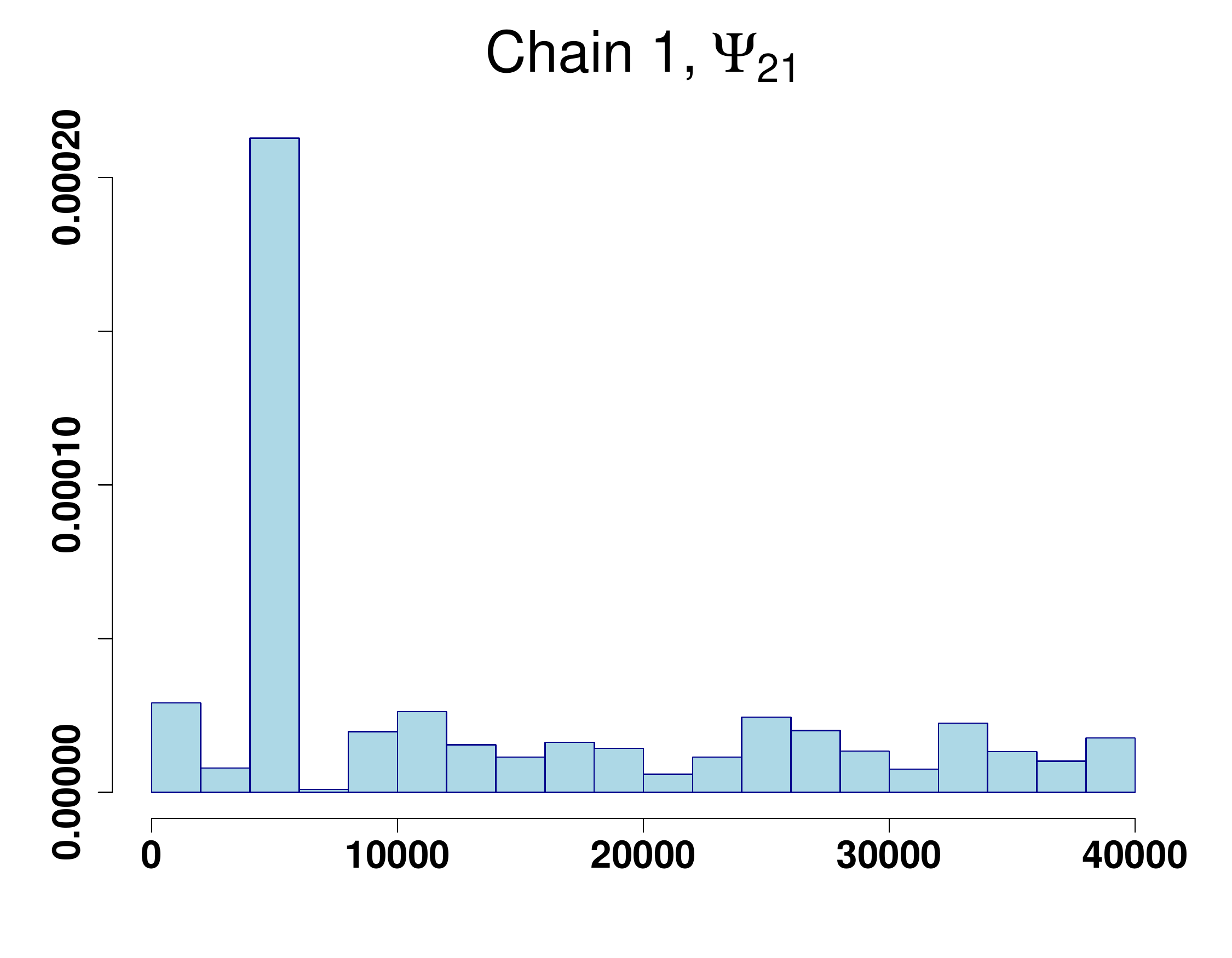}&\hspace{-0.5cm}\includegraphics[width=4.0cm]{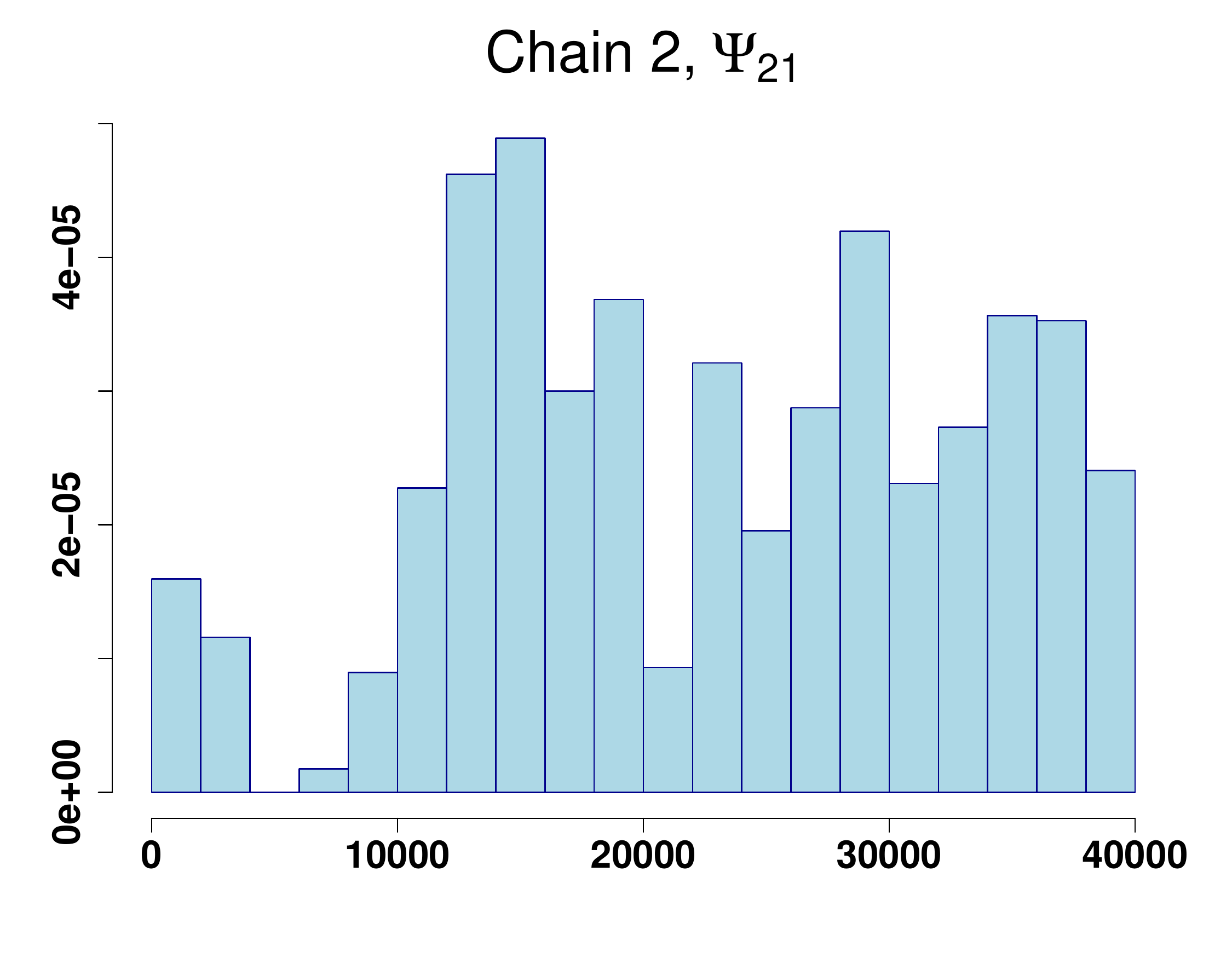}&
\hspace{-1.0cm}\includegraphics[width=4.0cm]{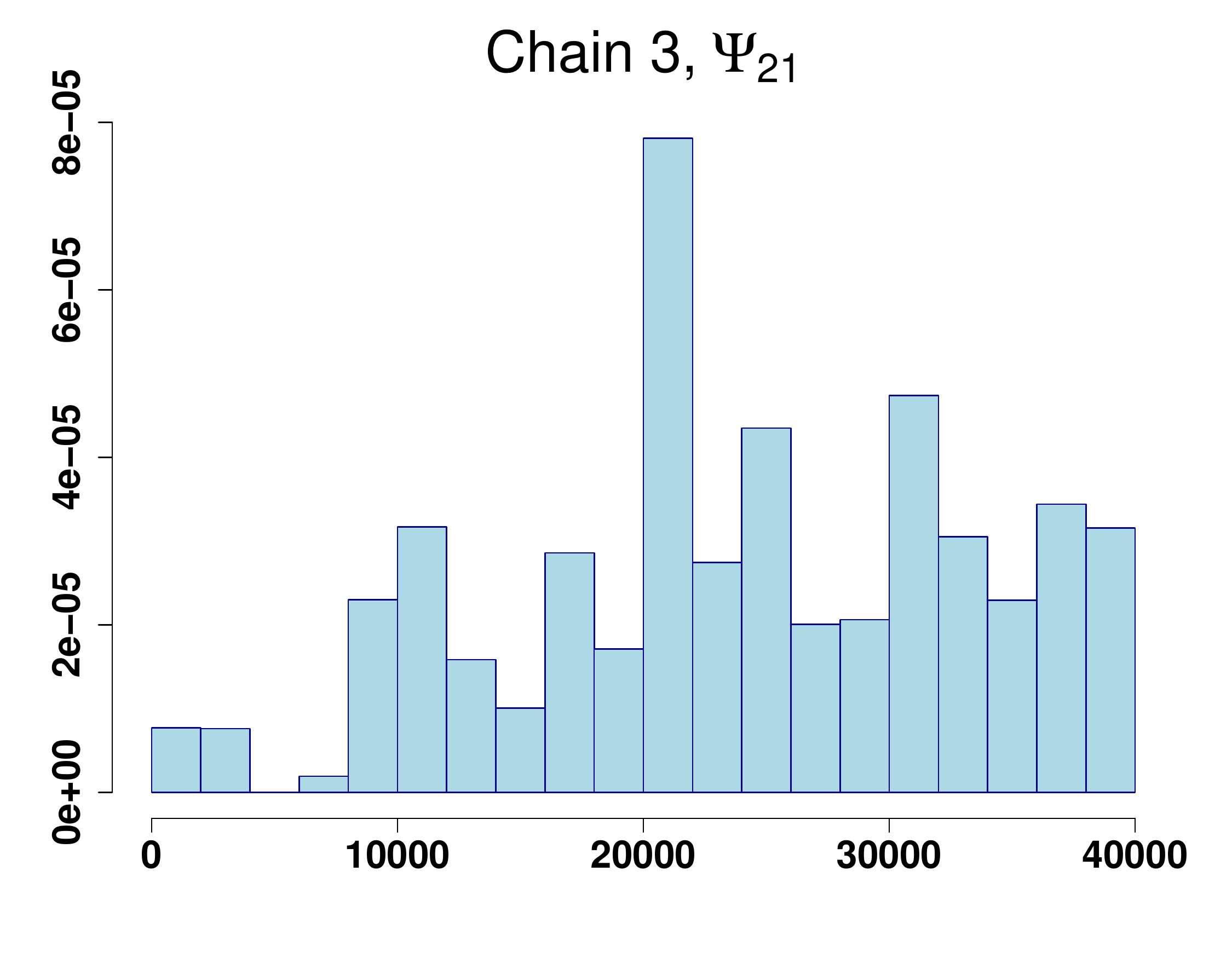}&\hspace{-1.5cm}\includegraphics[width=4.0cm]{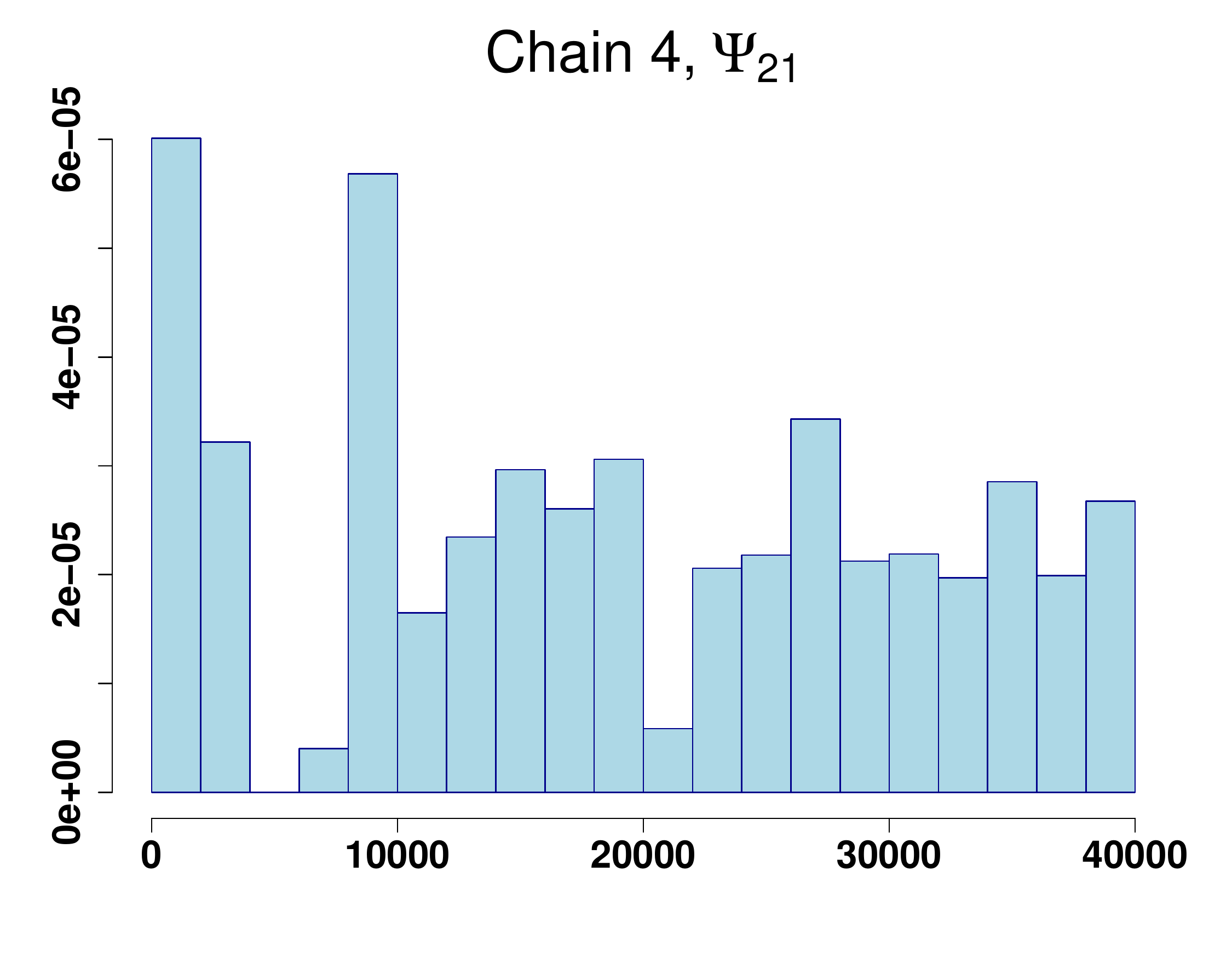}\\
\hspace{0.0cm}\includegraphics[width=4.0cm]{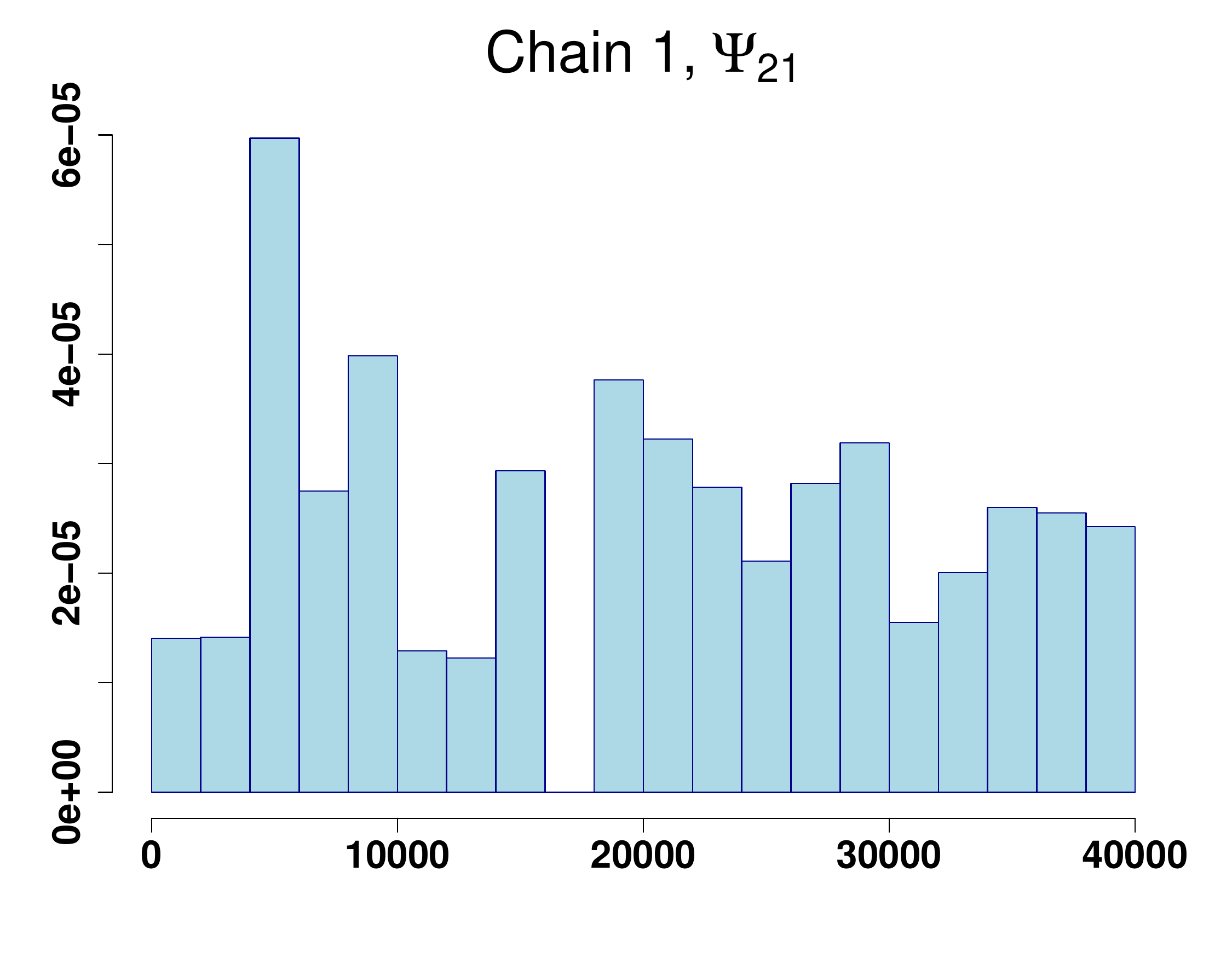}&\hspace{-0.5cm}\includegraphics[width=4.0cm]{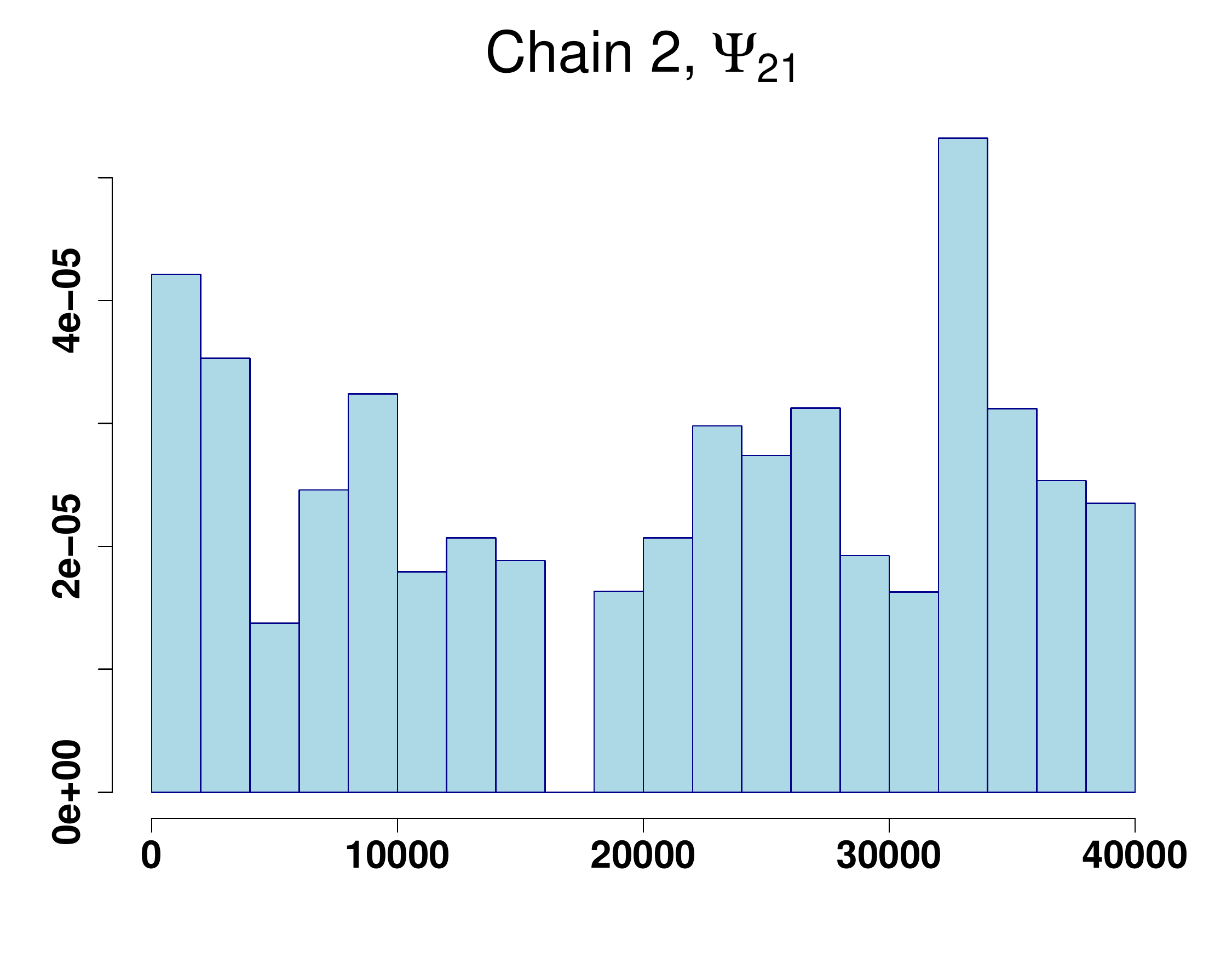}&
\hspace{-1.0cm}\includegraphics[width=4.0cm]{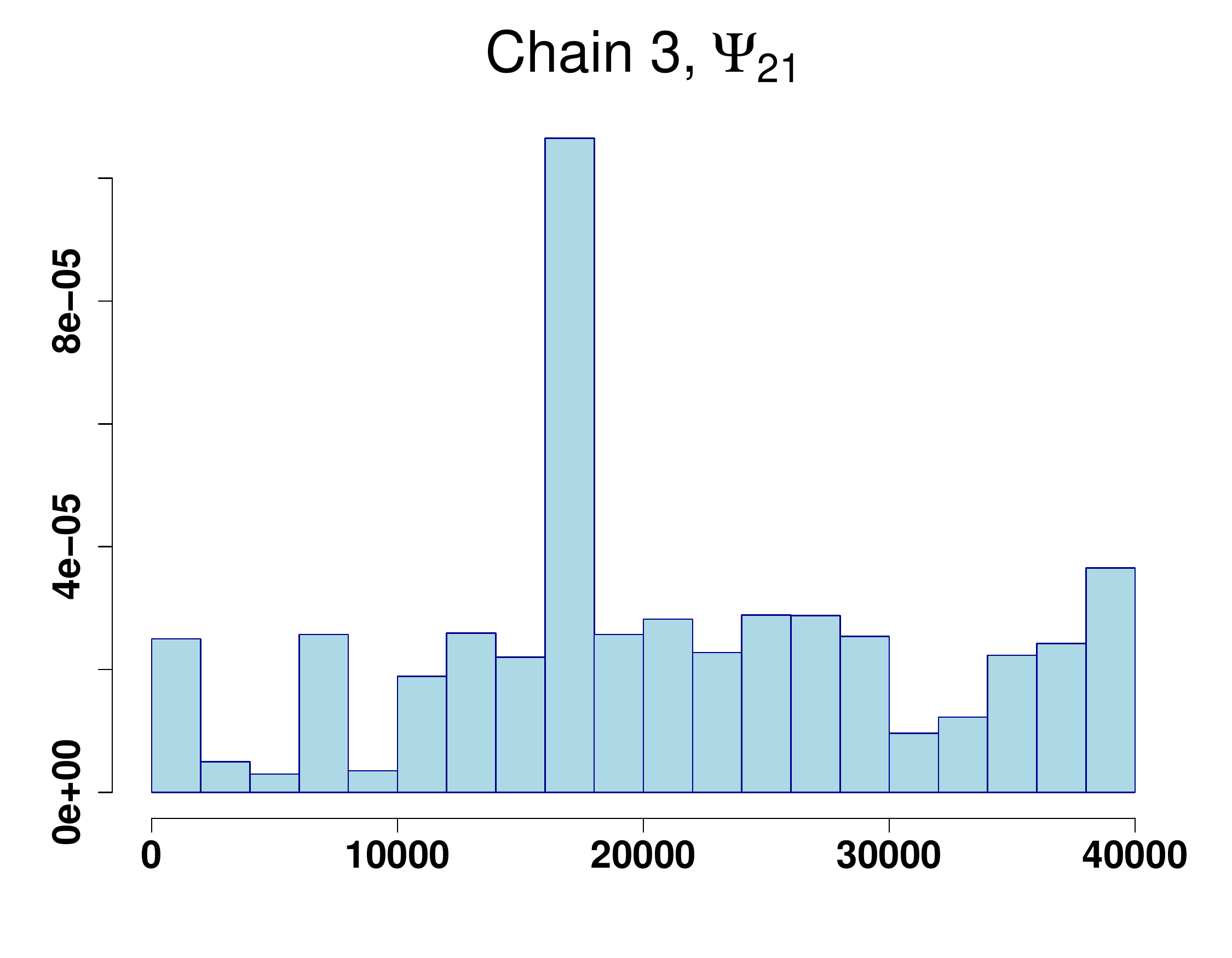}&\hspace{-1.5cm}\includegraphics[width=4.0cm]{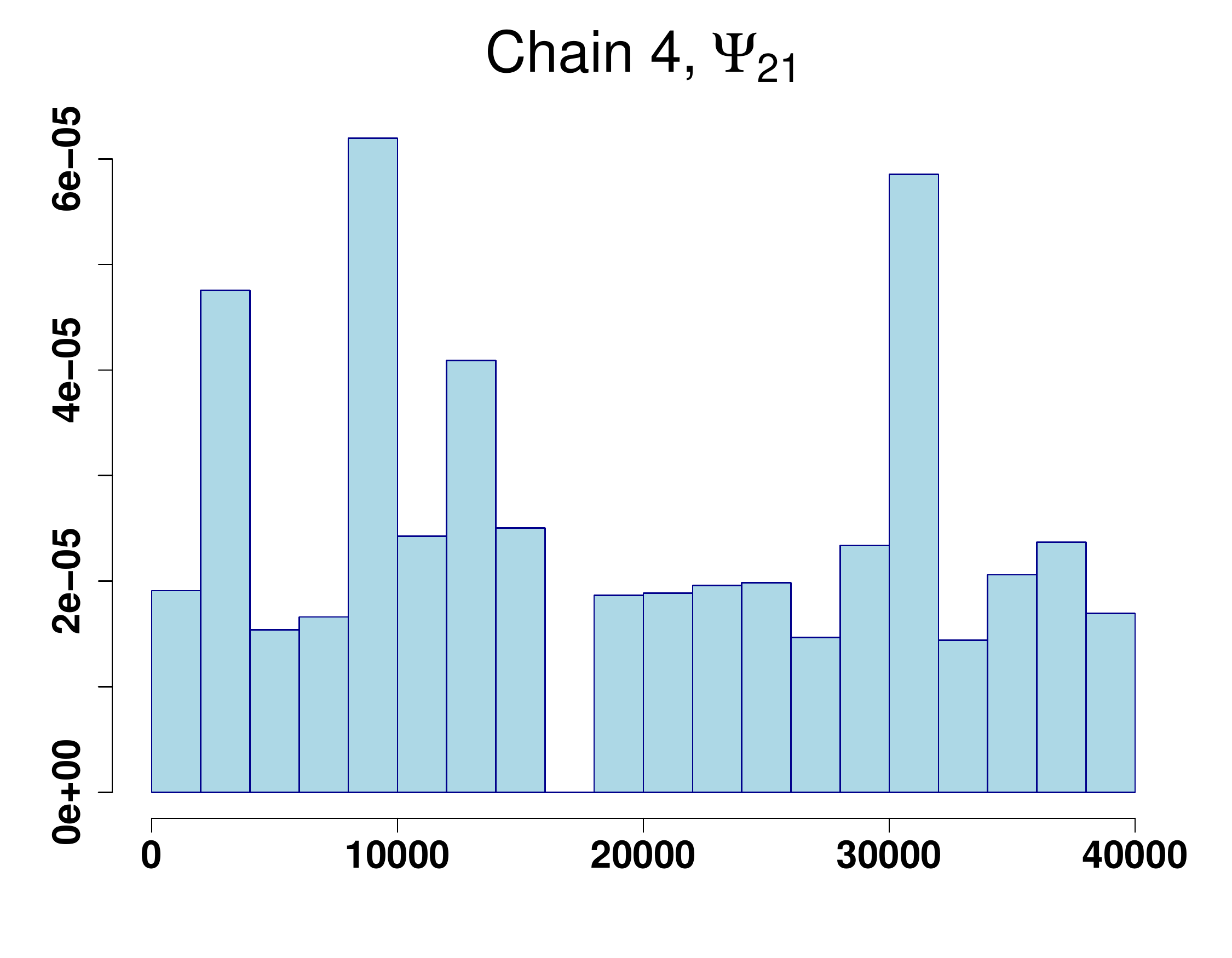}\\
\hspace{0.0cm}\includegraphics[width=4.0cm]{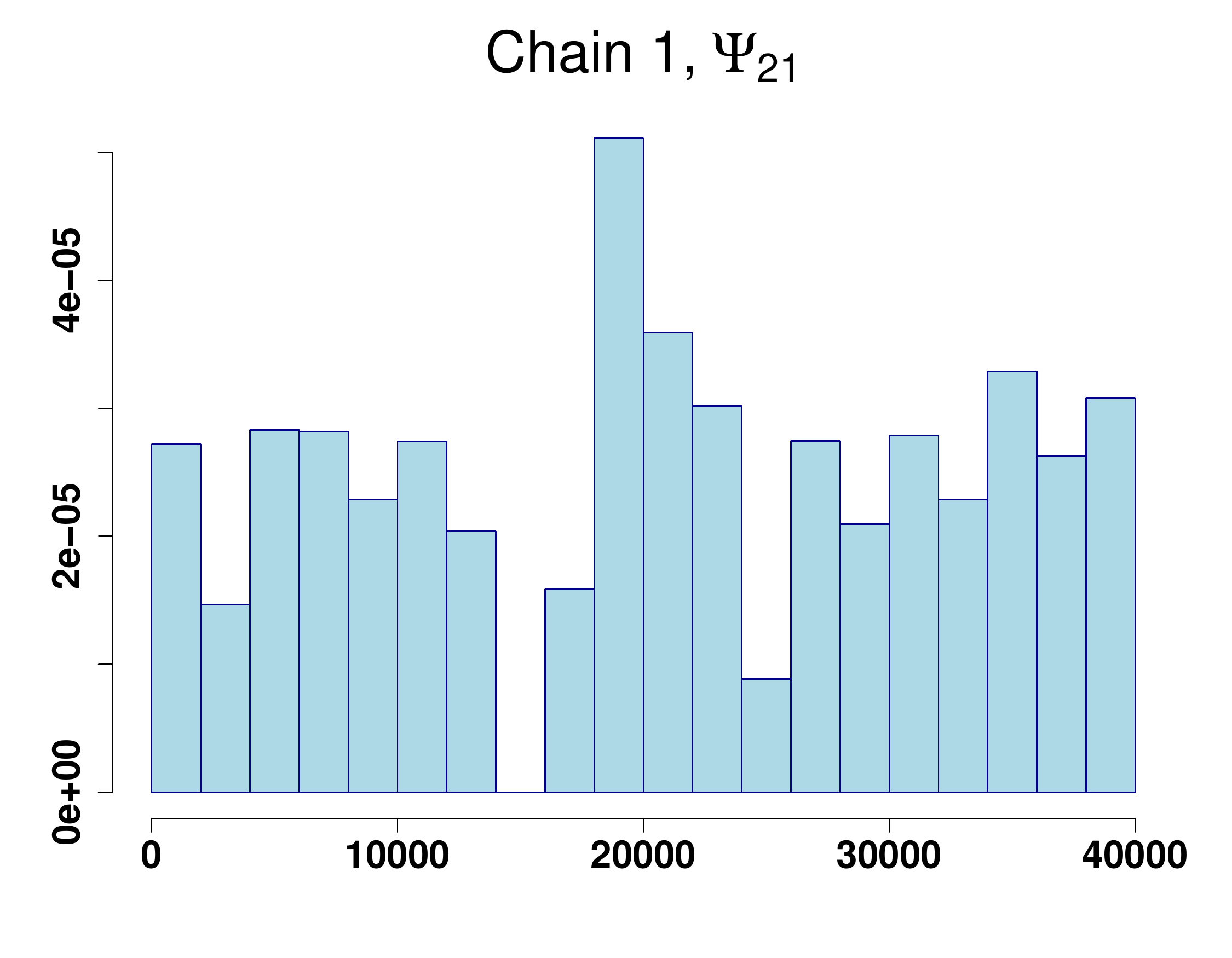}&\hspace{-0.5cm}\includegraphics[width=4.0cm]{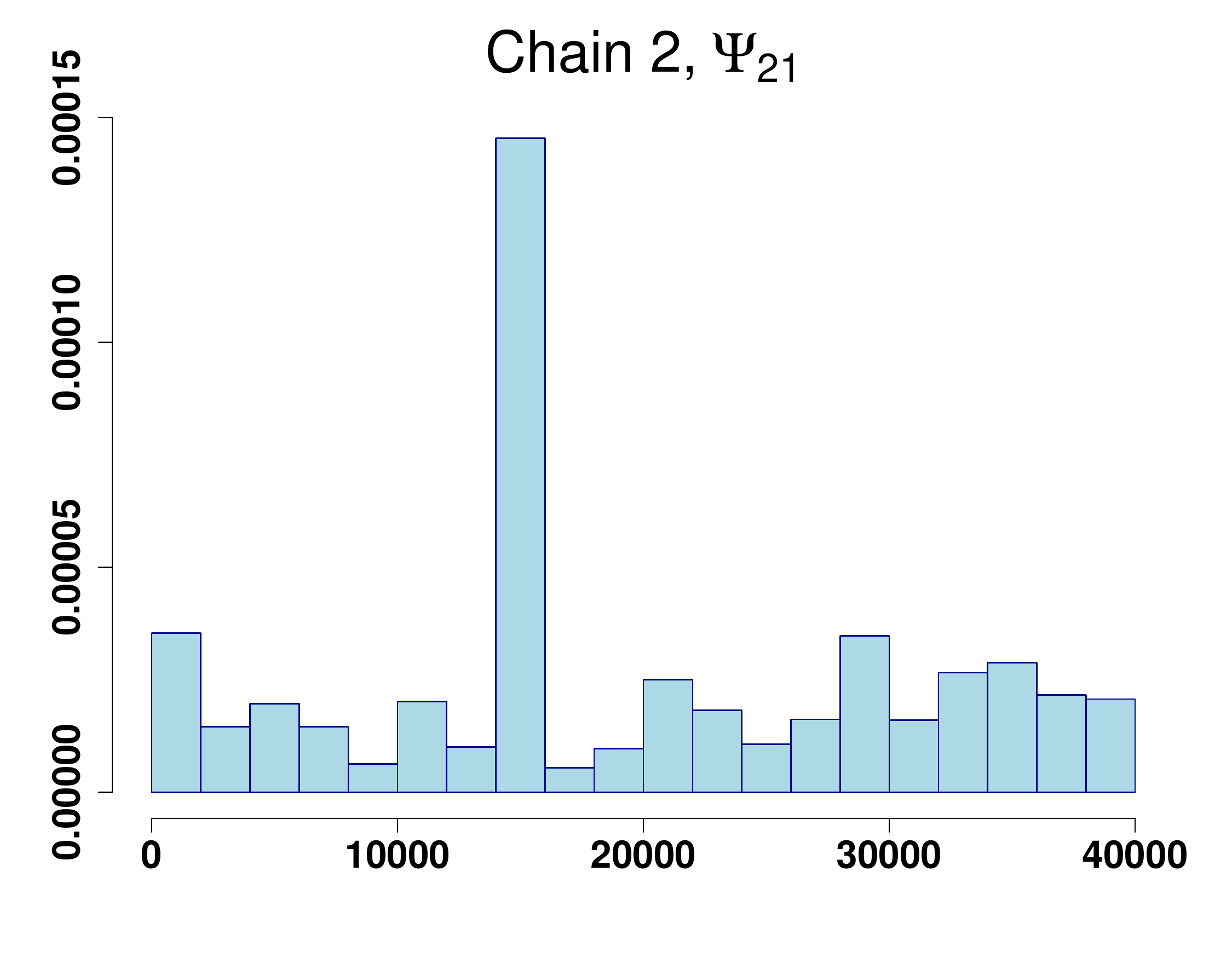}&
\hspace{-1.0cm}\includegraphics[width=4.0cm]{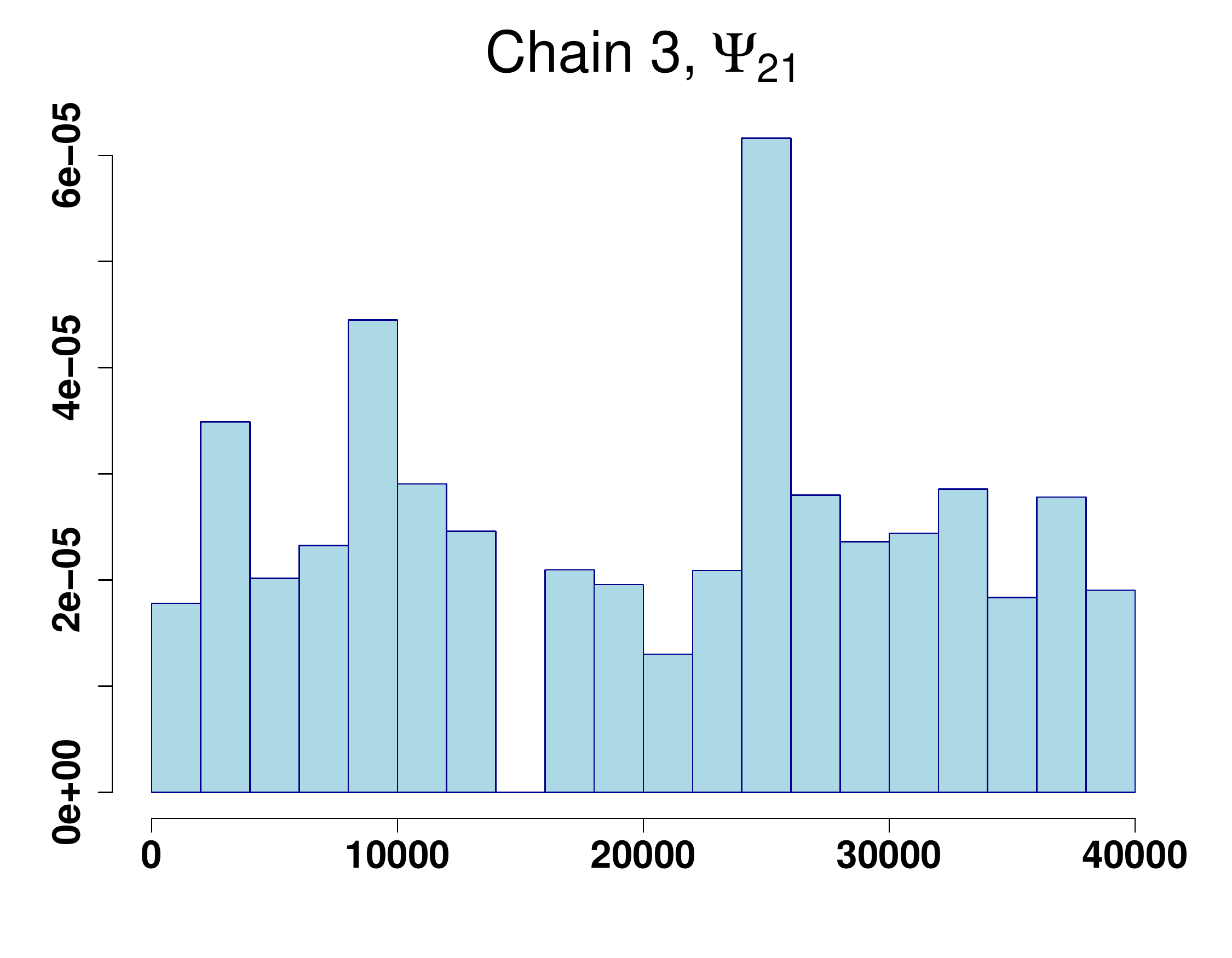}&\hspace{-1.5cm}\includegraphics[width=4.0cm]{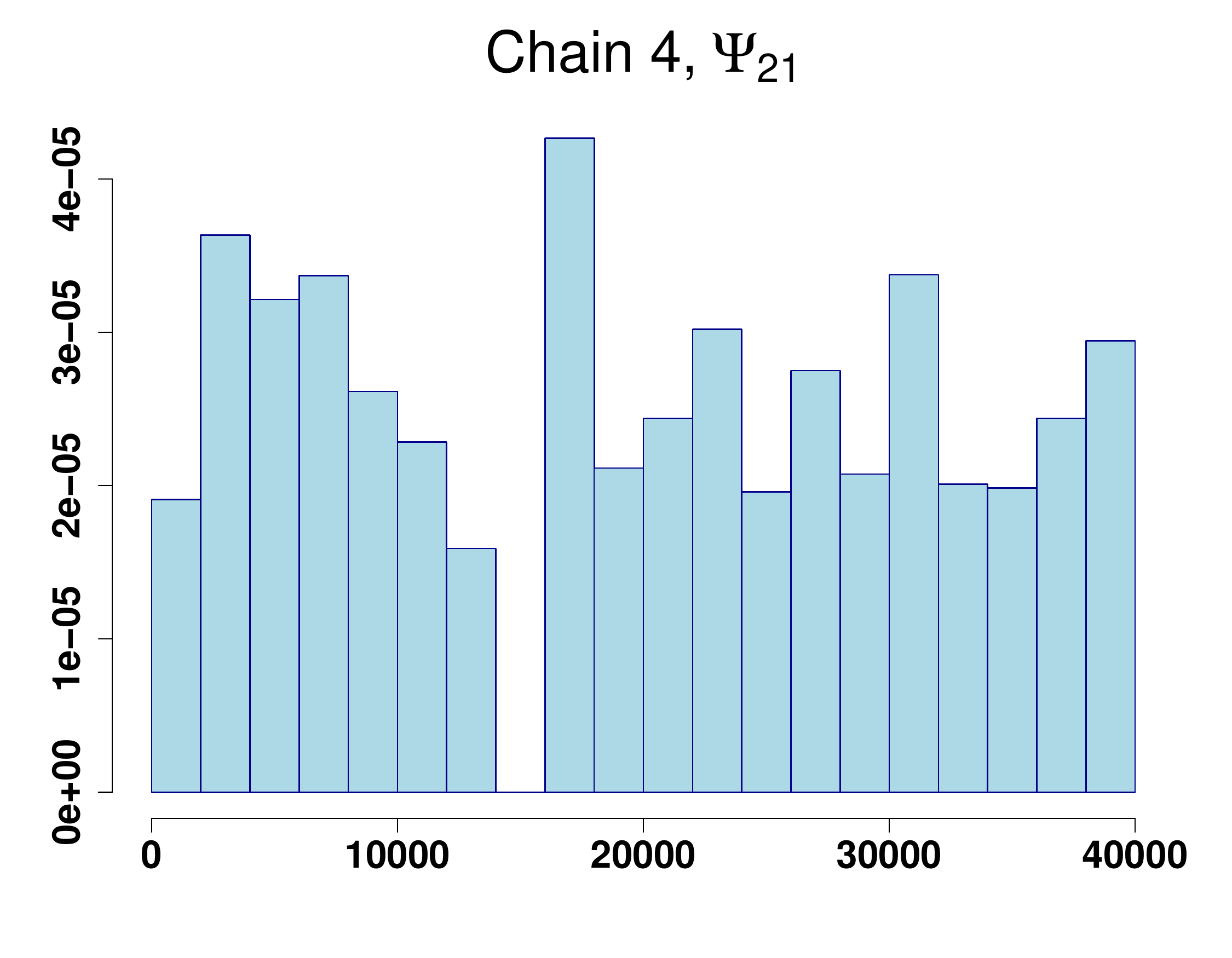}\\
\end{tabular}
 \caption{Rank plots of posterior draws from four chains in the case of the parameter $\Psi_{21}$ (DBP, SBP) of the normal multivariate random effects model by employing the Jeffreys prior (first to third rows) and the Berger and Bernardo reference prior (fourth to sixth rows). The samples from the posterior distributions are drawn by Algorithm A (first and fourth rows), Algorithm B (second and fifth rows) and Algorithm C (third and sixth rows).}
\label{fig:emp-study-rank-Psi21-nor}
 \end{figure}

\begin{figure}[h!t]
\centering
\begin{tabular}{p{4.0cm}p{4.0cm}p{4.0cm}p{4.0cm}}
\hspace{0.0cm}\includegraphics[width=4.0cm]{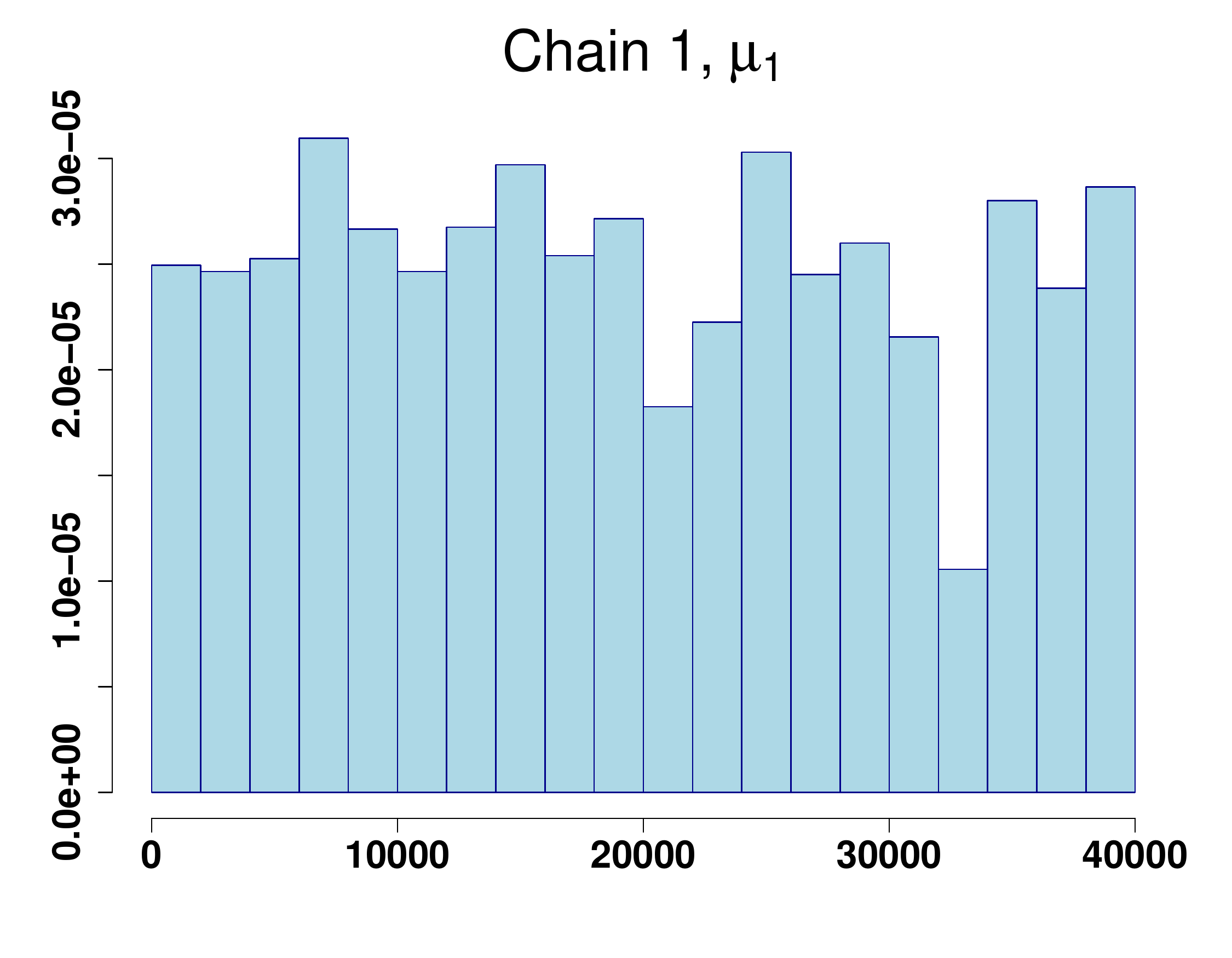}&\hspace{-0.5cm}\includegraphics[width=4.0cm]{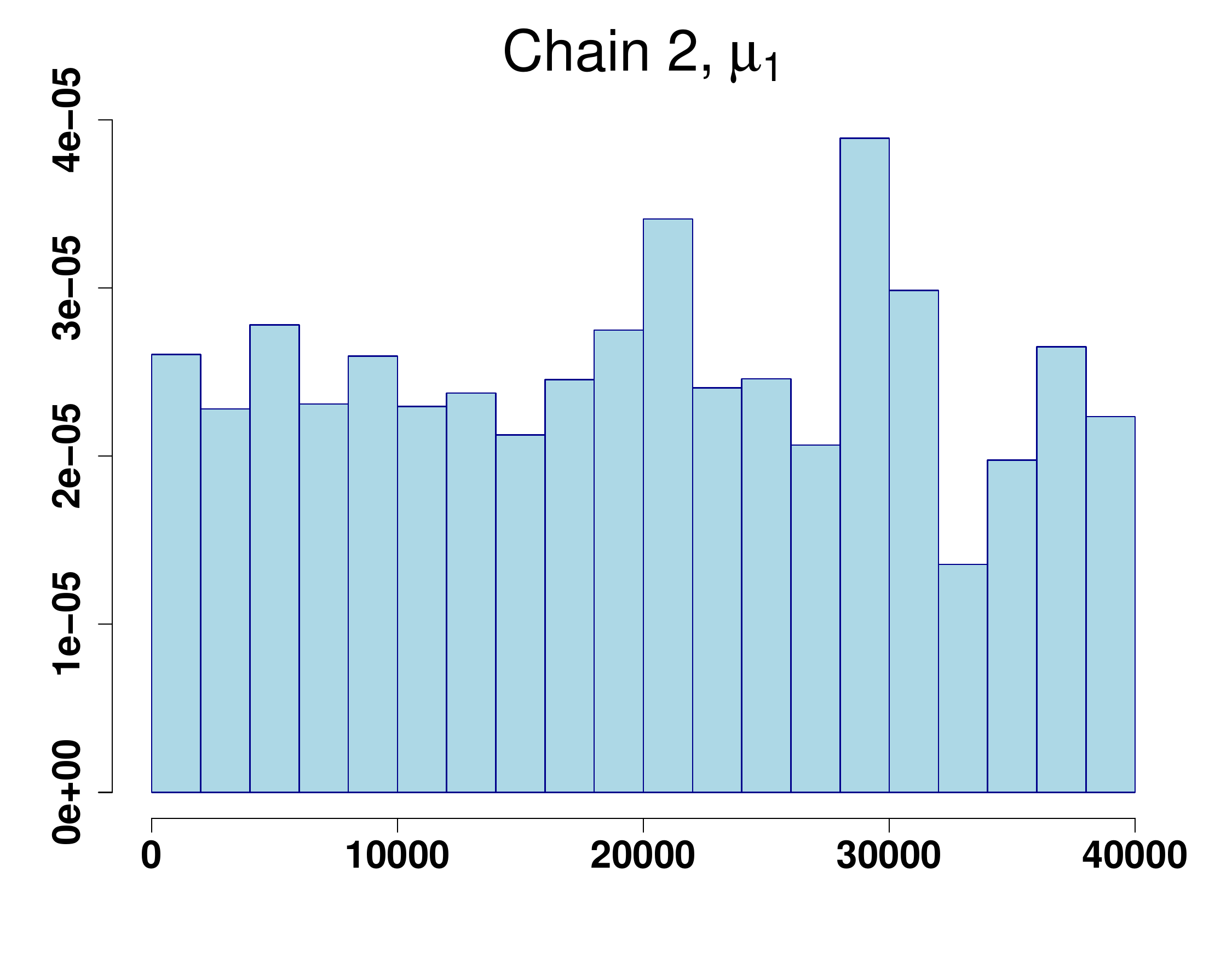}&
\hspace{-1.0cm}\includegraphics[width=4.0cm]{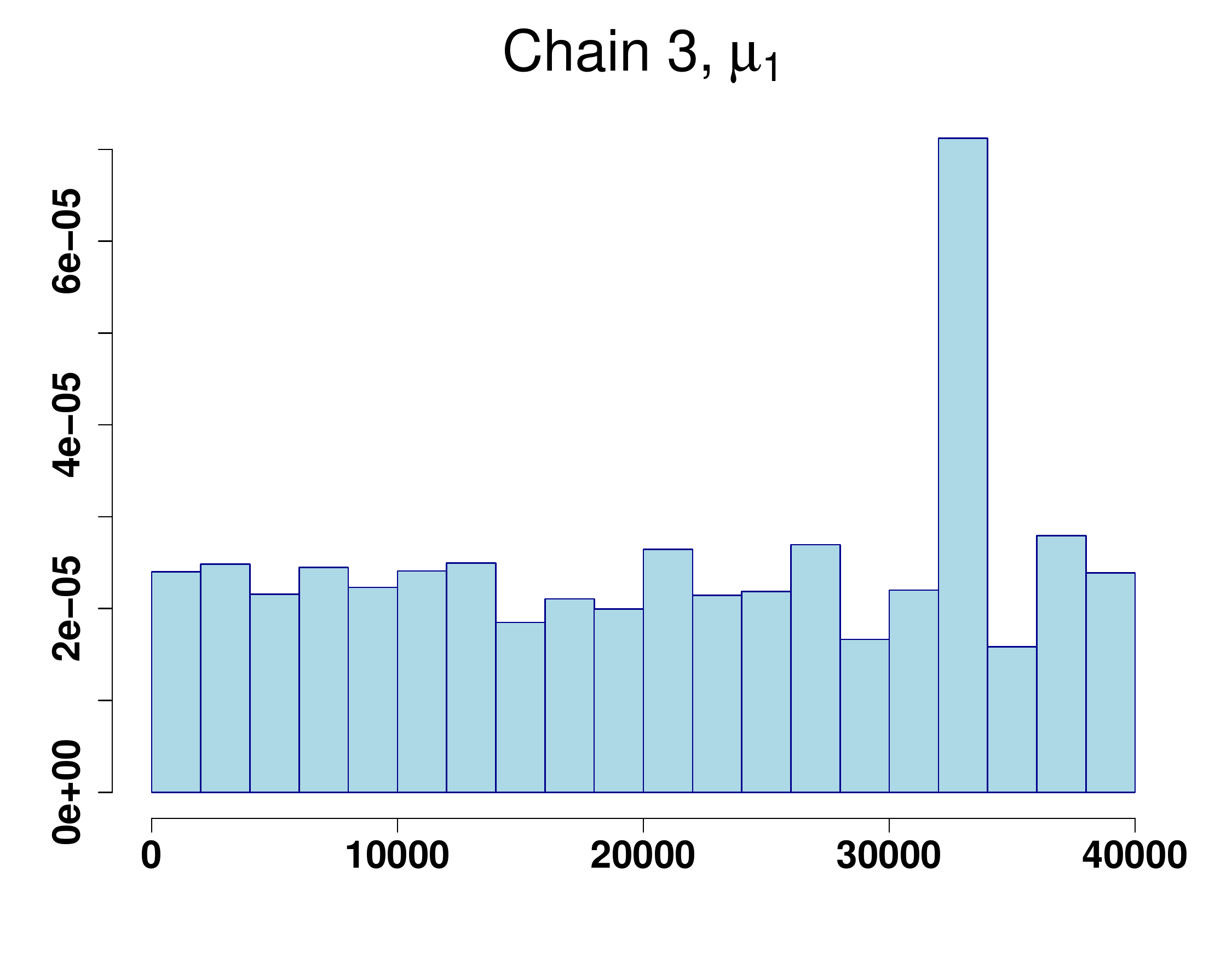}&\hspace{-1.9cm}\includegraphics[width=4.0cm]{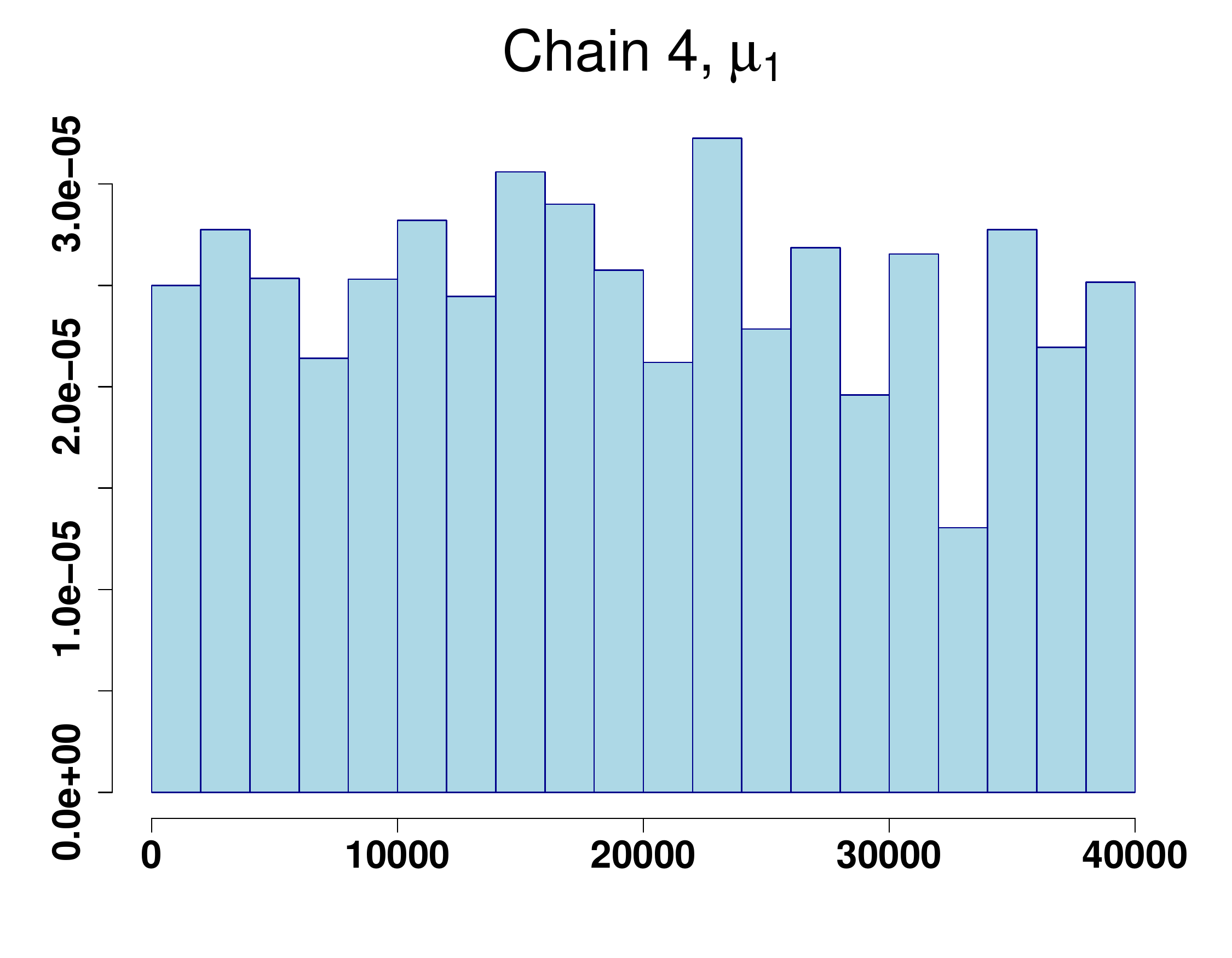}\\
\hspace{0.0cm}\includegraphics[width=4.0cm]{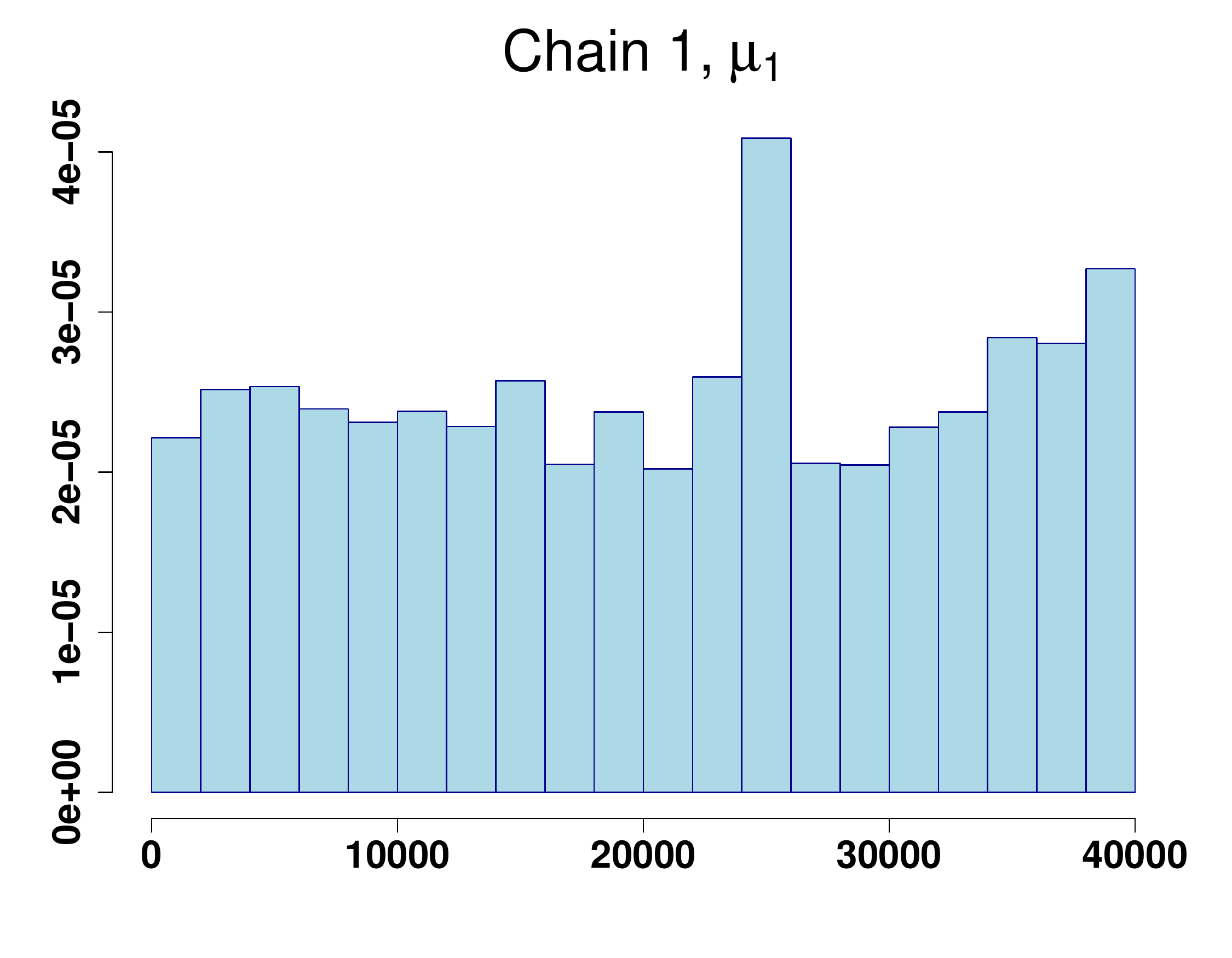}&\hspace{-0.5cm}\includegraphics[width=4.0cm]{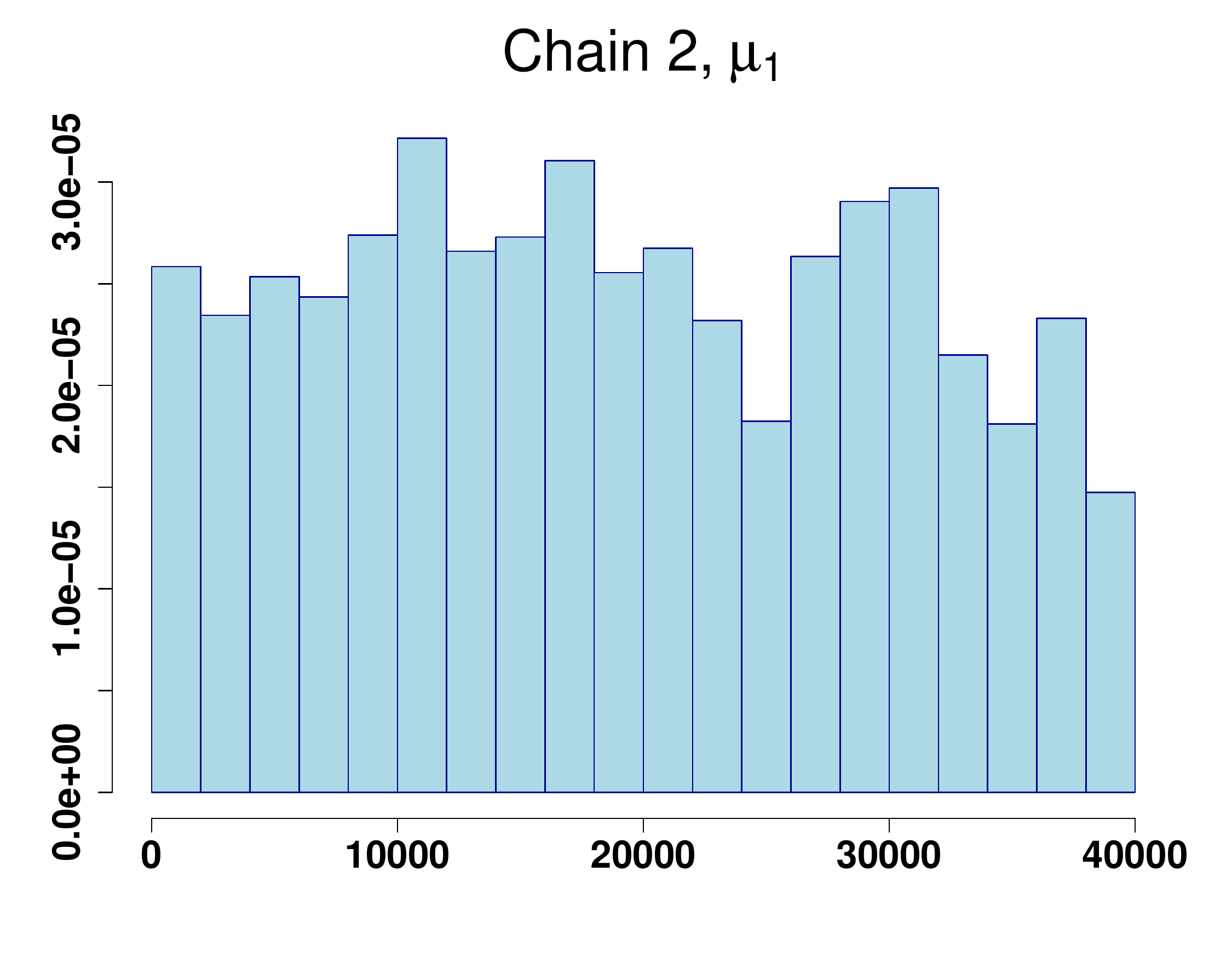}&
\hspace{-1.0cm}\includegraphics[width=4.0cm]{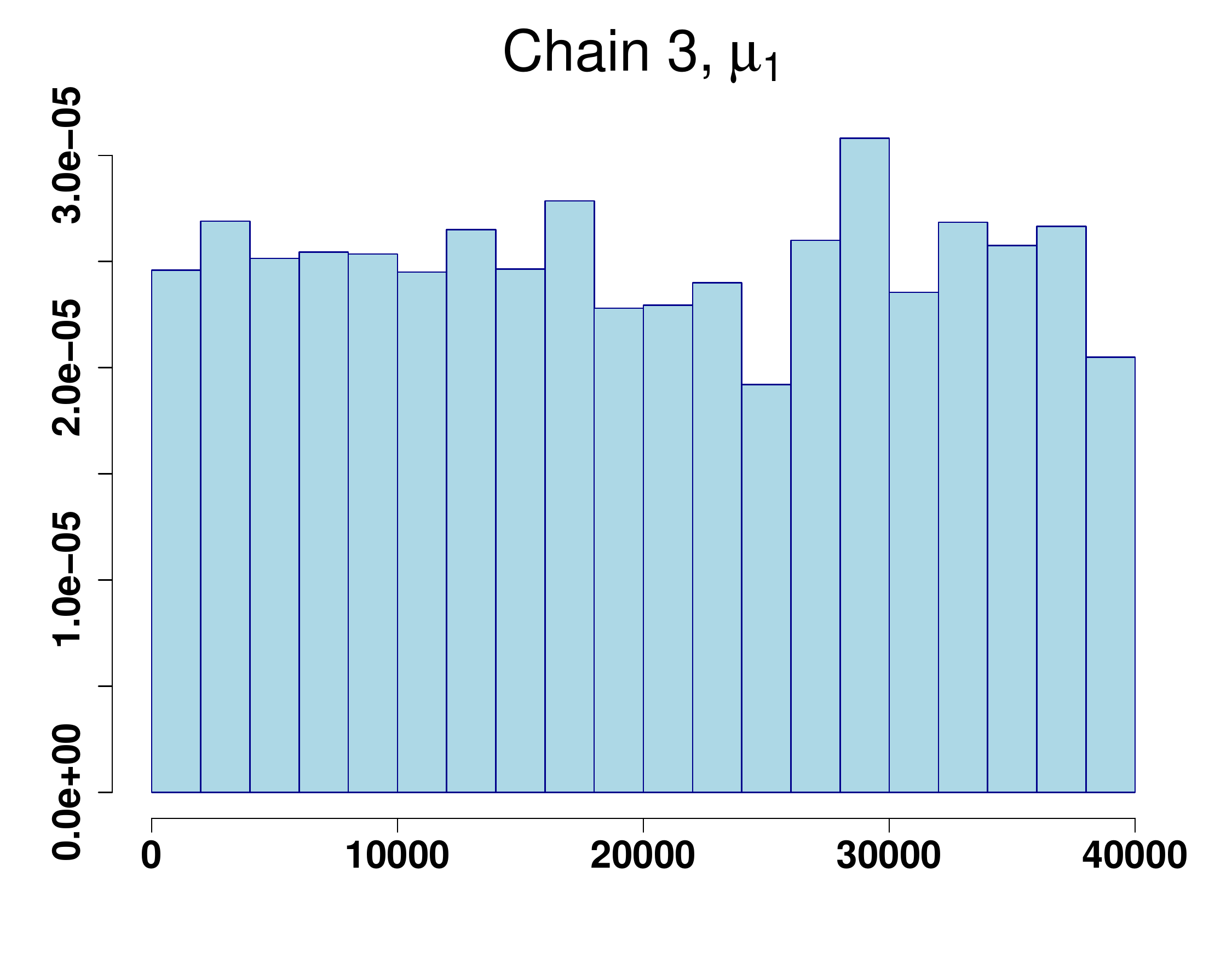}&\hspace{-1.9cm}\includegraphics[width=4.0cm]{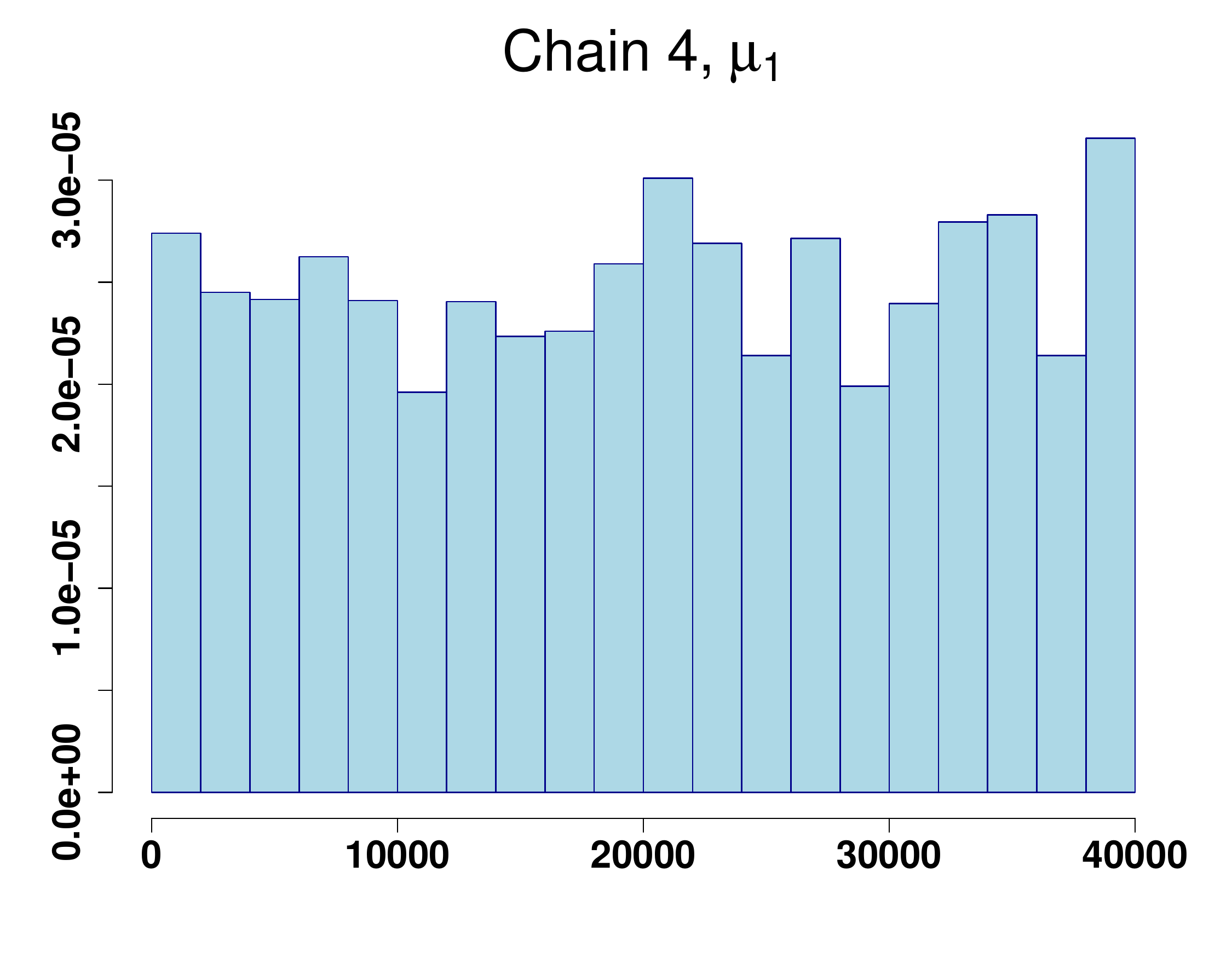}\\
\hspace{0.0cm}\includegraphics[width=4.0cm]{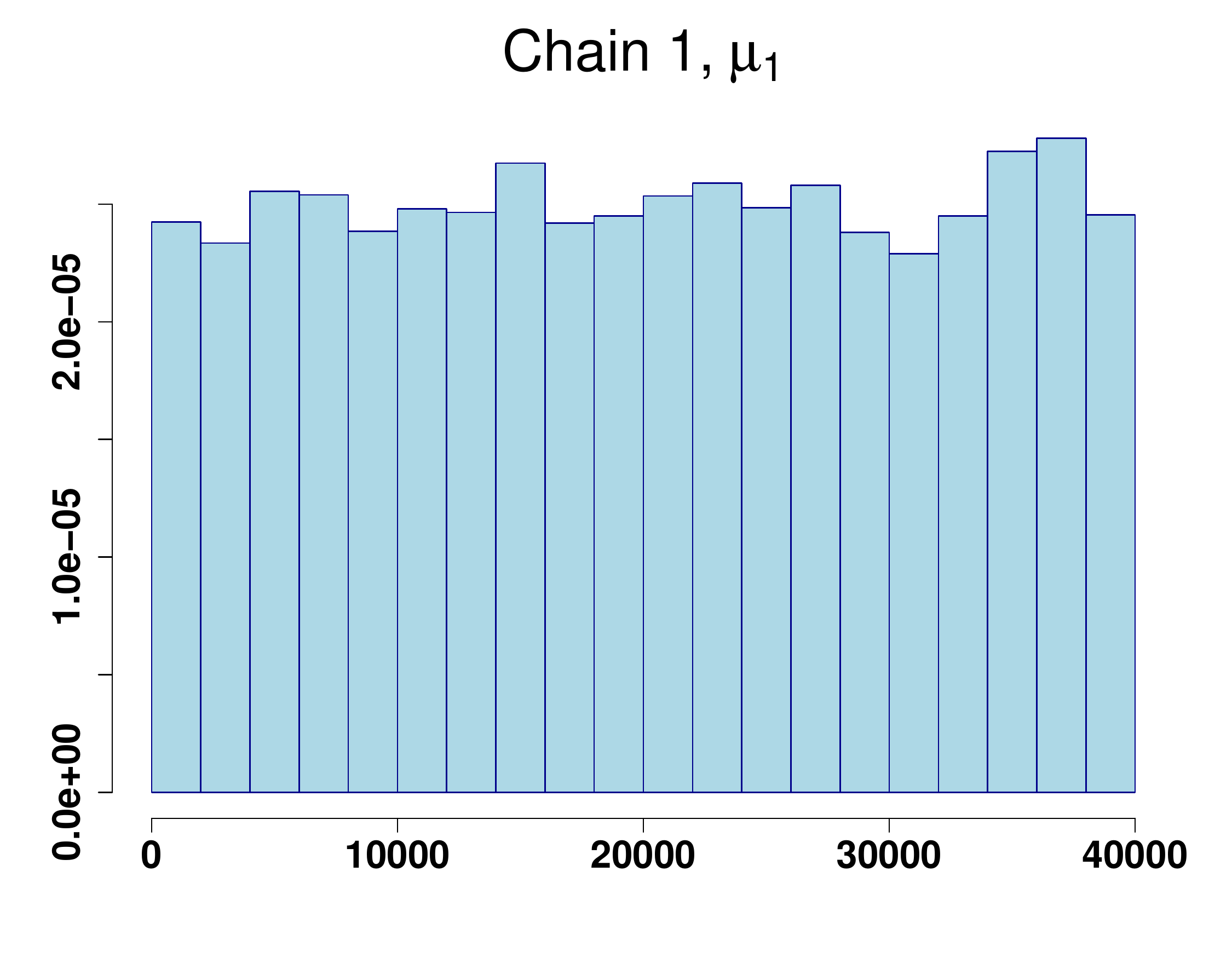}&\hspace{-0.5cm}\includegraphics[width=4.0cm]{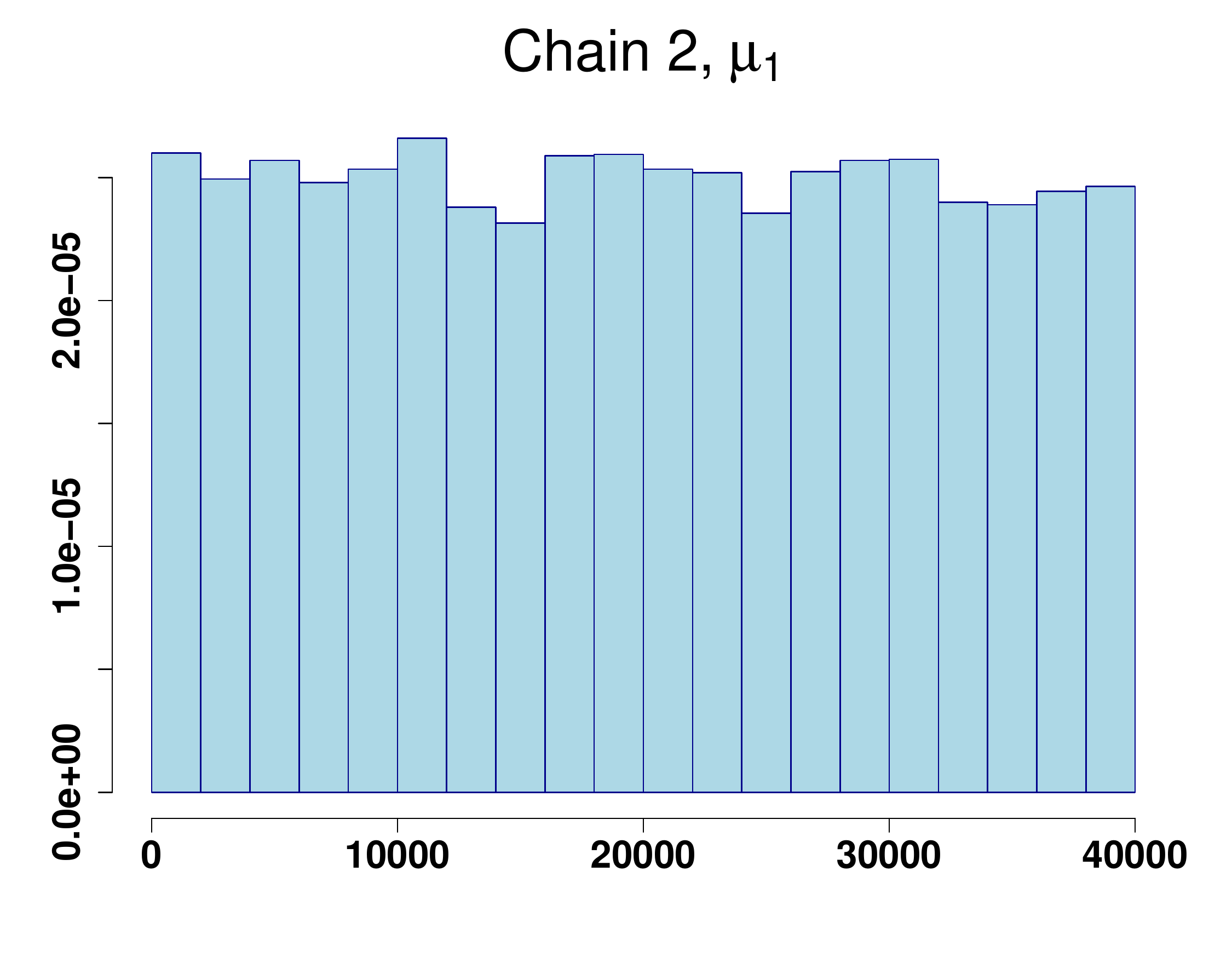}&
\hspace{-1.0cm}\includegraphics[width=4.0cm]{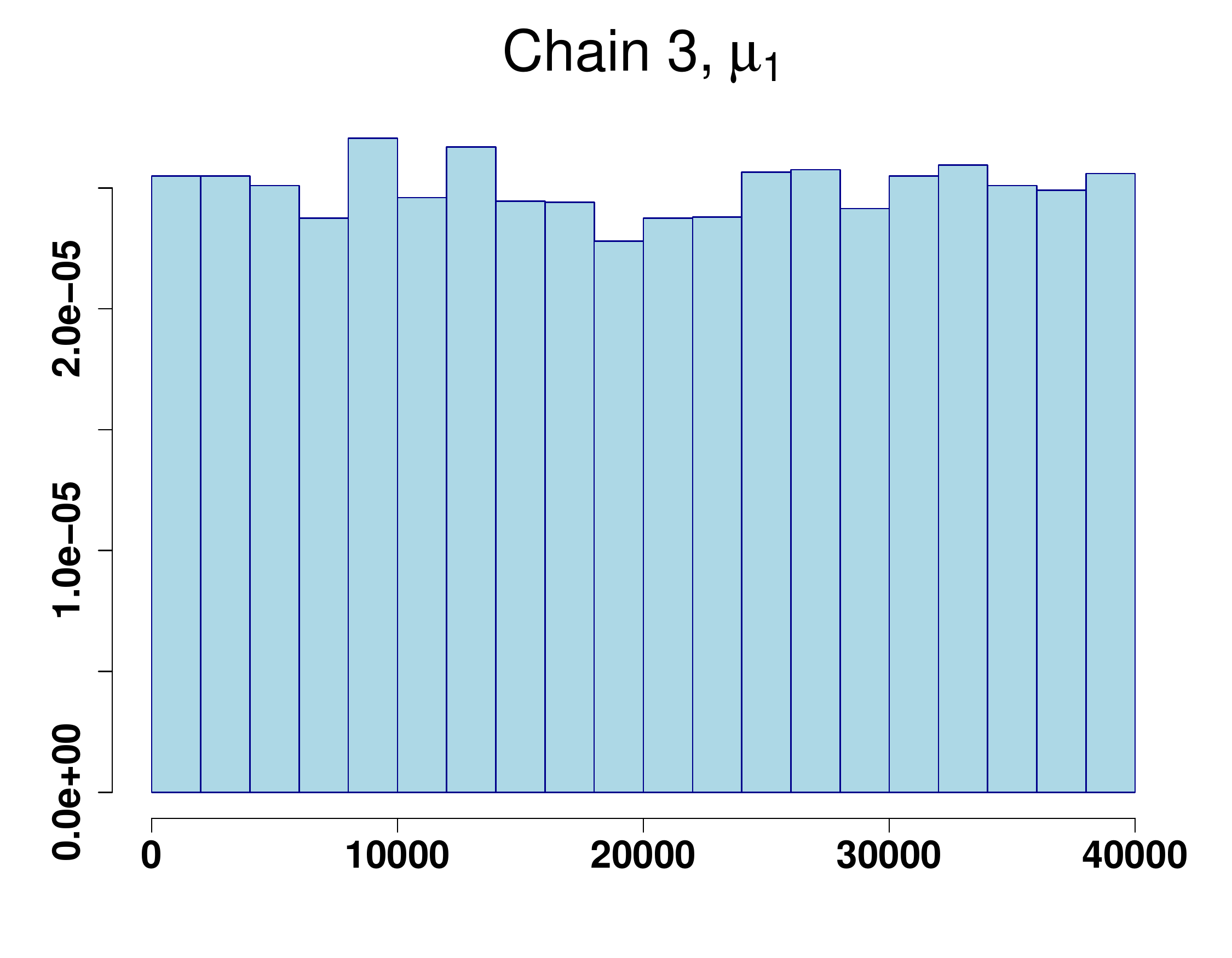}&\hspace{-1.9cm}\includegraphics[width=4.0cm]{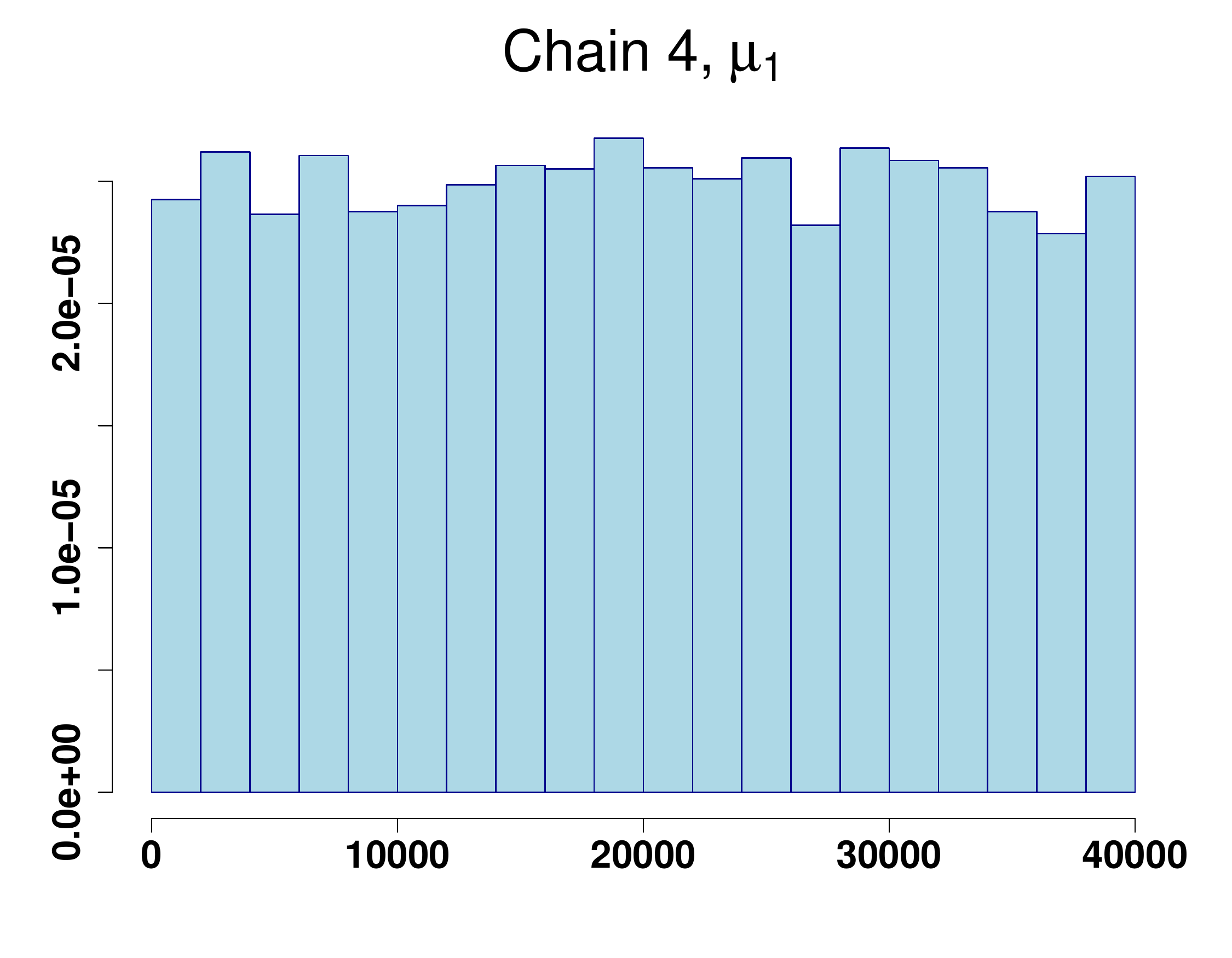}\\
\hspace{0.0cm}\includegraphics[width=4.0cm]{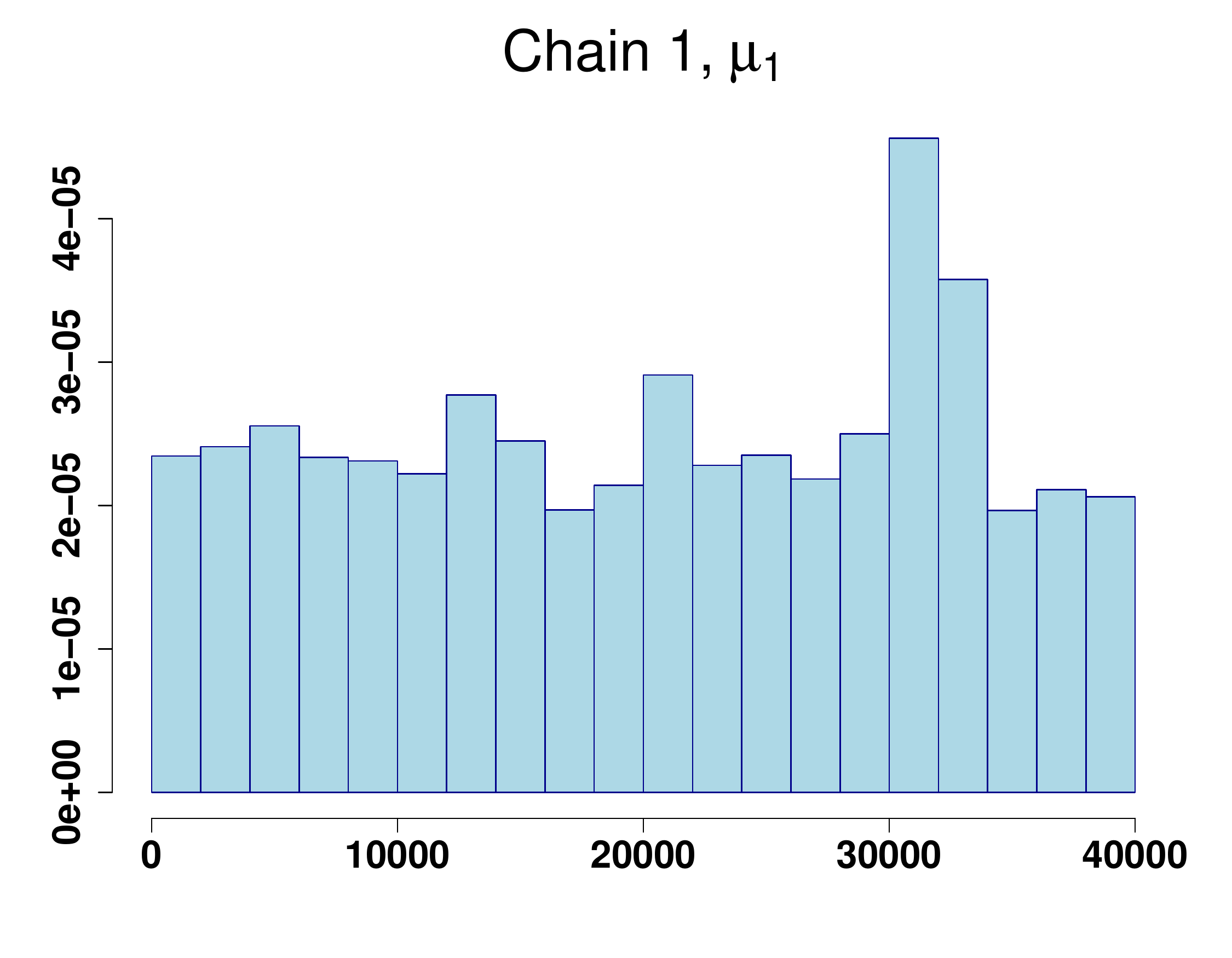}&\hspace{-0.5cm}\includegraphics[width=4.0cm]{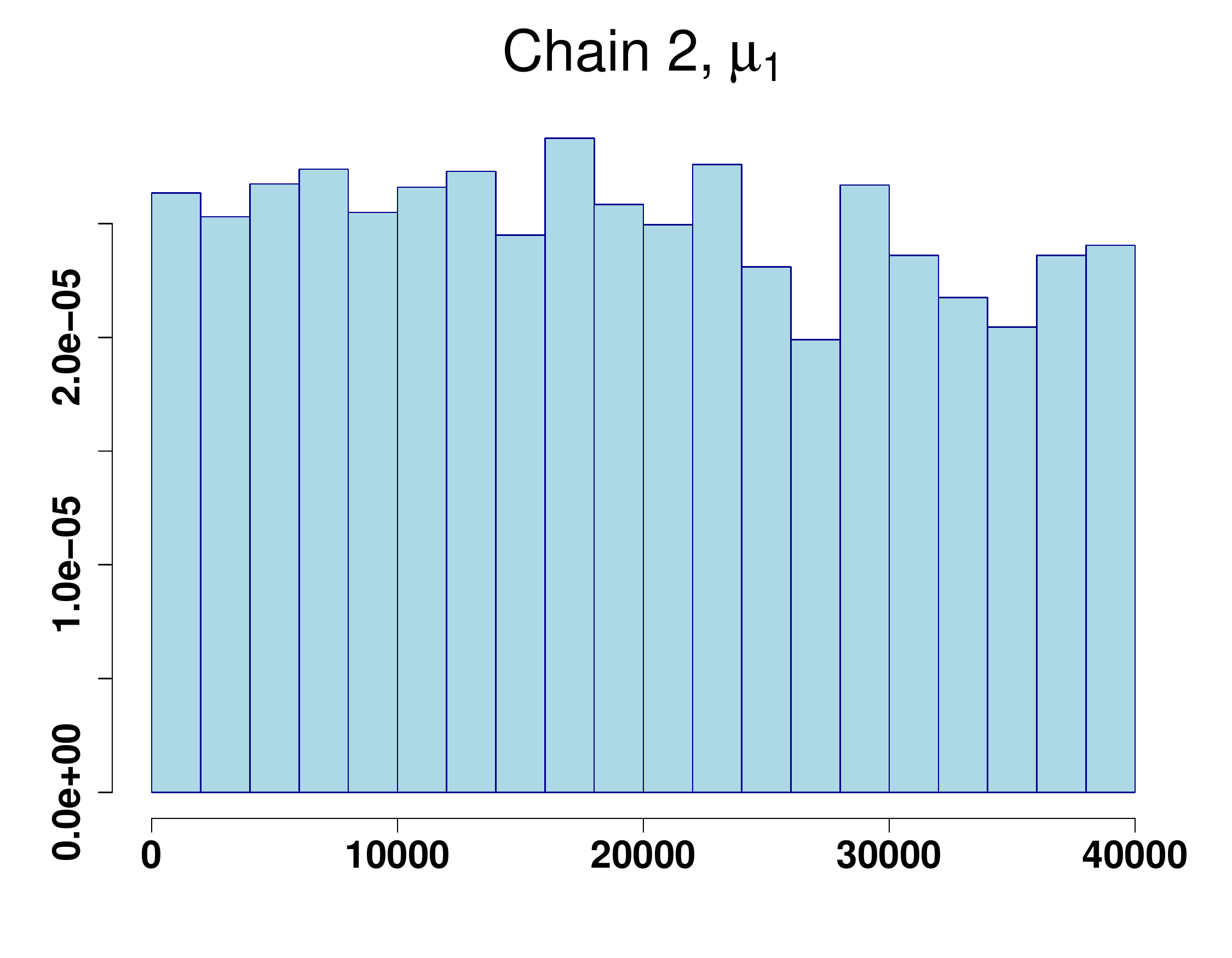}&
\hspace{-1.0cm}\includegraphics[width=4.0cm]{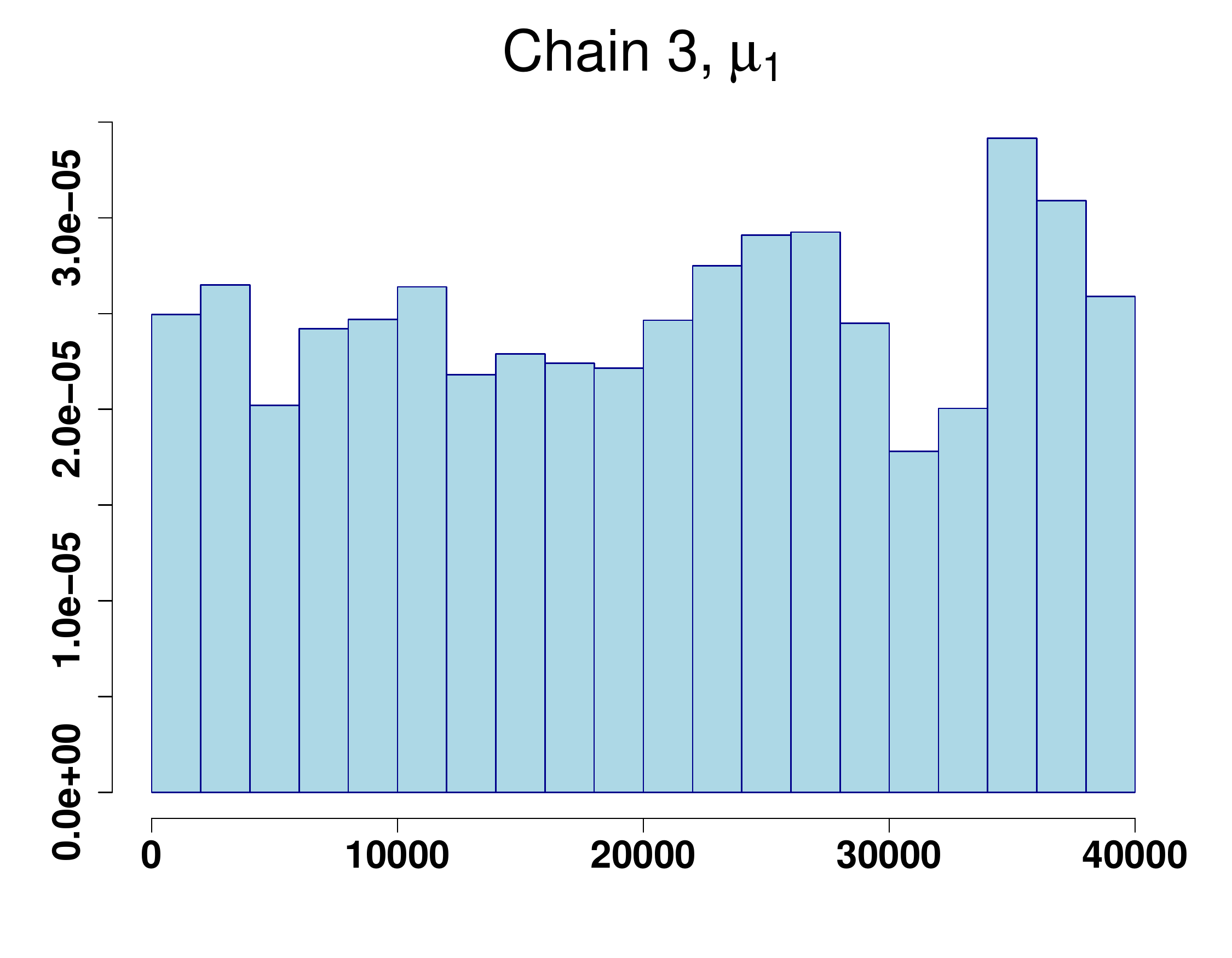}&\hspace{-1.9cm}\includegraphics[width=4.0cm]{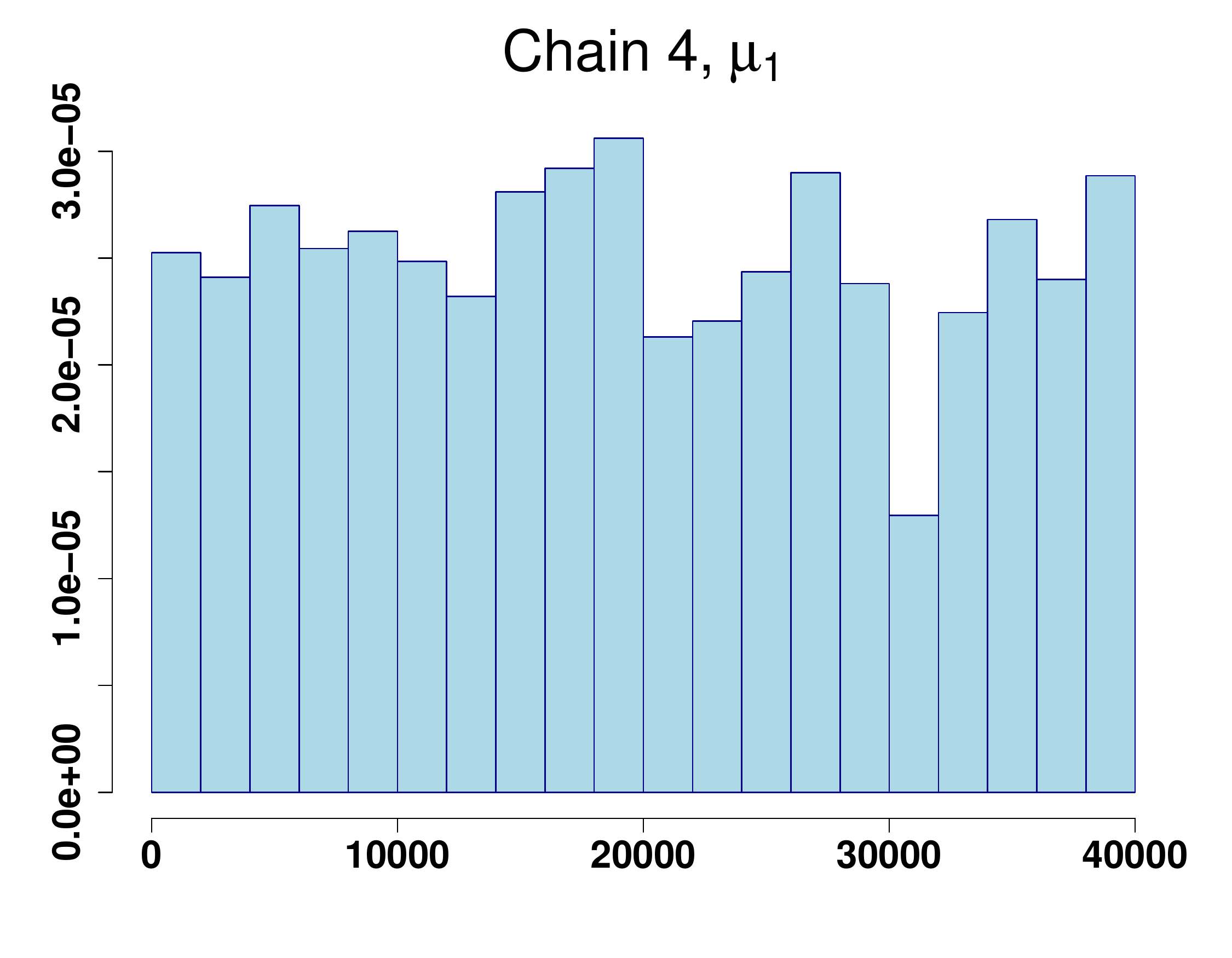}\\
\hspace{0.0cm}\includegraphics[width=4.0cm]{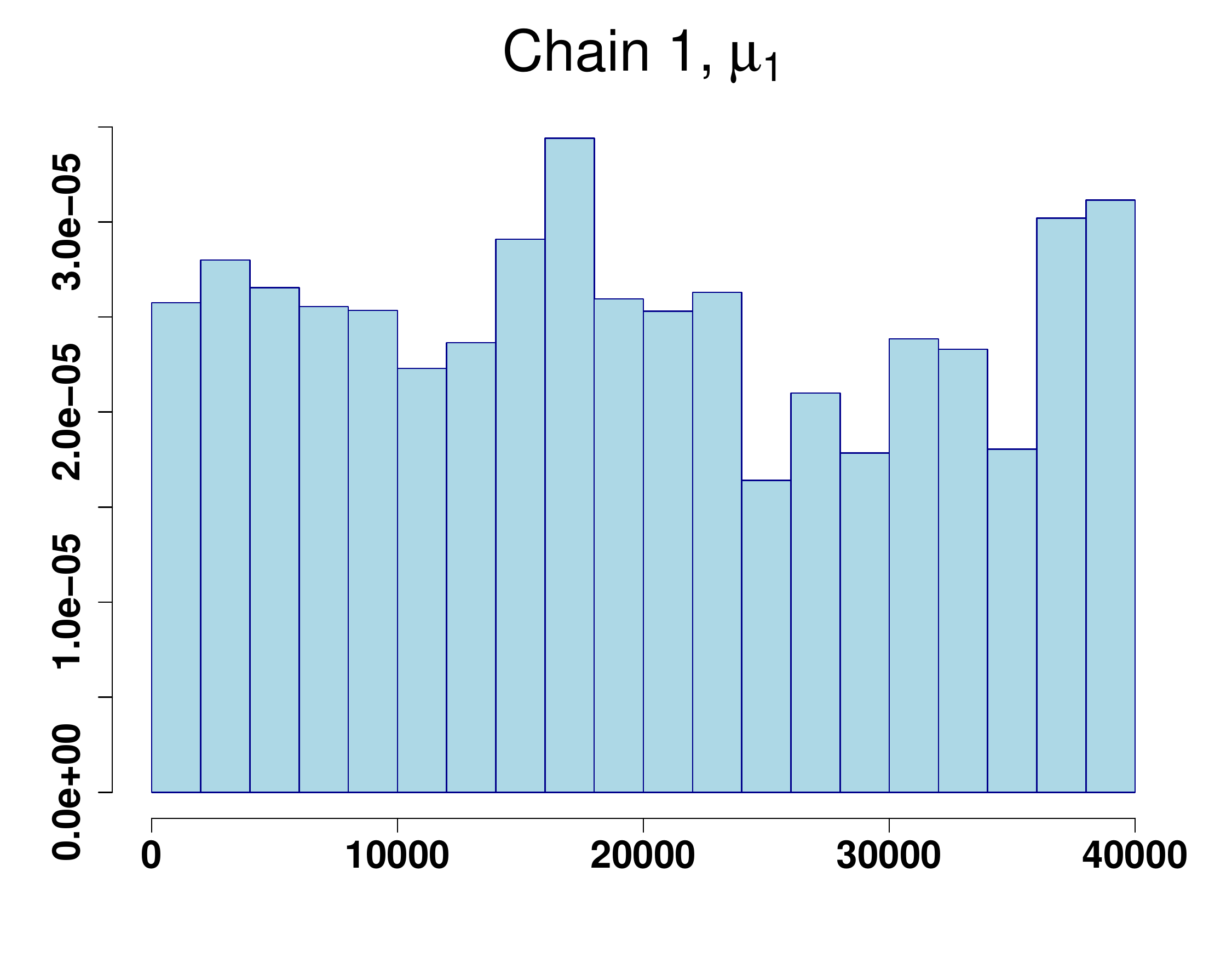}&\hspace{-0.5cm}\includegraphics[width=4.0cm]{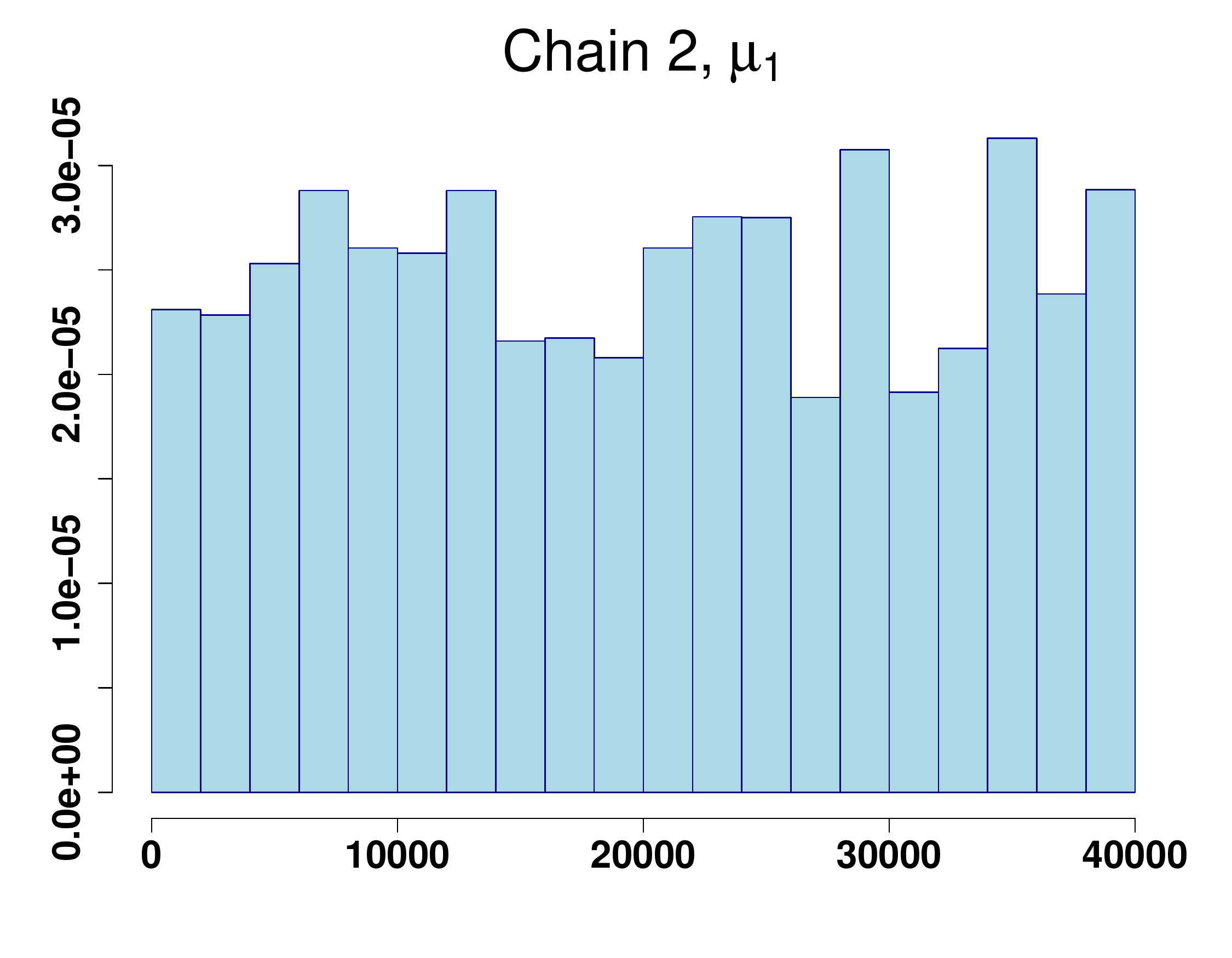}&
\hspace{-1.0cm}\includegraphics[width=4.0cm]{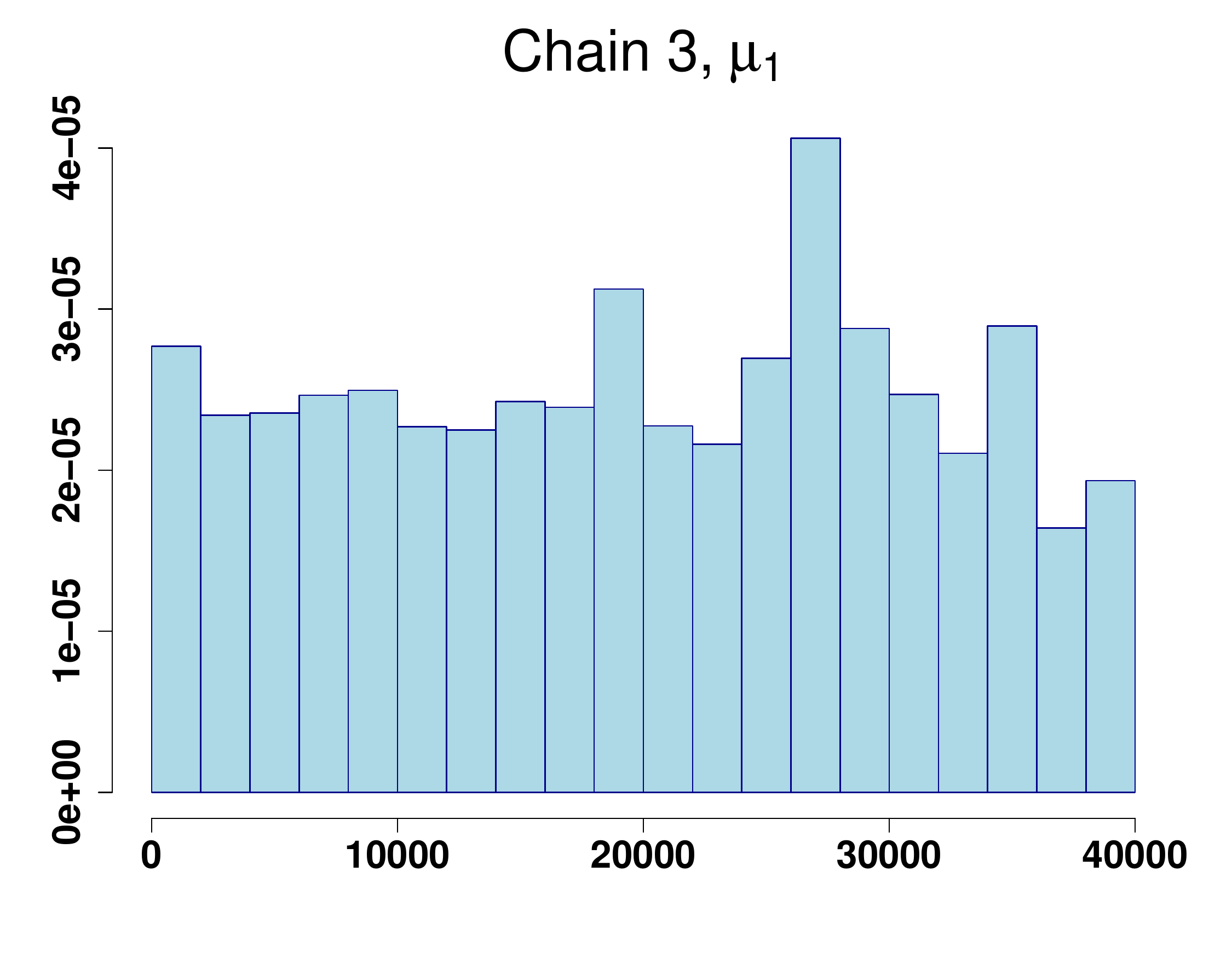}&\hspace{-1.9cm}\includegraphics[width=4.0cm]{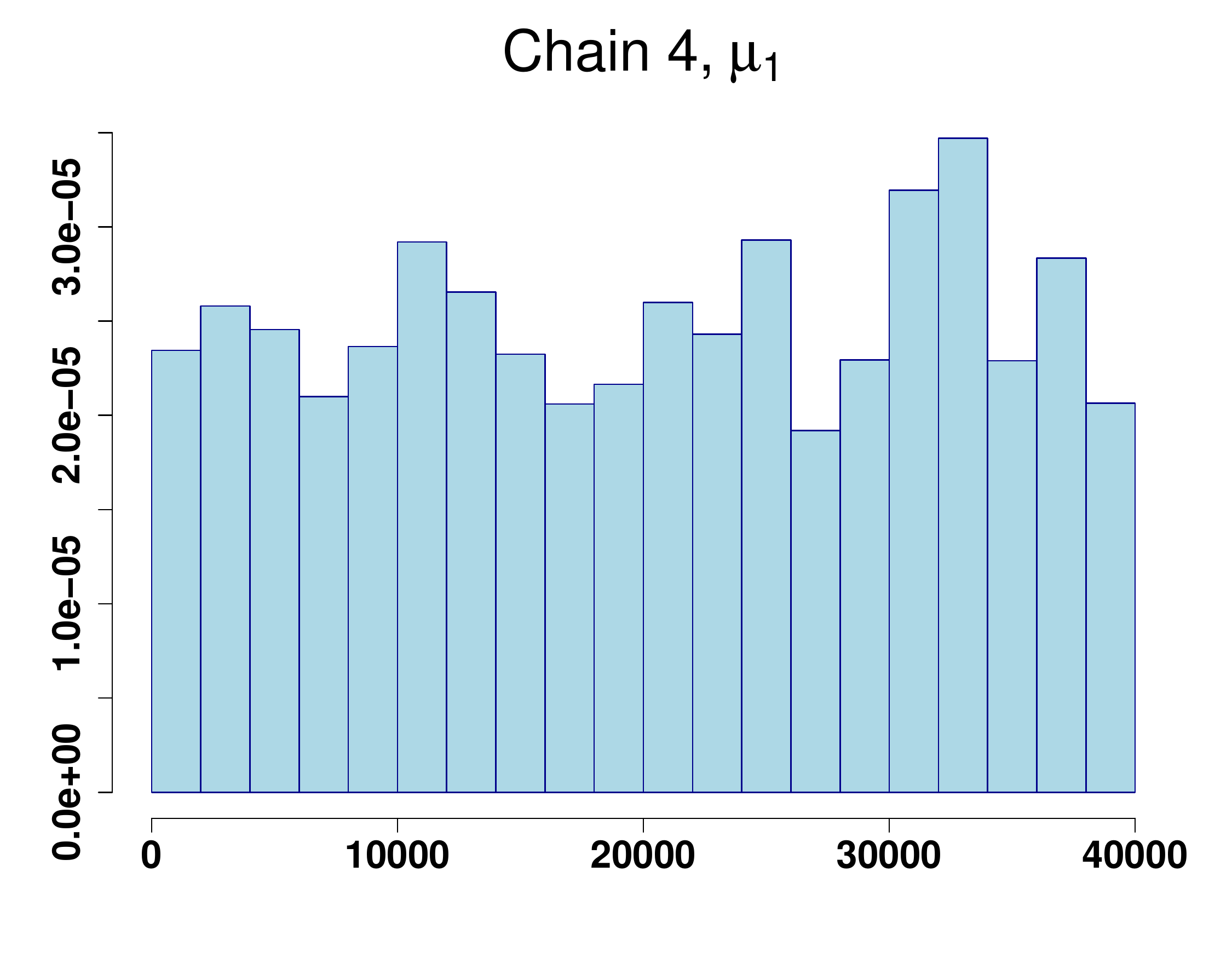}\\
\hspace{0.0cm}\includegraphics[width=4.0cm]{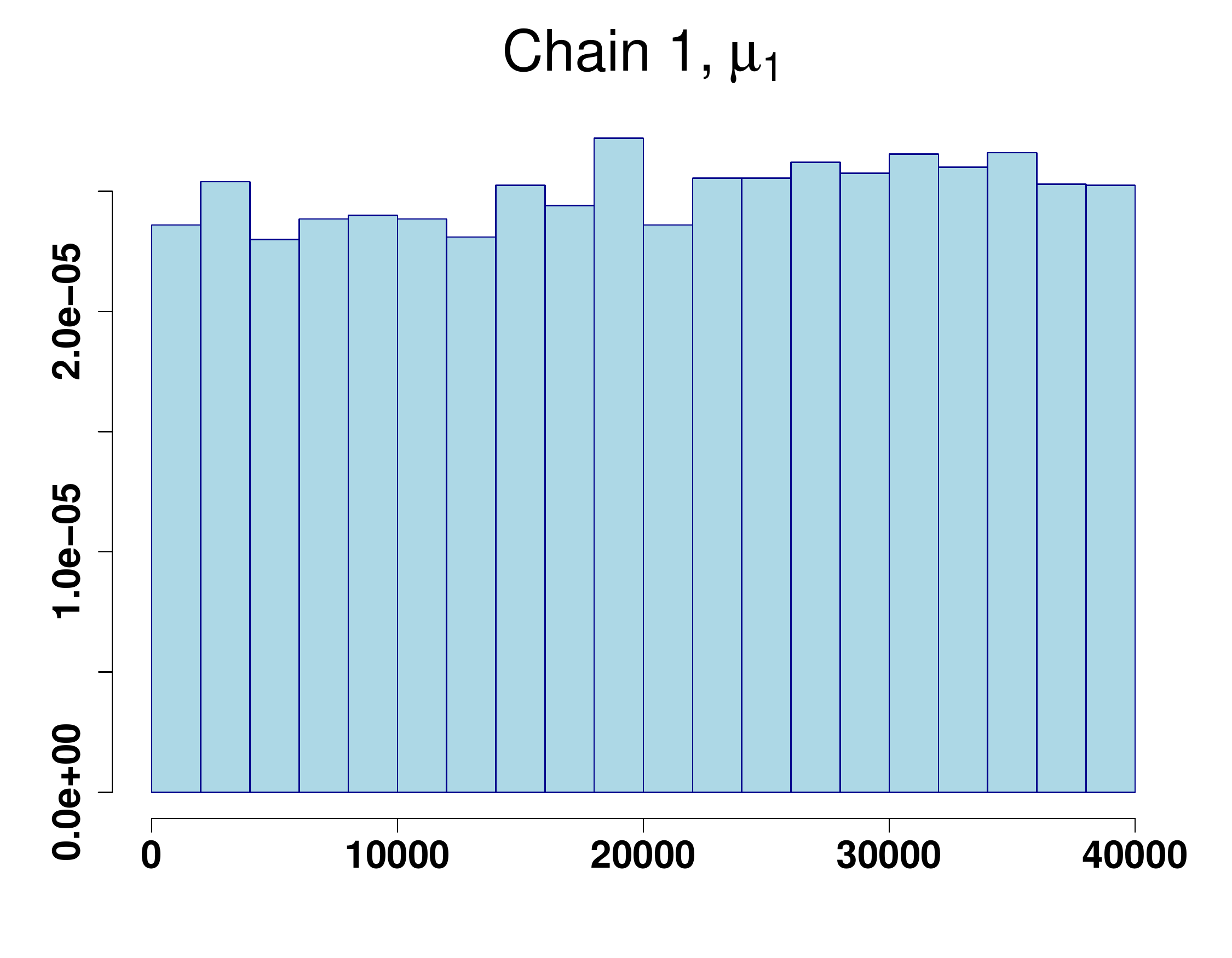}&\hspace{-0.5cm}\includegraphics[width=4.0cm]{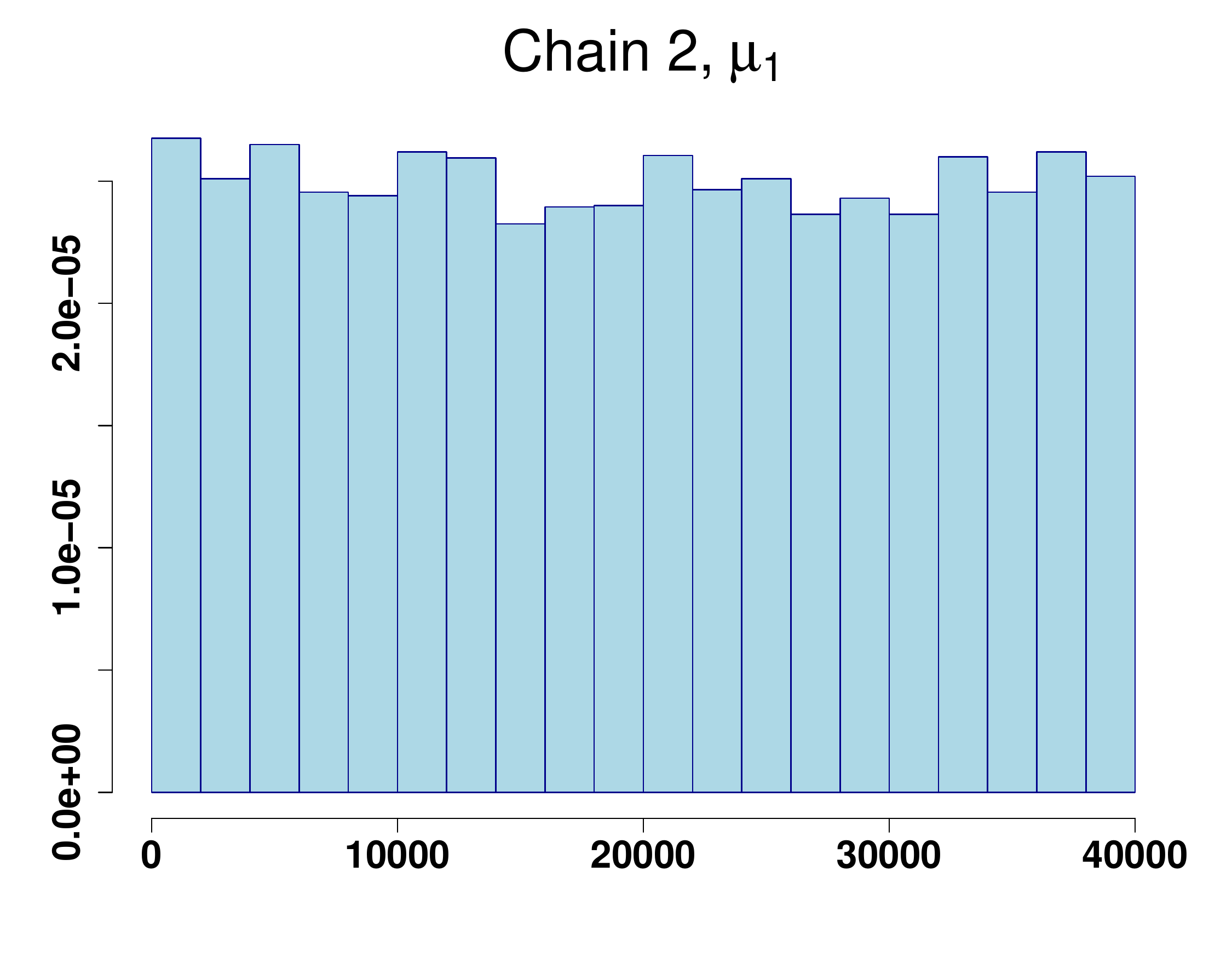}&
\hspace{-1.0cm}\includegraphics[width=4.0cm]{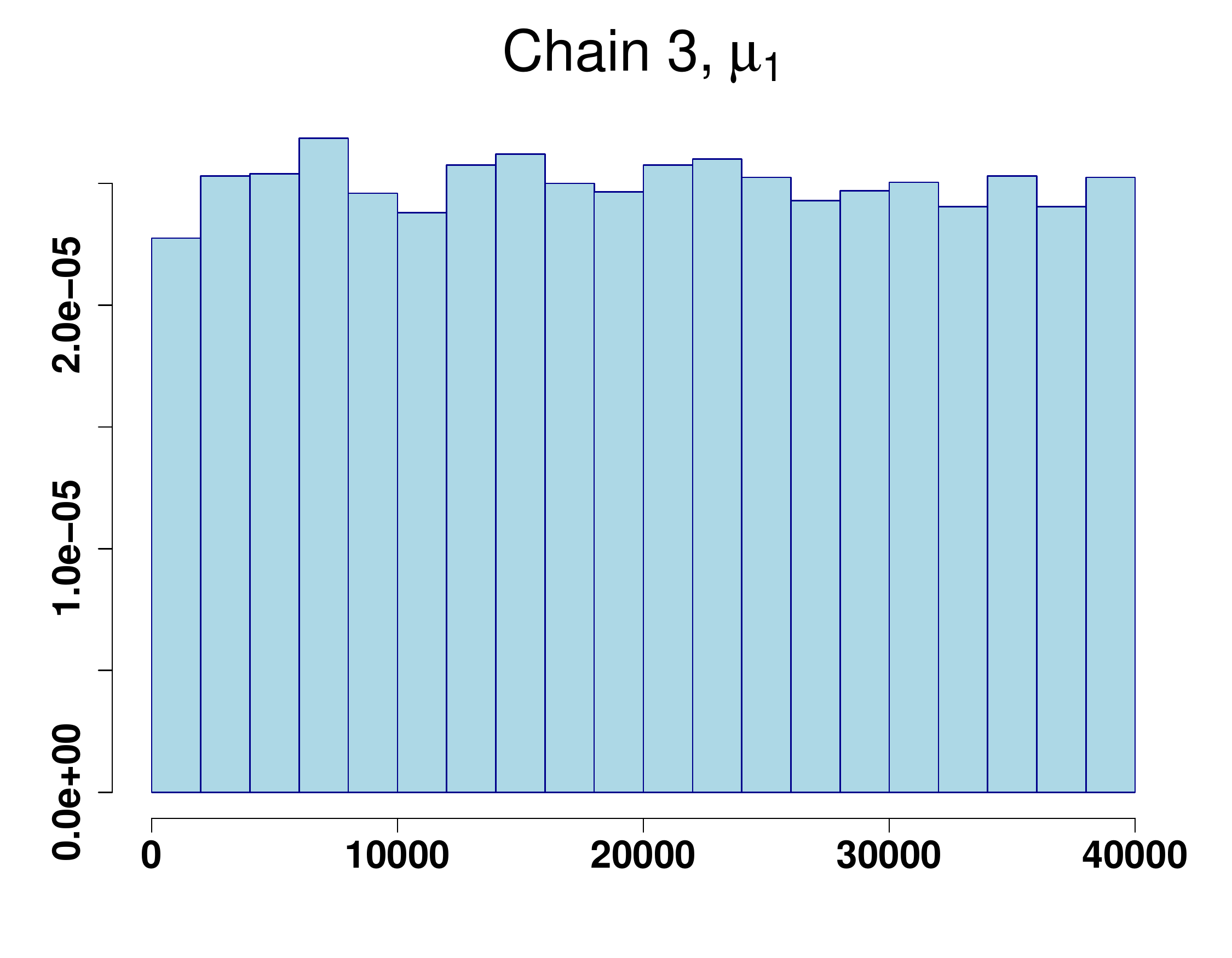}&\hspace{-1.9cm}\includegraphics[width=4.0cm]{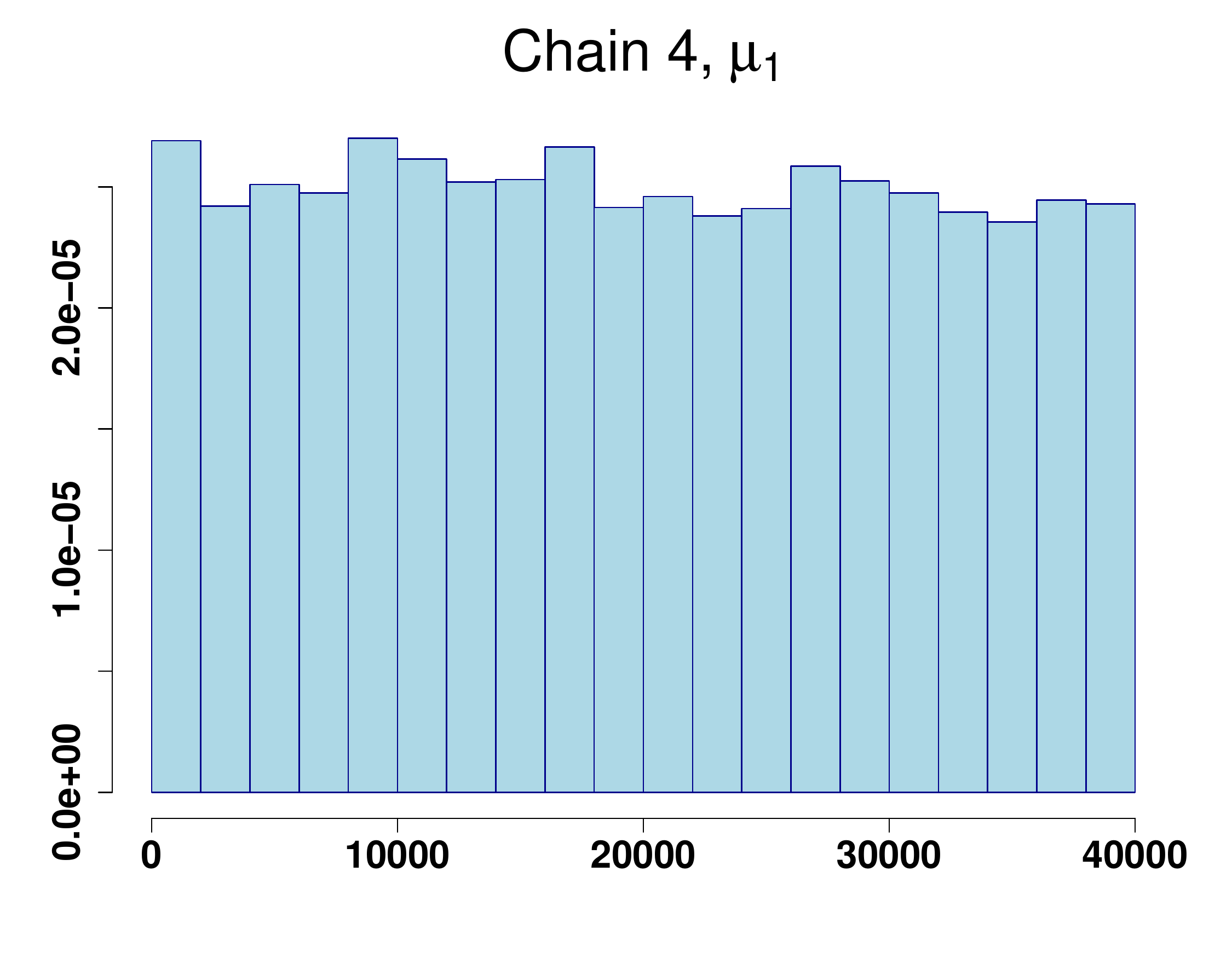}\\
\end{tabular}
 \caption{Rank plots of posterior draws from four chains in the case of the parameter $\mu_{1}$ (SBP) of the $t$ multivariate random effects model by employing the Jeffreys prior (first to third rows) and the Berger and Bernardo reference prior (fourth to sixth rows). The samples from the posterior distributions are drawn by Algorithm A (first and fourth rows), Algorithm B (second and fifth rows) and Algorithm C (third and sixth rows).}
\label{fig:emp-study-rank-mu1-t}
 \end{figure}

\begin{figure}[h!t]
\centering
\begin{tabular}{p{4.0cm}p{4.0cm}p{4.0cm}p{4.0cm}}
\hspace{0.0cm}\includegraphics[width=4.0cm]{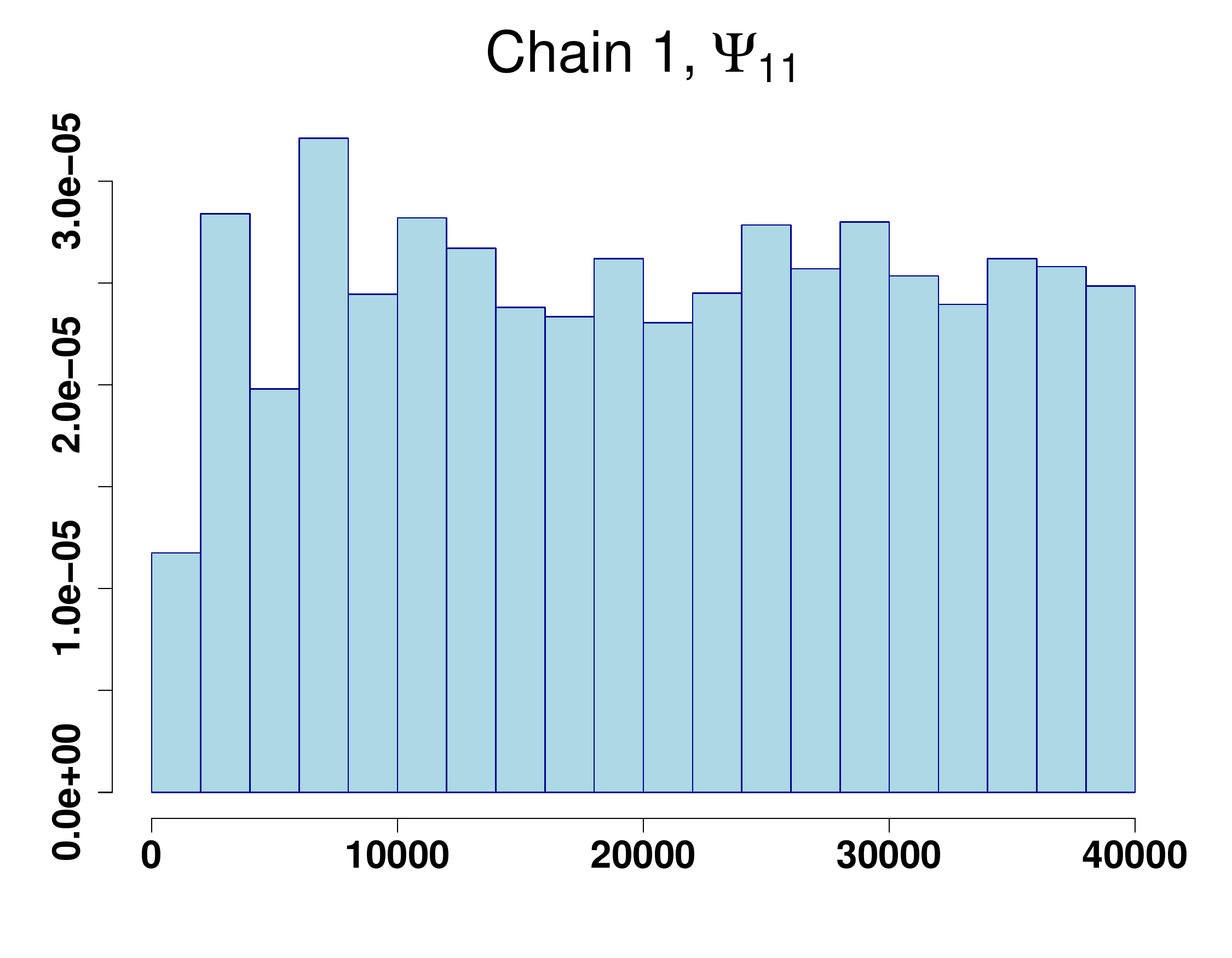}&\hspace{-0.5cm}\includegraphics[width=4.0cm]{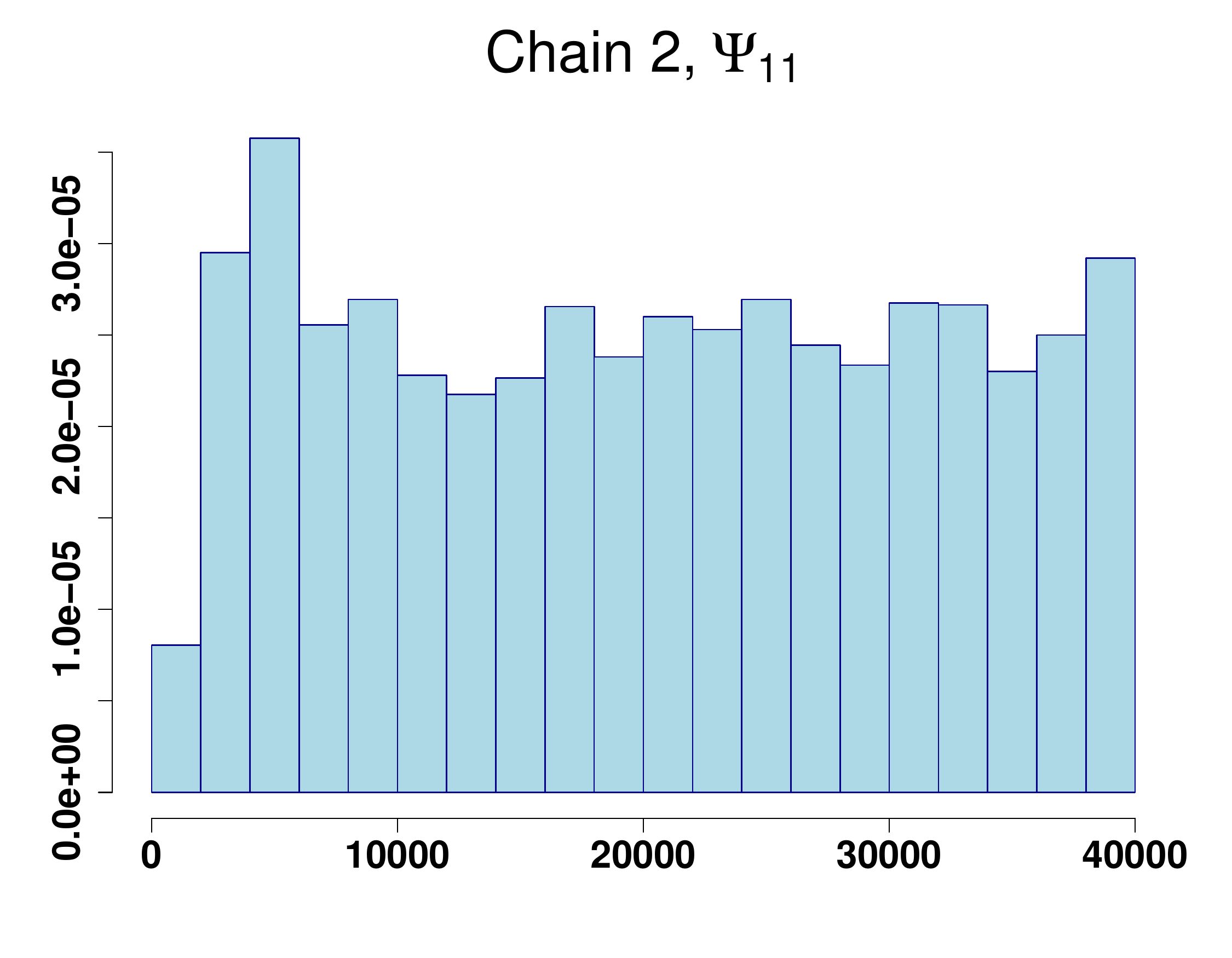}&
\hspace{-1.0cm}\includegraphics[width=4.0cm]{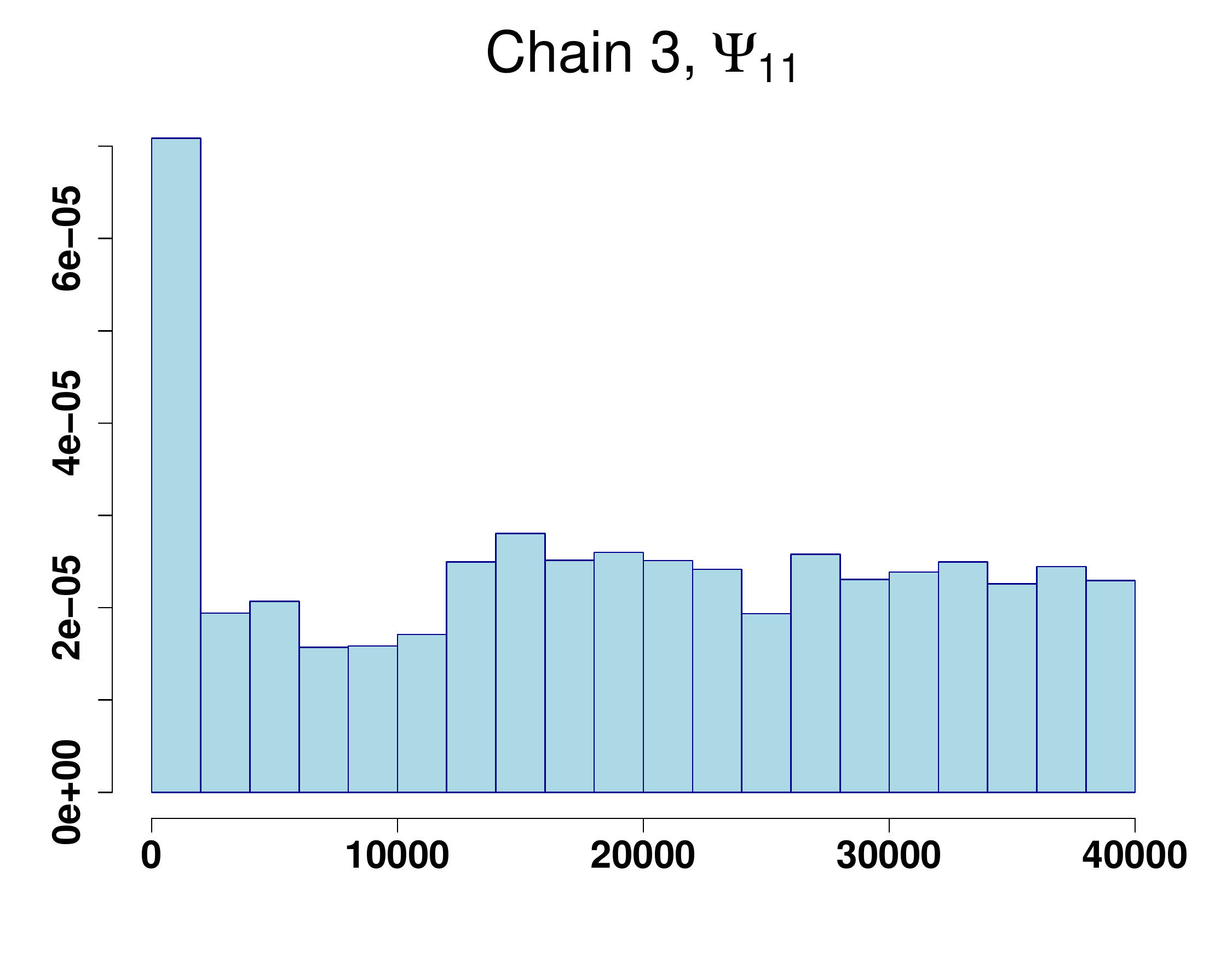}&\hspace{-1.5cm}\includegraphics[width=4.0cm]{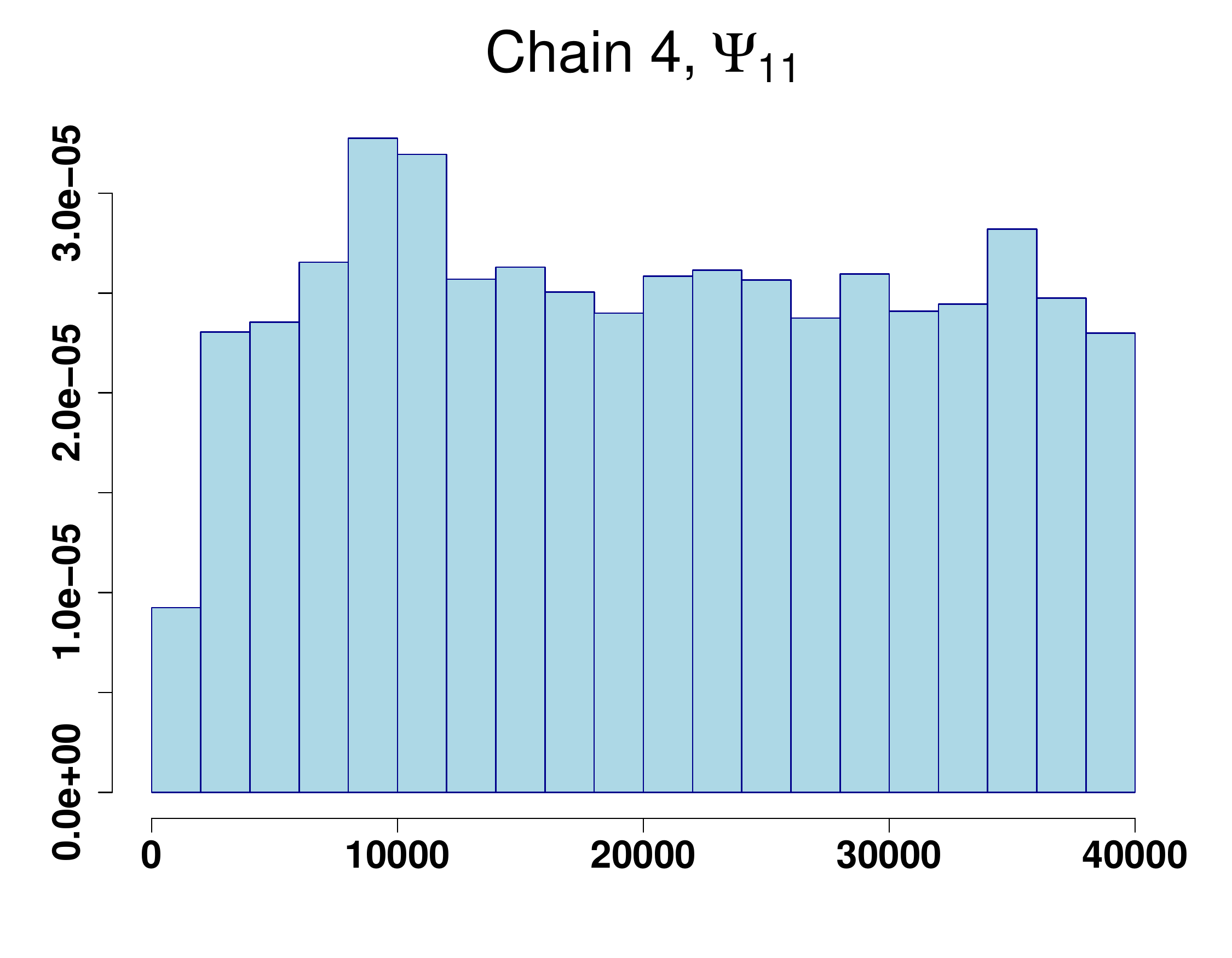}\\
\hspace{0.0cm}\includegraphics[width=4.0cm]{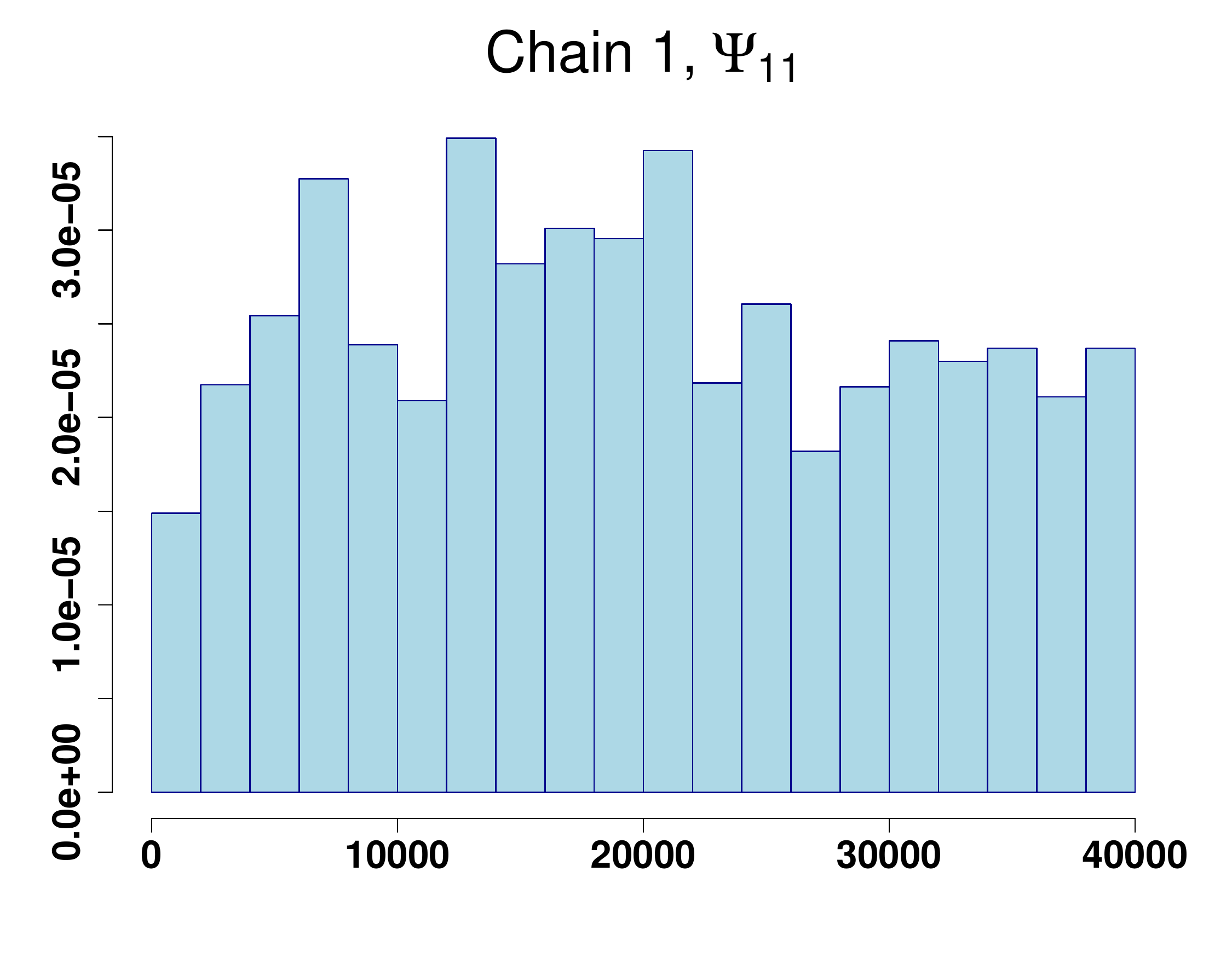}&\hspace{-0.5cm}\includegraphics[width=4.0cm]{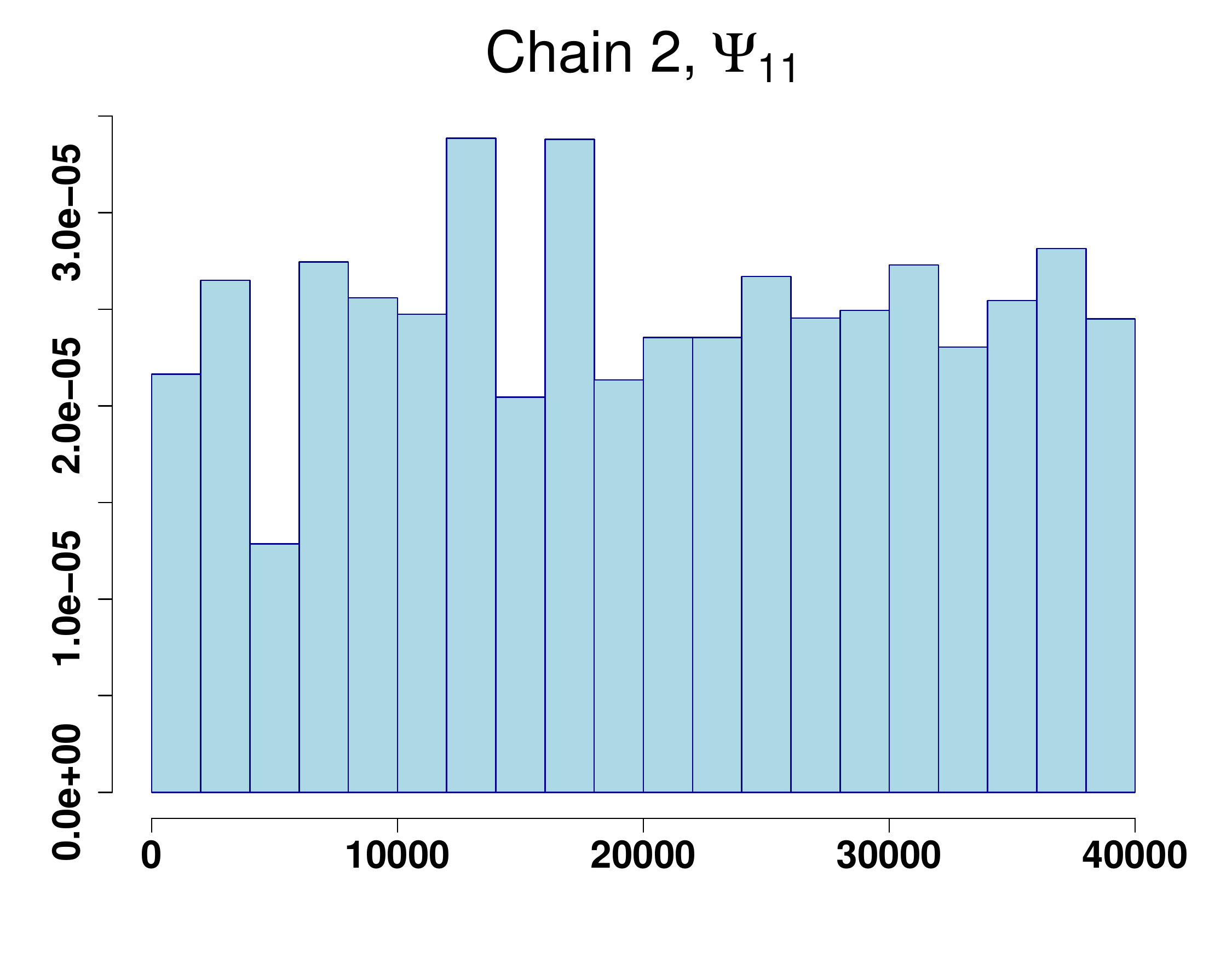}&
\hspace{-1.0cm}\includegraphics[width=4.0cm]{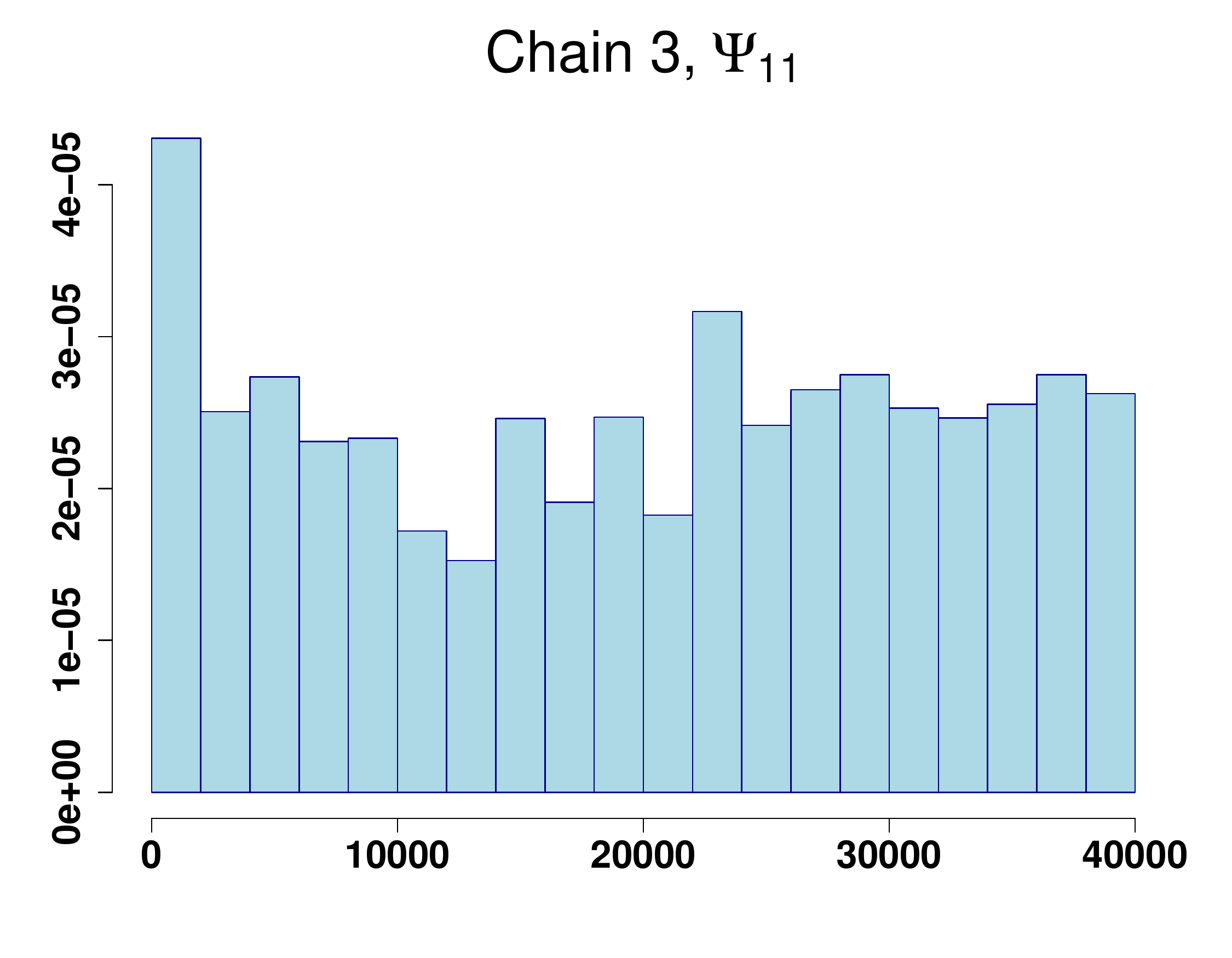}&\hspace{-1.5cm}\includegraphics[width=4.0cm]{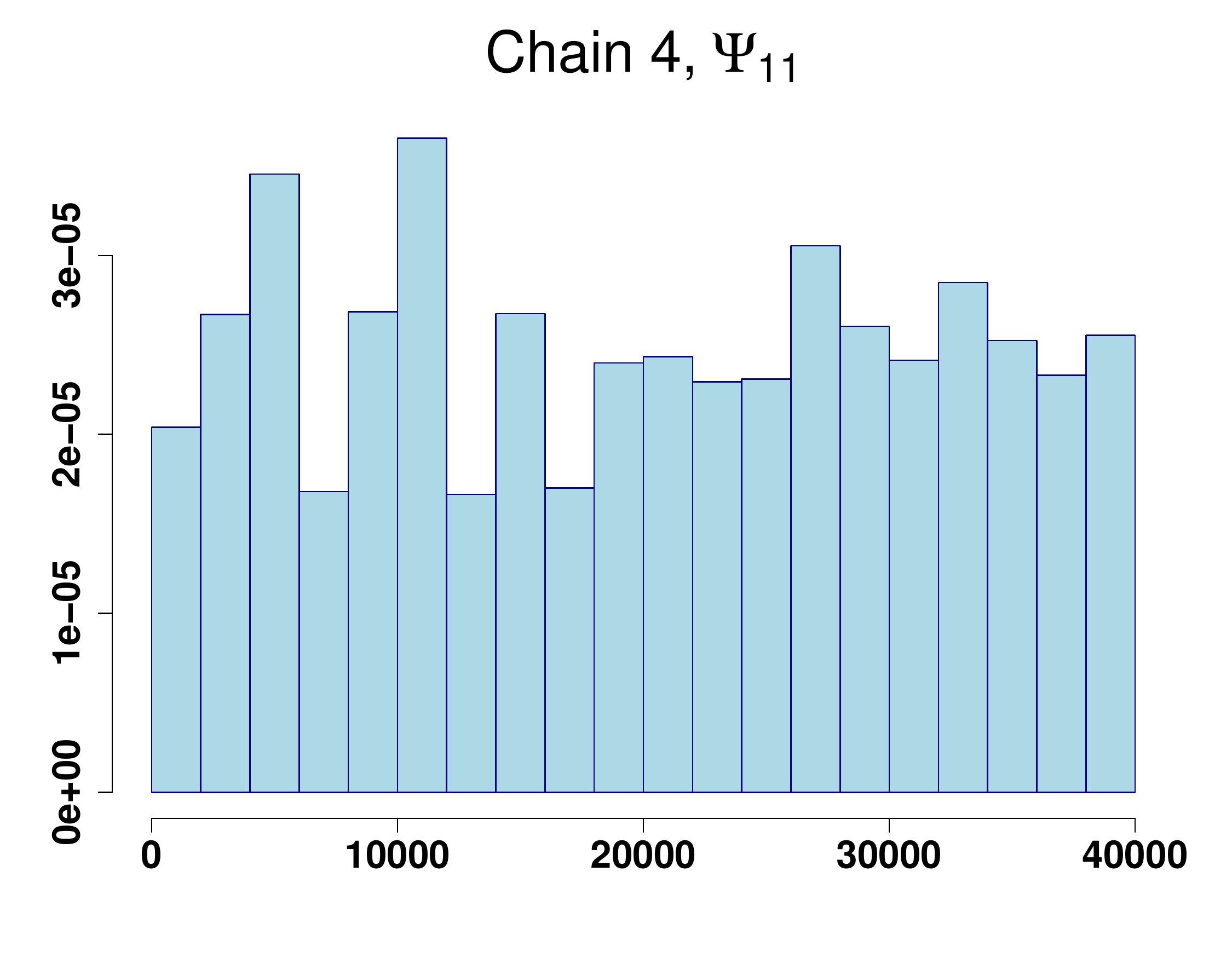}\\
\hspace{0.0cm}\includegraphics[width=4.0cm]{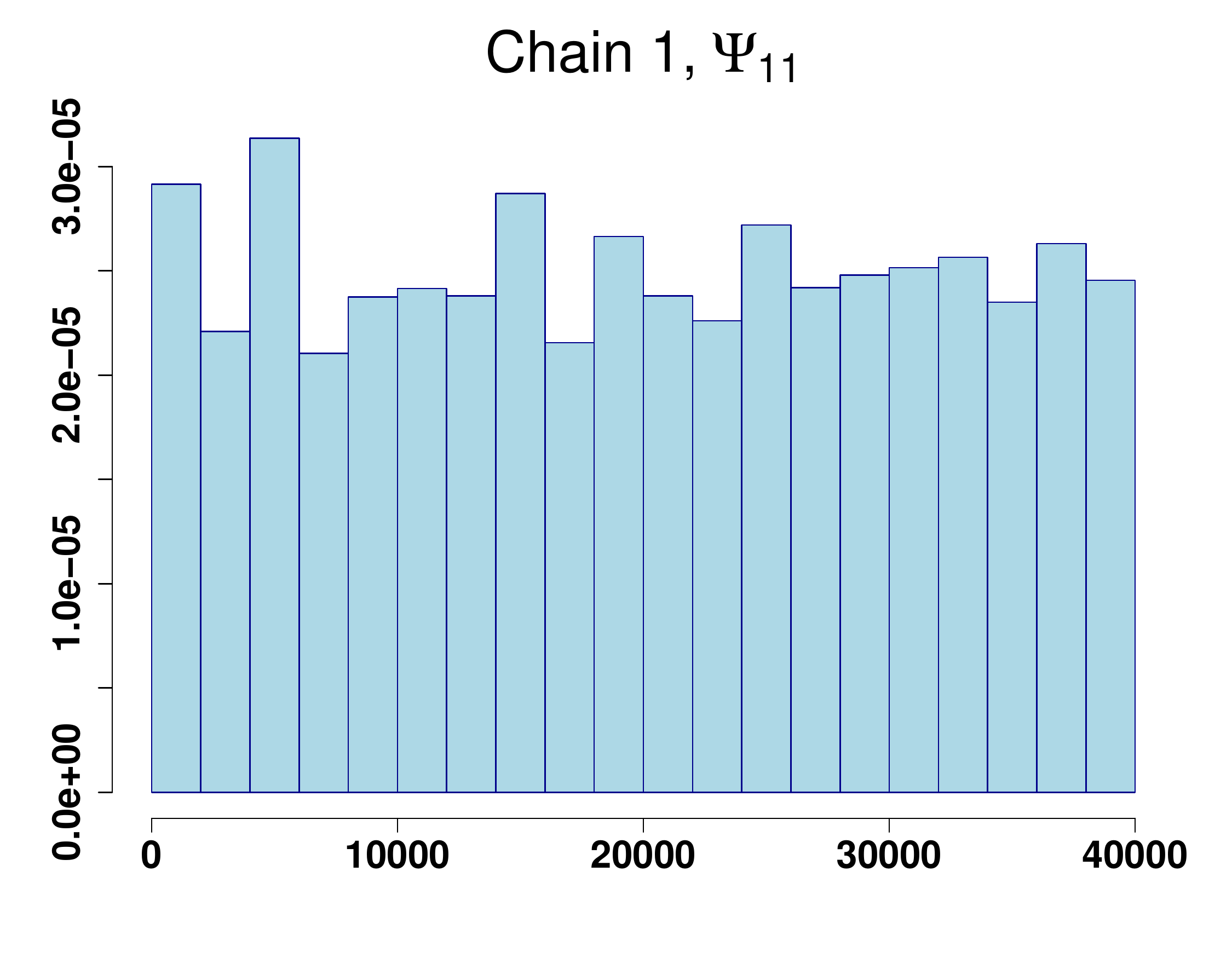}&\hspace{-0.5cm}\includegraphics[width=4.0cm]{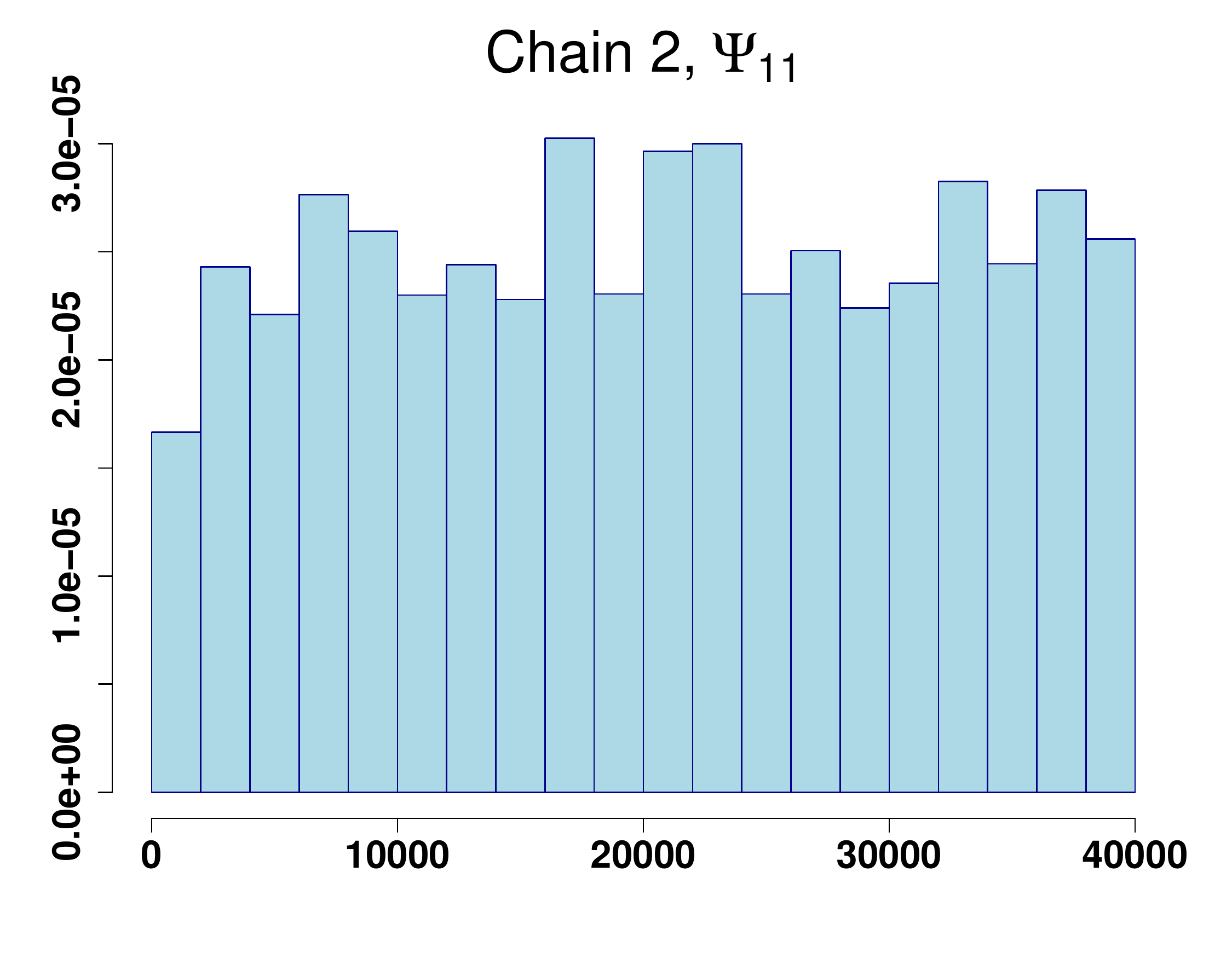}&
\hspace{-1.0cm}\includegraphics[width=4.0cm]{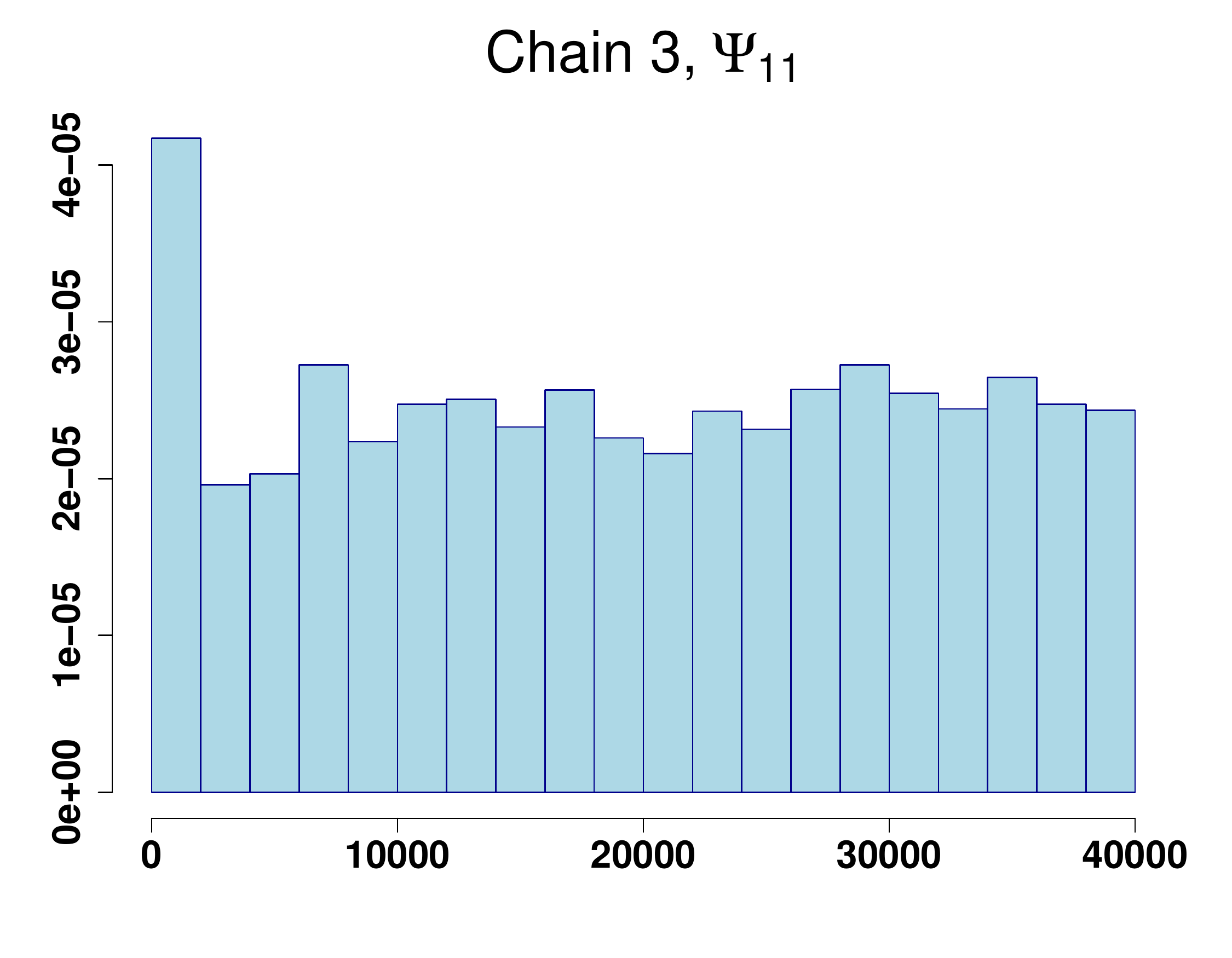}&\hspace{-1.5cm}\includegraphics[width=4.0cm]{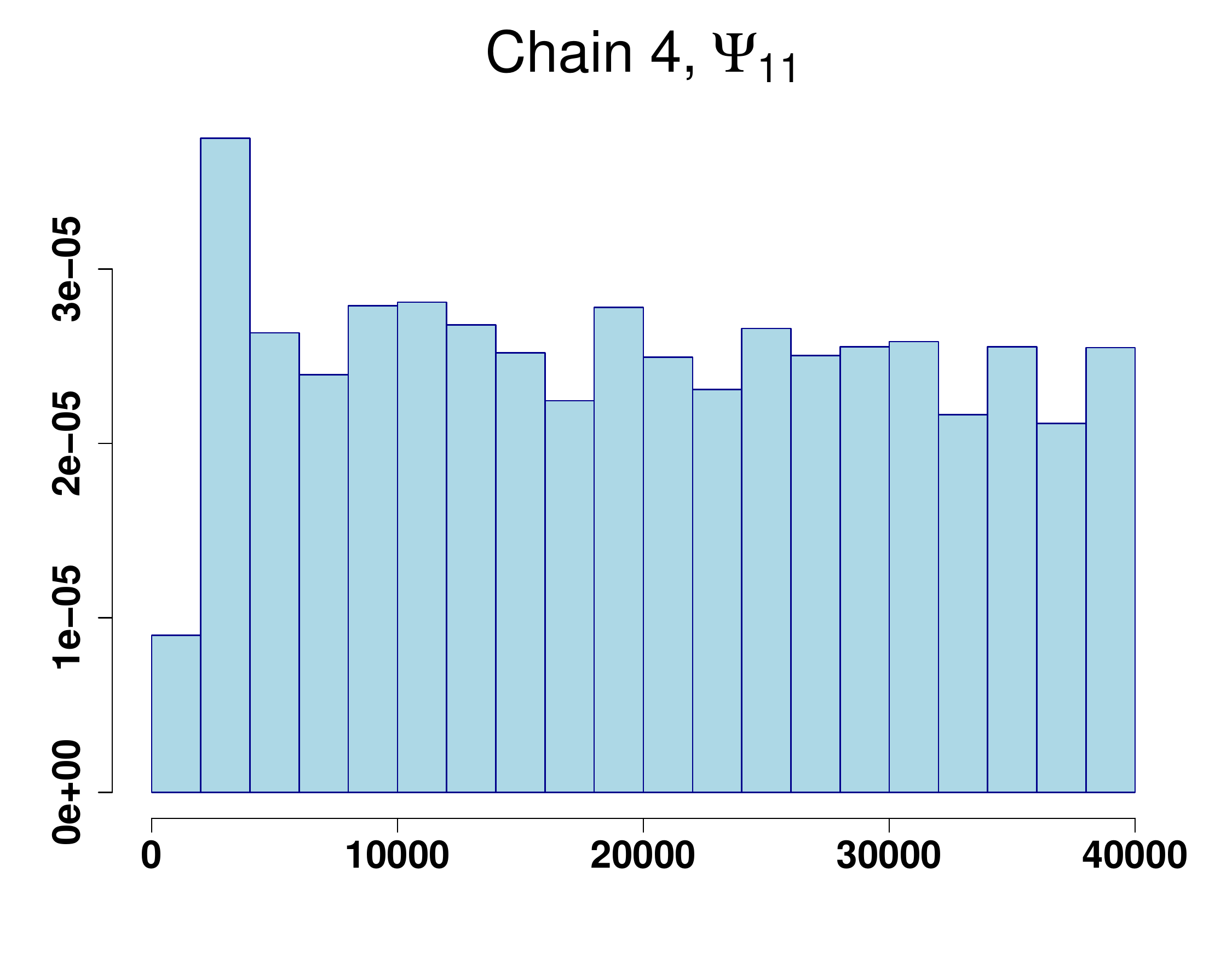}\\
\hspace{0.0cm}\includegraphics[width=4.0cm]{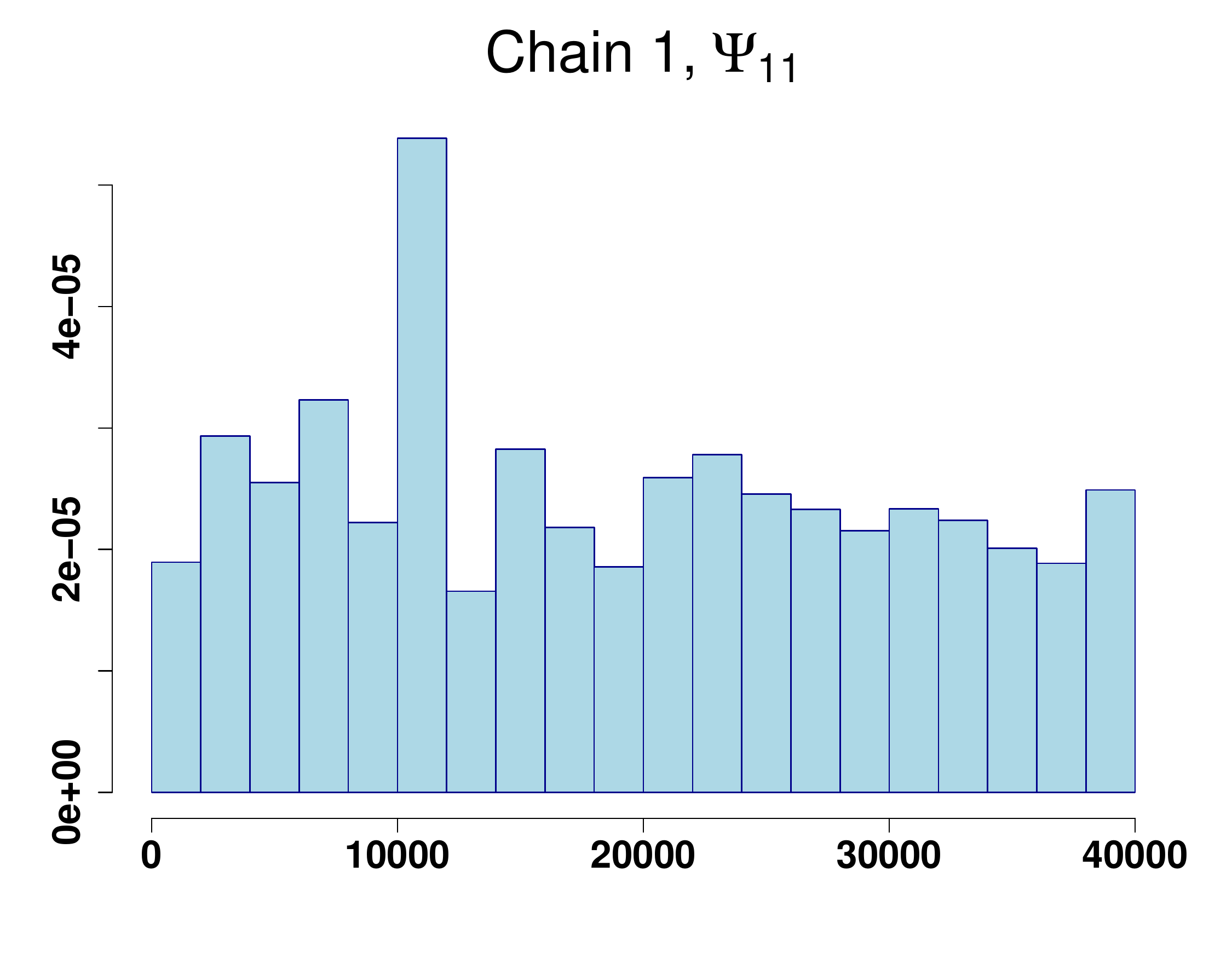}&\hspace{-0.5cm}\includegraphics[width=4.0cm]{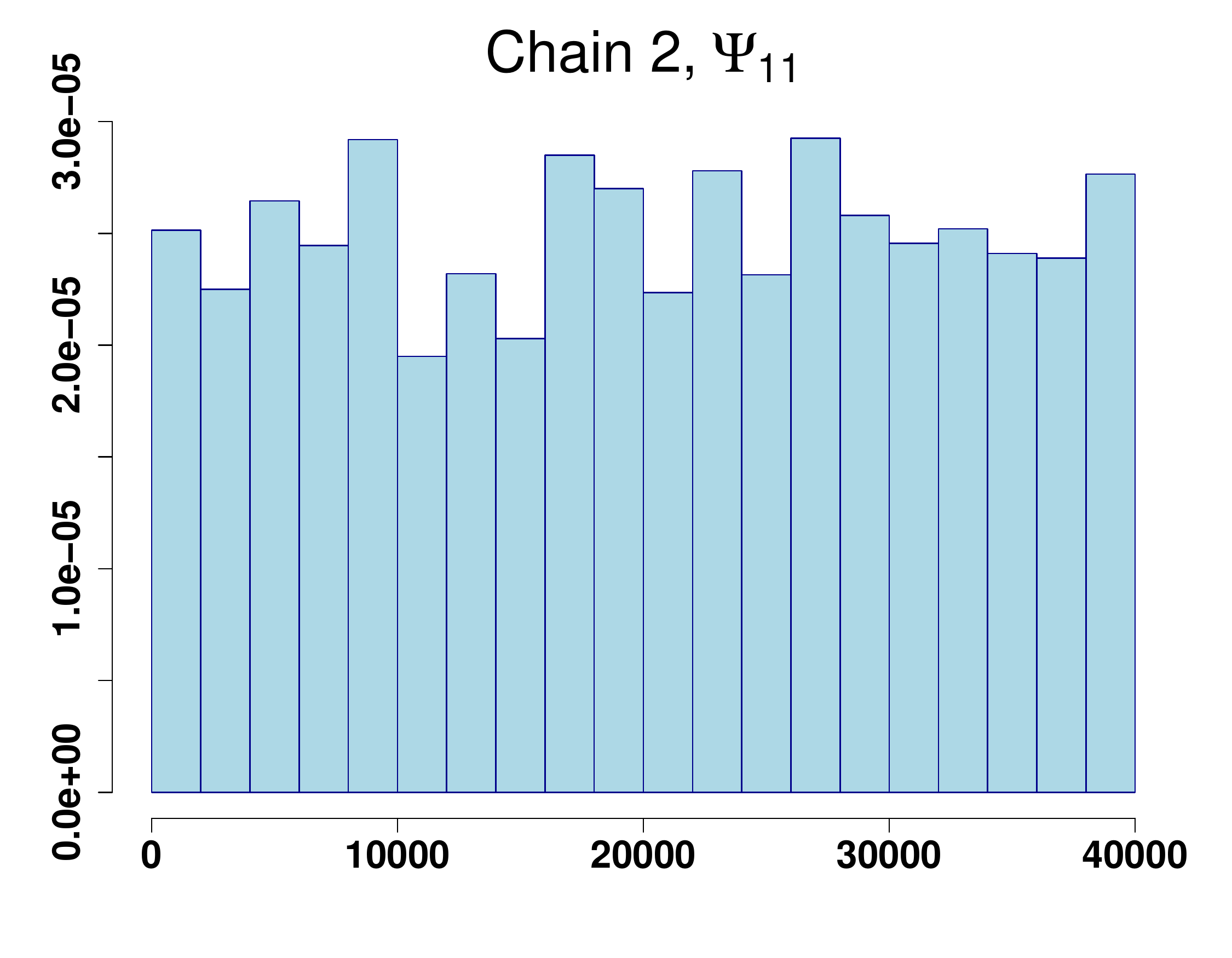}&
\hspace{-1.0cm}\includegraphics[width=4.0cm]{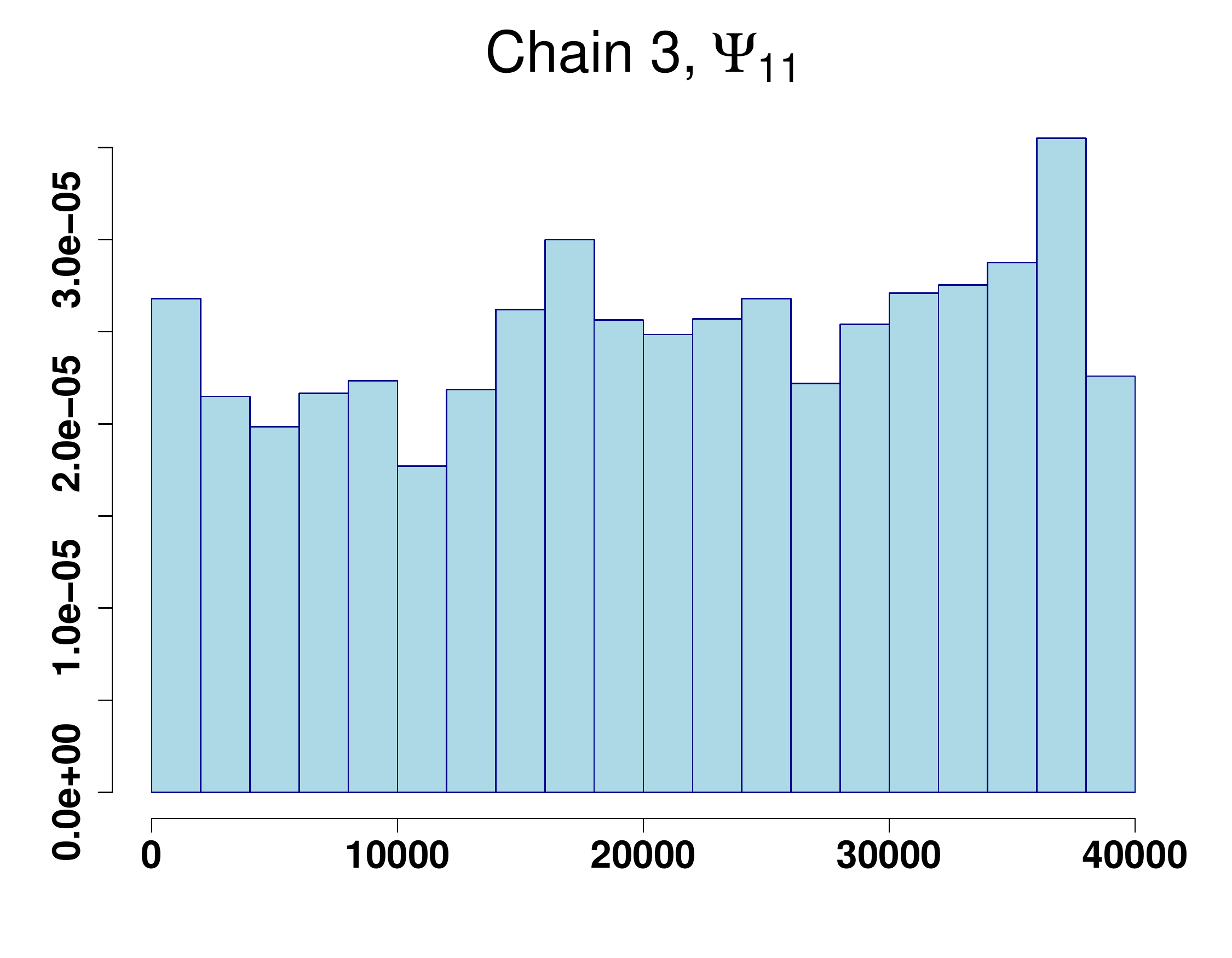}&\hspace{-1.5cm}\includegraphics[width=4.0cm]{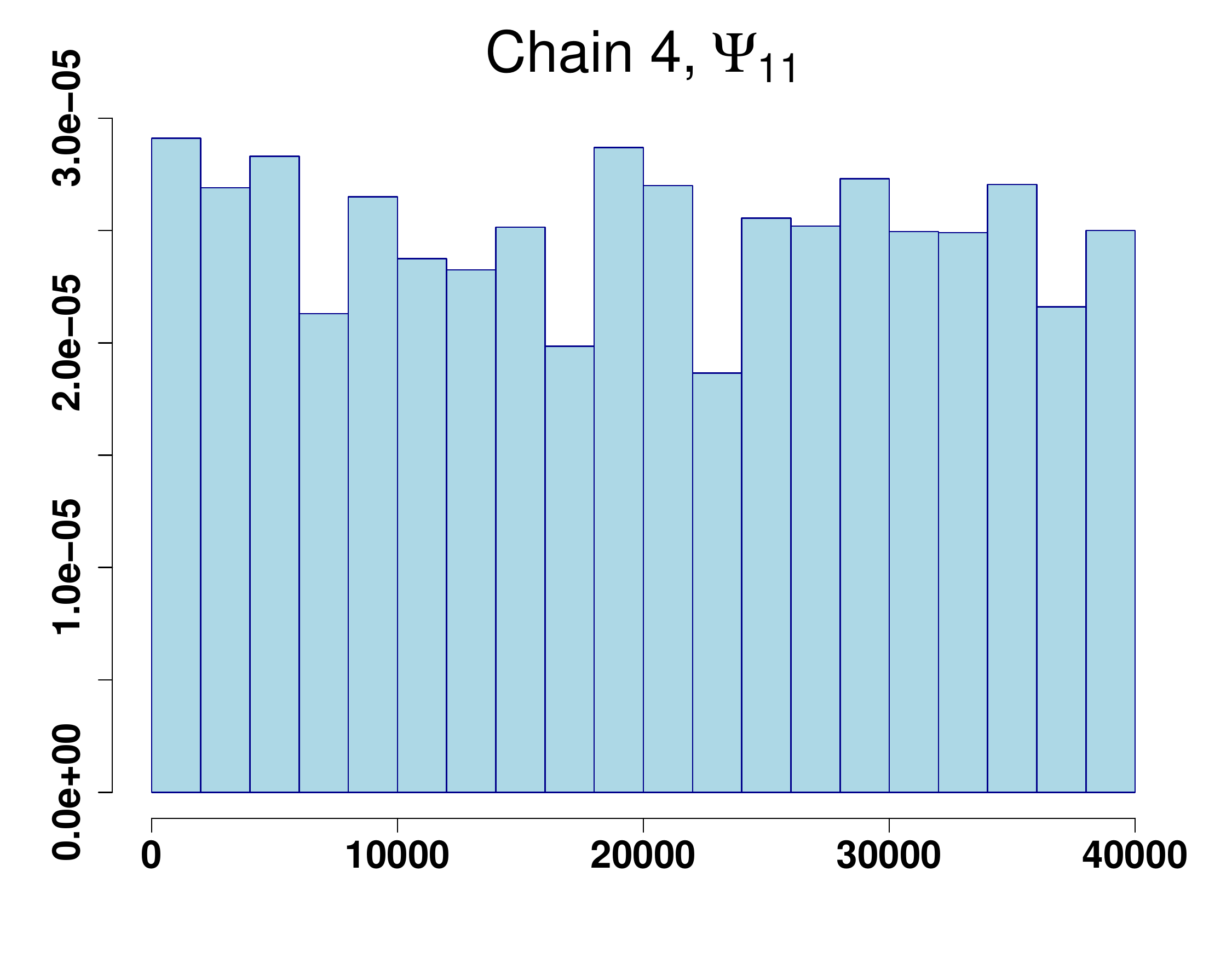}\\
\hspace{0.0cm}\includegraphics[width=4.0cm]{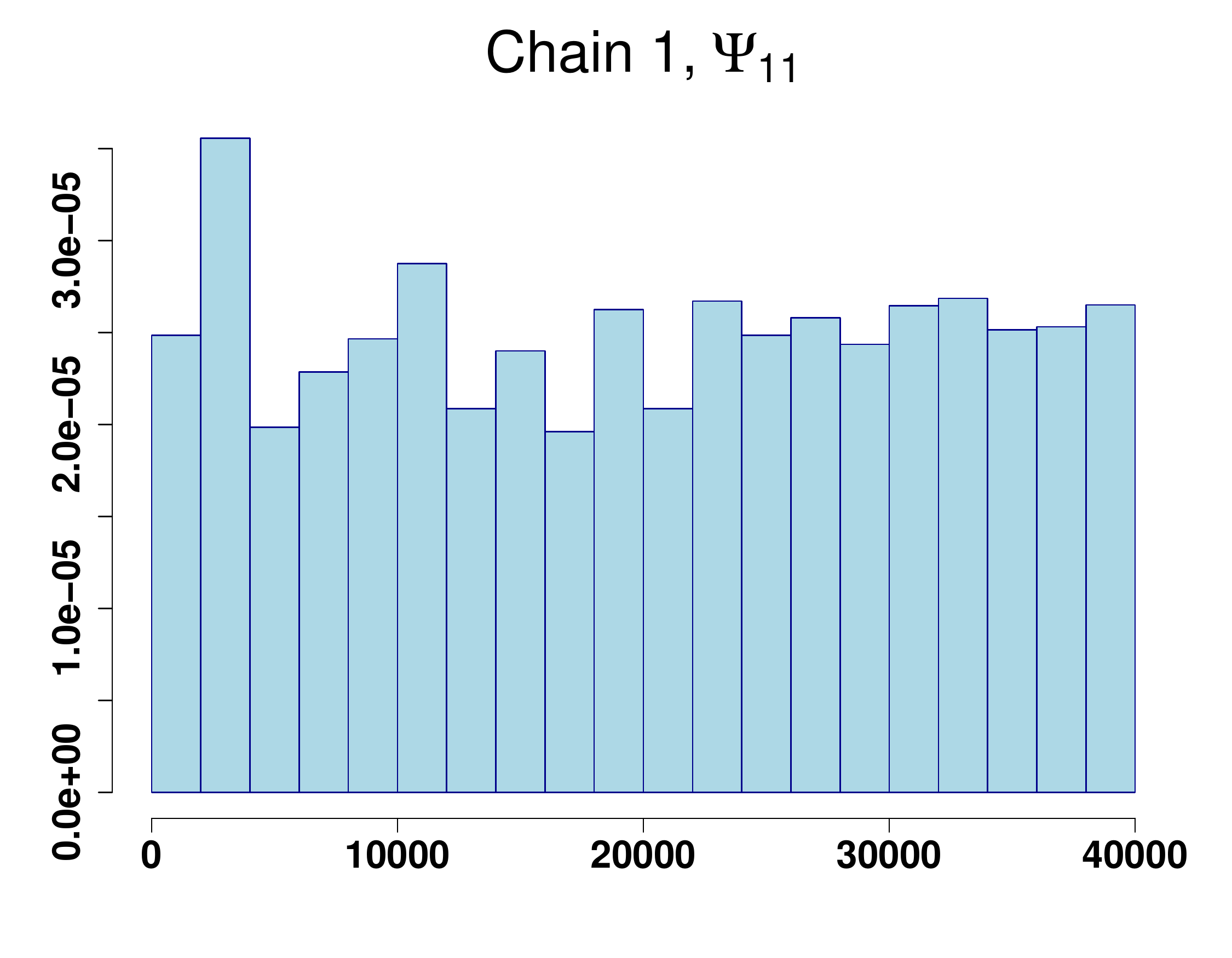}&\hspace{-0.5cm}\includegraphics[width=4.0cm]{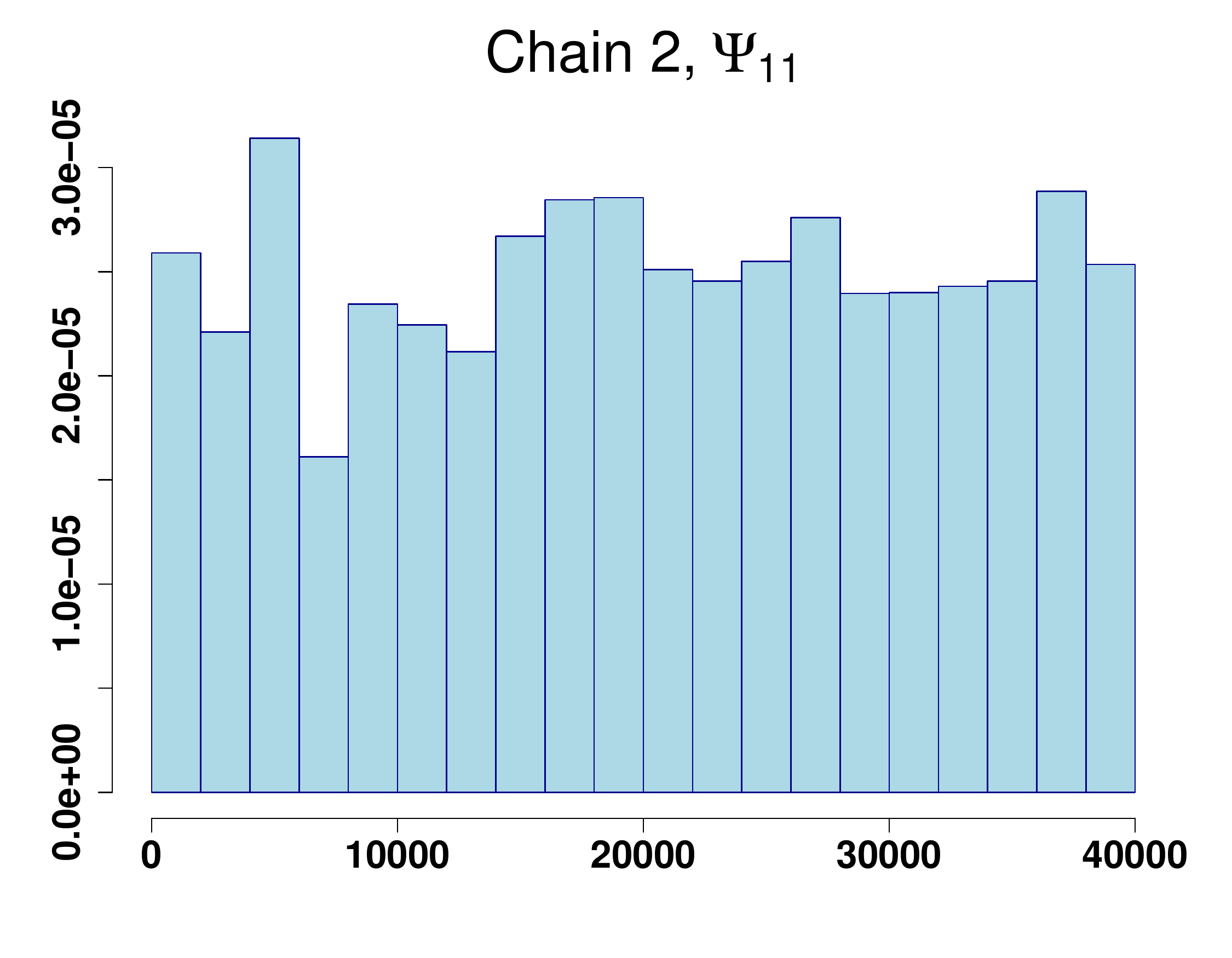}&
\hspace{-1.0cm}\includegraphics[width=4.0cm]{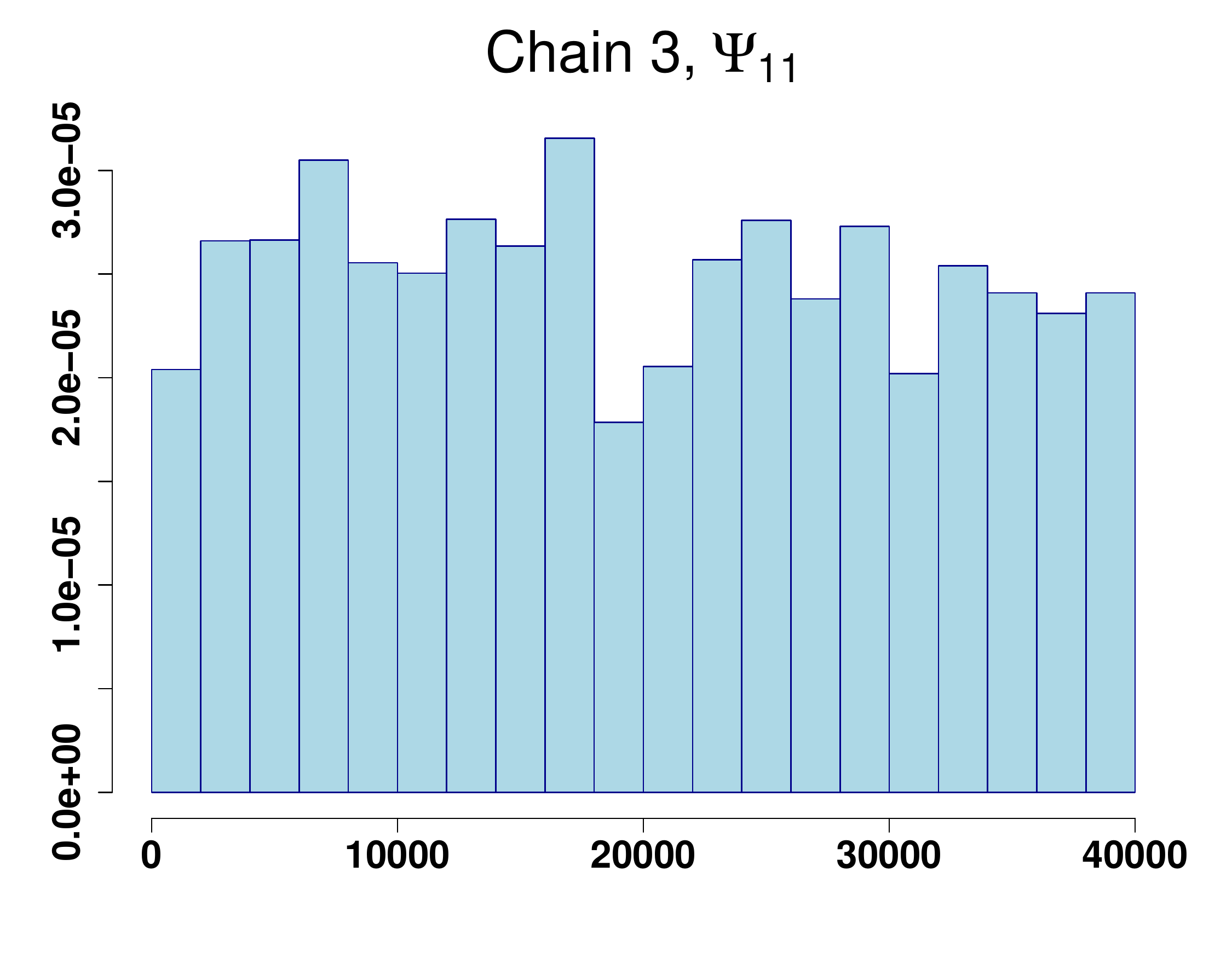}&\hspace{-1.5cm}\includegraphics[width=4.0cm]{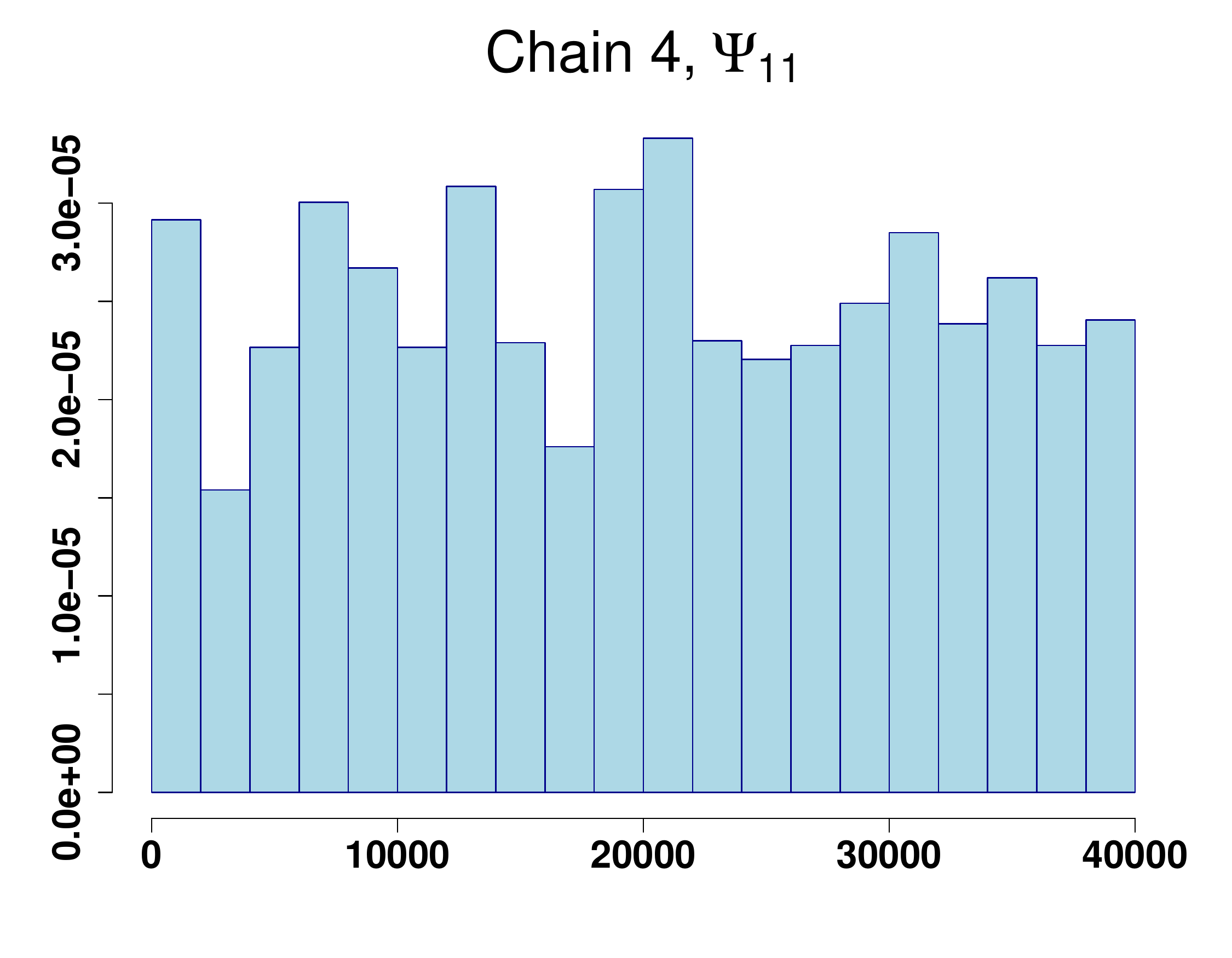}\\
\hspace{0.0cm}\includegraphics[width=4.0cm]{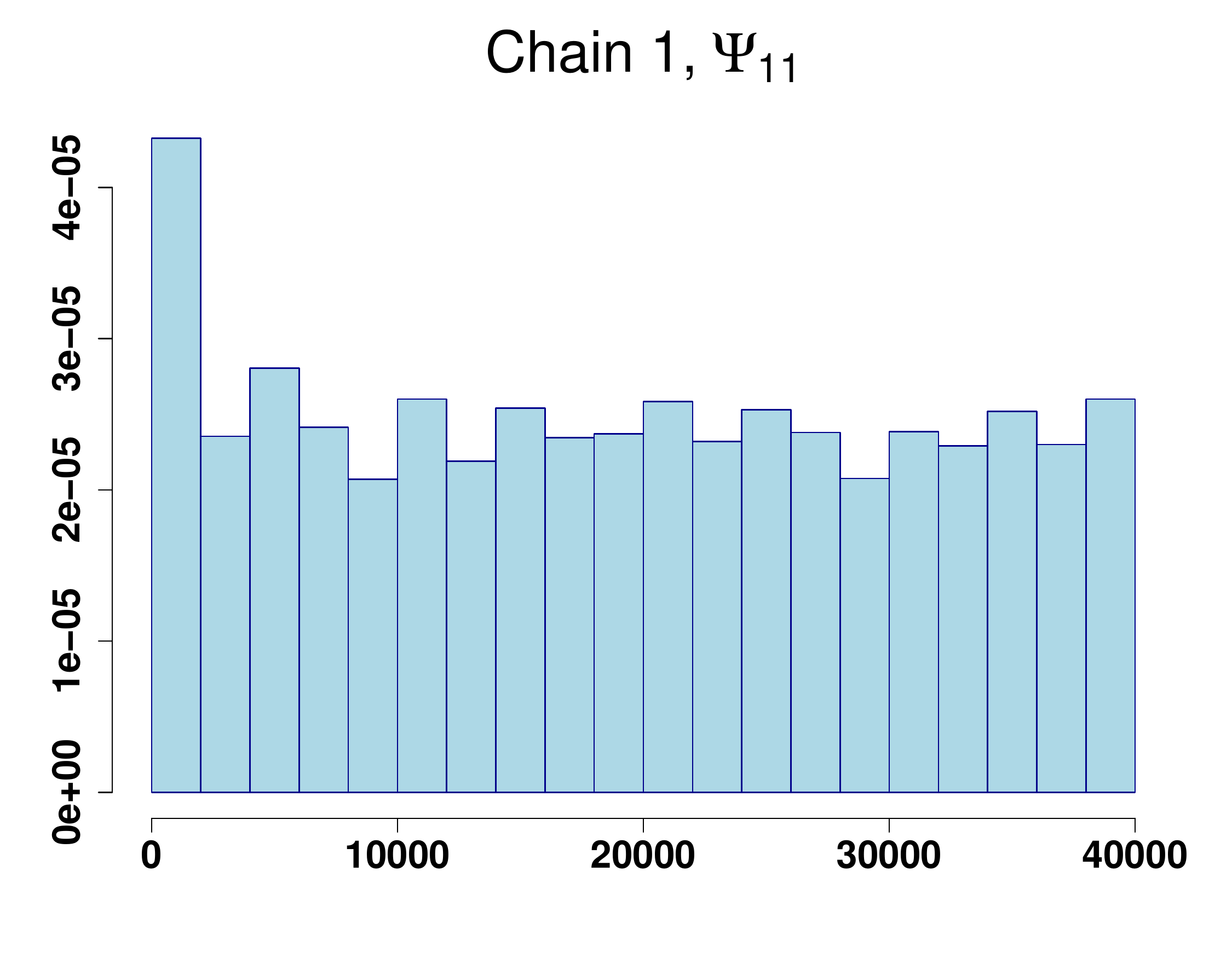}&\hspace{-0.5cm}\includegraphics[width=4.0cm]{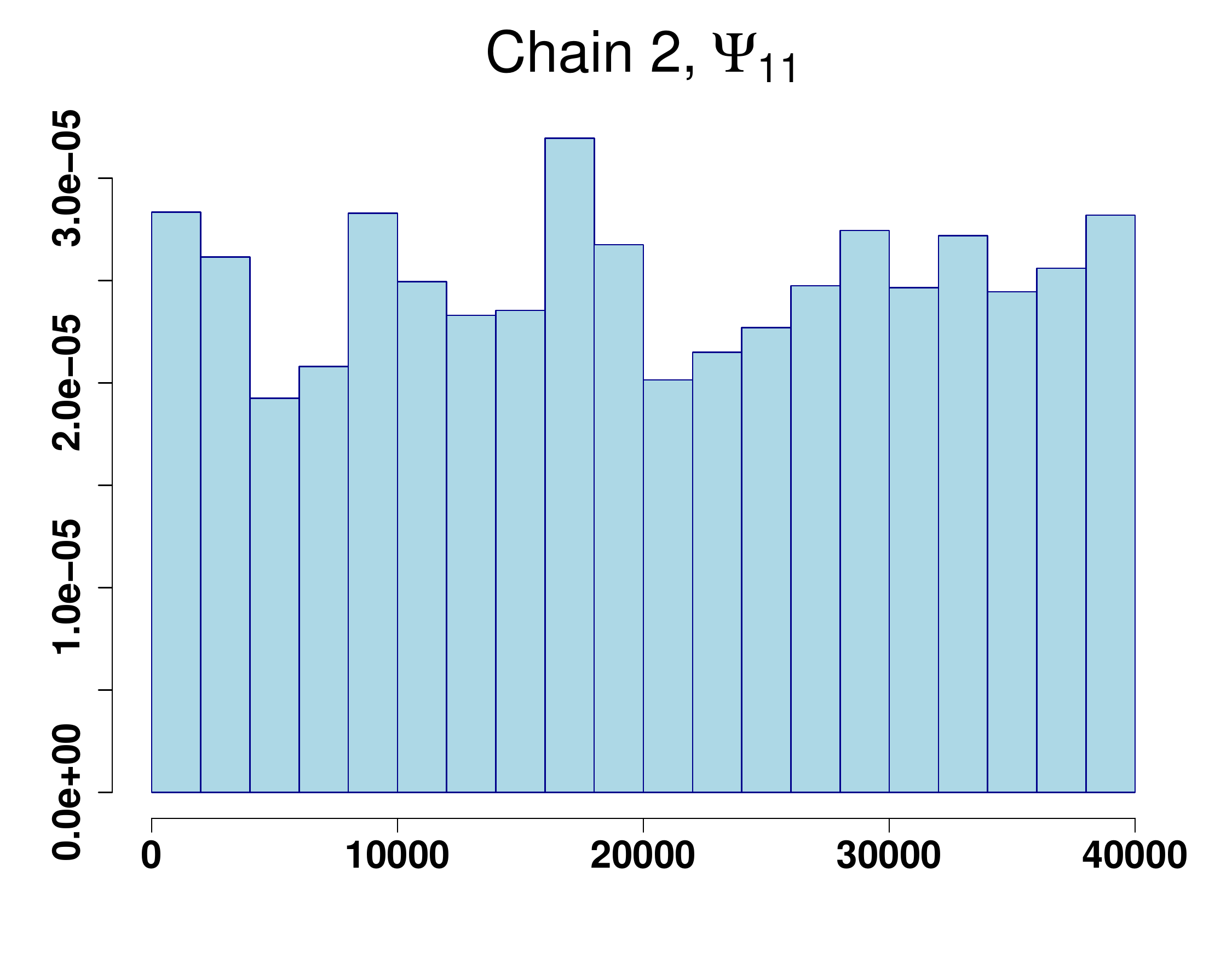}&
\hspace{-1.0cm}\includegraphics[width=4.0cm]{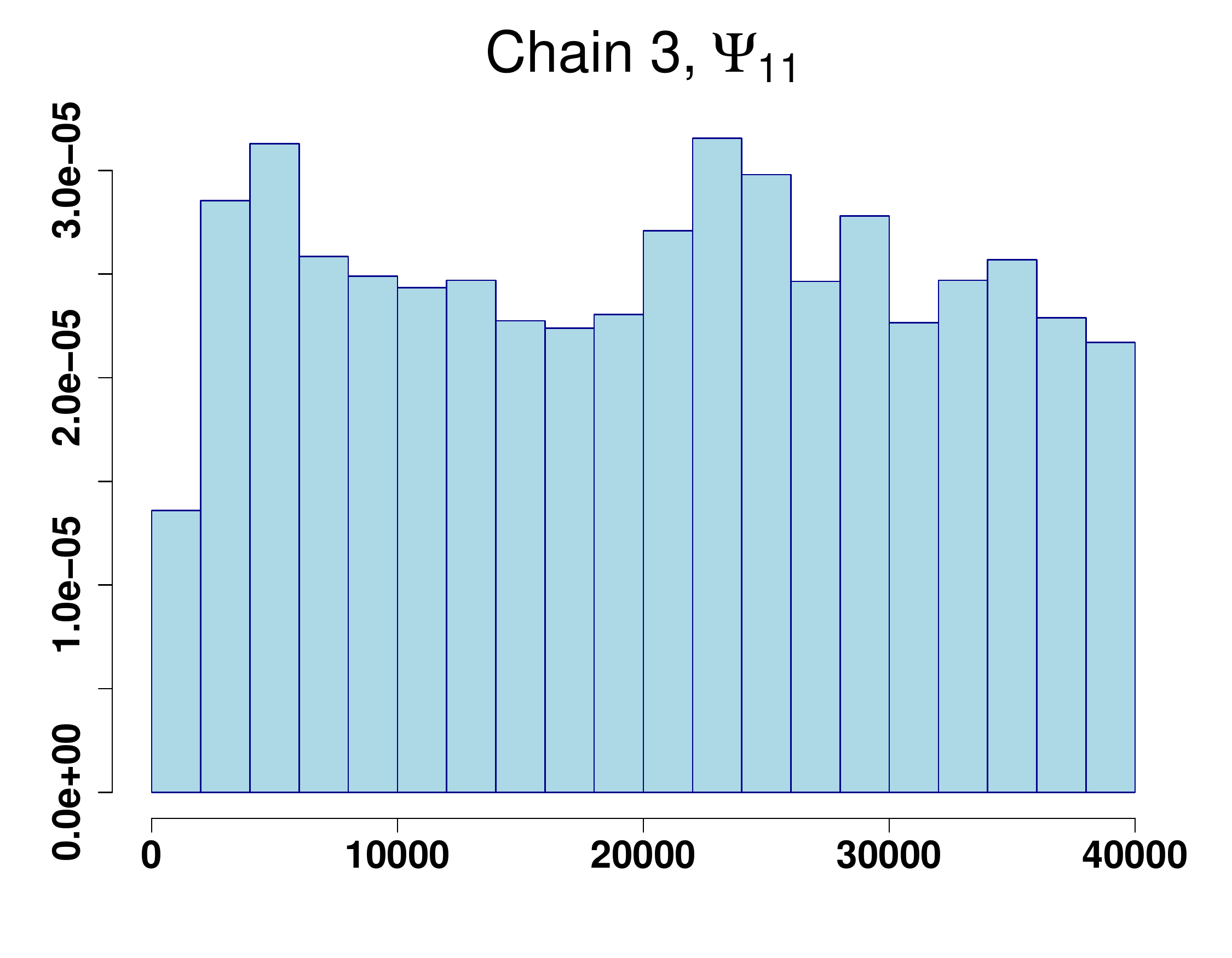}&\hspace{-1.5cm}\includegraphics[width=4.0cm]{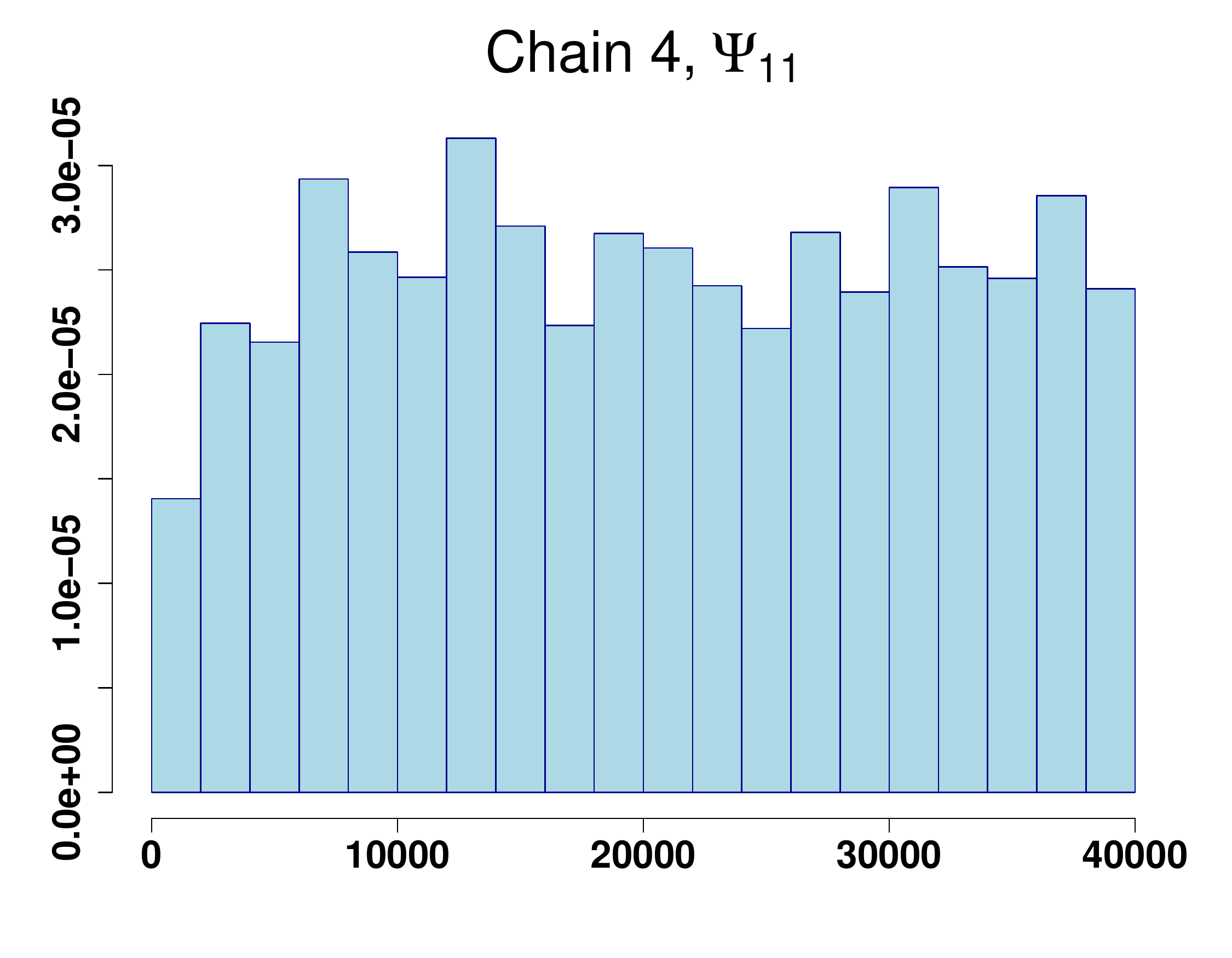}\\
\end{tabular}
 \caption{Rank plots of posterior draws from four chains in the case of the parameter $\Psi_{11}$ (SBP) of the $t$ multivariate random effects model by employing the Jeffreys prior (first to third rows) and the Berger and Bernardo reference prior (fourth to sixth rows). The samples from the posterior distributions are drawn by Algorithm A (first and fourth rows), Algorithm B (second and fifth rows) and Algorithm C (third and sixth rows).}
\label{fig:emp-study-rank-Psi11-t}
 \end{figure}

\begin{figure}[h!t]
\centering
\begin{tabular}{p{4.0cm}p{4.0cm}p{4.0cm}p{4.0cm}}
\hspace{0.0cm}\includegraphics[width=4.0cm]{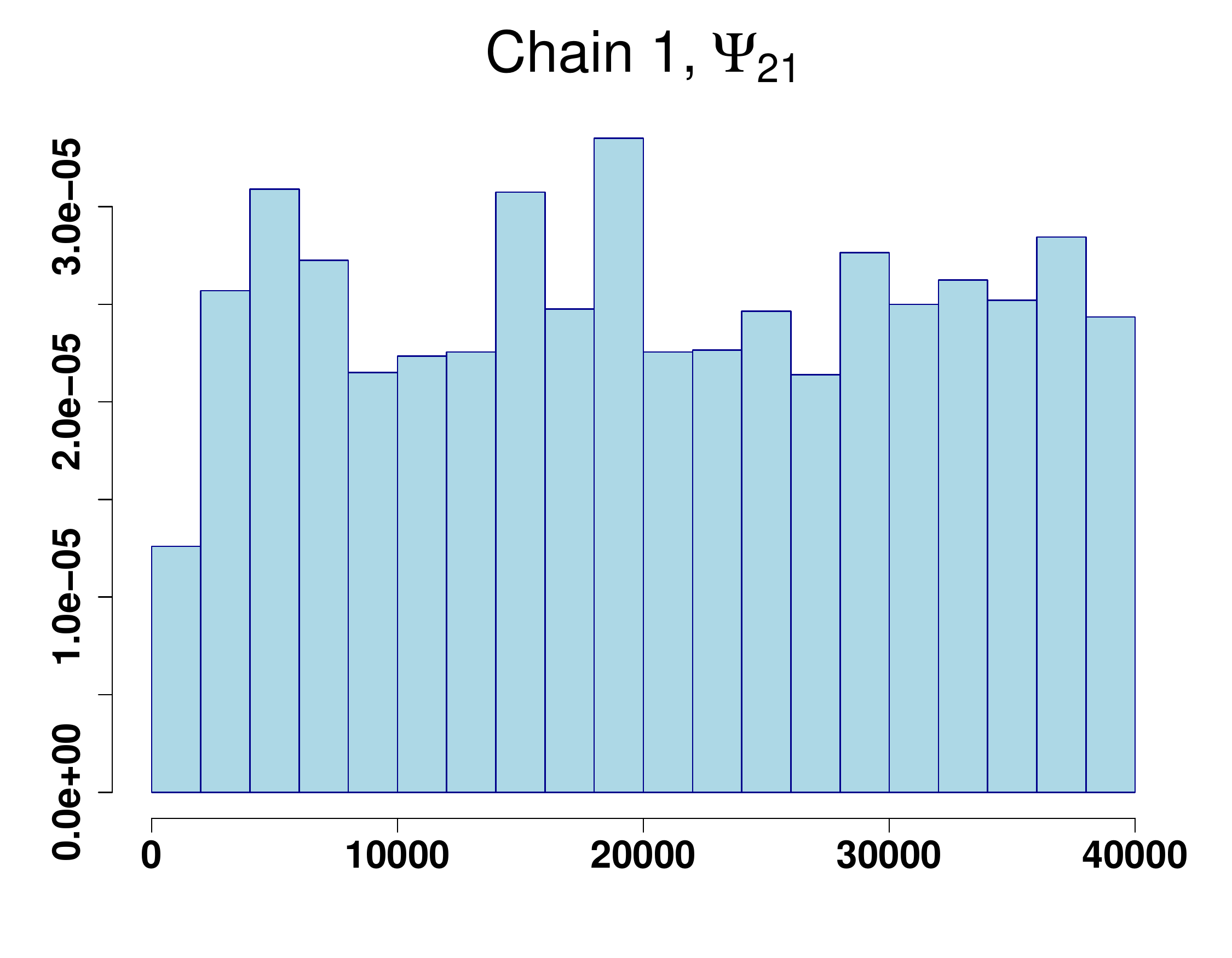}&\hspace{-0.5cm}\includegraphics[width=4.0cm]{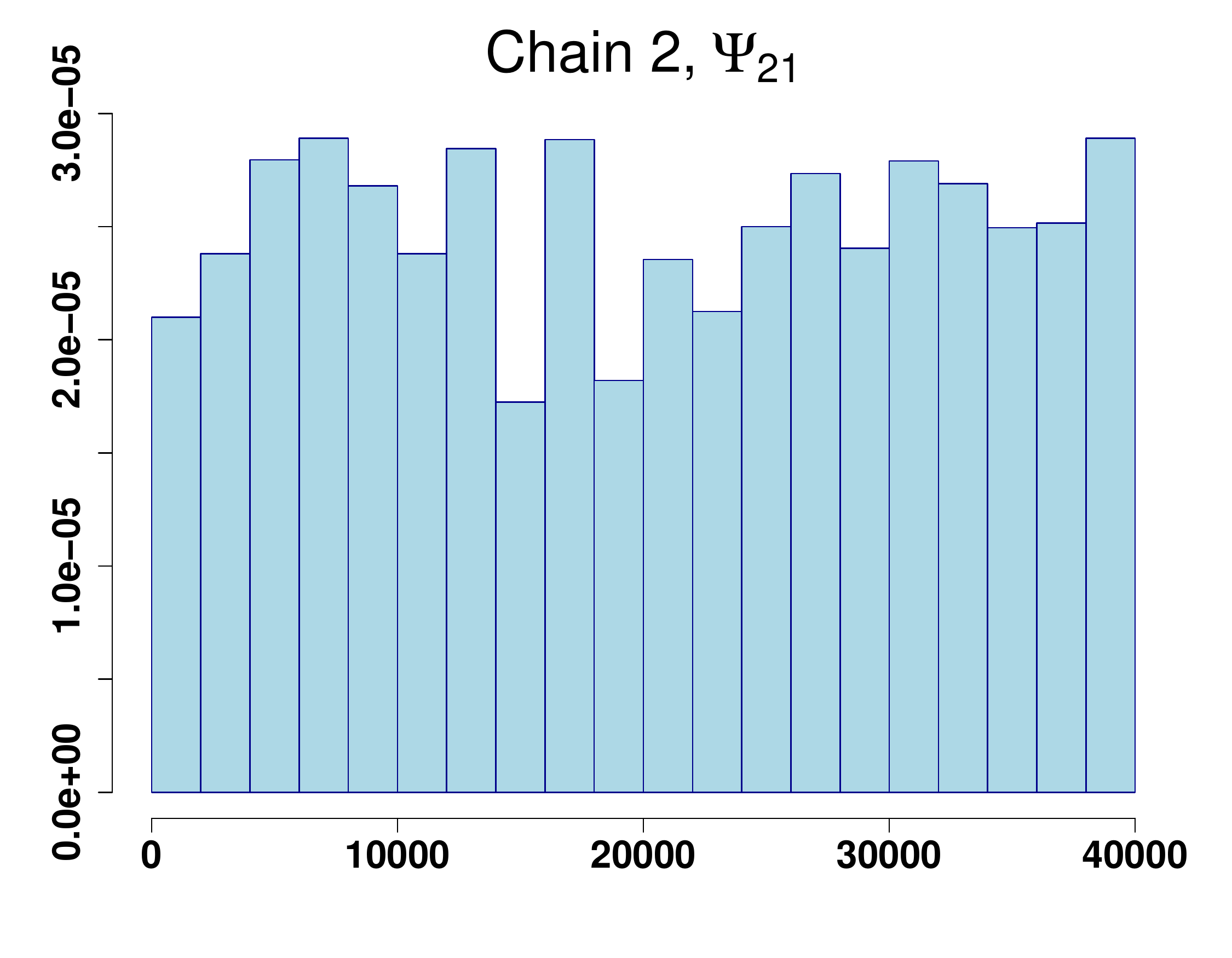}&
\hspace{-1.0cm}\includegraphics[width=4.0cm]{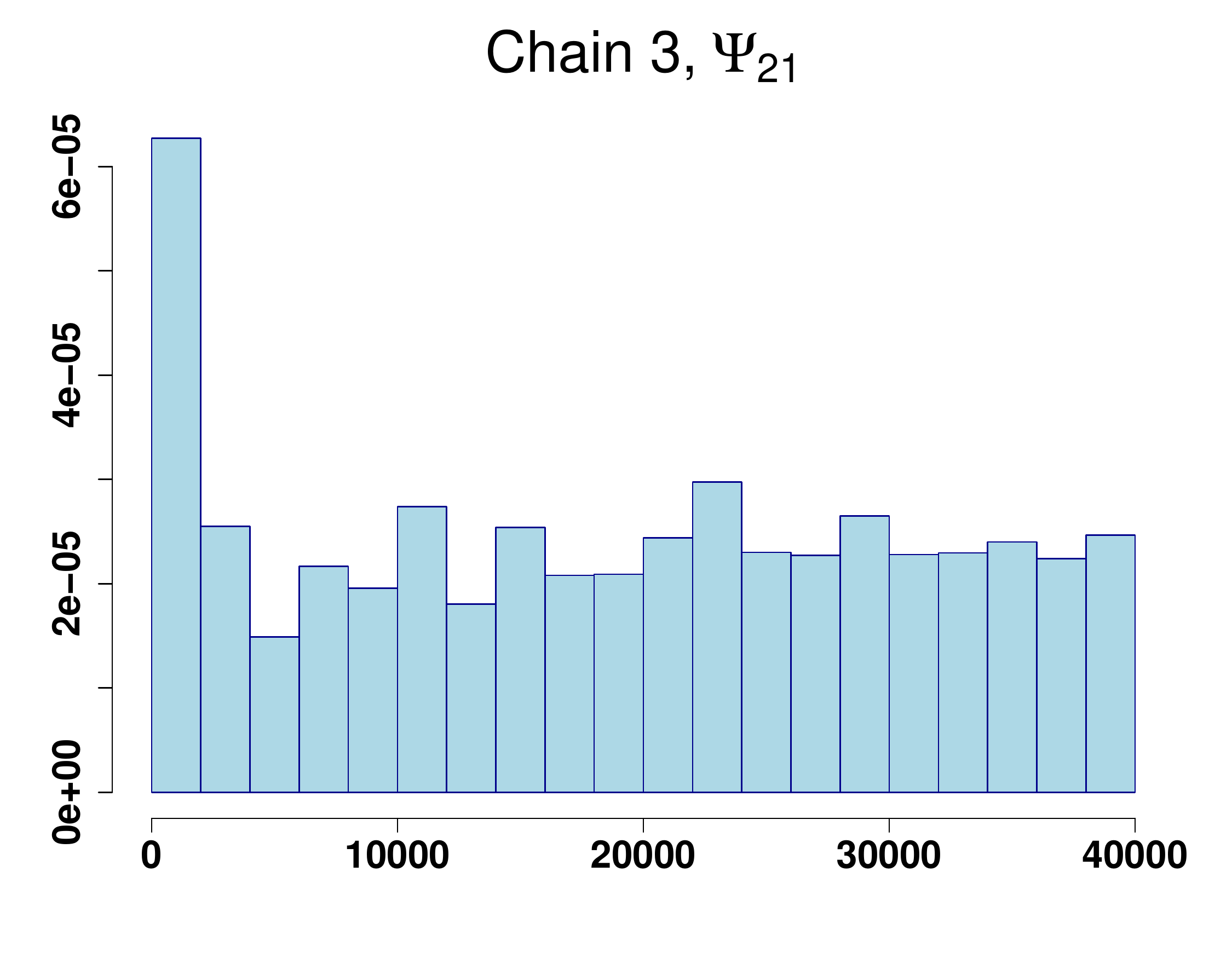}&\hspace{-1.5cm}\includegraphics[width=4.0cm]{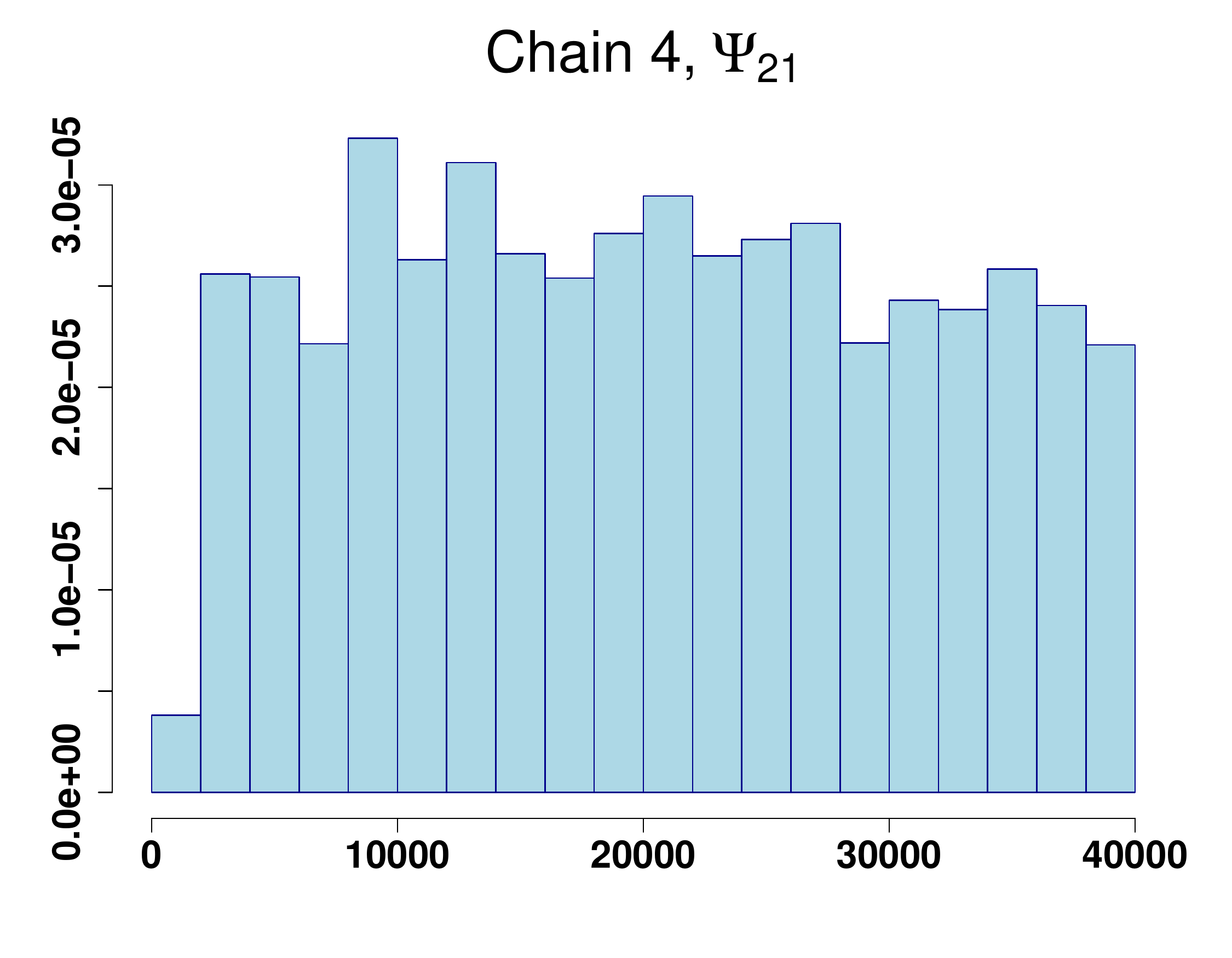}\\
\hspace{0.0cm}\includegraphics[width=4.0cm]{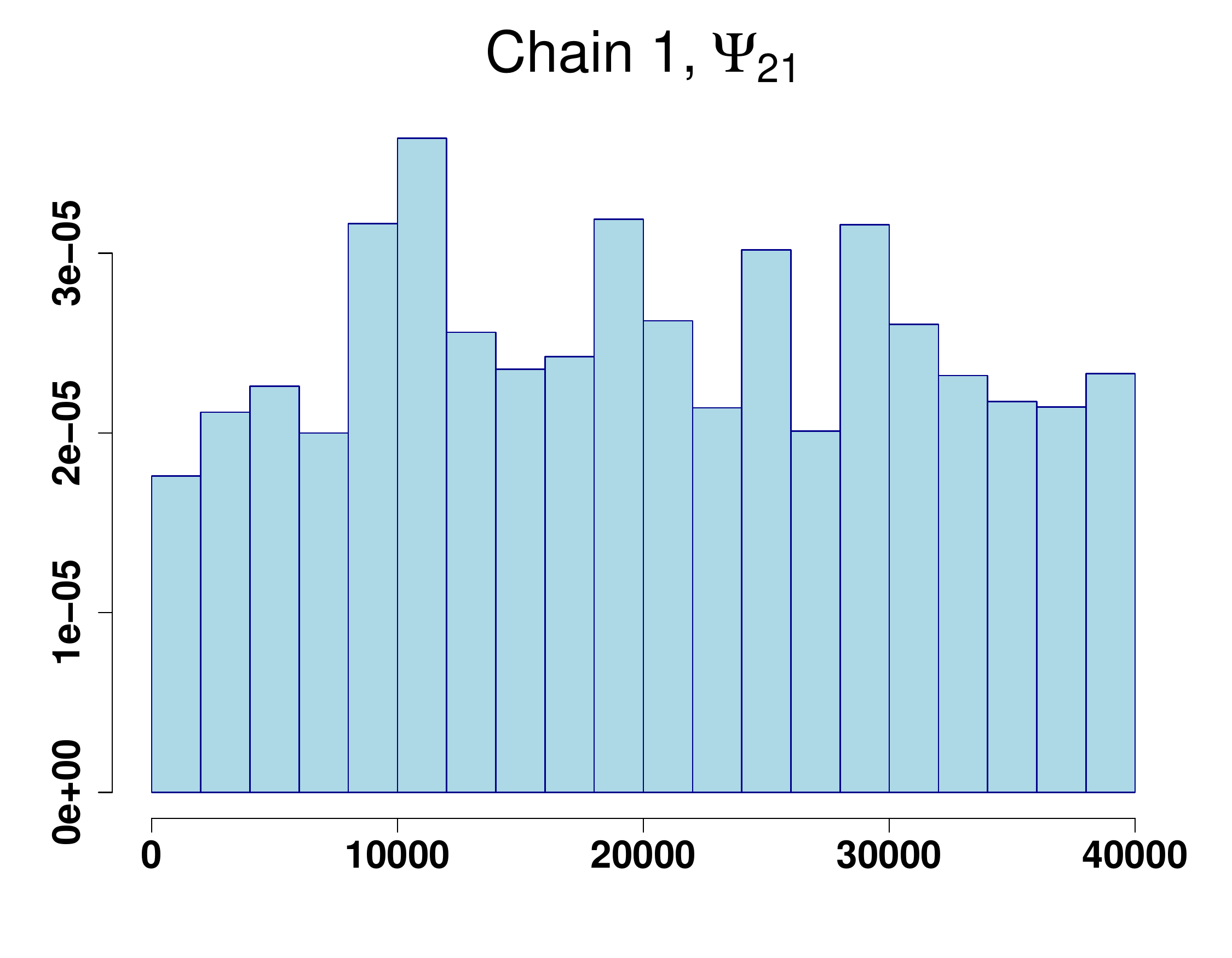}&\hspace{-0.5cm}\includegraphics[width=4.0cm]{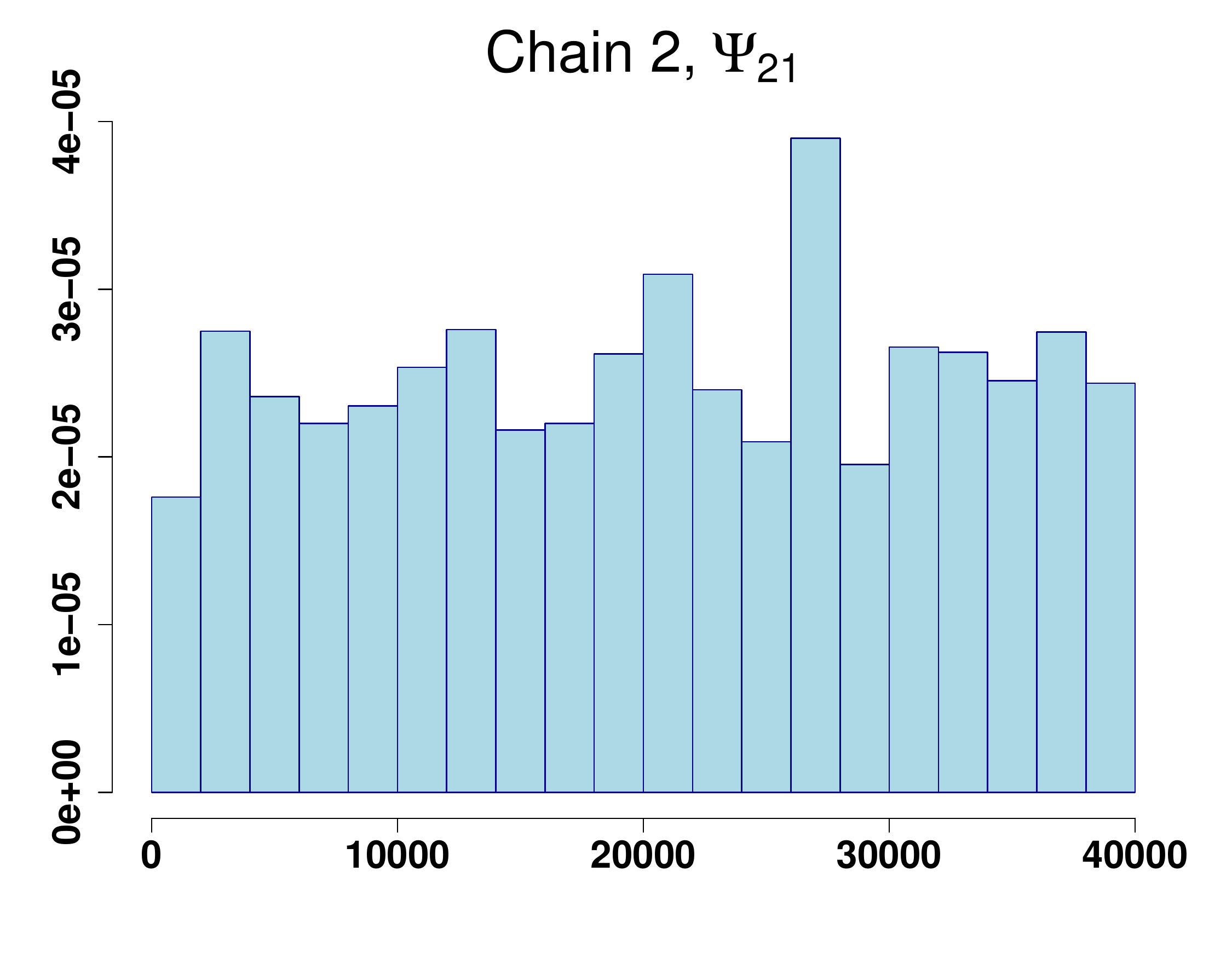}&
\hspace{-1.0cm}\includegraphics[width=4.0cm]{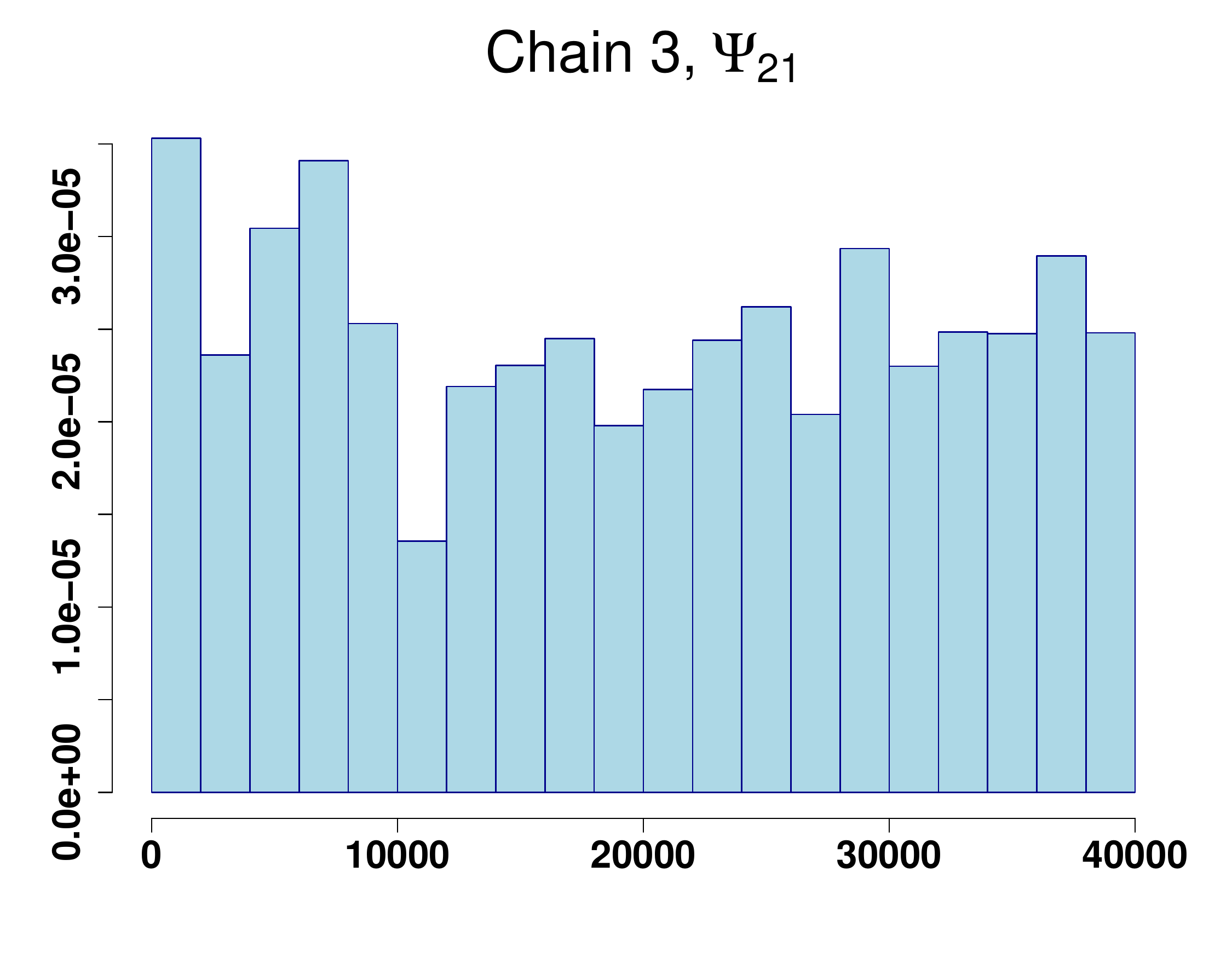}&\hspace{-1.5cm}\includegraphics[width=4.0cm]{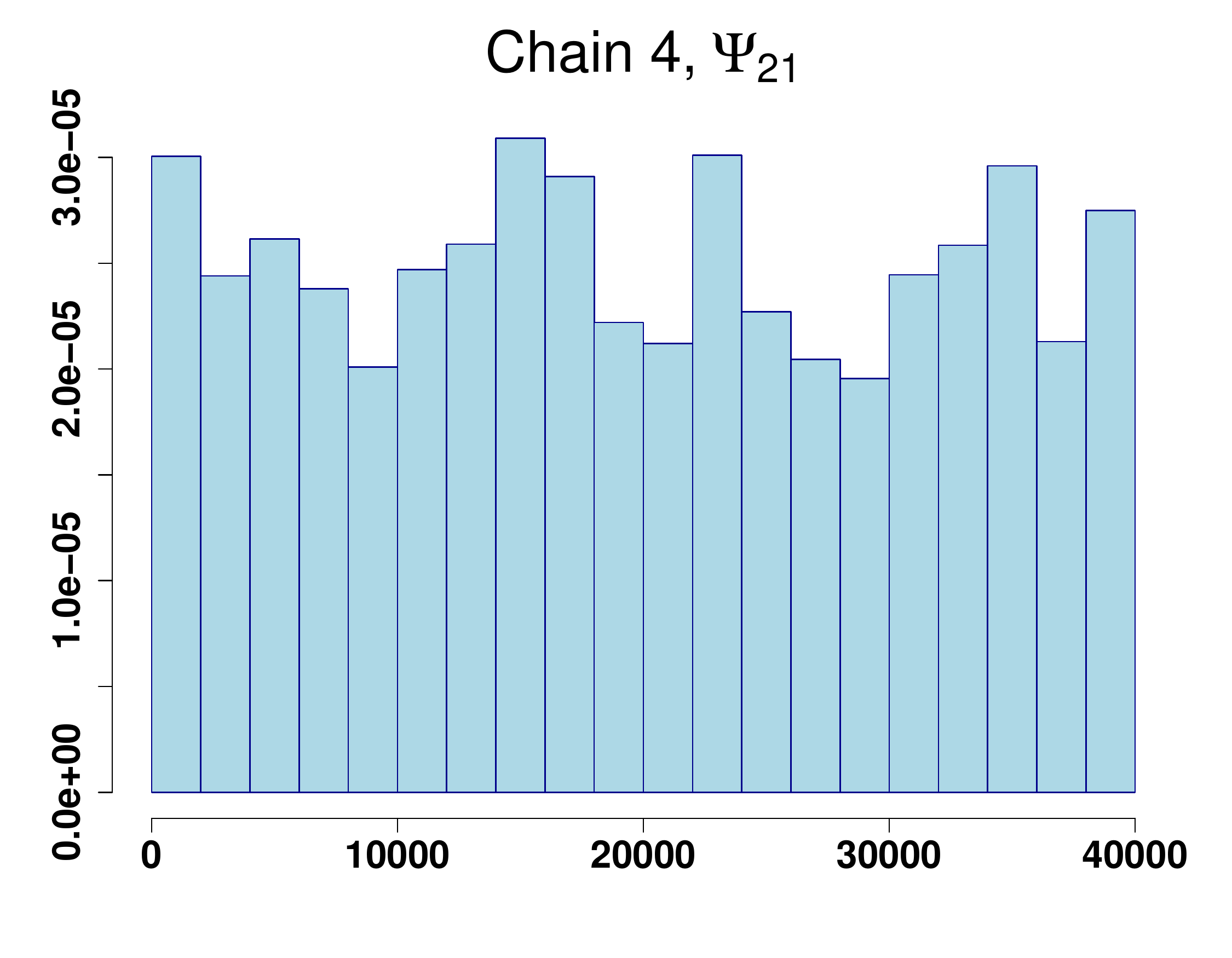}\\
\hspace{0.0cm}\includegraphics[width=4.0cm]{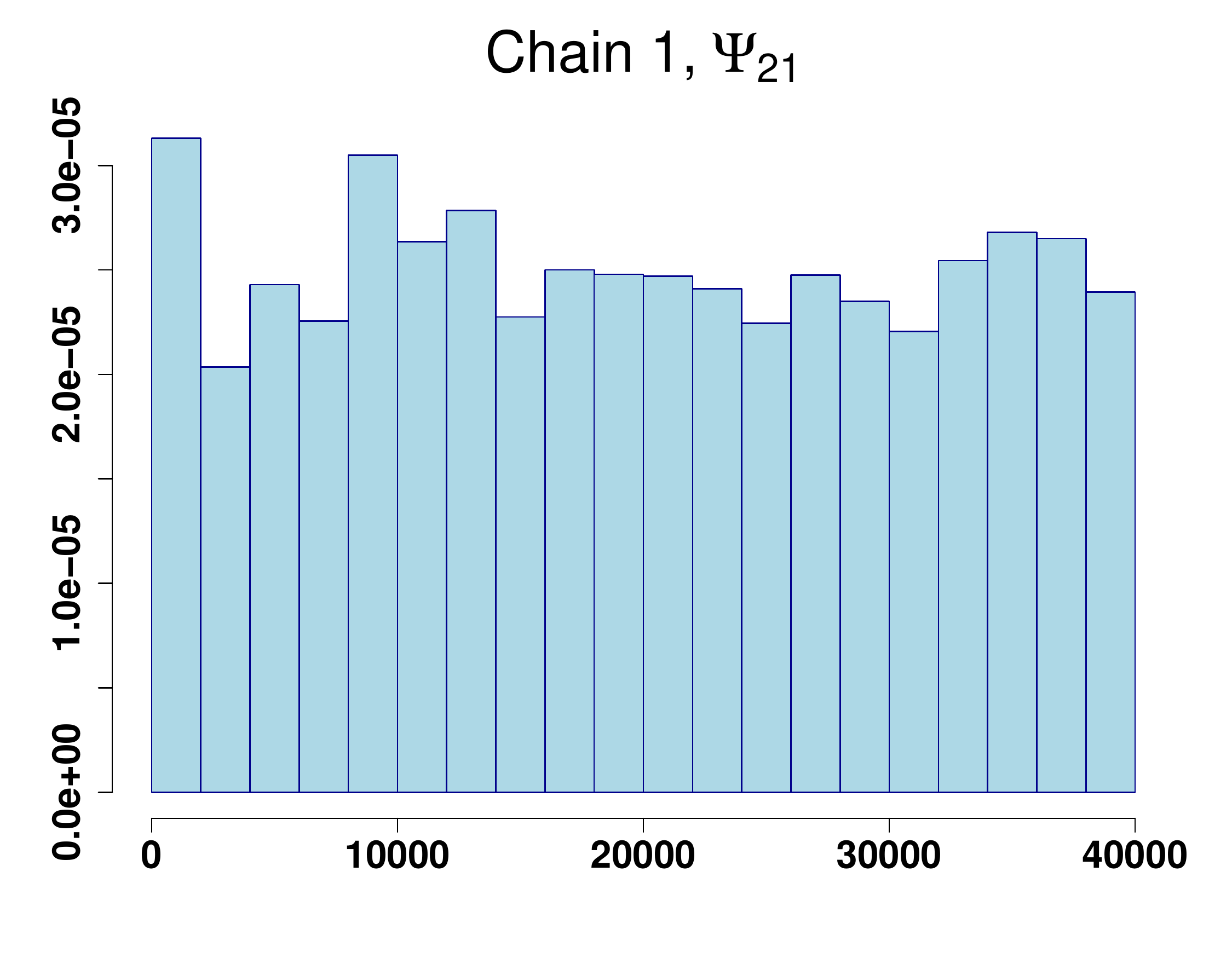}&\hspace{-0.5cm}\includegraphics[width=4.0cm]{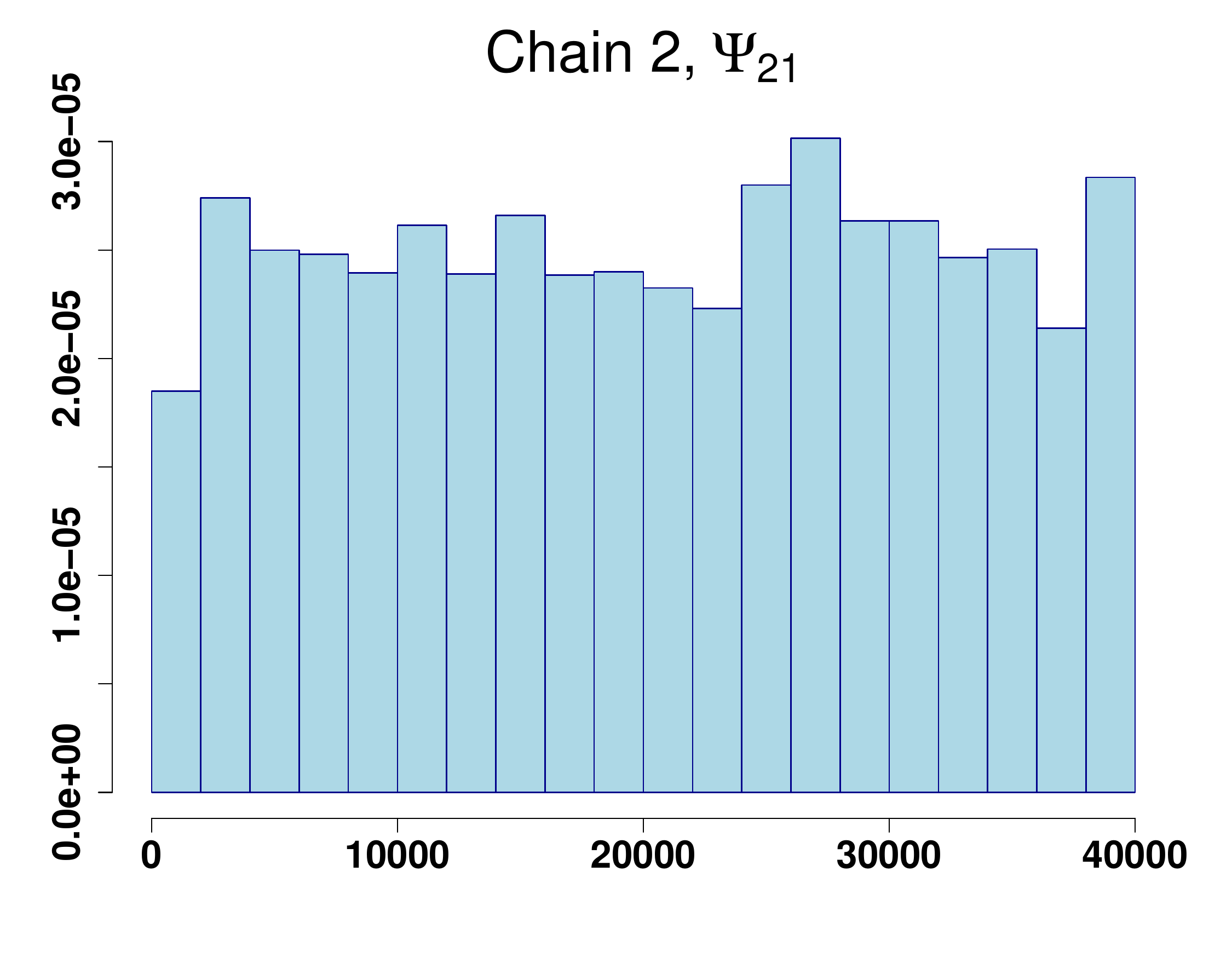}&
\hspace{-1.0cm}\includegraphics[width=4.0cm]{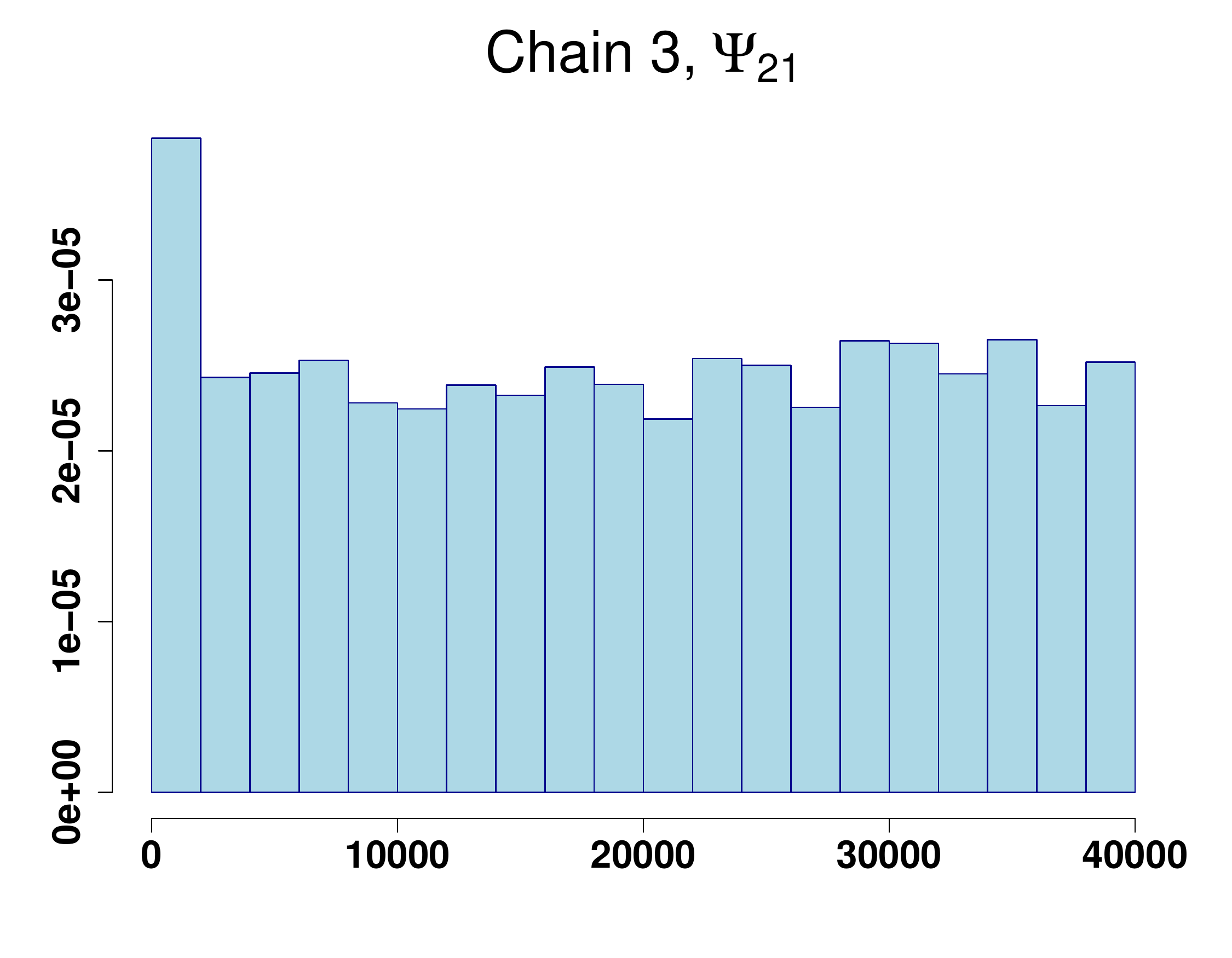}&\hspace{-1.5cm}\includegraphics[width=4.0cm]{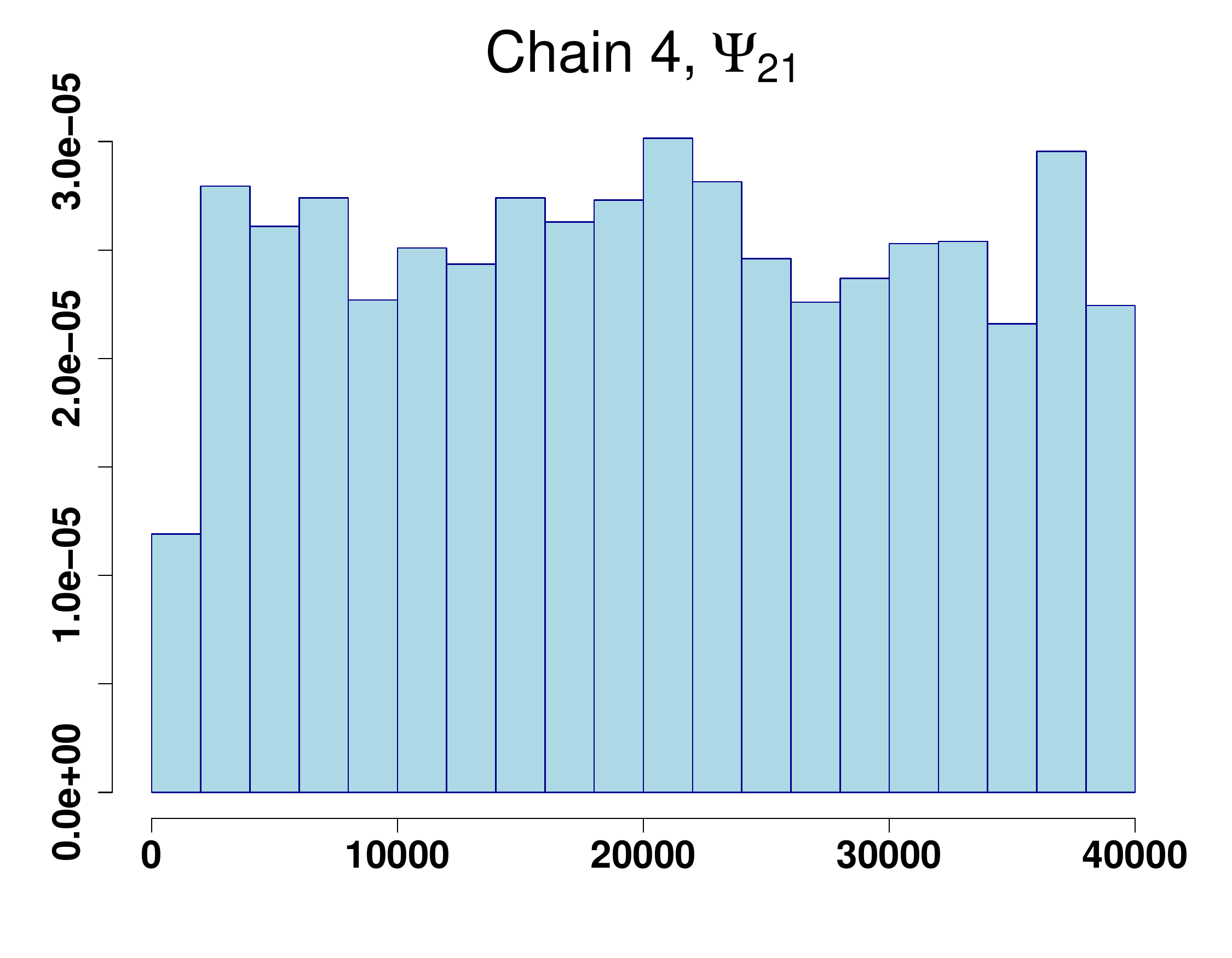}\\
\hspace{0.0cm}\includegraphics[width=4.0cm]{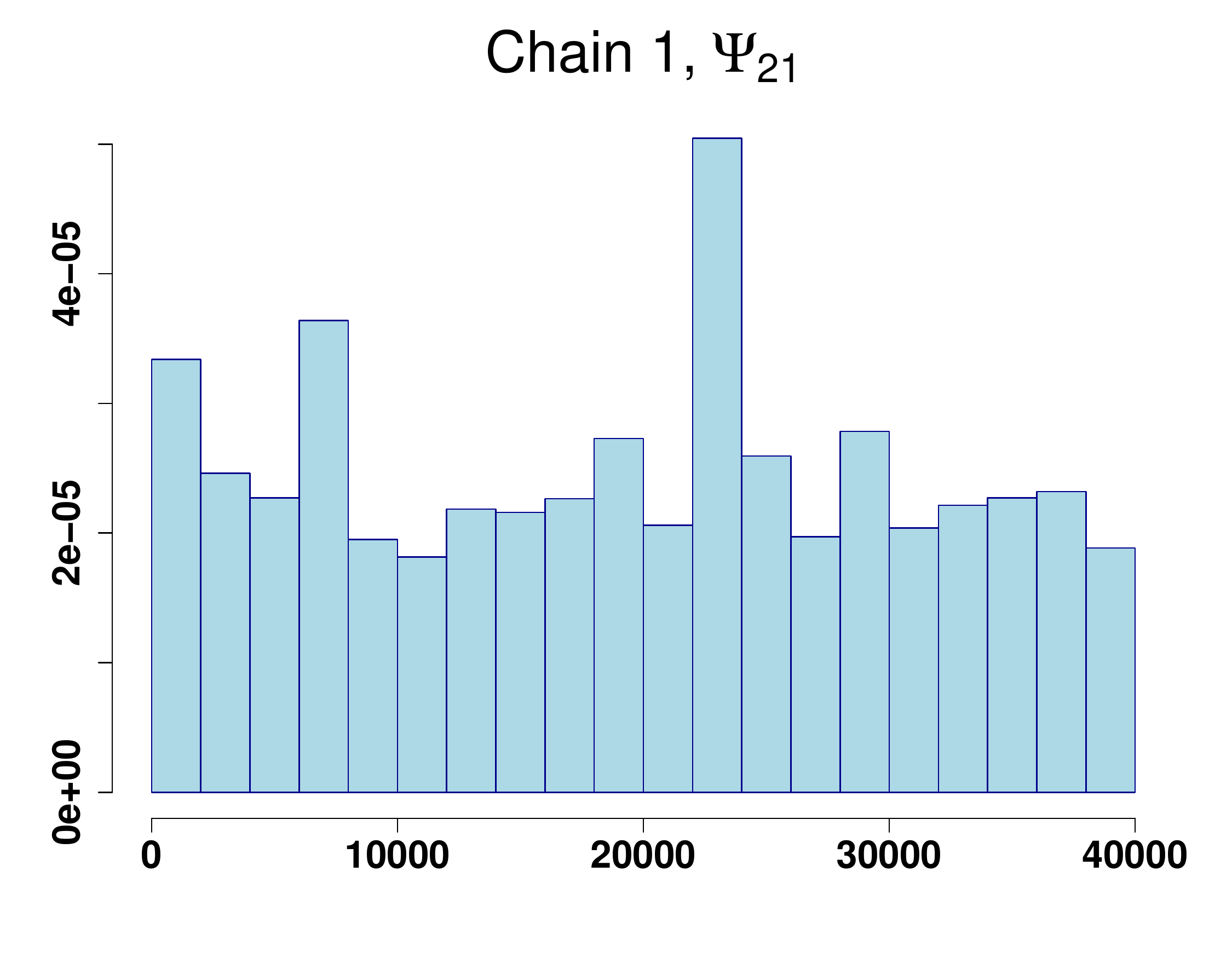}&\hspace{-0.5cm}\includegraphics[width=4.0cm]{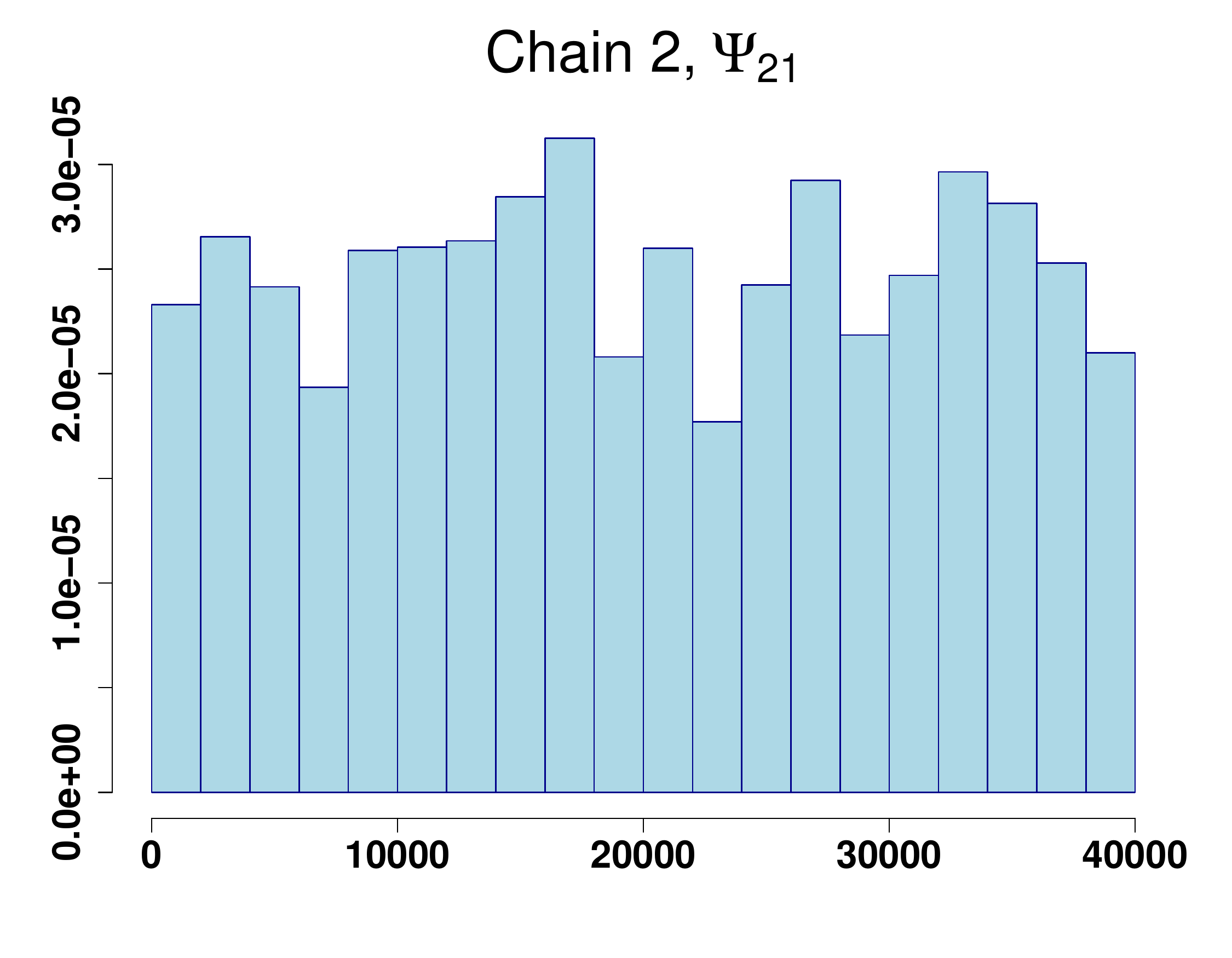}&
\hspace{-1.0cm}\includegraphics[width=4.0cm]{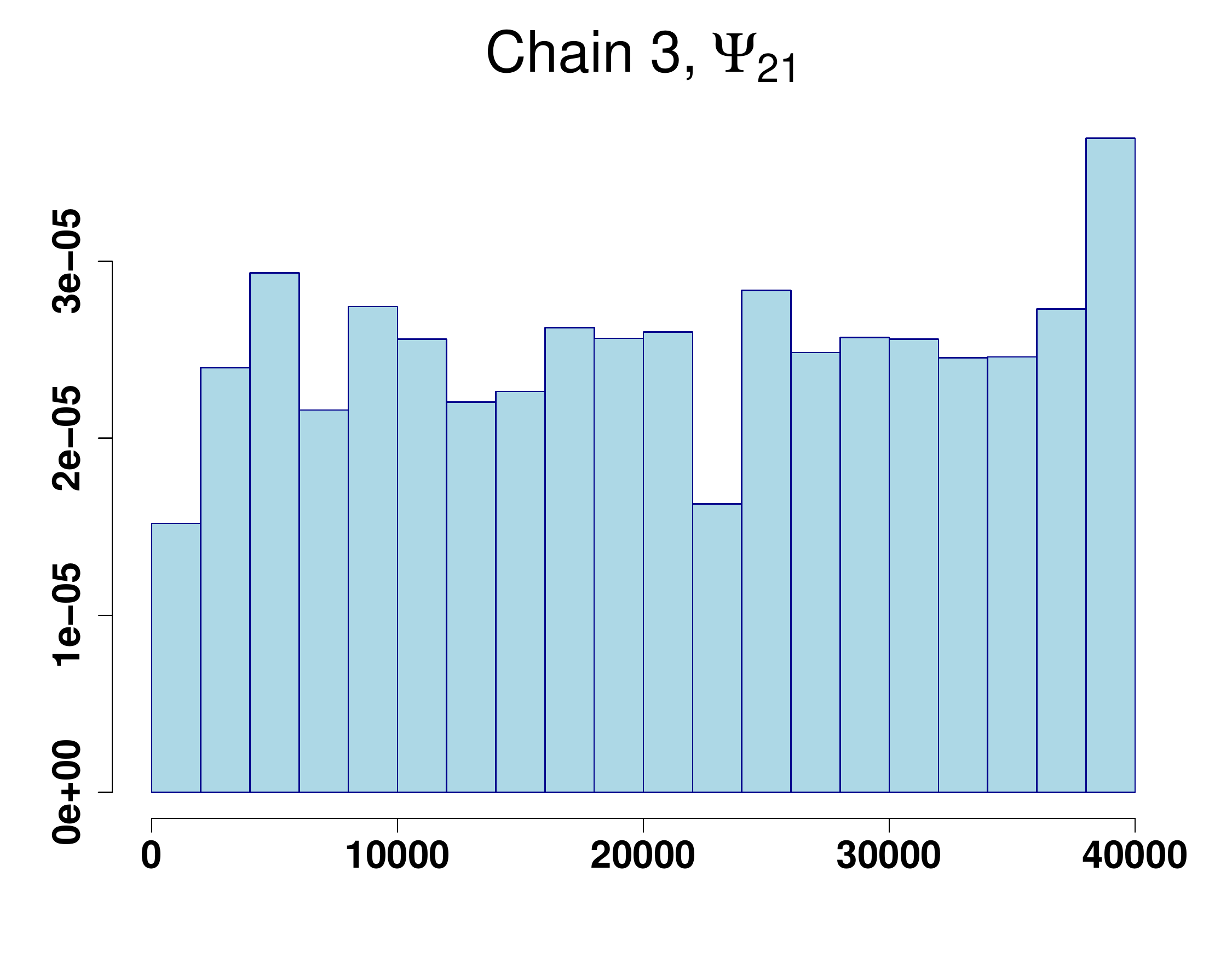}&\hspace{-1.5cm}\includegraphics[width=4.0cm]{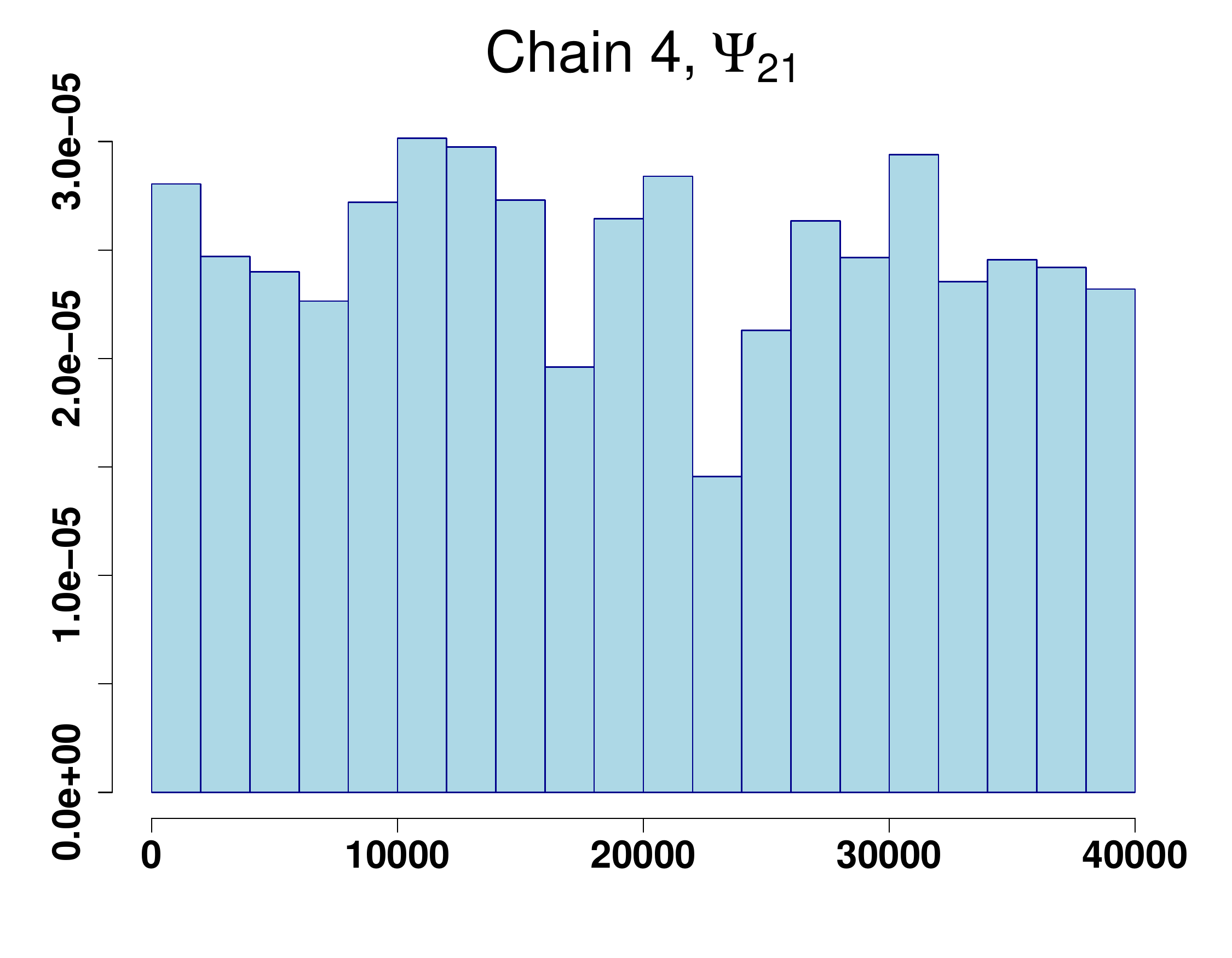}\\
\hspace{0.0cm}\includegraphics[width=4.0cm]{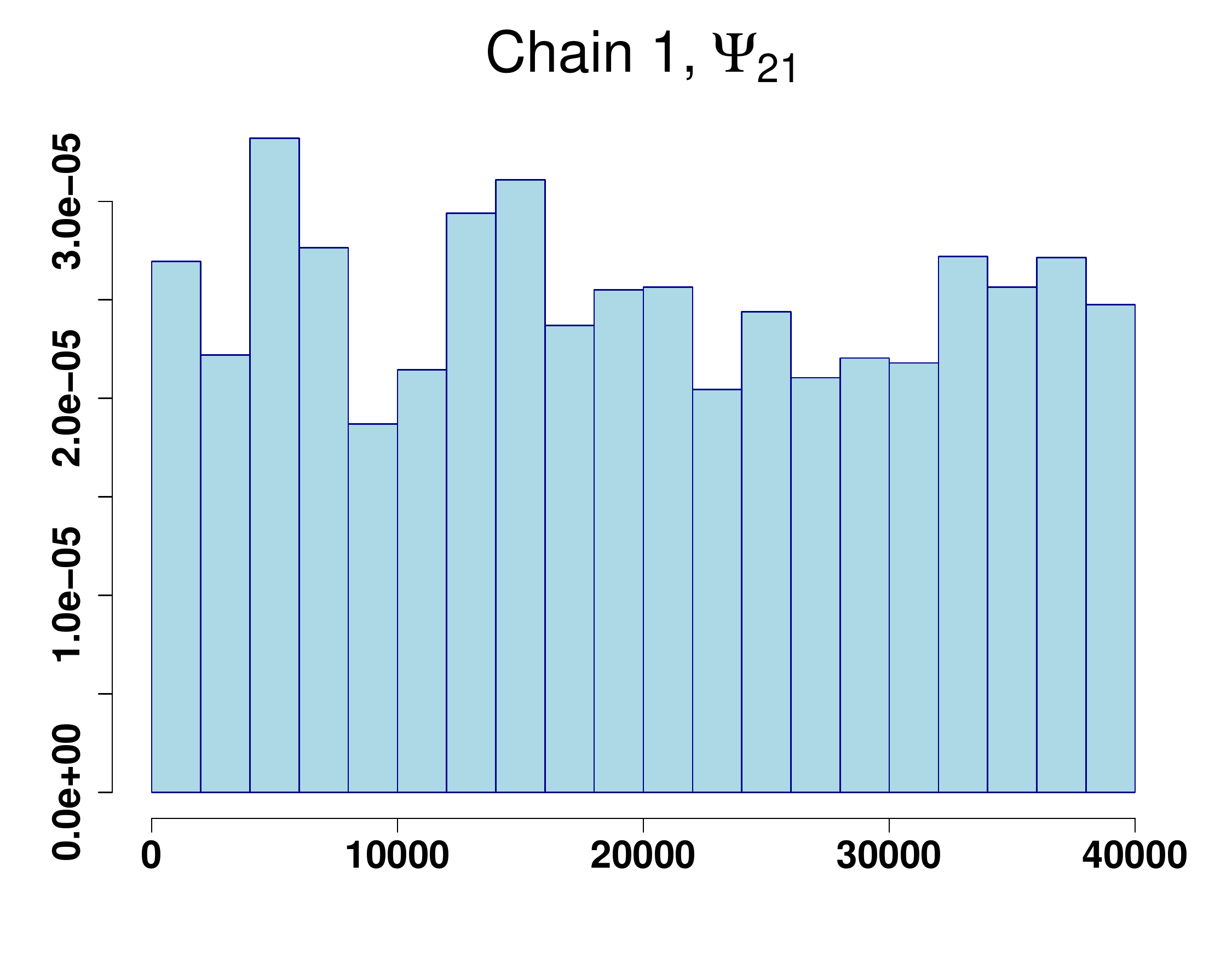}&\hspace{-0.5cm}\includegraphics[width=4.0cm]{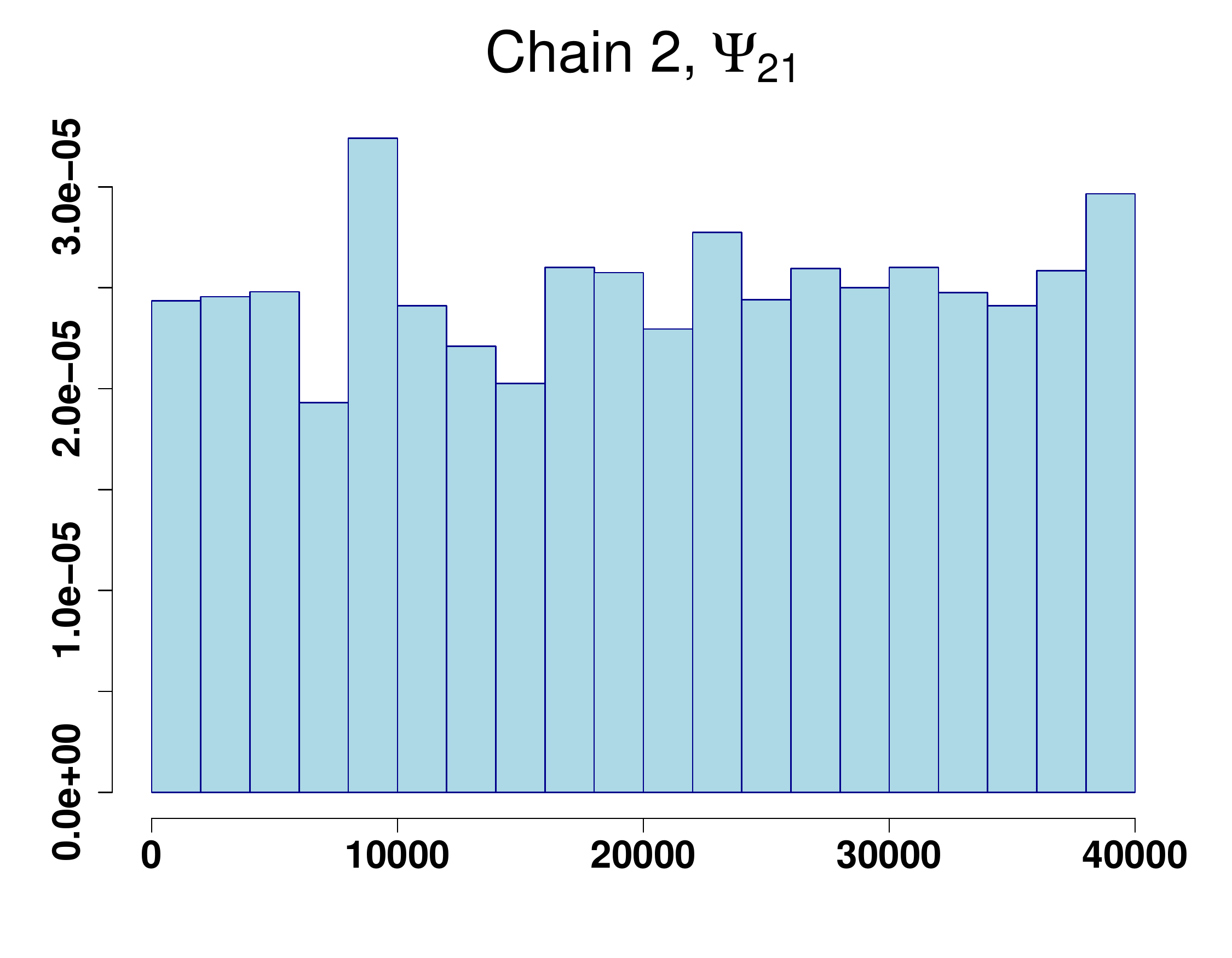}&
\hspace{-1.0cm}\includegraphics[width=4.0cm]{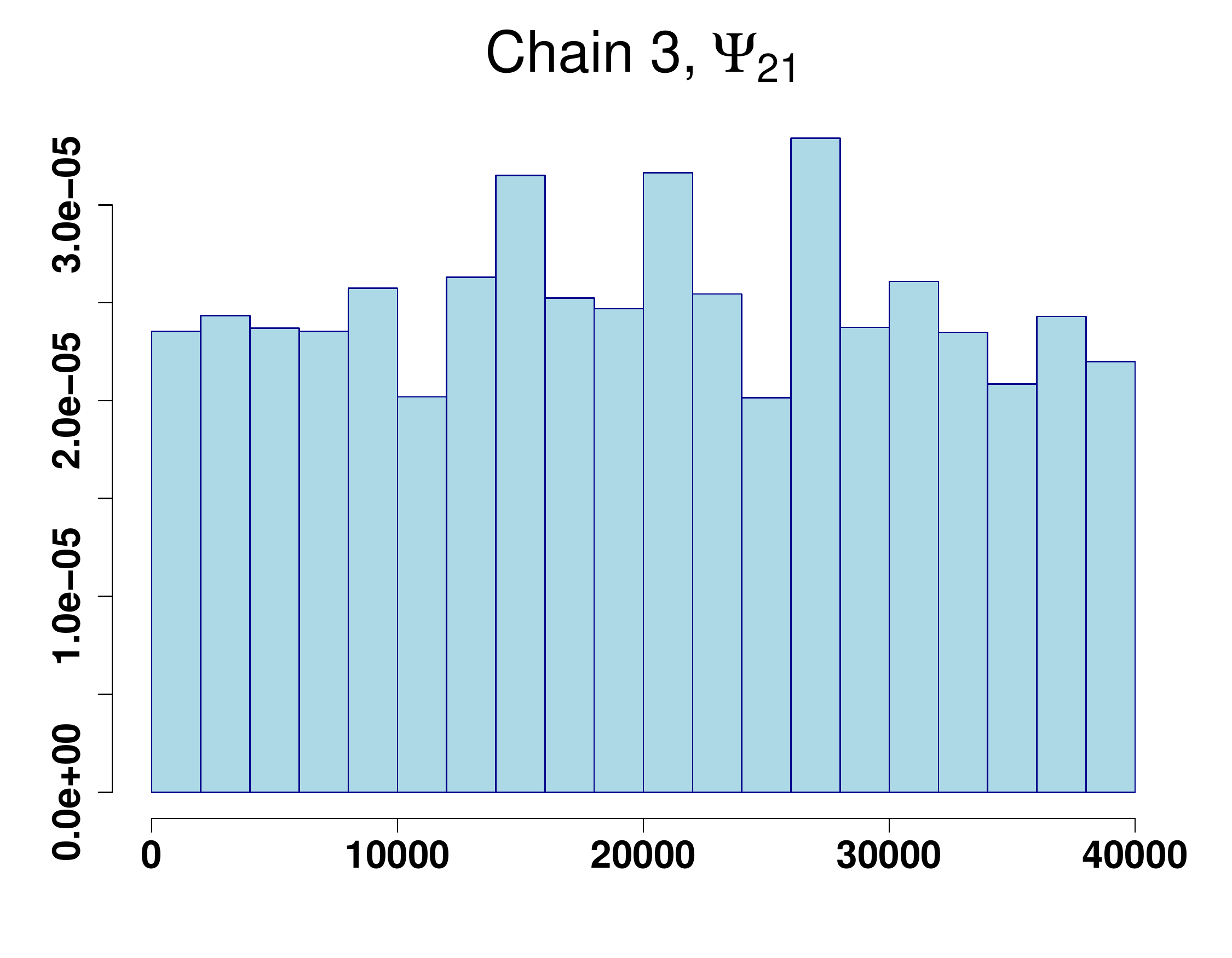}&\hspace{-1.5cm}\includegraphics[width=4.0cm]{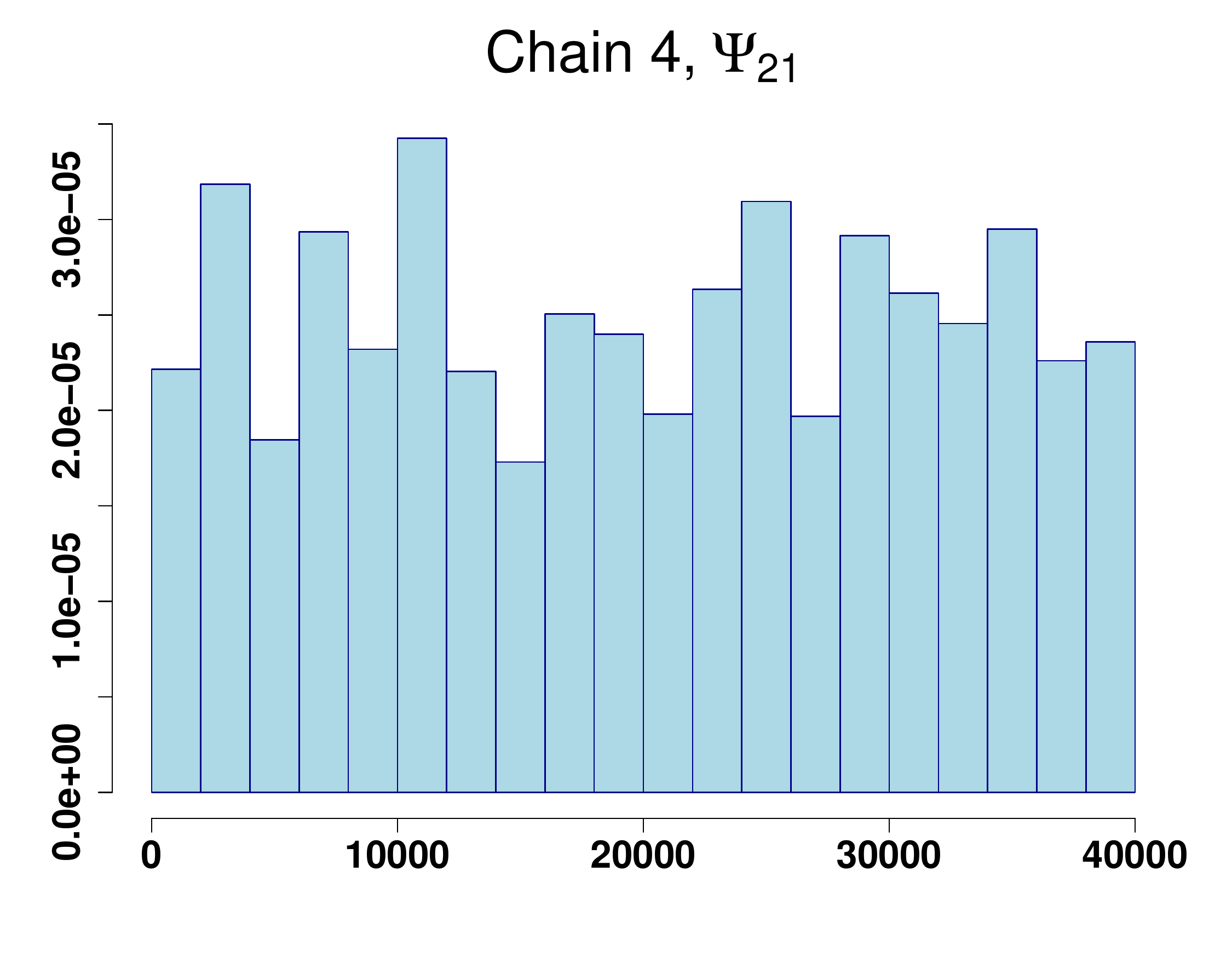}\\
\hspace{0.0cm}\includegraphics[width=4.0cm]{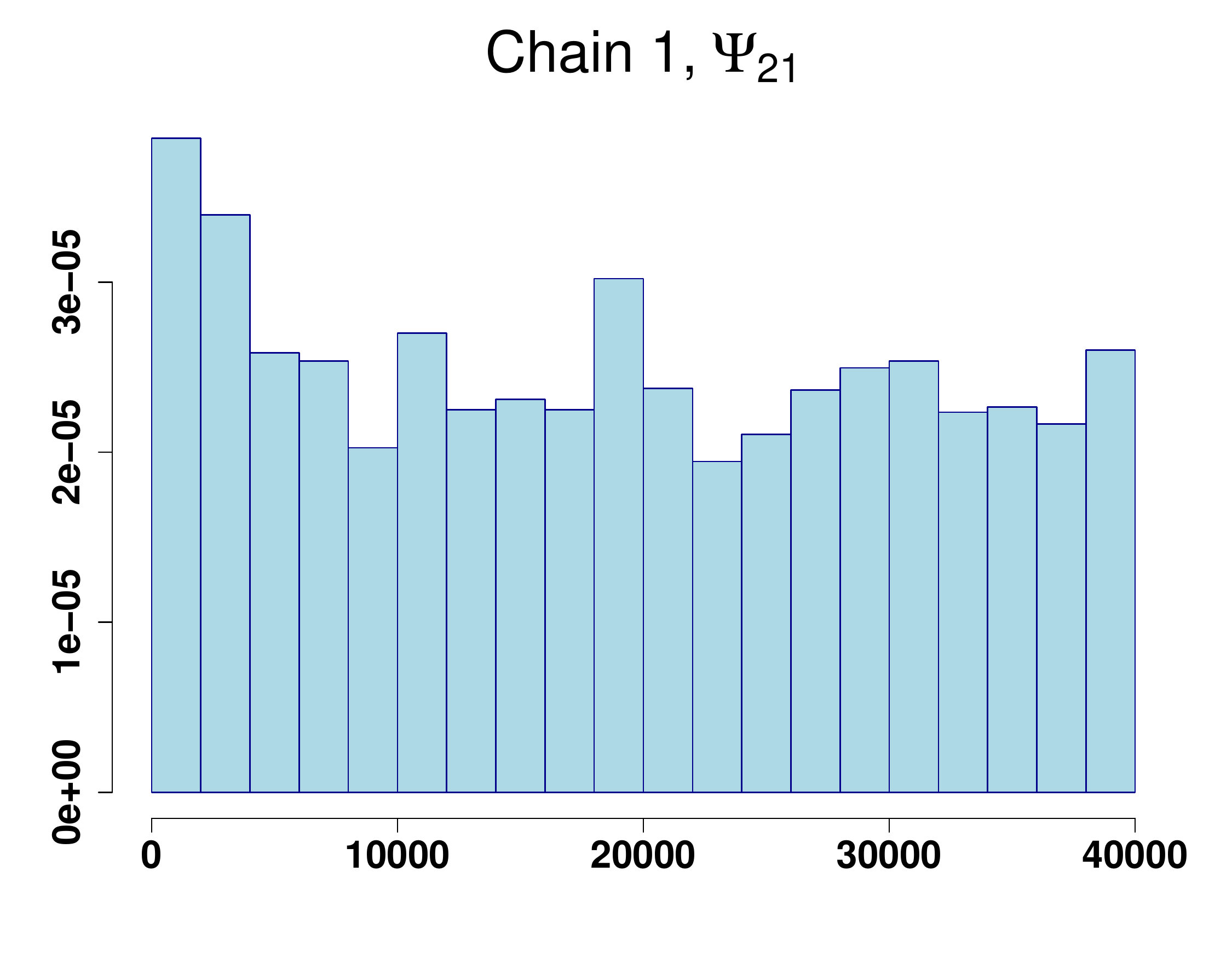}&\hspace{-0.5cm}\includegraphics[width=4.0cm]{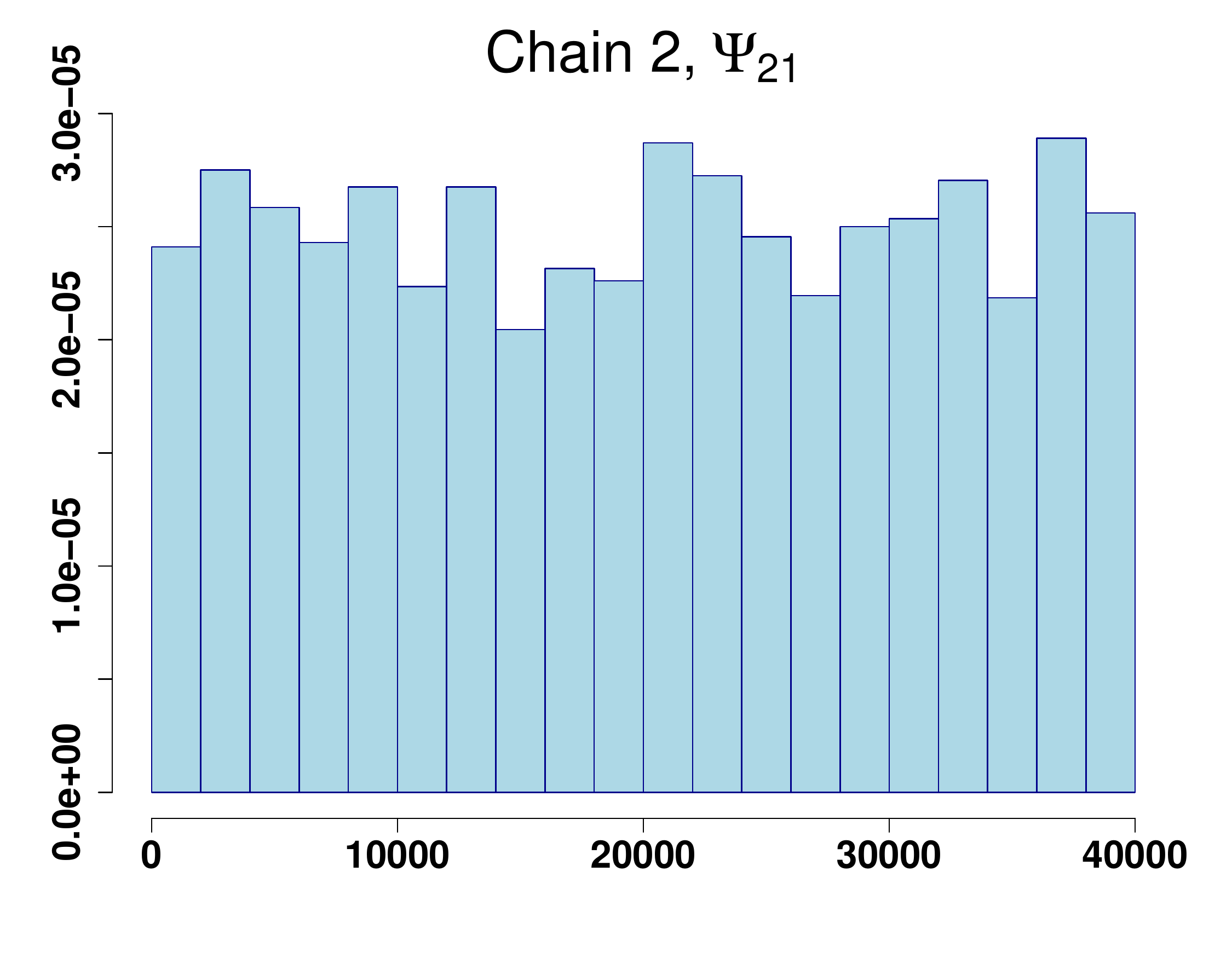}&
\hspace{-1.0cm}\includegraphics[width=4.0cm]{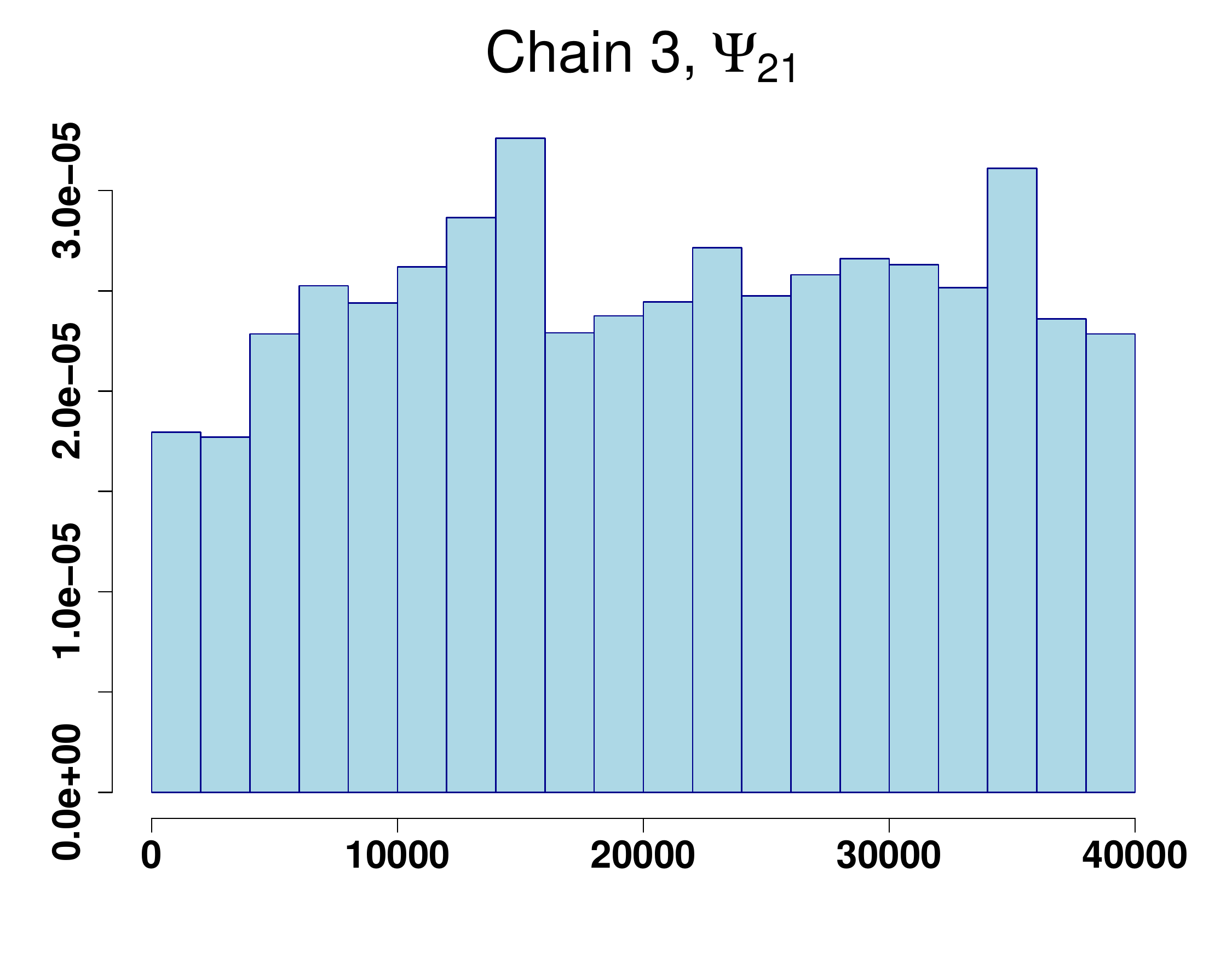}&\hspace{-1.5cm}\includegraphics[width=4.0cm]{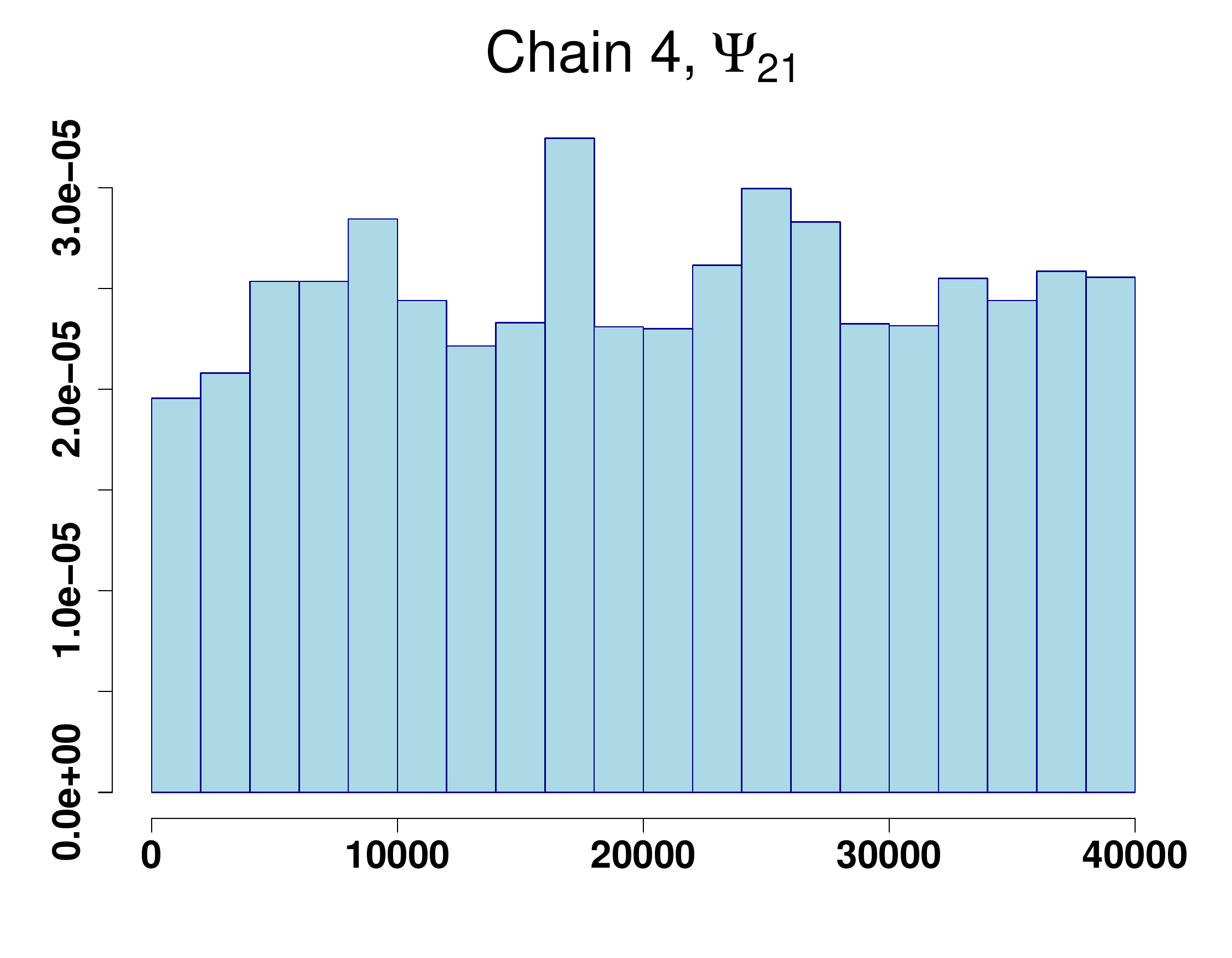}\\
\end{tabular}
 \caption{Rank plots of posterior draws from four chains in the case of the parameter $\Psi_{21}$ (DBP, SBP) of the $t$ multivariate random effects model by employing the Jeffreys prior (first to third rows) and the Berger and Bernardo reference prior (fourth to sixth rows). The samples from the posterior distributions are drawn by Algorithm A (first and fourth rows), Algorithm B (second and fifth rows) and Algorithm C (third and sixth rows).}
\label{fig:emp-study-rank-Psi21-t}
 \end{figure}

Figures \ref{fig:emp-study-rank-mu1-nor} to \ref{fig:emp-study-rank-Psi21-nor} present the results obtained under the assumption of the normal multivariate random effects model, while Figures \ref{fig:emp-study-rank-mu1-t} to \ref{fig:emp-study-rank-Psi21-t} depict the plots under the $t$ multivariate random effects model. The figures provide the rank plots of posterior draws, which are obtained from the constructed four Markov chains for each algorithm and prior. If good mixing properties of the constructed Markov chains are present, then the plots should be similar to the histograms corresponding to the uniform distribution. As such, the algorithm with the best performance should results in the rank plot which is closest to the histogram of the uniform distribution. In the figures the results are provided in the case of $\mu_1$, $\Psi_{11}$ and $\Psi_{21}$, while the results for $\mu_2$ and $\Psi_{22}$ are similar; they are available from the authors by request.

The rank plots of posterior draws created in the case of the parameter $\mu_1$ by the hybrid Gibbs sampler are indistinguishable from the histogram corresponding to the uniform distributions, independently whether the normal multivariate random effects model or the $t$ multivariate random effects model is fitted to the data. It is interesting that the performance of the two Metropolis-Hastings algorithms cannot be clearly ranked. While Algorithm A performs better under the Jeffreys prior, Algorithm B is better under the Berger and Bernardo reference prior. When the rank plots are constructed for the components of the between-study covariance matrix $\bPsi$, then better results are achieved under the Jeffreys prior, while the differences in plots related to the application of different algorithms to draw samples from the posterior distribution are not large. Finally, we note that better mixing properties of the constructed Markov chains are present under the $t$ multivariate random effects model.

\begin{table}[h!t]
\centering
{
\begin{tabular}{c||c|c|c|c|c}
  \hline \hline
& $\mu_{1}$ & $\mu_{2}$ & $\Psi_{11}$& $\Psi_{21}$ &$\Psi_{22}$\\
  \hline
  \multicolumn{6}{c}{Jeffreys prior, Algorithm A}\\
  \hline
normal distribution     &1.0064 &1.0096 &1.0071 &1.0023 &1.0027\\
$t$-distribution        &1.0014 &1.0004 &1.0053 &1.0066 &1.0028\\
  \hline
  \multicolumn{6}{c}{Jeffreys prior, Algorithm B}\\
  \hline
normal distribution     &1.0781 &1.0438 &1.0246 &1.0168 &1.0263\\
$t$-distribution        &1.0020 &1.0023 &1.0009 &1.0012 &1.0027\\
  \hline
    \multicolumn{6}{c}{Jeffreys prior, Algorithm C}\\
  \hline
normal distribution     &1.0000 &1.0000 &1.0097 &1.0030 &1.0051\\
$t$-distribution        &1.0002 &1.0002 &1.0030 &1.0024 &1.0030\\
  \hline
\multicolumn{6}{c}{Berger and Bernardo reference prior, Algorithm A}\\
  \hline
normal distribution     &1.0058 &1.0133 &1.0951 &1.0680 &1.1080\\ 
$t$-distribution        &1.0009 &1.0005 &1.0050 &1.0044 &1.0026\\
  \hline
  \multicolumn{6}{c}{Berger and Bernardo reference prior, Algorithm B}\\
  \hline
normal distribution     &1.0196 &1.0307 &1.0458 &1.0205 &1.0308\\
$t$-distribution        &1.0010 &1.0006 &1.0017 &1.0025 &1.0039\\
    \hline
  \multicolumn{6}{c}{Berger and Bernardo reference prior, Algorithm C}\\
  \hline
normal distribution     &1.0000 &1.0001 &1.0379 &1.0010 &1.0035\\
$t$-distribution        &1.0004 &1.0001 &1.0033 &1.0023 &1.0027\\
\hline\hline
\end{tabular}
}
\caption{Split-$\hat{R}$ estimates based on the rank normalization (see, \cite{vehtari2021rank}) computed for the normal multivariate random effects model and for the $t$ multivariate random effects model by employing the Berger and Bernardo reference prior and the Jeffreys prior. The samples from the posterior distributions are drawn by Algorithm A (first and fourth rows), Algorithm B (second and fifth rows) and Algorithm C (third and sixth rows). }
\label{tab:hatR}
\end{table}

Table \ref{tab:hatR} provides the values of the split-$\hat{R}$ estimates based on the rank normalization, another robust measure for studying the mixing properties of the constructed Markov chains as suggested in \cite{vehtari2021rank}. The values close to one indicates better mixing properties and, as such, the algorithm, which produces the values of the split-$\hat{R}$ estimate closest to one, might be considered to be the approach with the best performance. The results provided in Table \ref{tab:hatR} are in line with the rank pots presented in Figures \ref{fig:emp-study-rank-mu1-nor} to \ref{fig:emp-study-rank-Psi21-t}. The smallest values are obtained when the Gibbs sampler (Algorithm C) is used to draw samples form the posterior distribution independently of the distributional assumption used in the definition of the multivariate random effects model. The application of the $t$ multivariate random effects model reduces the split-$\hat{R}$ estimates considerably. The exception is the case when Algorithm C is used for the parameter $\mu_1$  where very small increases are observed and some cases corresponding to parameters $\Psi_{21}$ and $\Psi_{22}$. Moreover, the values computed under the $t$ multivariate random effects model are always below $1.01$, the target value suggested in \cite{vehtari2021rank}.

\begin{figure}[H]
\centering
\begin{tabular}{p{7.0cm}p{7.0cm}}
\includegraphics[width=6.5cm]{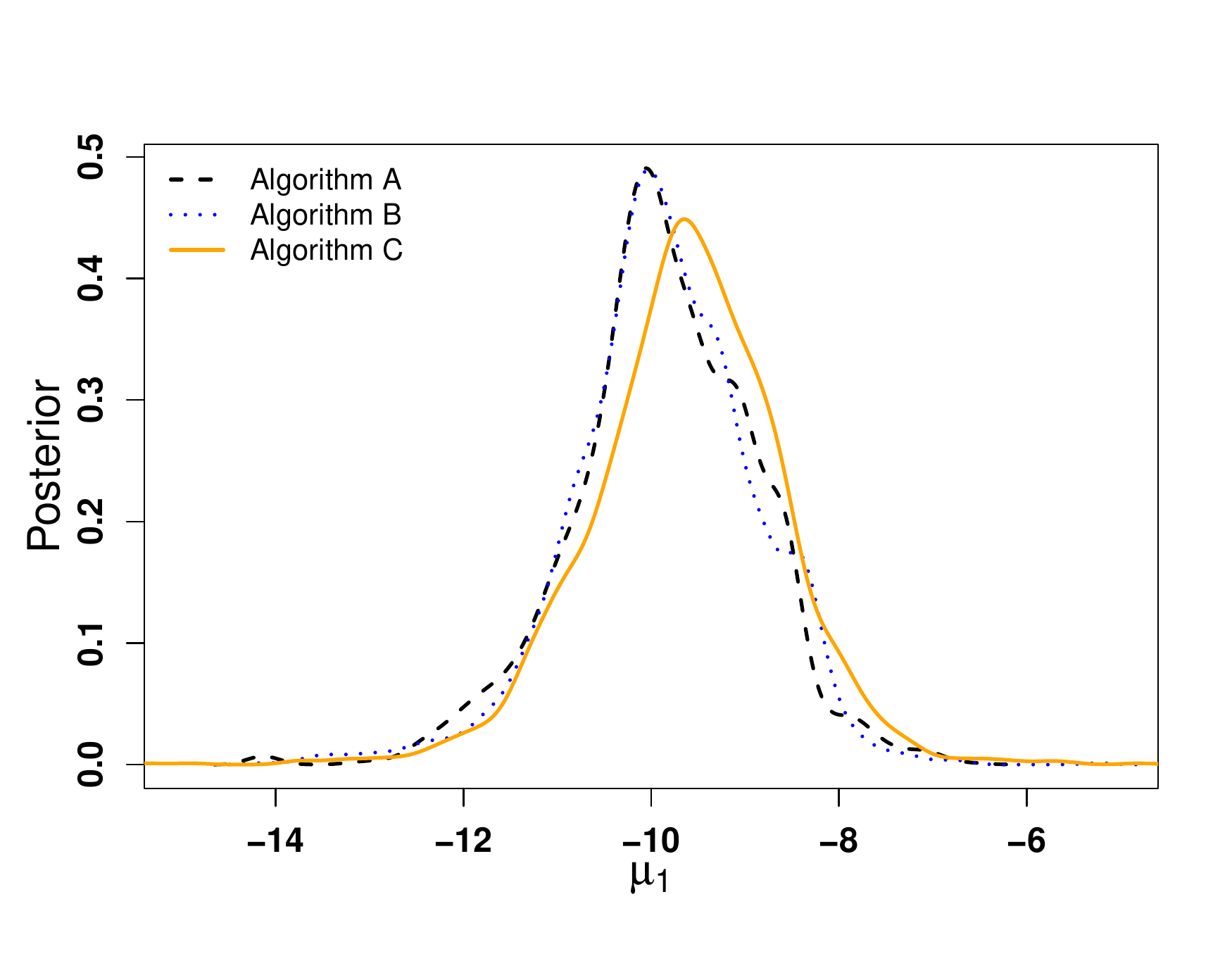}&\includegraphics[width=6.5cm]{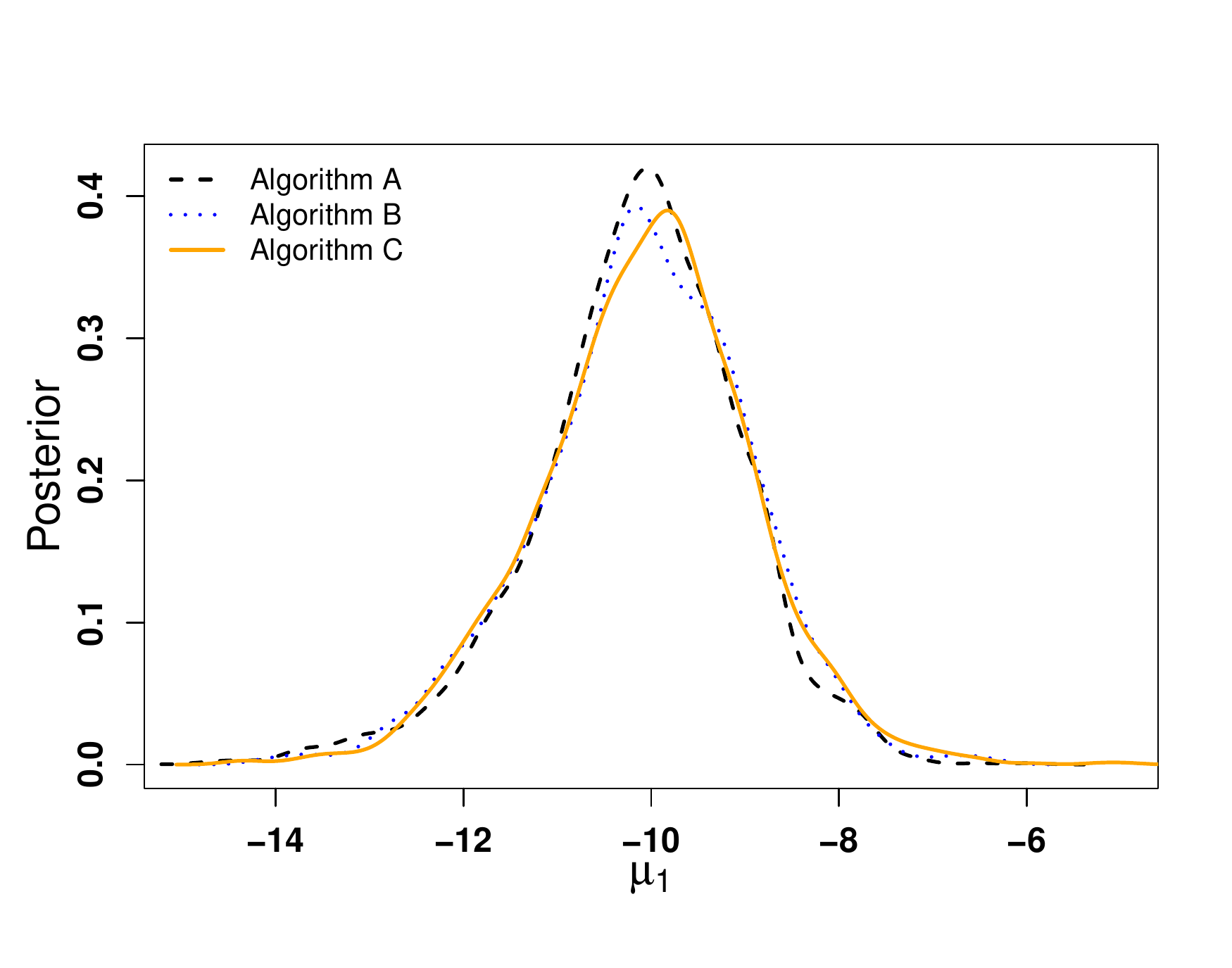}\\[-1cm]
\includegraphics[width=6.5cm]{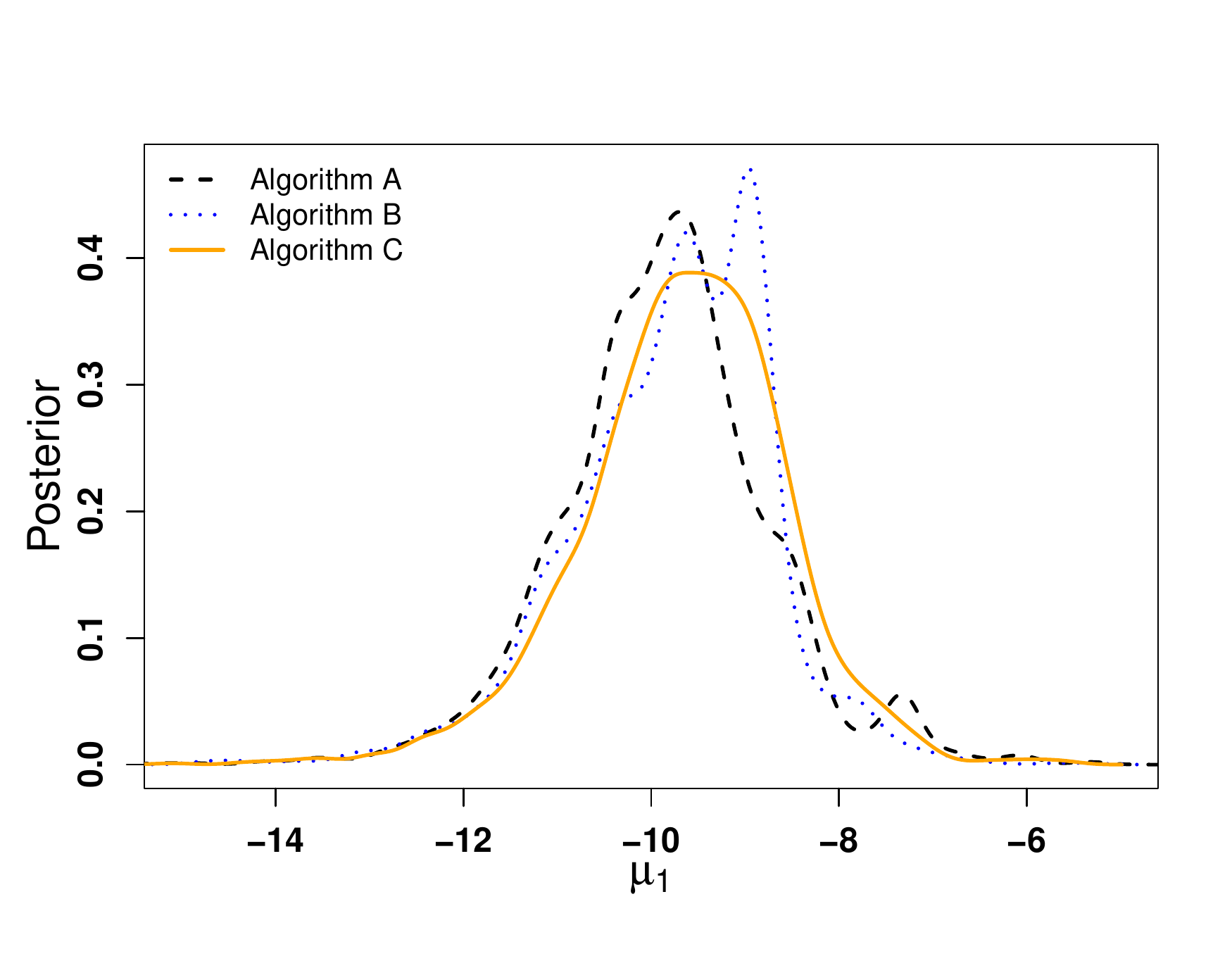}&\includegraphics[width=6.5cm]{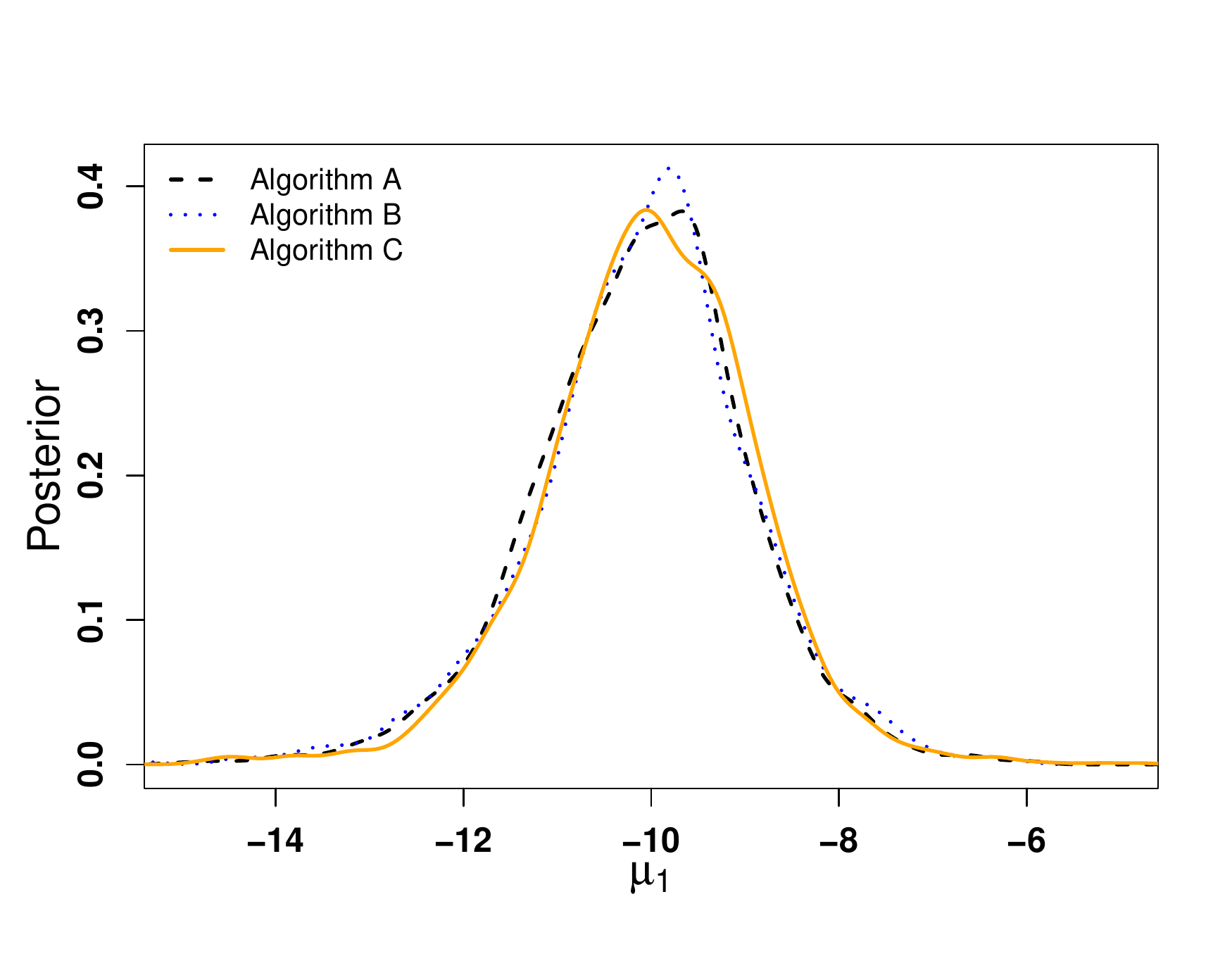}\\[-0.7cm]
\end{tabular}
 \caption{Kernel density estimators of the marginal posterior for $\mu_1$ (SBP) of the normal multivariate random effects model (left column) and $t$ multivariate random effects model (right column), obtained for the data from Table \ref{tab:data} by employing the Berger and Bernardo reference prior and the Jeffreys prior and using Algorithms A, B and C to draw samples from the posterior distribution.}
\label{fig:emp-study-posterior-mu1}
 \end{figure}
 
\begin{figure}[H]
\centering
\begin{tabular}{p{7.0cm}p{7.0cm}}
\includegraphics[width=6.5cm]{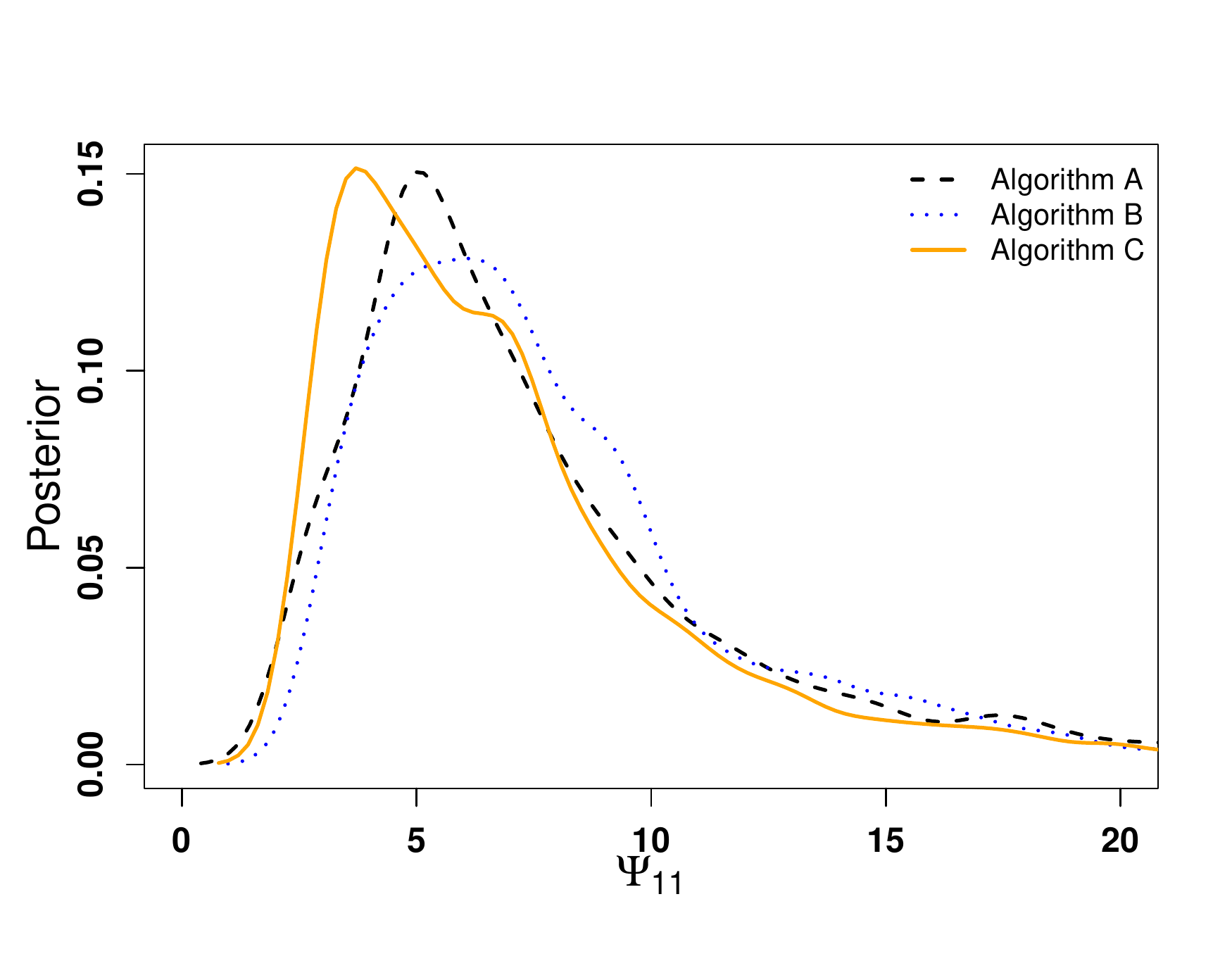}&\includegraphics[width=6.5cm]{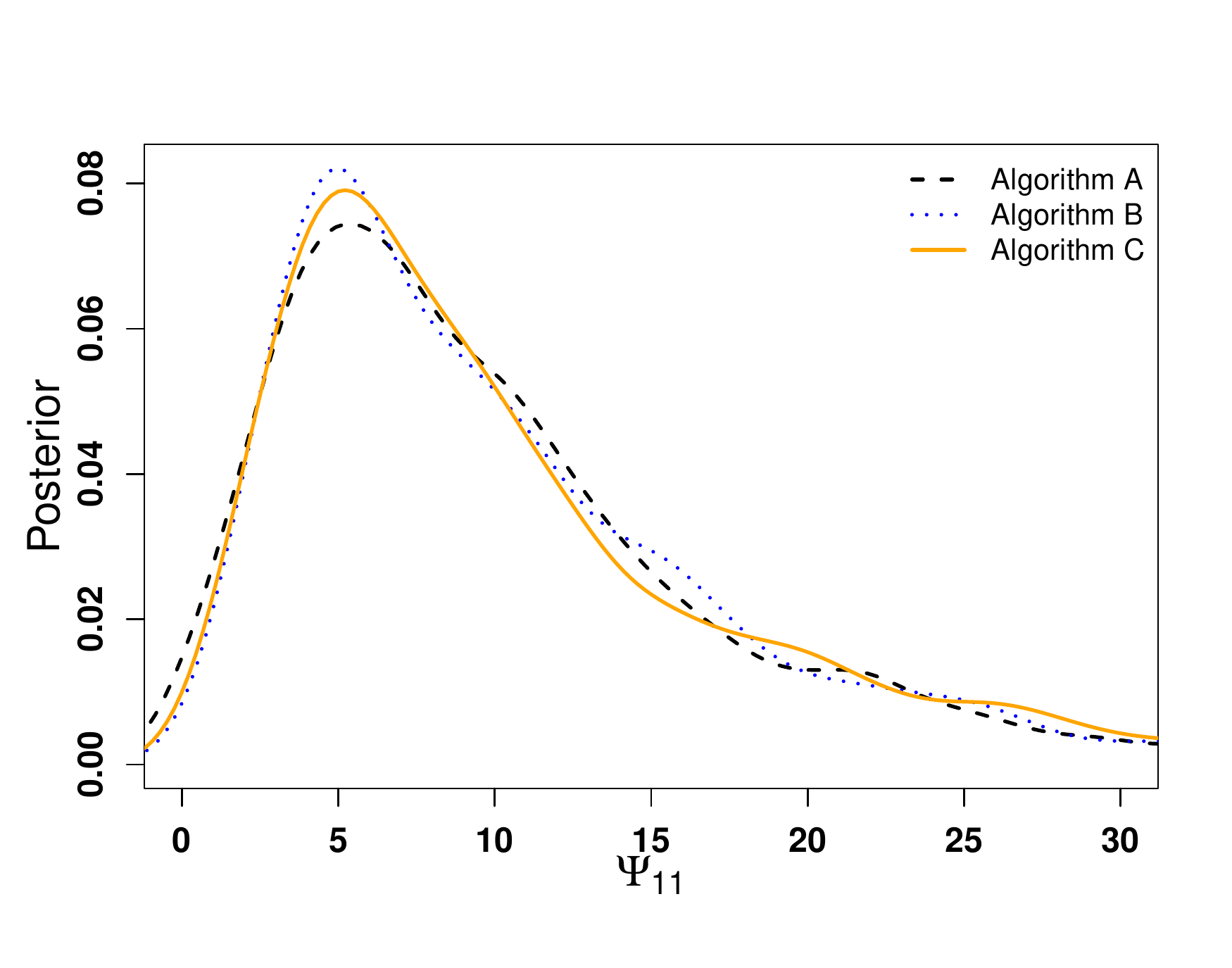}\\[-1cm]
\includegraphics[width=6.5cm]{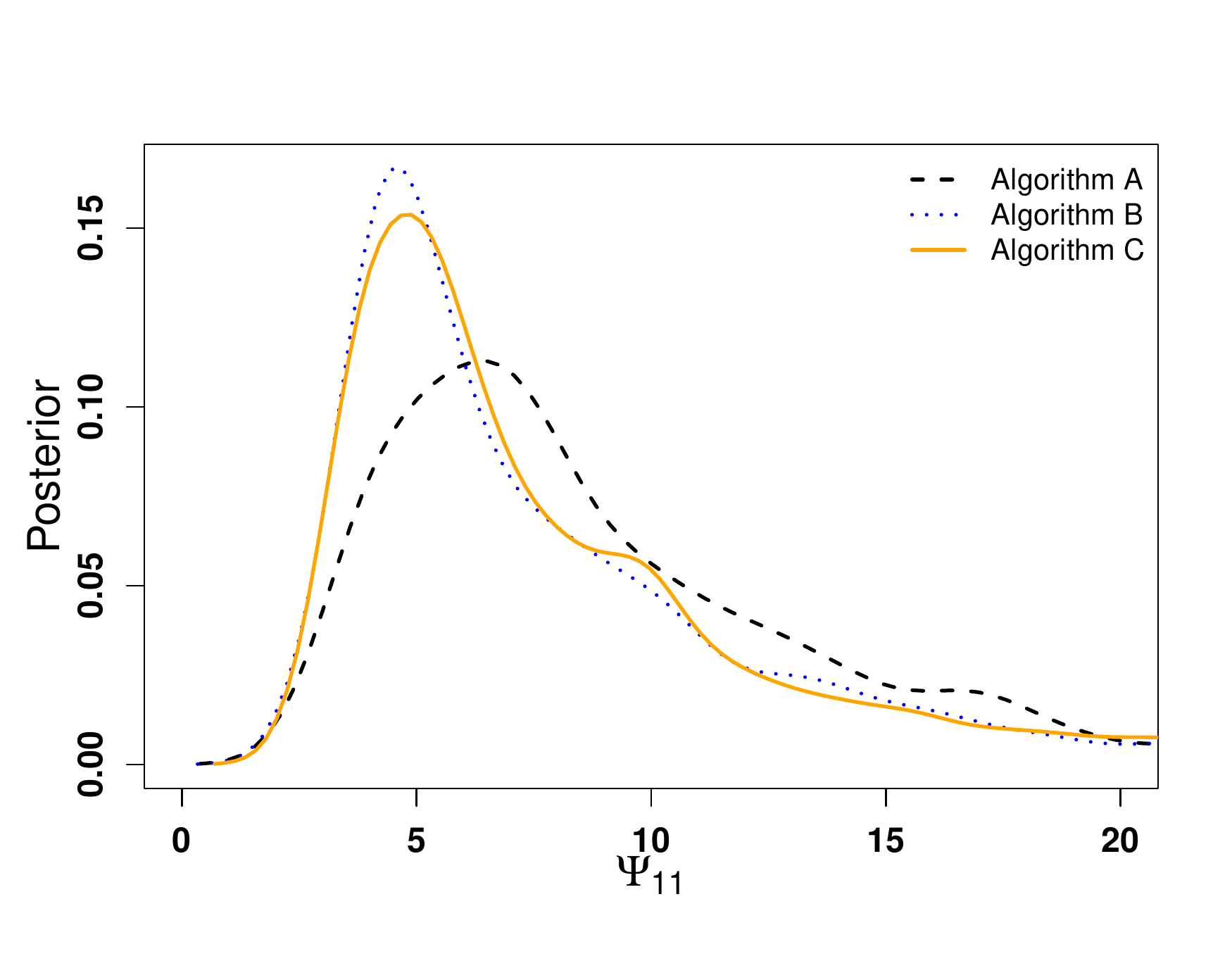}&\includegraphics[width=6.5cm]{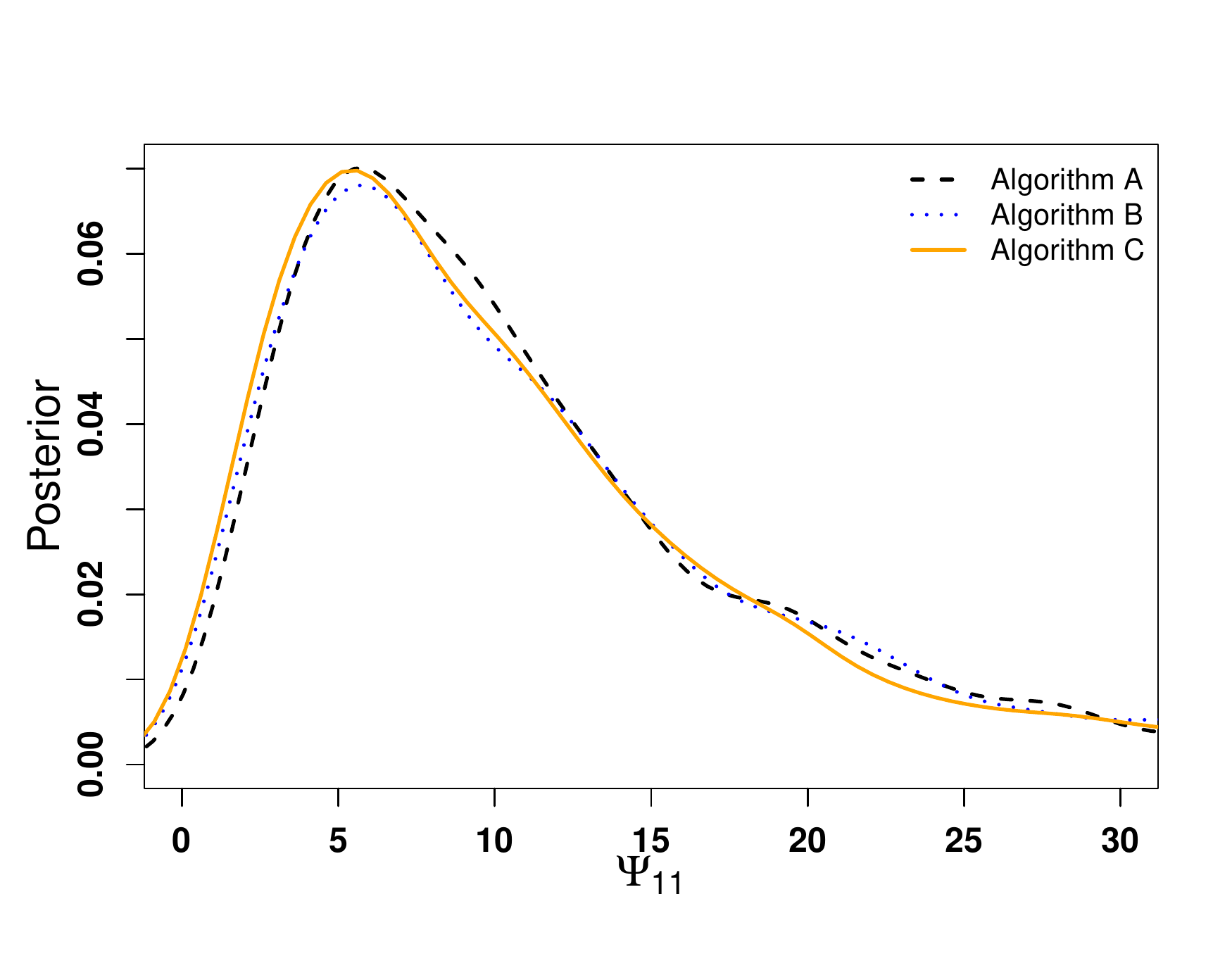}\\[-0.7cm]
\end{tabular}
 \caption{Kernel density estimators of the marginal posterior for $\Psi_{11}$ (SBP) of the normal multivariate random effects model (left column) and $t$ multivariate random effects model (right column), obtained for the data from Table \ref{tab:data} by employing the Berger and Bernardo reference prior and the Jeffreys prior and using Algorithms A, B and C to draw a sample from the posterior distribution.}
\label{fig:emp-study-posterior-Psi11}
 \end{figure}

\begin{figure}[H]
\centering
\begin{tabular}{p{7.0cm}p{7.0cm}}
\includegraphics[width=6.5cm]{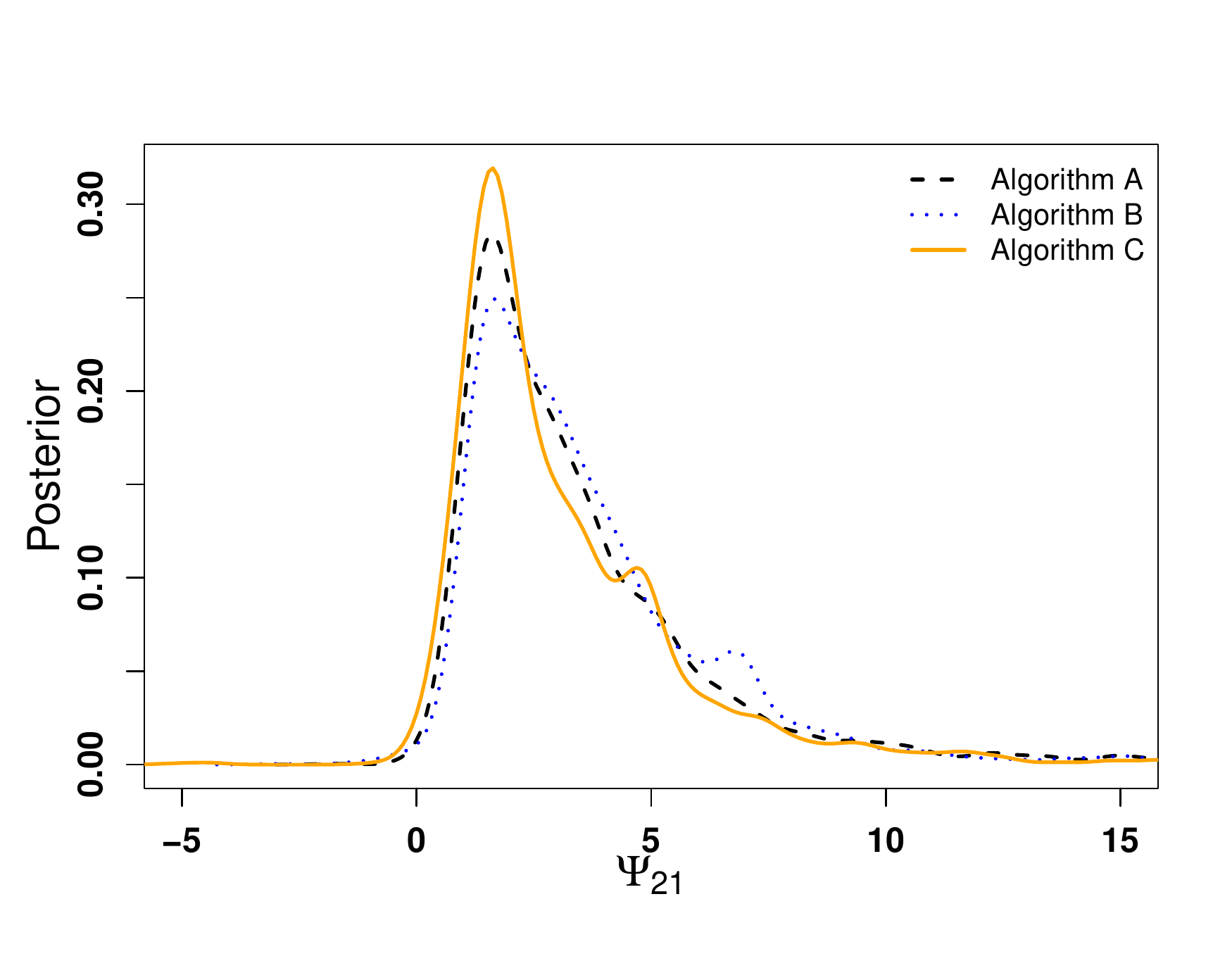}&\includegraphics[width=6.5cm]{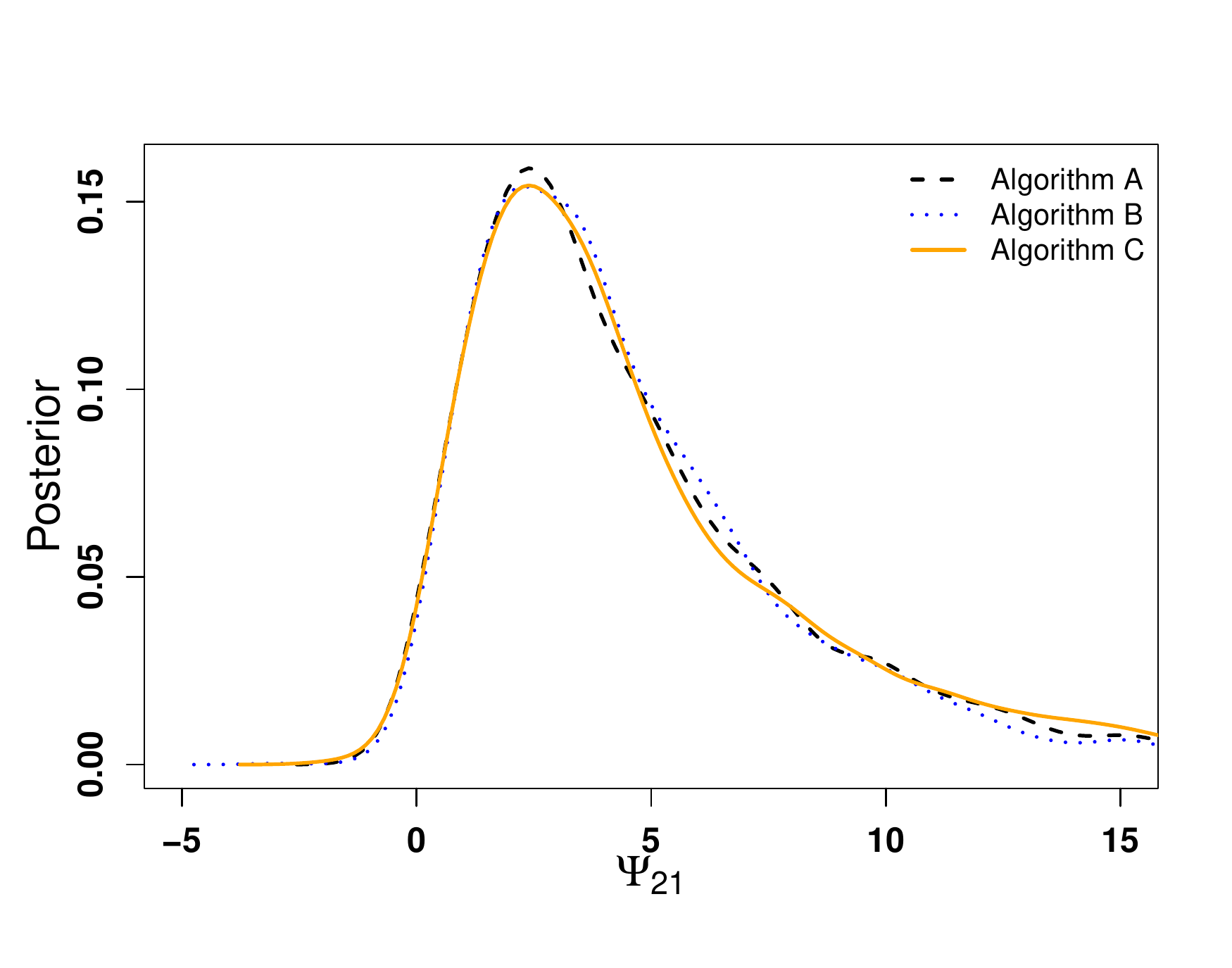}\\[-1cm]
\includegraphics[width=6.5cm]{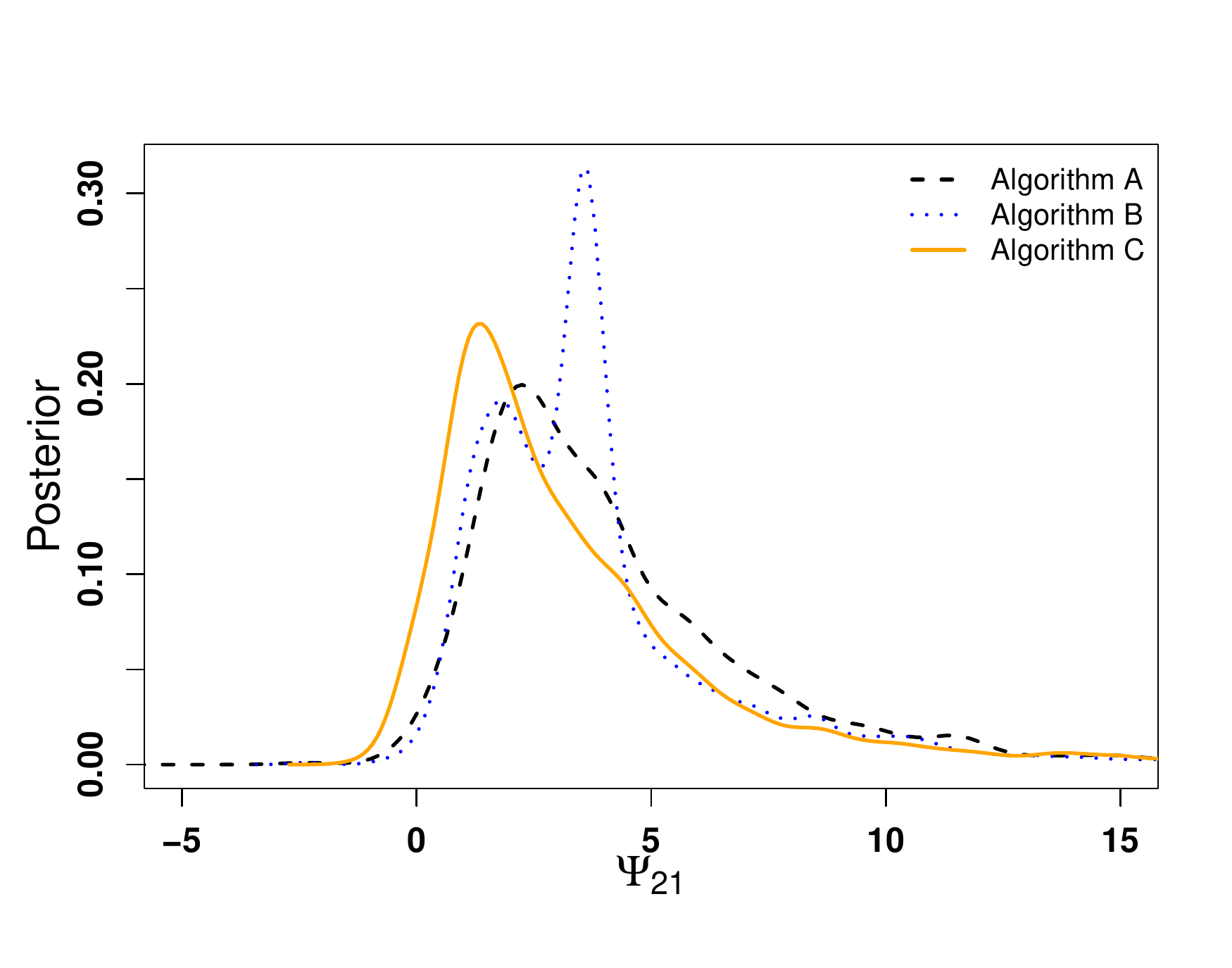}&\includegraphics[width=6.5cm]{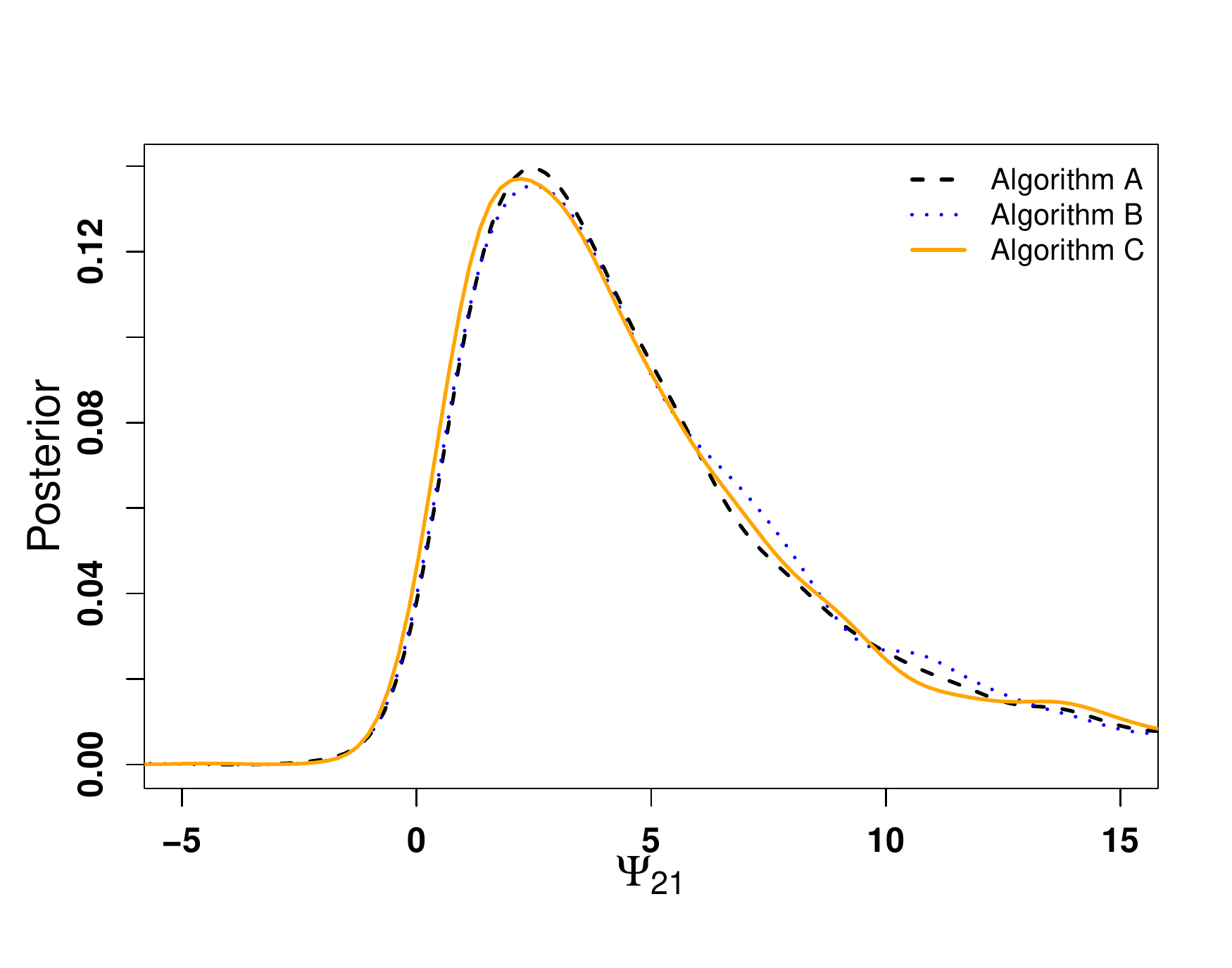}\\[-0.7cm]
\end{tabular}
 \caption{Kernel density estimators of the marginal posterior for $\Psi_{21}$ (DBP, SBP) of the normal multivariate random effects model (left column) and $t$ multivariate random effects model (right column), obtained for the data from Table \ref{tab:data} by employing the Berger and Bernardo reference prior and the Jeffreys prior and using Algorithms A, B and C to draw a sample from the posterior distribution.}
\label{fig:emp-study-posterior-Psi21}
 \end{figure}

Figures \ref{fig:emp-study-posterior-mu1}-\ref{fig:emp-study-posterior-Psi21} depict kernel density estimators of the marginal posteriors of $\mu_1$, $\Psi_{11}$ and $\Psi_{21}$ computed under the normal multivariate random effects model (left column) and $t$ multivariate random effects model (right column) when the Berger and Bernardo reference prior and the Jeffreys prior are assigned to the model parameters. For each prior and each algorithm, a Markov chain of length $B=110000$ is generated, where the first 10000 observations are used for as burn-in period. From the remaining 100000 observations we use each 50th one in the construction of the plots.

Larger differences between the curves are observed in the case of the normal multivariate random effects model, while the plots almost coincide under the assumption of the $t$-distribution. Moreover, the kernel density estimators obtained under the $t$-distribution are more smooth and are less peaked around the mode. Similarly, independently of the imposed distributional assumption, the application of the Jeffreys prior leads to the posterior with higher peaks. While the posteriors for $\mu_1$ are roughly symmetric, the marginal posteriors deduced for the elements of the between-study covariance matrix are skewed to the right. Furthermore, the kernel densities constructed by using Algorithm A and B under the assumption of normality have several modes in the case of $\mu_1$ and $\Psi_{21}$, while it is not the case when a Markov chain is constructed by employing the Gibbs sampler (Algorithm C). Finally, we note that the posteriors deduced under the Berger-Bernrado reference prior are more flat than the ones obtained when the Jeffreys prior is used.

Finally, the Bayesian point estimators together with the credible intervals obtained under the three algorithms are compared in Tables \ref{tab:emp-mu} and \ref{tab:emp-Psi} for the overall mean vector $\bmu$ and the between-study covariance matrix $\bPsi$. Since the marginal posteriors for $\Psi_{11}$, $\Psi_{21}$ and $\Psi_{22}$ are skewed to the right the credible intervals of the form $[q_{\beta}, q_{1-(\alpha-\beta)}]$ are constructed with $\alpha=0.05$ and $\beta=0.0001$, motivated by the results presented in Figure \ref{fig:sim-study-cov-prob-beta} and obtained for the between-study variance $\Psi_{11}$. The lengths of this type of credible intervals appear to be considerably smaller than the lengths of the probability symmetric credible intervals. In the case of $\mu_1$ and $\mu_2$, probability symmetric credible intervals are presented. For the completeness of the presentation, the estimators for these parameters obtained by employing the methods of the frequentist statistics are provided as well. The considered methods are the maximum likelihood estimator, the restrictive maximum likelihood estimator described in \citet{gasparrini2012multivariate}, and the method of moment estimators from \citet{jackson2013matrix}. 

The results of in Tables \ref{tab:emp-mu} and \ref{tab:emp-Psi} are in line with the findings obtained in Figures \ref{fig:emp-study-posterior-mu1} to \ref{fig:emp-study-posterior-Psi21}. In particular, wider credible intervals are obtained under the $t$ random effects model when the Berger and Bernardo reference prior is employed. This phenomenon becomes even more pronounced when credible intervals for the elements of the between-study covariance matrix $\bPsi$ are constructed. Also, the application of the Berger-Bernardo reference prior leads to larger values of the estimated elements of the between-study covariance matrix, independently of the distributional assumption used in the definition of the random effects model. Interestingly, despite larger difference obtained for the values of the between-study covariance matrix, the differences between the Bayesian estimators constructed for the elements of the overall mean vector are not large, which is explained by the fact that the whole posterior of $\bPsi$ is used, when Bayesian inference procedures for $\bmu$ are conducted. Finally, it is noted that the confidence intervals obtained under the three frequentist approaches are considerably narrower than the credible intervals delivered by the Bayesian procedures. This is explained by the fact, that the frequentist methods ignore the uncertainty related to the estimation of the between-study covariance matrix, while the Bayesian approaches take it into account. 

\renewcommand{\baselinestretch}{1.0}
\begin{table}[h!t]
\centering
\begin{tabular}{c||c|c||c|c}
  \hline \hline
  & \multicolumn{2}{c||}{Normal random effects model}&\multicolumn{2}{c}{t random effects model}\\
  \cline{2-5}
  & $\mu_{1}$ (SBP) & $\mu_{2}$ (DBP) & $\mu_{1}$ (SBP) & $\mu_{2}$ (DBP)\\
  \hline
  \multicolumn{5}{c}{Jeffreys prior, Algorithm A}\\
  \hline
  post. mean/ med.& -9.87/-9.93 & -4.52/-4.50 & -10.15/-10.08 & -4.68/-4.65  \\
  post. sd. &  0.98 & 0.58 & 1.10 & 0.59  \\
  cred. inter. & [-11.95,-7.96] & [-5.74,-3.56] & [-12.61,-8.11] & [-5.92,-3.64]\\
  \hline
  \multicolumn{5}{c}{Jeffreys prior, Algorithm B}\\
  \hline
  post. mean/med. & -9.86/-9.89 & -4.54/-4.54 & -10.09/-10.05 & -4.68/-4.67 \\
  post. sd. & 0.97 & 0.55 & 1.13 & 0.63  \\
  cred. inter. &  [-11.90,-8.18] & [-5.66,-3.48] & [-12.42,-8.04] & [-5.95,-3.49]  \\
  \hline
  \multicolumn{5}{c}{Jeffreys prior, Algorithm C}\\
  \hline
  post. mean/med. & -9.63/-9.61 & -4.45/-4.44 & -10.06/-10.00 & -4.64/-4.65  \\
  post. sd. & 1.03 & 0.57 & 1.13 & 0.62  \\
  cred. inter. &[-11.74,-7.68] & [-5.60,-3.40] & [-12.39,-7.92] & [-5.85,-3.38]  \\
  \hline
  \multicolumn{5}{c}{Berger and Bernardo reference prior, Algorithm A}\\
  \hline
  post. mean/med. & -9.84/-9.83 & -4.56/-4.52 & -10.10/-10.05 & -4.66/-4.66  \\
  post. sd. & 1.14 & 0.58 & 1.14 & 0.64  \\
  cred. inter. &[-12.06,-7.32] & [-5.77,-3.38] & [-12.52,-7.87] & [-5.91,-3.36] \\
  \hline
  \multicolumn{5}{c}{Berger and Bernardo reference prior, Algorithm B}\\
  \hline
  post. mean/med. & -9.77/-9.63 & -4.40/-4.39 &  -10.08/-9.99 & -4.66/-4.63 \\
  post. sd. & 1.07 & 0.63 & 1.18 & 0.67 \\
  cred. inter. & [-12.18,-7.75] & [-5.65,-3.17] & [-12.66,-7.79] & [-6.08,-3.34] \\
  \hline
  \multicolumn{5}{c}{Berger and Bernardo reference prior, Algorithm C}\\
  \hline
  post. mean/med. & -9.66/-9.60 & -4.45/-4.40 & -10.02/-9.99 & -4.64/-4.64\\
  post. sd. & 1.10 & 0.60 &1.15 & 0.63 \\
  cred. inter. & [-12.03,-7.58] & [-5.74,-3.36]  &[-12.33,-7.86] & [-5.93,-3.41] \\
  \hline
  \multicolumn{5}{c}{ML, \citet{gasparrini2012multivariate}}\\
  \hline
  estimator & -9.47 & -4.41 & -- & -- \\
  stand. error & 0.68 & 0.44 & -- & -- \\
  conf. inter. & [-10.79, -8.14] & [-5.26, -3.55] & -- & -- \\
  \hline
  \multicolumn{5}{c}{REML, \citet{gasparrini2012multivariate}}\\
  \hline
  estimator & -9.51 & -4.43 & -- & --  \\
  stand. error & 0.73 & 0.47 & -- & -- \\
  conf. inter. & [-10.95, -8.07] & [-5.35, -3.51] & -- & -- \\
  \hline
  \multicolumn{5}{c}{Method of moments, \citet{jackson2013matrix}}\\
  \hline
  estimator & -9.17 & -4.31 & -- & --  \\
  stand. error & 0.55 & 0.36 & -- & -- \\
  conf. inter. & [-10.26, -8.08] & [-5.02, -3.60] & -- & -- \\
\hline \hline
\end{tabular}
\caption{Posterior mean, posterior median, posterior standard deviation and 95\% credible interval) for the parameter $\bmu$ of the multivariate random effects model obtained for the data from Table \ref{tab:data} by employing the Berger and Bernardo reference prior and the Jeffreys prior. The last three panels of the table include the results of the maximum likelihood estimator and the restrictive maximum likelihood estimator described in \citet{gasparrini2012multivariate}, and the method of moment estimators from \citet{jackson2013matrix}. }
\label{tab:emp-mu}
\end{table}

\begin{table}[h!t]
\centering
{\footnotesize
\begin{tabular}{c||c|c|c||c|c|c}
  \hline \hline
  & \multicolumn{3}{c||}{Normal random effects model}&\multicolumn{3}{c}{t random effects model}\\
  \cline{2-7}
  & $\psi_{11}$ (SBP) & $\psi_{21}$ &$\psi_{22}$ (DBP) & $\psi_{11}$ (SBP) & $\psi_{21}$ &$\psi_{22}$ (DBP)\\
  \hline
  \multicolumn{7}{c}{Jeffreys prior, Algorithm A}\\
  \hline
  post. mean/med. &8.37/6.52& 3.57/2.73 &  2.76/2.18 & 11.04/8.37 & 5.17/3.71 & 3.88/2.89 \\
  post. sd. &6.43 & 3.06 & 1.94  & 10.37 & 5.41 & 3.64 \\
  cred. inter. & [2.56, 19.75] & [-1.42,9.28] & [0.76,6.43] & [0.51,28.44] & [-0.61,13.65] & [0.23,9.79]\\
  \hline
  \multicolumn{7}{c}{Jeffreys prior, Algorithm B}\\
  \hline
  post. mean/med. &8.68/7.04 & 3.76/2.96 & 3.02/2.44 &  11.06/8.46 & 5.08/3.69 & 3.85/2.96 \\
  post. sd. & 6.31 & 2.96 & 2.07 & 9.62 & 4.83 & 3.19 \\
  cred. inter. &  [2.96, 18.79] & [-2.70, 8.82] & [0.74,7.29] & [0.71,27.53] & [-2.49,13.63] & [0.23,9.58]  \\
  \hline
  \multicolumn{7}{c}{Jeffreys prior, Algorithm C}\\
  \hline
  post. mean/med. &7.51/6.11 & 3.20/2.32  & 2.60/1.99 & 11.37/8.42 & 5.21/3.75 & 3.91/3.01  \\
  post. sd. & 5.85 & 2.81 & 1.85 & 10.34 & 5.07 & 3.31\\
  cred. inter. & [2.78,17.38] & [-5.05,8.13] & [0.68,5.94] & [0.47,29.49] & [-1.75,14.28] & [0.17,9.93]  \\
  \hline
  \multicolumn{7}{c}{Berger and Bernardo reference prior, Algorithm A}\\
  \hline
  post. mean/med. &10.41/7.72 & 4.47/3.48  & 3.61/2.76& 13.04/9.39 & 6.09/4.12 & 4.62/3.31 \\
  post. sd. & 9.13 & 4.09 & 2.88 & 13.54 & 6.77 & 4.33 \\
  cred. inter. & [3.24,25.38] & [-7.10,11.49] & [0.82,9.36] & [0.57,34.55] & [-1.83,17.59] & [0.19,12.36] \\
  \hline
  \multicolumn{7}{c}{Berger and Bernardo reference prior, Algorithm B}\\
  \hline
  post. mean/med. &8.95/6.22 & 4.03/3.60 & 4.18/3.04 &  12.77/9.39 & 5.79/4.20 & 4.30/3.45 \\
  post. sd. & 8.36 & 3.71 & 3.04 &  12.03 & 5.74 & 3.62 \\
  cred. inter. & [2.95,22.46] & [-2.34,10.21] & [0.71,9.14] & [0.40,34.86 ] & [-5.06,16.41] & [0.17,11.15]  \\
  \hline
  \multicolumn{7}{c}{Berger and Bernardo reference prior, Algorithm C}\\
  \hline
  post. mean/med. & 9.06/6.39 & 3.62/2.42 & 2.96/2.24 & 12.38/9.05 & 5.71/4.00 & 4.40/3.33\\
  post. sd. & 8.11 & 4.04 & 2.74  &  13.48 & 6.64 & 4.65 \\
  cred. inter. & [3.00,23.32] & [-1.17,10.85] & [0.61,7.80] &[0.52,33.91] & [-3.90,15.85] & [0.12,11.43]\\
  \hline
  \multicolumn{7}{c}{ML, \citet{gasparrini2012multivariate}}\\
  \hline
  estimator & 3.29 & 1.51 & 1.57 & -- & -- & -- \\
  \hline
  \multicolumn{7}{c}{REML, \citet{gasparrini2012multivariate}}\\
  \hline
  estimator & 3.92 & 1.81& 1.83 & -- & -- & --  \\
  \hline
  \multicolumn{7}{c}{Method of moments, \citet{jackson2013matrix}}\\
  \hline
  estimator & 2.03 & 0.2 & 1.04 & -- & -- & --  \\
\hline \hline
\end{tabular}
}
\caption{Posterior mean, posterior median, posterior standard deviation and 95\% credible interval) for the parameter $\bPsi$ of the multivariate random effects model obtained for the data from Table \ref{tab:data} by employing the Berger and Bernardo reference prior and the Jeffreys prior. The last three panels of the table include the results of the maximum likelihood estimator and the restrictive maximum likelihood estimator described in \citet{gasparrini2012multivariate}, and the method of moment estimators from \citet{jackson2013matrix}.}
\label{tab:emp-Psi}
\end{table}

\renewcommand{\baselinestretch}{1.4}

\section{Summary}\label{sec:sum}

Multivariate random effects model is used as a common quantitative tool to pool the results of the individual study together when several features are measured in each individual study. The model is widely used in different fields of science. While the inference procedures for the parameters of the model were first derived by the methods of the frequentist statistics, the Bayesian approaches have increased its popularity. 

Even though the joint posterior for the parameters of the multivariate random effects model is derived in the analytical form, it is a challenging task to construct Bayesian inference procedures due to the complicated structure which is present in the marginal posterior of the between-study covariance matrix. To deal with the problem, two Metropolis-Hastings algorithms to draw samples from the posterior distribution were suggested in the literature. In the present paper, an alternative approached based on the hybrid Gibbs sampler is proposed. Via a simulation study and empirical illustration we showed that the new numerical approach performs similarly to the existent ones or even outperforms in terms of the convergence properties of the constructed Markov chain, which are assessed by using the rank plots and the split-$\hat{R}$ estimates based on the rank normalization (see, \cite{vehtari2021rank}). Moreover, the application of the $t$ multivariate random effects model appears to provide more reliable results in the conducted empirical study in comparison the model based on the assumption of normality.

\section*{Acknowledgement}
Olha Bodnar acknowledges valuable support from the internal grand (R\"{o}rlig resurs) of the \"{O}rebro University. This research is a part of the project \emph{Statistical Models and Data Reductions to Estimate Standard Atomic Weights and Isotopic Ratios for the Elements, and to Evaluate the Associated Uncertainties} (No. 2019-024-1-200), IUPAC (International Union of Pure and Applied Chemistry).

{\footnotesize
\bibliography{Bibfile2022-10-22}
}

\end{document}